\documentclass[%
 reprint,
nofootinbib,
 amsmath,amssymb,
 aps,
]{revtex4-2}

\usepackage{graphicx}
\usepackage{dcolumn}
\usepackage{bm}
\usepackage{mathrsfs}
\usepackage{xcolor}

\begin{document}

\preprint{APS/123-QED}

\title{
Paraxial diffusion-field retrieval. II.  
Fokker-Planck generalization of the transport-of-intensity equation
}

\author{David M.~Paganin}
 \affiliation{School of Physics and Astronomy, Monash University, Clayton, Victoria, 3800, Australia}

\author{Kaye S.~Morgan}
 \affiliation{School of Physics and Astronomy, Monash University, Clayton, Victoria, 3800, Australia}

\date{\today}

\begin{abstract}
The transport-of-intensity equation (TIE), namely the continuity equation associated with a coherent paraxial optical wavefield, has been very widely employed for phase retrieval.  In particular, the TIE is an elliptic second-order partial differential equation which may be solved for the phase of a coherent paraxial field such as a monochromatic scalar optical beam, given both the intensity and longitudinal intensity derivative in a plane perpendicular to the optical axis.  This paper shows how the coherent flow associated with the TIE may be augmented by a diffusive flow associated  with a scalar or tensor diffusion field.  Such diffusive flow can arise via scattering from unresolved spatially random microstructure within an illuminated sample, the blurring effects of an extended chaotic source that illuminates the sample, the resolution-reducing effect of shot noise in detected intensity images of the sample, and the sharpening effect (negative diffusion) associated with scattering from sharp sample edges.  Augmenting the TIE's modeling of coherent flow with a diffuse-flow channel leads to a Fokker-Planck extension to this equation.  Two different TIE augmentations are obtained, using several complementary derivations.  The inverse problems of phase retrieval and diffusion-field retrieval are then considered, for both defocus-based imaging and mask-based imaging (structured-illumination imaging).  When symmetric overfocus and underfocus images are employed for the purposes of phase retrieval, the diffusive term drops out and our Fokker-Planck formalism implies that any ensuing TIE-based phase-retrieval method needs no modification in light of our formalism.  However, the same focal-series dataset---which is typically an infocus image, a weakly overfocused image, and a weakly underfocused image---may also be employed to access the additional channel of information associated with the Fokker-Planck diffusion field.  Our formalism is applicable to visible-light microscopy, x-ray imaging, electron microscopy, and neutron imaging.   
\end{abstract}

\maketitle


\section{Introduction}

This article is a direct continuation of our previous paper \cite{PaganinPellicciaMorgan2023} (henceforth Paper I). 
{ 
In our earlier publication the position-dependent blur of a propagating paraxial optical scalar beam was associated with a position-dependent dimensionless diffusion coefficient (diffusion field).  Several distinct blurring mechanisms were considered, including (i) the influence of spatially random sample structures too small to be directly resolved by a given imaging system, (ii) blur associated with an extended chaotic source of sample illumination in the imaging system, and (iii) the resolution-reducing effects of noise sources such as those associated with image detection.  Our core focus in both papers is the inverse problem \cite{Sabatier2000} of ``diffusion-field retrieval'', namely means for recovering the diffusion field from intensity measurements in one or more planes perpendicular to the optical axis of a paraxial beam. 

For almost all of Paper I, the position-dependent diffusion field was taken to be the {\em only} factor (besides the unpropagated position-dependent beam intensity itself) governing the beam's intensity evolution with respect to propagation distance along the optical axis.  It was only in the final (short) subsection, namely Sec.~V\,H, that the refractive influence of wavefield phase was taken into account.  This led to equations of a Fokker-Planck form \cite{Risken1989}, namely Eqs.~(146) and (147) of Paper I, which augment the nondirected diffuse flow with a directed coherent flow (transverse ``drift'') associated with refraction.  In turn, these Fokker-Planck equations may be viewed as augmenting the transport-of-intensity equation (TIE) \cite{Teague1983}---which is very widely employed in coherent paraxial-optics phase retrieval\footnote{``Phase retrieval'' may be broadly defined as the inverse problem of reconstructing wavefield phase from (noninterferometric) intensity measurements \cite{gerchberg1972practical,paganin1998}.  Paraxial ``diffusion-field retrieval'' is a deliberately similar term, for reconstructing paraxial-beam diffusion fields from intensity measurements alone.} \cite{TIE==LongReviewArticle2020}---with the effects of the paraxial diffusion field treated in Paper I.   
A number of existing publications study the Fokker-Planck extension to the TIE in the context of coherent paraxial x-ray wave propagation. This existing thread in the x-ray imaging research literature was initiated by a pair of theory papers we published in 2019 \cite{PaganinMorgan2019,MorganPaganin2019}.  These were followed by a number of experimental applications and extensions using both synchrotron and laboratory x-ray sources \cite{MIST,Gureyev2020, MISTdirectional,PaganinPelliccia2021,alloo2022dark, alloo2023SciRep, Leatham2023,Beltran2023,Esposito2023,Magnin2023,Ahlers2024,Leatham2024,CelestreReview2025,Alloo2025,Ahlers2025,LiuAllooLangerPavlovPhysRevA2025,AllooPhysicaScripta2025,YuanDas25,Alloo2026JSynchRad,alloo2026arXiv,allan2026arXiv,Magnin2026,Lindberg2026PMB,Zboray2026OpticaOpenPreprint}.\footnote{Applications beyond x-ray optics were given cursory mention in an earlier paper \cite{PaganinMorgan2019}, but not developed in any level of detail.  Also, a form of Fokker-Planck extension to the TIE was written at the end of a recent paper on neutron optics \cite{PaganinNeutron2023}, but again there was no further development beyond such a brief mention.}  

Encouraged by such extensive uptake within the x-ray imaging community, it is timely to consider diffusion-field retrieval in the primary context of coherent visible-light imaging. That goal, which includes a development and significant extension of foundations that is not tied to assumptions which are specific to x-ray imaging, is the core focus of the present paper.  In this setting we note that the diffusion-field-retrieval problem of Paper I is considerably complicated by the additional coherent-flow effects taken into account by the Fokker-Planck extension to the TIE.  Conversely, the Fokker-Planck extension to the TIE is clarified by Paper I's focus on the role played by the dimensionless diffusion field.

}

Our paper is structured as follows.  Section~\ref{sec:BriefReviewOfTIE} briefly reviews the theory underpinning the TIE \cite{Teague1983}, namely the continuity equation associated with the paraxial equation of coherent scalar wave optics.  It is pointed out that the TIE has been the basis for a very large number of papers on phase retrieval (noninterferometric intensity measurement) using visible light, x rays, electrons, and neutrons.   Section~\ref{sec:RoughDerivationOfFPE} gives two informal derivations for the Fokker-Planck generalization of the TIE \cite{PaganinMorgan2019,MorganPaganin2019}, in the form that was stated without proof at the end of Paper I.  This generalization augments the coherent-flow term in the TIE with a position-dependent diffusive flow associated with a paraxial diffusion field. The first informal derivation, which is given in Sec.~\ref{sec:First
RoughDerivation}, manually bifurcates the optical flow into coherent and diffuse channels.  Section~\ref{sec:Second
RoughDerivation} gives a second derivation where this bifurcation is seen to emerge in a natural way, from a geometrical-optics model wherein we average over an ensemble of beams that are each continuously distorted in a random manner.  These complementary informal derivations aim to explain key aspects of the Fokker-Planck extension to the TIE in a manner that is both mathematically simple and conceptually clear. Section~\ref{sec:Critique} explains that the two derivations are incomplete in that both neglect the influence of a coherent-flow correction term, which is of the same order in sample-to-detector propagation distance (defocus distance) as the diffuse-flow term.  This correction term may be associated with a higher-order version of the TIE. Such critique of our two informal derivations motivates the development of Sec.~\ref{sec:InfinityOfTIEs}, which obtains an infinite hierarchy of higher-order TIEs, namely expressions for the $m$th longitudinal derivative of the intensity associated with a paraxial monochromatic scalar beam that obeys the paraxial wave equation, where $m$ is any positive integer. This result enables Sec.~\ref{sec:FPEandExtendedFPE}, which considers the simultaneous presence of both coherent and diffuse energy flow in a paraxial beam, in a manner that overcomes the preceding critique. We thereby obtain the corrected paraxial-optics Fokker-Planck generalization of the TIE. Section \ref{sec:FPE--GenericRemarks} explains how the Fokker-Planck extension to the TIE is indeed a special case of the form for the Fokker-Planck equation that is typically employed in its more usual context of nonequilibrium statistical mechanics.  Section \ref{sec:PhaseAndDiffusionFieldRetrieval}, which is broken into two parts, considers the inverse problems \cite{Sabatier2000} of phase retrieval [Sec.~\ref{sec:PhaseRetrieval}] and diffusion-field retrieval [Sec.~\ref{sec:DiffusionFieldRetrieval}].  Both defocus-based schemes and mask-based (structured illumination) schemes are considered.  For defocus-based phase retrieval employing positive and negative defoci that are equal in magnitude, the diffusive term drops out in the Fokker-Planck generalization of the TIE; this implies that {\em existing TIE-based phase-retrieval schemes, employing symmetric overfocus and underfocus images, need no modification} in light of the Fokker-Planck generalization of the TIE.  More importantly, {\em an additional channel of information may be
recovered from such TIE focal-series datasets, namely the scalar or tensor diffusion field}.  Section \ref{sec:DiffusionFieldModels} augments the physical models for diffusion-field generation that were presented in Paper I, by considering { (i) a first-principles derivation which also serves to illustrate the relation between the diffusion-field concept and the notion of local small-angle diffuse scatter (Sec.~\ref{sec:DiffusionFieldFromFirstPrinciples}), (ii)} the influence of noise on the positive component of the diffusion field [Sec.~\ref{sec:NoiseContributionToParaxialDiffusionField}], and { (iii)} negative diffusion fields that may be associated with scattering from sharp edges in a sample or from coherent defocus [Sec.~\ref{sec:DiffusionFieldModels:NegativeCases}].  { 
Several points of practical importance are treated in Sec.~\ref{secs:PracticalAspects}: reconstruction uniqueness (Sec.~\ref{subsec:Practicalities1}) and stability (Sec.~\ref{subsec:Practicalities2}), sensitivity with respect to noise (Sec.~\ref{subsec:Practicalities3}), the number of measurements required (Sec.~\ref{subsec:Practicalities4}), and conditions for the ignorability of effects associated with either the higher-order TIE (Sec.~\ref{subsec:Practicalities6}) or the  diffusion field (Sec.~\ref{subsec:Practicalities7}).} Three closely related topics lead the discussion in Sec.~\ref{sec:Discussion}, namely a link between the optical-physics concept of a paraxial diffusion field and the statistical-physics notion of a renormalization group [Sec.~\ref{sec:RunningCoupling}], a simple model for the dependence upon imaging-system spatial resolution of the partitioning of the paraxial flow into coherent and diffuse channels [Sec.~\ref{sec:HowDiffusionFieldVariesWithResolution}], and the role of Liouville's theorem, spatiotemporal coarse graining, and unresolved speckle, in the context of the formalism developed in the present paper [Sec.~\ref{sec:Liouville&UnresolvedSpeckle}].  We also discuss some implications of the diffusion-field concept in the context of the theory of partially coherent optical fields [Sec.~\ref{sec:LinkWithPartialCoherence}], followed by application-specific remarks concerning the use of our paper's formalism for visible-light imaging, x-ray imaging, electron-microscope imaging, and neutron imaging [Sec.~\ref{sec:FPEforElectronsNeutronsPhotonsEtc}]. Concluding remarks are given in Sec.~\ref{sec:Conclusion}.

\section{Background: Transport-of-intensity equation}\label{sec:BriefReviewOfTIE}

\begin{figure}[ht!]
\centering
\includegraphics[width=0.9\columnwidth]{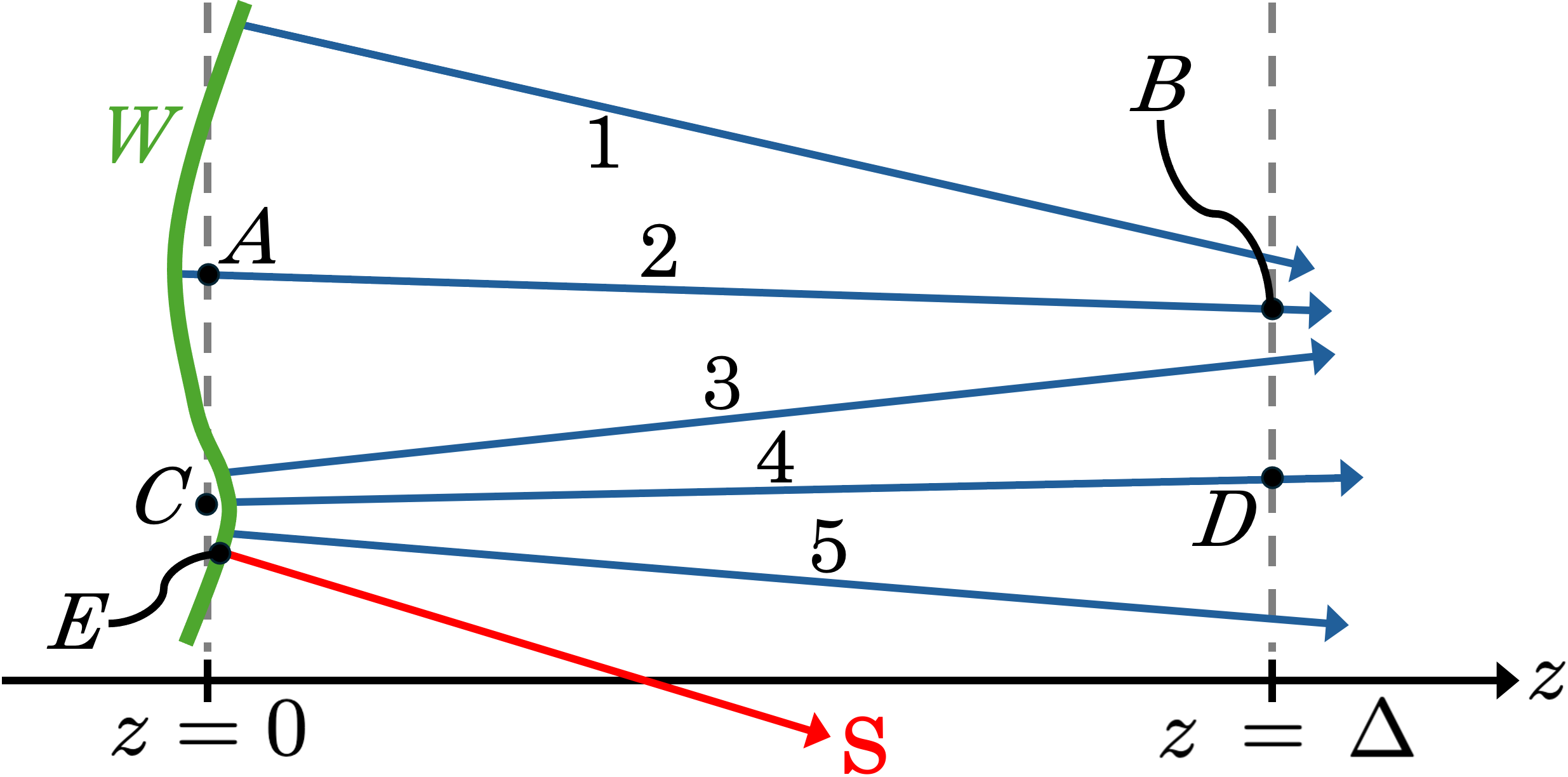}
\caption{Coherent flow of a paraxial scalar optical field, associated with the transport-of-intensity equation \cite{Teague1983}. According to geometrical optics, when propagating through the vacuum that is assumed to lie between the planes $z=0$ and $z=\Delta>0$, the flow may be modeled in terms of straight rays such as those labeled $1$ to $5$.  According to wave optics, we may instead model the flow via an intensity and phase, with wavefronts (such as that indicated by the curved line $W$) associated with surfaces of constant phase. The transport-of-intensity equation is consistent with both viewpoints.}
\label{Fig:CoherentFlow}
\end{figure}

As shown in Fig.~\ref{Fig:CoherentFlow}, consider a paraxial monochromatic scalar light field, such as a propagating laser beam that is deformed by passage through an illuminated sample.  If we picture this field in terms of geometrical optics, we may consider the set of rays associated with the wavefront\footnote{In the language of geometrical optics, the wavefront is a ``surface of constant action'' (Ref.~\cite{Lanczos1970}, p.~273) or ``surface of constant eikonal'' (Ref.~\cite{BornWolf}, p. 119).  In wave-optical language, the wavefront is a ``surface of constant phase''.  The action $S$, eikonal $\mathscr{S}$, and phase $\phi$ are linked via $S/\hbar=2\pi\mathscr{S}/\lambda=\phi$, where $\hbar=h/(2\pi)$, $h$ is the Planck constant, and $\lambda$ is the vacuum wavelength (Ref.~\cite{Lanczos1970}, p.~278; Ref.~\cite{BornWolf}, pp.~118-119).} indicated by the curved line $W$.  Each of these rays is perpendicular to the wavefront. Since the wavefront is locally converging in the vicinity of point $A$, the rays labeled $1,2,3$ will get closer together as they propagate in the direction of the optical axis $z$.  This local convergence of rays implies that the intensity will increase as we move from $A$ to $B$.  Similarly, in the vicinity of point $C$, the rays $3,4,5$ are locally diverging, hence the intensity will decrease when moving from $C$ to $D$. In both cases, we assume that no rays cross one another in the vacuum-filled slab of space between $z=0$ and $z=\Delta > 0$, where $z$ is a Cartesian coordinate associated with the optical axis; for our context of paraxial optics, this condition corresponds to $\Delta$ being sufficiently small.  

A mathematical model, for the qualitative observations of the previous paragraph, proceeds as follows. Begin by noting that the intensity $I(x,y,z)$ and phase $\Phi(x,y,z)$, of a scalar monochromatic field that is not necessarily paraxial, can be combined into the complex wavefunction
\begin{equation}
\widetilde{\psi}(x,y,z)=\sqrt{I(x,y,z)}\exp[i\Phi(x,y,z)]
\label{eq:MadelungTransformation}
\end{equation}
that obeys the Helmholtz equation
\begin{equation}
    (\nabla^2+k^2)\widetilde{\psi}(x,y,z)=0.
\label{eq:HelmholtzEquationPlainForm}
\end{equation}
Here, 
\begin{equation}
\nabla^2=\nabla\cdot\nabla=\frac{\partial^2}{\partial x^2}+\frac{\partial^2}{\partial y^2}+\frac{\partial^2}{\partial z^2}
\label{eq:3DLaplacian}
\end{equation}
is the three-dimensional Laplacian, $\nabla$ is the three-dimensional gradient operator,  
\begin{equation}
    k=2\pi/\lambda
\label{eq:waveNumber}
\end{equation}
is the wavenumber associated with radiation vacuum wavelength $\lambda$, and $(x,y,z)$ is a Cartesian coordinate system with the $(x,y)$ planes being transverse to the optical axis $z$. Substitute Eq.~(\ref{eq:MadelungTransformation}) into Eq.~(\ref{eq:HelmholtzEquationPlainForm}), expand the resulting expression, cancel a common factor of $\exp(i\Phi)$, and then separate real from imaginary parts. One of the equations so obtained is the (nonparaxial) continuity equation
\begin{equation}
\nabla\cdot[I(x,y,z)\nabla\Phi(x,y,z)]=0,
\label{eq:nonparaxialContinuityEquation}
\end{equation}
which expresses local conservation of energy. This equation also shows that $I\nabla\Phi$ is proportional to the flow vector (Poynting vector) of the propagating field.\footnote{{To make this identification, note that Eq.~(\ref{eq:nonparaxialContinuityEquation}) is a special case of the Poynting theorem for steady-state electromagnetic field flow in vacuum (see, e.g., Sec.~6.7 of \citet{JacksonElectrodynamicsBook}).}} For simplicity, work in a system of units where the constant of proportionality is unity, hence we speak of
\begin{equation}
    \mathbf{S}(x,y,z)=I(x,y,z)\nabla\Phi(x,y,z)
\label{eq:PoyntingVector}
\end{equation}
as {\em the} Poynting vector for the flow \cite{BerryFiveMomenta2013}.  As shown for the point labeled $E$ in Fig.~\ref{Fig:CoherentFlow}, the Poynting vector is perpendicular to the wavefronts (surfaces of constant phase).

Since we are interested in the case of paraxial beams, it is natural to decompose the wavefield phase $\Phi(x,y,z)$ as a sum of the phase $kz$ associated with a plane wave that travels in the direction of the optical axis $z$, and a comparatively slowly-varying modulation (envelope) phase $\phi(x,y,z)$ that quantifies the manner in which the plane-wave ``carrier wave'' phase $kz$ is distorted.  Thus
\begin{equation}
  \Phi(x,y,z)=kz+\phi(x,y,z),  
\label{eq:relationBetweenBigAndLittlePhi}
\end{equation}
where 
\begin{equation}
    \vert \nabla \phi(x,y,z) \vert \ll k.
\label{eq:paraxialityCondition}
\end{equation}
Next, return to Eq.~(\ref{eq:nonparaxialContinuityEquation}) and split it into transverse and longitudinal contributions as
\begin{eqnarray}
\nonumber\nabla_{\perp}\cdot[I(x,y,z)\nabla_{\perp}\Phi(x,y,z)]\quad\quad\quad\quad\quad \\ +\frac{\partial}{\partial z}\left[I(x,y,z)\frac{\partial}{\partial z}\Phi(x,y,z)\right]=0,
\label{eq:nonparaxialContinuityEquationSplitInTwoPieces}
\end{eqnarray}
where $\nabla_{\perp}$ denotes the gradient operator in the $(x,y)$ plane.  Equation~(\ref{eq:relationBetweenBigAndLittlePhi}) implies that $\nabla_{\perp}\Phi=\nabla_{\perp}\phi$, with Eqs.~(\ref{eq:relationBetweenBigAndLittlePhi}) and (\ref{eq:paraxialityCondition}) together implying that $\partial\Phi/\partial z$ may be approximated by $k$.  Hence Eq.~(\ref{eq:nonparaxialContinuityEquationSplitInTwoPieces}) becomes the following paraxial continuity equation known as the ``transport-of-intensity equation'' (TIE) \cite{Teague1983}:
\begin{equation}
    -\nabla_{\perp}\cdot[I(x,y,z)\nabla_{\perp}\phi(x,y,z)]=k\frac{\partial}{\partial z} I(x,y,z).
    \label{eq:TIE}
\end{equation}

The TIE states that the ``convergence'' (i.e., the negative of the divergence) of the transverse component of the flow vector (Poynting vector) is proportional to the longitudinal rate of change of the intensity. This accords with the intuitive remarks made earlier. Indeed, upon making the asymmetric first-order finite-difference approximation 
\begin{equation}
\left[\frac{\partial}{\partial z} I(x,y,z)\right]
_{z=0}\approx\frac{I(x,y,z=\Delta)-I(x,y,z=0)}{\Delta}
\label{eq:IntermediateResult02}
\end{equation}
to the longitudinal intensity derivative, the TIE leads to 
\begin{align}
\label{eq:outOfFocusConstrastVanillaForm}      
\nonumber I(x, &y,z=\Delta) \approx I(x,y,z=0) \\ &-\frac{\Delta}{k}\nabla_{\perp}\cdot\left[I(x,y,z=0)\nabla_{\perp}\phi(x,y,z=0)\right].
\end{align}
The contribution of the second line of the above equation may be spoken of as ``out of focus contrast'' \cite{Zernike1942,Bremmer1952,Cowley1959,CowleyBook} induced by the locally converging or diverging character of the flow.  Moreover, if the infocus intensity $I(x,y,z=0)$ is approximately uniform, with a value of $I_{\textrm{in}}$, this out-of-focus contrast is proportional to the transverse Laplacian 
\begin{equation}
\nabla_{\perp}^2=\frac{\partial^2}{\partial x^2}+\frac{\partial^2}{\partial y^2}
\label{eq:TransverseLaplacian}
\end{equation}
of the phase of the infocus field \cite{Bremmer1952}:
\begin{equation}
\label{eq:LaplacianContrast}      
\frac{I(x, y,z=\Delta)}{I_{\textrm{in}}} \approx  1-\frac{\Delta}{k}\nabla_{\perp}^2\phi(x,y,z=0).
\end{equation}

The TIE has been very widely utilized for phase retrieval.  In particular, Eq.~(\ref{eq:TIE}) is an elliptic second-order partial differential equation with nonconstant coefficients \cite{GilbargTrudingerBook}, for which the measured intensity and longitudinal intensity derivative, over a given plane of fixed $z$, may be used to solve for the unknown phase over that plane \cite{Gureyev1995}.  Since such an approach to phase measurement does not require the formation of interference fringes, its coherence requirements are rather lax \cite{paganin1998}.  The vast research literature on TIE-based phase retrieval will not be reviewed here, with the reader being referred to the comprehensive review article by \citet{TIE==LongReviewArticle2020} together with its extensive list of references.

\section{Fokker-Planck extension to the TIE: two informal derivations}\label{sec:RoughDerivationOfFPE}

The distinction between specular and diffuse reflection is a key concept in optical physics \cite{HechtOpticsBook}. For example, if light is reflected from a roughened plane mirror, the specular reflection (for which the angle of incidence is equal in magnitude to the angle of reflection) will be accompanied by a diffuse component that travels in a continuum of different directions.  

\begin{figure}[ht!]
\centering
\includegraphics[width=0.9\columnwidth]{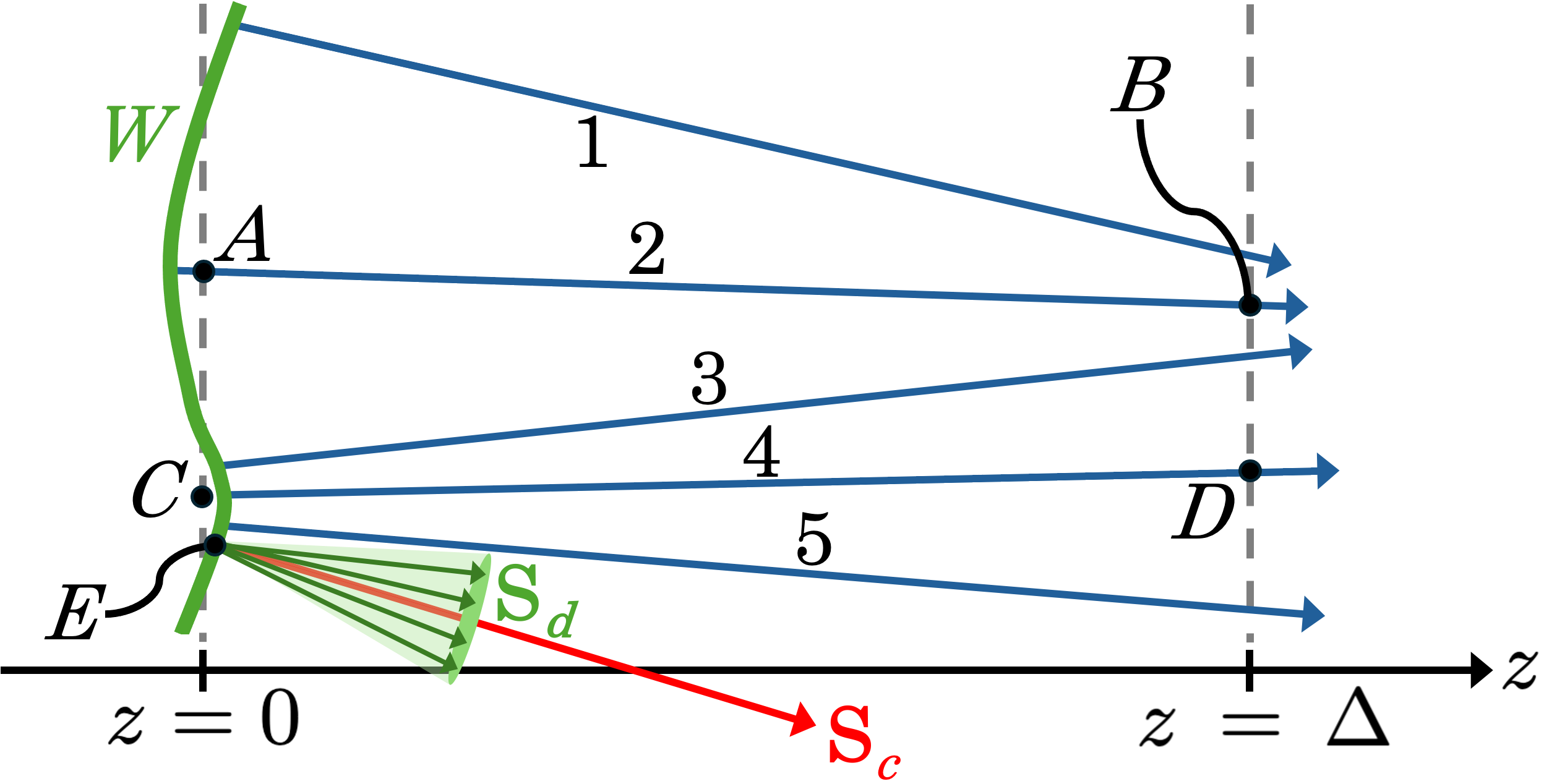}
\caption{Combined coherent and diffuse flow of a paraxial scalar optical field, associated with Fokker-Planck generalizations of the transport-of-intensity equation.}
\label{Fig:CoherentAndDiffuseFlow}
\end{figure}

As shown in Fig.~\ref{Fig:CoherentAndDiffuseFlow}, a similar phenomenon occurs when paraxial quasimonochromatic light \cite{BornWolf} passes through a refracting sample that contains ``internal roughness'' associated with spatially-random unresolved microstructure \cite{PaganinMorgan2019} [cf.~Figs.~3, 6, 7, and 10(b) in Paper I]. In this case, suppose the plane $z=0$ to correspond to the nominally planar exit surface of a thin sample (not shown) that lies immediately upstream of this plane. At each point over this plane, the Poynting vector $\mathbf{S}$ may be split into two components $\mathbf{S}_c$ and $\mathbf{S}_d$ that are respectively associated with the coherent (specular) and diffuse components of the flow.  The coherent component of the flow corresponds to a fraction $1-F(x,y,z)$ of the rays, streaming through the point $E$ in Fig.~\ref{Fig:CoherentAndDiffuseFlow}, in the direction of the local phase gradient according to Eq.~(\ref{eq:PoyntingVector}).  The diffuse component of the flow has a different character.  It may be associated with a fraction $F(x,y,z)$ of the rays, streaming through the point $E$, having a random spread of directions indicated by the ray ensemble filling the cone that is shown in the diagram with green vectors.\footnote{While we here emphasize the contribution to the cone-like diffusive ray ensemble that is associated with ``internal roughness'' of the sample, another contribution to this diffusive flow is due to the degree of spatial coherence of the illuminating beam (e.g., via an extended incoherent source).  For further information, see Sec.~II of Paper I, especially Fig.~3 and the upper line of Eq.~(23).} { Note also that the decomposition of an optical beam's flow into coherent and diffuse components is not unique in the sense that it depends on both the spatial and temporal resolution of the intensity detection, with progressively finer spatial or temporal resolution in general leading to a decrease in the fraction of the flow that is assigned to the diffuse channel.}\footnote{This point is further developed in Secs.~\ref{sec:RunningCoupling},  \ref{sec:HowDiffusionFieldVariesWithResolution}, and \ref{sec:LinkWithPartialCoherence}.}  

The presence of combined coherent and diffuse flow leads to a modified form of the TIE \cite{PaganinMorgan2019,MorganPaganin2019,PaganinPelliccia2021,PaganinPellicciaMorgan2023}, with the diffuse flow leading to position-dependent blur which augments the defocus-induced phase contrast associated with the coherent flow. This yields a modified version of Eq.~(\ref{eq:outOfFocusConstrastVanillaForm}), as very briefly sketched in Sec.~V\,H near the end of Paper I; this modification is a form of Fokker-Planck equation \cite{Risken1989} since it is a continuity equation that simultaneously describes both coherent and diffusive flow. 

In Secs.~\ref{sec:First
RoughDerivation} and \ref{sec:Second
RoughDerivation} below, we give two crude derivations that each emphasize different aspects of the optical physics associated with the Fokker-Planck generalization of the TIE in the forms previously published \cite{PaganinMorgan2019,MorganPaganin2019,PaganinPelliccia2021,PaganinPellicciaMorgan2023}.  We critique these derivations in Sec.~\ref{sec:Critique}, thereby motivating the more involved development given later in the paper (Secs.~\ref{sec:InfinityOfTIEs} and \ref{sec:FPEandExtendedFPE}). 

\subsection{First informal derivation}\label{sec:First
RoughDerivation}

Return consideration to the intensity distribution $I(x,y,z)$ associated with Fig.~\ref{Fig:CoherentAndDiffuseFlow}. Compressing our notation by writing $I(x,y,z=0)$ as $I_0(x,y)$, we decompose the ``unpropagated intensity'' into the diffuse and coherent channels via the identity\footnote{{Cf.~the first-principles derivation in Sec.~\ref{sec:DiffusionFieldFromFirstPrinciples}.}} 
\begin{equation} 
I_0(x,y)  = \underbrace{F(x,y) I_0(x,y)}_\text{diffuse channel} 
+ \underbrace{[1-F(x,y)] I_0(x,y)}_\text{coherent channel} .
\label{eq:AlternativeFPEderivation1}
\end{equation}
For the moment, suppress explicit functional dependence on $(x,y)$ for clarity. In propagating through a small distance $\Delta$, we consider the diffuse-channel intensity distribution $FI_0$ to be locally blurred via application of the diffusive operator 
\begin{equation}
    \mathscr{D}\equiv\mathbb{I}()+\tfrac{1}{2}\nabla_{\perp}^2[\sigma_{\Delta}^2()],
\end{equation}
where $\sigma_{\Delta}$ is a position-dependent blur-kernel standard deviation [see Eq.~(8) in Paper I], $\mathbb{I}$ denotes the identity operator, and empty brackets indicate where the acted-upon function should be inserted.\footnote{For an alternative derivation of this diffusive operator, see Eqs.~(25)-(31) in Ref.~\cite{GPM2020}.}  Conversely, the coherent-channel intensity $(1-F)I_0$ is locally sharpened using the TIE \cite{Teague1983}, according to the operator 
\begin{equation}
    \mathscr{T}\equiv\mathbb{I}()-\tfrac{\Delta}{k}\nabla_{\perp}\cdot [()\nabla_{\perp}\phi]
\end{equation}
given by Eq.~(\ref{eq:outOfFocusConstrastVanillaForm}). We may therefore write the defocus-induced intensity by separately evolving the two flows:

\begin{widetext}
\begin{equation} 
\begin{split}
I_0  \quad\quad &= \quad \quad\quad\quad F I_0 \quad\quad\quad\quad\,\,\,+ \quad\quad\quad\quad\quad (1-F) I_0
\\ &\quad\quad\quad\quad\quad\quad \Bigg\downarrow \scalebox{0.65}{$z=\Delta\ge 0$} \quad\quad\quad\quad\quad\quad\quad\quad\quad\quad
\Bigg\downarrow \scalebox{0.65}{$z=\Delta\ge 0$}
\\  &\quad\quad\quad F I_0 + \tfrac{1}{2}\nabla_{\perp}^2(\sigma_{\Delta}^2F I_0) \quad + \quad (1-F) I_0 - \tfrac{\Delta}{k}\nabla_{\perp}\cdot[(1-F)I_0\nabla_{\perp}\phi].
\end{split}
\label{eq:AlternativeFPEderivation2}
\end{equation}
\end{widetext}
Now recall Fig.~3 in Paper I, but simplify the scenario presented there by ignoring the influence of source-size blur. This simplification implies that we may take
\begin{equation}
\sigma_{\Delta}(x,y)=\theta_s(x,y) \Delta, 
\end{equation}
where $\theta_s(x,y)$ is a position-dependent blur angle associated with the position-dependent diffuse scatter induced by the unresolved sample microstructure (again see Fig.~3 of Paper I), and we are here reinstating explicit functional dependence upon transverse coordinates $(x,y)$. Also introduce the dimensionless diffusion coefficient (diffusion field)\footnote{Note that Eq.~(\ref{eq:DimensionlessScalarDiffusionCoefficient}) corresponds to a special case of Eq.~(24) in Paper I, with the angular extended-source size $\theta_0$ (see Figs.~2, 3 and 5 of Paper I) being set to zero.}
\begin{equation}
D(x,y)=\frac{1}{2} F(x,y) \, \left[\theta_s(x,y)\right]^2.
\label{eq:DimensionlessScalarDiffusionCoefficient}
\end{equation}
With the indicated changes, the lower line of Eq.~(\ref{eq:AlternativeFPEderivation2}) leads to the finite-difference form 
\begin{align}
\nonumber &I(x,y,z=\Delta \ge 0) =I(x,y,z=0) 
\\  \nonumber &-\frac{\Delta}{k} \nabla_{\perp}\cdot \{[1-F(x,y)]I(x,y,z=0) \nabla_{\perp}\phi(x,y,z=0)\}
\\ &+\Delta^2\nabla_{\perp}^2[D(x,y) I(x,y,z=0)]
\label{eq:SecIII--FPE-scalar-case}
\end{align}
of the isotropic-diffusion Fokker-Planck generalization of the TIE given in Eq.~(146) of Paper I.\footnote{See, also, Refs.~\cite{MorganPaganin2019,PaganinMorgan2019,PaganinPelliccia2021,Leatham2023,Leatham2024}.  Cf.~Refs.~\cite{alloo2022dark,Beltran2023,alloo2023SciRep,Alloo2025} and the earlier papers by Davis \cite{Davis1991,Davis1992,Davis1993,Davis1994}.}

\subsection{Second informal derivation}\label{sec:Second
RoughDerivation}

Here we provide a second informal derivation of the Fokker-Planck TIE generalization. This calculation is inspired by, and partly analogous to, the derivation of Zitterbewegung-induced smearing of the effective potential (``Darwin term'') for an electron bound to a hydrogen nucleus in relativistic quantum theory.\footnote{See, e.g., p.~119 of \citet{SakuraiAdvancedQuantumMechanics}.} Geometrical optics is assumed throughout this derivation.

For the moment, we work in one transverse dimension $x$.  Hence Eq.~(\ref{eq:outOfFocusConstrastVanillaForm}) becomes
\begin{align}
\nonumber I(x,z=\Delta) &\approx I(x,z=0) \\ &-\frac{\Delta}{k}\frac{\partial}{\partial x}\left[I(x,z=0)\frac{\partial}{\partial x}\phi(x,z=0)\right].
\label{eq:AppB02}      
\end{align}
On the right side of this equation, make the replacement 
\begin{equation}
\label{eq:AppB03}      
x \longrightarrow x - \delta x(x).
\end{equation}
Here $\delta x$ is an $x$-dependent smooth random function which continuously deforms the propagated intensity $I(x,z=\Delta)$, in a manner consistent with the added presence of unresolved random refractive structure \cite{ThinWolfBook} { within a thin sample placed immediately upstream of the plane $z=0$ in Fig.~\ref{Fig:CoherentAndDiffuseFlow}}.  This unresolved structure may also be associated with temporally unresolved fluctuations that are due to an extended chaotic source \cite{BornWolf,LoudonBook} or the non-zero temperature of the sample \cite{FrozenPhononModel}.   

For any one particular instance of $\delta x (x)$, drawn from the statistical ensemble $\mathscr{E}$ of all possible realizations for this stochastic function, the replacement in Eq.~(\ref{eq:AppB03}) implies that Eq.~(\ref{eq:AppB02}) tentatively becomes
\begin{align}
\label{eq:AppB04}
&I(x,z=\Delta) \stackrel{?}{\longrightarrow} I(x-\delta x(x),z=0) \\ \nonumber &-\frac{\Delta}{k}\frac{\partial}{\partial x}\left[I(x-\delta x(x),z=0)\frac{\partial}{\partial x}\phi(x-\delta x(x),z=0)\right].
\end{align}
In writing the preceding expression, we have made the strong but sometimes reasonable assumption that the previously mentioned unresolved { random refractive} sample structure is sufficiently slowly varying, such that it merely {\em geometrically distorts the defocused intensity that would exist in the absence of such fluctuations}.

Taylor expand all functions on the right side of Eq.~(\ref{eq:AppB04}), working to second order in $\delta x(x)$, and note that---for the purposes of this expansion---the final line of Eq.~(\ref{eq:AppB04}) may be considered as a single function.  Ensemble average over $\mathscr{E}$, as denoted using an overline (cf.~Refs.~\cite{PaganinMorgan2019} and \cite{Nesterets2008}).  By assumption, $\overline{\delta x(x)}$ vanishes at any transverse position $x$, hence
\begin{align}
\nonumber
&I(x,z=\Delta) \stackrel{?}{\longrightarrow} \overline{I(x-\delta x(x),z=\Delta)} 
\\ \nonumber &\approx I(x,z=0) + \frac{\overline{[\delta x(x)]^2}}{2!}\frac{\partial^2}{\partial x^2}I(x,z=0)
\\  \nonumber &-\frac{\Delta}{k}\frac{\partial}{\partial x}\left[I(x,z=0)\frac{\partial}{\partial x}\phi(x,z=0)\right] 
\\  &- \frac{\Delta}{k}\frac{\overline{[\delta x(x)]^2}}{2!}\frac{\partial^3}{\partial x^3}\left[I(x,z=0)\frac{\partial}{\partial x}\phi(x,z=0)\right].
\label{eq:AppB05}
\end{align}

Consistent with a paraxial small-defocus distortion in the regime of geometrical optics, we now assume that
\begin{equation}
\label{eq:AppB06}      
\delta x(x)=\varepsilon(x)\Delta,
\end{equation}
where $\varepsilon(x)$ is a position-dependent refraction angle associated with any one instance (``realization'') of the unresolved sample fluctuations (see Fig.~6 in Paper I).  Retaining terms that are no higher than second order in $\Delta$, the final term in Eq.~(\ref{eq:AppB05}) drops out since it is of third order with respect to $\Delta$.  We are left with
\begin{align}
\nonumber
I(x, z=\Delta) &\stackrel{?}{\longrightarrow} \overline{I(x-\delta x(x),z=\Delta)} 
\\ \nonumber &\approx I(x,z=0) + \frac{\Delta^2}{2}[\theta_s(x)]^2\frac{\partial^2}{\partial x^2}I(x,z=0)
\\  &-\frac{\Delta}{k} \frac{\partial}{\partial x}\left[I(x,z=0)\frac{\partial}{\partial x}\phi(x,z=0)\right],
\label{eq:AppB07}
\end{align}
where
\begin{equation}
\label{eq:AppB08}      
\theta_s(x)=\sqrt{\,\overline{[\varepsilon(x)]^2}}
\end{equation}
is the root-mean-square half-angle associated with the sample-induced fluctuations in ray angle [cf.~Eq.~(73) in Paper I, together with Fig.~6 in that paper]. 

To complete the present derivation, an additional physical factor must be taken into account. 
 This extra factor is directly related to the previously mentioned fact that, when a light beam illuminates a plane mirror that contains some degree of surface roughness, the reflected beam is split into specular and diffuse components \cite{HechtOpticsBook}.  In our case, we may instead speak of (i) ``specular refraction by the sample'' which manifests as the coherent flow associated with the phase gradient, coupled with (ii) position-dependent diffuse scatter by the sample that is associated with diffusive flow.  To refine our calculation so as to take this factor into account, let us again bifurcate the propagated intensity into a sum of coherent-flow and diffuse-flow channels, respectively, via 
\begin{align}
\nonumber
I(x,z=\Delta)\longrightarrow & [1-F(x)]I(x,z=\Delta)
\\ &+ F(x) \overline{I(x-\delta x(x),z=\Delta)}.
\label{eq:AppB09}      
\end{align}
If Eq.~(\ref{eq:AppB02}) is used to estimate the first term in the preceding sum, with { the middle line of} Eq.~(\ref{eq:AppB07}) being used for the ensemble average in the second term,\footnote{{We here ignore the bottom line of Eq.~(\ref{eq:AppB07}), since we wish to exclude the influence of coherent-flow effects when calculating a diffuse-flow contribution to the propagated intensity. See Sec.~\ref{sec:DiffusionFieldFromFirstPrinciples} for first-principles justification of this step.}} we obtain
\begin{align}
\nonumber
I(x,&\,z=\Delta) \approx I(x,z=0)
\\ \nonumber &-[1-F(x)]\frac{\Delta}{k}\frac{\partial}{\partial x}\left[I(x,z=0)\frac{\partial}{\partial x}\phi(x,z=0)\right]
\\ &+\frac{1}{2}F(x)[\theta_s(x)]^2\Delta^2\frac{\partial^2}{\partial x^2}I(x,z=0).
\label{eq:AppB10}      
\end{align}

Reinstate the second transverse dimension {\em under the simplifying assumption that the unresolved sample fluctuations are rotationally symmetric in their statistical character}.  Stated differently, we here assume that the diffuse flow at each point may be associated with a local smearing function that is invariant under rotations about that point in the $(x,y)$ plane.  Recall the form for the dimensionless diffusion coefficient (scalar diffusion field) $D(x,y)$ in Eq.~(\ref{eq:DimensionlessScalarDiffusionCoefficient}),
then assume $D(x,y)$ to be sufficiently slowly spatially varying that it approximately commutes with the transverse Laplacian operator $\nabla_{\perp}^2$.\footnote{Alternatively, one may follow similar logic to Sec.~III\,C of Paper I, and consider this interchange in order---of the scalar diffusion field and the transverse Laplacian---to amount to a {\em change in representation for the diffusion field}, rather than implying any additional approximation.} { Similarly, $F(x,y)$ is taken to be sufficiently slowly varying that it approximately commutes with the transverse divergence operator. We thereby arrive once more at Eq.~(\ref{eq:SecIII--FPE-scalar-case}).}

We now generalize the result of the previous paragraph.  In particular, we now treat the case where rotational symmetry cannot be assumed for the statistical character of the sample-induced stochastic wavefield fluctuations.  Each instance of the random-microstructure stochastic distortion field is now represented by $\delta x (x,y)$ for the geometric distortions in the $x$ transverse direction and $\delta y (x,y)$  for the geometric distortions in the $y$ transverse direction.  Similarly, we generalize Eq.~(\ref{eq:AppB06}) to
\begin{equation}
\label{eq:AppB06==generalisedTo2D}      
\delta x(x,y)=\varepsilon_x(x,y)\Delta
\quad\textrm{and}\quad
\delta y(x,y)=\varepsilon_y(x,y)\Delta,
\end{equation}
where $\varepsilon_x(x,y)$ is a position-dependent refraction (deflection) angle in the $x$ direction due to unresolved random sample fluctuations in refractive index, and $\varepsilon_y(x,y)$ is the corresponding angle in the $y$ direction.  With these definitions in place, and again discarding Taylor-series terms in $\Delta$ that are of higher than second order, the chain of logic leading from Eq.~(\ref{eq:AppB04}) to (\ref{eq:AppB07}) generalizes to [cf.~Eqs.~(105) and (147) in Paper I]
\begin{align}
\nonumber
I(x, &y, z=\Delta) \stackrel{?}{\longrightarrow} \overline{I(x-\delta x(x,y), y-\delta y(x,y), z=\Delta)} 
\\ \nonumber \approx& I(x,y,z=0) +\frac{\Delta^2}{2}\nabla_{\perp}\cdot[\mathbf{C}(x,y)\nabla_{\perp}I(x,y,z=0)]
\\ &-\frac{\Delta}{k} \nabla_{\perp}\cdot[I(x,y,z=0)\nabla_{\perp}\phi(x,y,z=0)],
\label{eq:AppB07==2D-version}
\end{align}
where we have introduced the position-dependent angular-deflection covariance matrix field\footnote{(a) Note also that in writing Eq.~(\ref{eq:AppB07==2D-version}) we have assumed the elements of $\mathbf{C}(x,y)$ to be sufficiently slowly varying functions of spatial position that they approximately commute with $\partial/\partial x$ and $\partial/\partial y$. {(b) In the middle line of Eq.~(\ref{eq:AppB07==2D-version}), $\nabla_{\perp}I$ is considered to be a two-component column vector (cf.~p.~16 of Paper I).}}
\begin{equation}
    \mathbf{C}(x,y)\equiv \begin{pmatrix}
\overline{[\varepsilon_x(x,y)]^2} & \overline{[\varepsilon_x(x,y)][\varepsilon_y(x,y)]} \\
\overline{[\varepsilon_x(x,y)][\varepsilon_y(x,y)]} & \overline{[\varepsilon_y(x,y)]^2} 
\end{pmatrix}.
\label{eq:CovarianceMatrix}
\end{equation}
To refine the calculation, by taking into account the bifurcation into coherent and diffuse flows, we invoke the generalization of Eq.~(\ref{eq:AppB09}) to two transverse dimensions, namely
\begin{align}
\nonumber
I(x, &y,z=\Delta)\longrightarrow  [1-F(x,y)]I(x,y,z=\Delta)
\\ &+ F(x,y) \overline{I(x-\delta x(x,y), y-\delta y(x,y), z=\Delta)}.
\label{eq:AppB09==GenerliseTo2D}      
\end{align}
If Eq.~(\ref{eq:outOfFocusConstrastVanillaForm}) is used to estimate the first term in the preceding sum, with { the middle line of} Eq.~(\ref{eq:AppB07==2D-version}) being used for the ensemble average in the second term, we arrive at
\begin{align}
\nonumber
&I(x,y,\,z=\Delta) \approx I(x,y,z=0)
\\ \nonumber &-\frac{\Delta}{k}[1-F(x,y)]\nabla_{\perp}\cdot\left[I(x,y,z=0)\nabla_{\perp}\phi(x,y,z=0)\right]
\\ &+\frac{1}{2}\Delta^2 F(x,y)\nabla_{\perp}\cdot [\mathbf{C}(x,y)\nabla_{\perp}I(x,y,z=0)].
\label{eq:AlmostThereWithZitterbewegung-likeCalc}   
\end{align}
Now enforce both local and global energy conservation by assuming that the diffuse-scatter fraction $F(x,y)$ is a sufficiently slowly varying function of position that multiplication by $F(x,y)$ may be interchanged in order with application of the transverse divergence operator.\footnote{``Putting $F$ inside the divergence operation'' is consistent with both local and global energy conservation, in a similar sense to that described in the paragraph of Paper I which contains Eq.~(26).  In effect, what we are doing is (i) to acknowledge that our approximate analysis spoils the energy conservation ``$\iint I(x,y,z=\Delta) dx\,dy={\textrm{constant}}$'' that we know must hold for any $\Delta$, and then (ii) modifying the resulting expression to restore energy conservation, by ensuring that the diffuse Fick-law flow \cite{CrankBook} is a conserved current, in the sense of Noether's theorem \cite{MandlShawBook,PaganinMorgan2019}. A similar comment applies to the coherent-flow component of the paraxial intensity transport. Cf.~use of the ``Cayley form'' to restore unitarity, in a finite-difference approximation that otherwise spoils unitary evolution, by \citet{Goldberg1967}.} In addition, introduce the dimensionless tensor diffusion field
\begin{equation}
\widetilde{\mathbf{D}}(x,y)=\frac{1}{2} F(x,y) \mathbf{C}(x,y).
\label{eq:NiceNewEquationForTensorDiffusionField}
\end{equation}
We thereby arrive at the anisotropic-diffusion finite-difference form 
\begin{align}
\nonumber
&I(x,y,\,z=\Delta) \approx I(x,y,z=0)
\\ \nonumber &-\frac{\Delta}{k}\nabla_{\perp}\cdot\left\{[1-F(x,y)]I(x,y,z=0)\nabla_{\perp}\phi(x,y,z=0)\right\}
\\ &+\Delta^2 \nabla_{\perp}\cdot [\widetilde{\mathbf{D}}(x,y)\nabla_{\perp}I(x,y,z=0)]
\label{eq:EndofZitterbewegung-likeCalc}   
\end{align}
of the Fokker-Planck TIE extension, which coincides with Eq.~(147) in Paper I.\footnote{See, also, Refs.~\cite{MorganPaganin2019,PaganinMorgan2019,PaganinPelliccia2021,PaganinNeutron2023}. Cf.~Ref.~\cite{MISTdirectional}.} 

We close this section with five miscellaneous remarks. (i) The TIE can be crudely summarized by the statement that ``phase gradients reveal coherent flow'' for a paraxial beam whose transverse intensity distribution evolves as it propagates along the optical axis $z$ (which is considered to be an evolution parameter analogous to time, in this context).  In the Fokker-Planck extensions of the TIE, such a summary may be generalized to ``intensity gradients reveal diffusive flow\footnote{The statement that ``intensity gradients reveal diffusive flow'' is quantified by Fick's first law of diffusion in the presence of a position-dependent diffusion coefficient (see, e.g., pp.~2 and 4 of \citet{CrankBook}).  It is the position-dependent diffusion coefficient, namely the ``diffusion field'' and its associated diffuse current, which is revealed through the manner in which the intensity distribution diffuses.  Cf.~Fig.~1 in Paper I.} and phase gradients reveal coherent flow''.\footnote{This simple statement needs some clarification.  (a) Spatially rapid {\em phase} gradients (namely phase fluctuations occurring at length scales finer than the detector resolution) can also lead to diffusive flow, not only intensity gradients. However, in the present paper such unresolved rapid phase gradients are considered to have been coarse grained into an associated diffusion field associated with intensity gradients. See Ref.~\cite{PaganinMorgan2019} for an example of this process. Cf.~Secs.~{\ref{sec:DiffusionFieldFromFirstPrinciples} and} \ref{sec:RunningCoupling}. (b) In the absence of any intensity gradients, ``edges'' of areas of diffusive flow (i.e., regions where the diffusion field varies rapidly with transverse position) can still be observed.  For example, the final line of Eq.~(\ref{eq:SecIII--FPE-scalar-case}) is nonzero if the diffusing intensity is uniform and the diffusion field is nonuniform. Cf.~Fig.~4 in Ref.~\cite{MorganPaganin2019}.} (ii) It is interesting to compare the expressions for the scalar and tensor diffusion fields that are given in Eqs.~(\ref{eq:DimensionlessScalarDiffusionCoefficient}) and (\ref{eq:NiceNewEquationForTensorDiffusionField}), respectively.\footnote{See, also, the ``$\theta_0=0$'' special case of the scalar diffusion field in Eq.~(24) from Paper I, together with the tensor diffusion field in Eqs.~(106)-(109) from the same paper.}  The former expression factorizes the scalar diffusion field $D(x,y)$ as equal to one half of the product of the scatter-fraction $F(x,y)$ with the {\em variance} of the stochastic angular deflection associated with spatially unresolved sample microstructure [see, also, Eq.~(\ref{eq:AppB08})].  The latter expression factorizes the rank-two tensor diffusion field $\widetilde{\mathbf{D}}(x,y)$ as equal to one half of the product of the scatter-fraction $F(x,y)$ with the $2 \times 2$ {\em covariance matrix} of the stochastic angular deflection associated with spatially unresolved sample microstructure.  (iii) If terms of higher than second order in $\Delta$ are retained in the Taylor expansion that was employed earlier in the present section, then a hierarchy of additional diffusion fields is generated. This extends our Fokker-Planck equation into a Kramers-Moyal equation \cite{MorganPaganin2019,PaganinMorgan2019,Risken1989}. (iv) The same rotation-matrix field $\mathcal{R}_{\psi(x,y)}$ that was used to diagonalize $\widetilde{\mathbf{D}}(x,y)$ in Eqs.~(106)-(109) of Paper I may also be used to diagonalize $\mathbf{C}(x,y)$.\footnote{Note that $\psi(x,y)$ denotes a rotation angle in the sense given by Fig.~9 of Paper I, with $\psi(x,y,z)$ denoting a complex wavefield in the sense employed from Sec.~\ref{sec:InfinityOfTIEs} onwards in the present paper [cf.~Eq.~(\ref{eq:MadelungTransformation})].  The indicated functional dependence, which is used consistently, distinguishes the two quantities.\label{footnote:Disambiguate-psi-notation}}  The physical picture in Fig.~9 from that earlier paper still holds in the present context, with the only difference being that the locally-elliptical transverse section of the diffuse-scatter fan corresponds to a sheaf of rays rather than a sheaf of wavevectors. (v) Transition from tensor diffusion to scalar diffusion follows if we may make the approximation 
\begin{equation}
\widetilde{\mathbf{D}}(x,y)\approx \mathbf{I}_2 \widetilde{D}(x,y),
\label{eq:TransitionFromTensorToScalarDiffusion}
\end{equation}
where $\mathbf{I}_2$ is the $2 \times 2$ unit matrix and $\widetilde{D}(x,y)$ is a scalar function.\footnote{Cf.~the approximation made in Eq.~(115) from Paper I.}  In this case the final line of Eq.~(\ref{eq:EndofZitterbewegung-likeCalc}) reduces to $\Delta^2 \nabla_{\perp}\cdot [\widetilde{D}(x,y)\nabla_{\perp}I(x,y,z=0)]$, which agrees with the representation for a scalar diffusion field in Sec.~III\,C of Paper I.  Moreover, if $\widetilde{D}(x,y)$ is a sufficiently slowly varying function of position that it approximately commutes with the gradient operator---or, alternatively, if we employ the change of representation for a scalar diffusion field that is  given in Eq.~(50) of Paper I---then we obtain the functional form given in the final line of Eq.~(\ref{eq:SecIII--FPE-scalar-case}).

\subsection{Critique of our two informal derivations}\label{sec:Critique}

Equations (\ref{eq:SecIII--FPE-scalar-case}) and (\ref{eq:EndofZitterbewegung-likeCalc}) are Fokker-Planck extensions of the TIE that may be criticized on the following grounds.  Observe that the coherent-flow middle line of each equation is of first order with respect to the defocus distance $\Delta$, whereas the diffuse-flow third line of each equation is of second order with respect to $\Delta$.  If one considers $\Delta$ to be the only relevant expansion parameter, then there is evidently a missing term, namely a contribution to the coherent flow that is of second order with respect to $\Delta$.  This issue is clarified in Secs.~\ref{sec:InfinityOfTIEs} and \ref{sec:FPEandExtendedFPE} below.  Alternatively, one may instead consider there to be {\em two} relevant parameters, namely $\Delta$ and the scalar or tensor diffusion field; if the influence of the  diffusion field is sufficiently strong, then Eqs.~(\ref{eq:SecIII--FPE-scalar-case}) and (\ref{eq:EndofZitterbewegung-likeCalc}) may be used without need for further modification.\footnote{ Section~\ref{subsec:Practicalities6} develops the final sentence in this paragraph.}

{ More generally, when the ``missing term'' is restored [see Eq.~(\ref{eq:SecIII--ExtendedFPE}) below], the order-$\Delta^2$ diffusion contribution is accompanied by another order-$\Delta^2$ contribution that is due to coherent Fresnel diffraction.  In all of the earlier derivations \cite{PaganinMorgan2019,MorganPaganin2019,PaganinPellicciaMorgan2023} these higher-order coherent terms are neglected, with this being an enabling assumption for the separation of coherent from diffusive effects (i.e., the ability to perform both phase retrieval and diffusion-field retrieval).  When the previously mentioned ``missing term'' is restored, and the higher-order coherent term is not ignorably small, is the separation of coherent from diffusive effects still possible?  A positive answer to this question will be developed in subsequent sections of the present paper.  The key strategy, which underpins this answer and guides much of the associated chain of logic, is as follows.  (i) Take a sufficient number of intensity measurements such that the order-$\Delta^2$ contribution may be algebraically eliminated. This removes the influence of both the diffusion-field and the second-order coherent contribution.  (ii) The resulting system of equations is a function of intensity and phase, and as we shall see, these equations can be solved for the phase (i.e., phase retrieval can be performed).  (iii) Since the phase has been retrieved and the intensity has been measured, the second-order coherent-Fresnel-diffraction contribution can be calculated.  (iv) The only remaining unknown function is the diffusion field, for which closed-form reconstruction formulae shall be written down. }

\section{An infinity of transport-of-intensity equations for coherent paraxial flow}\label{sec:InfinityOfTIEs}

As pointed out in Sec.~\ref{sec:Critique}, Eqs.~(\ref{eq:SecIII--FPE-scalar-case}) and (\ref{eq:EndofZitterbewegung-likeCalc}) should in general be augmented by a coherent-flow term that is of second order with respect to the defocus distance $\Delta$.  This augmentation is very closely related to the second-order longitudinal intensity derivative, for a coherent paraxial monochromatic scalar beam.  As shown below, this second-order derivative may be obtained by first calculating {\em all} positive-integer orders of the longitudinal intensity derivative.  This gives the ``infinity of TIEs'' in the title to this section. Our derivation is based on the previously unpublished analysis by \citet{PaganinPhDthesis1999}.  

We seek all longitudinal intensity derivatives $\partial^m I(x,y,z)/\partial z^m$ associated with the intensity 
\begin{equation}
I(x,y,z)=\vert\psi(x,y,z)\vert^2
\label{eq:InfinityOfTIEs01}
\end{equation}
of a paraxial complex scalar field $\psi(x,y,z)$.  Here, $m=1,2,\cdots$, the optical axis is $z$, and $(x,y)$ are Cartesian coordinates in planes perpendicular to $z$.  The complex field $\psi(x,y,z)$ is the spatial envelope of the full time-dependent complex wave field [cf.~Eq.~(\ref{eq:relationBetweenBigAndLittlePhi})]
\begin{equation}
\Psi(x,y,z,t)=\psi(x,y,z)\exp[i(kz-\omega t)],
\label{eq:InfinityOfTIEs02}
\end{equation}
where $k=2\pi/\lambda$ is the wavenumber [Eq.~(\ref{eq:waveNumber})] corresponding to the wavelength $\lambda$, $\omega$ is the angular frequency corresponding to the fixed energy $E=\hbar \omega$, $t$ denotes time, and $\hbar$ is Planck's constant $h$ divided by $2\pi$. 

This formalism is simultaneously applicable to a number of scenarios, including but not limited to (i) monochromatic complex scalar paraxial electromagnetic waves in situations where the polarization degree of freedom may be ignored, (ii) paraxial monoenergetic electron beams where the influence of spin may be neglected, and (iii) paraxial energy-filtered neutron beams where the influence of the neutron magnetic moment may be neglected.  In all of these cases, the Helmholtz equation (time-independent free-space Schr\"{o}dinger equation, time-independent free-space Klein-Gordon equation)
\begin{equation}
(\nabla^2+k^2)[\psi(x,y,z)\exp(ikz)]=0
\label{eq:HelmholtzEquation}
\end{equation}
reduces to the paraxial equation (parabolic equation)
\begin{equation}
\left(2ik\frac{\partial}{\partial z}+\nabla_\perp^2\right)\psi(x,y,z)=0
\label{eq:ParaxialEquation}
\end{equation}
under the paraxial approximation. Above, the three-dimensional Laplacian $\nabla^2$ is given by Eq.~(\ref{eq:3DLaplacian}), with the transverse Laplacian $\nabla_{\perp}^2$ corresponding to Eq.~(\ref{eq:TransverseLaplacian}).

A formal solution to the paraxial equation is \cite{Bremmer1952,Teague1983}
\begin{equation}
\psi(x,y,z=\Delta)=\exp\left(\frac{i \Delta \nabla_{\perp}^2}{2k}\right)\psi(x,y,z=0). 
\label{eq:FormalSolutionToParaxialEquation}
\end{equation}
Expand the exponential as a Taylor series, then form the squared magnitude of the ensuing expression.  Arrange terms in the resulting double summation so that the coefficients of increasing powers of $\Delta$ are isolated.  Hence 
\begin{align}
\nonumber    
I(x,y, &z=\Delta)=\sum_{m=0}^{\infty} \bigg\{ 
    \left(\frac{i}{2k}\right)^{\!\!m}
    \sum_{n=0}^m 
    \begin{pmatrix} m \\ n \end{pmatrix}
    \\ \nonumber
    &\times \left[\left(\nabla_{\perp}^2\right)^{m-n}\psi(x,y,z=0)\right]
\\ 
    &\times \left[\left(-\nabla_{\perp}^2\right)^{n}\psi^*(x,y,z=0)\right]
    \bigg\}\frac{\Delta^m}{m!},
\label{eq:InfinityOfTIEs03}  
\end{align}
where  
\begin{equation}
    \begin{pmatrix} m \\ n \end{pmatrix} = \frac{m!}{n!(m-n)!},
\label{eq:BinomialCoefficient}    
\end{equation}
is the binomial coefficient, and an asterisk superscript denotes complex conjugation. Term-by-term comparison of Eq.~(\ref{eq:InfinityOfTIEs03}) with the Taylor series expansion
\begin{equation}
I(x,y,z=\Delta)=\sum_{m=0}^{\infty} 
\frac{\Delta^m}{m!}
\left[\frac{\partial^m}{\partial z^m} I(x,y,z)\right]_{z=0}
\label{eq:LongitudinalTaylorSeriesExpansionOfIntensity}    
\end{equation}
for the intensity shows that 
\begin{align}
\nonumber    
\left[\frac{\partial^m}{\partial z^m} I(x, y, z)\right]_{z=0}
&= 
    \left(\frac{i}{2k}\right)^{\!\!m}
    \sum_{n=0}^m 
    \begin{pmatrix} m \\ n \end{pmatrix}
    \\ \nonumber
    &\times \left[\left(\nabla_{\perp}^2\right)^{m-n}\psi(x,y,z=0)\right]
\\ 
    &\times \left[\left(-\nabla_{\perp}^2\right)^{n}\psi^*(x,y,z=0)\right].
\end{align}
Dropping the ``$z=0$'' qualification then gives the required infinity of TIEs, namely \cite{PaganinPhDthesis1999}
\begin{align}
\nonumber    
&\frac{\partial^m}{\partial z^m} I(x, y, z)
= 
    \left(\frac{i}{2k}\right)^{\!\!m}
    \sum_{n=0}^m 
    \begin{pmatrix} m \\ n \end{pmatrix}
\\ &\quad\times
     \left[\left(\nabla_{\perp}^2\right)^{m-n}\psi(x,y,z)\right]
    \left[\left(-\nabla_{\perp}^2\right)^{n}\psi^*(x,y,z)\right].
\label{eq:InfinityOfTIEs--MainResult}  
\end{align}
This relates the indicated transverse derivatives of the complex field $\psi(x,y,z)$ to the $m$th derivative of intensity with respect to $z$ (where $m=0,1,2,\cdots$). 

For the trivial special case where $m=0$, Eq.~(\ref{eq:InfinityOfTIEs--MainResult}) reduces to Eq.~(\ref{eq:InfinityOfTIEs01}). When $m=1$, Eq.~(\ref{eq:InfinityOfTIEs--MainResult}) reduces to the standard form of the TIE in Eq.~(\ref{eq:TIE}). If $m=2$, we obtain the {\em second-order intensity transport equation}
\begin{align}
\nonumber 2k^2\frac{\partial^2}{\partial z^2} &I(x,y,z)  = \left\vert\nabla_{\perp}^2\psi(x,y,z)\right\vert^2 
\\  &-\textrm{Re}\left[\psi^*(x,y,z) (\nabla_{\perp}^2)^2  \psi(x,y,z)\right].
\label{eq:InfinityOfTIEs--2ndIntensityDerivative}
\end{align}
Hence, if $\Delta$ is sufficiently small that only the first three summation terms in Eq.~(\ref{eq:LongitudinalTaylorSeriesExpansionOfIntensity}) need to be retained, the propagated intensity associated with Fresnel diffraction may be approximated by
\begin{widetext}
\begin{align}
\label{eq:InfinityOfTIEs--SecondOrderExpansionOfIntensity}      
&I(x,y,z=\Delta) \approx I(x,y,z=0) -\frac{\Delta}{k}\nabla_{\perp}\cdot\left[I(x,y,z=0)\nabla_{\perp}\phi(x,y,z=0)\right]
    \\  \nonumber &+\frac{\Delta^2}{4k^2}\left\{\left\vert\nabla_{\perp}^2\left[\sqrt{I(x,y,z=0)}e^{i\phi(x,y,z=0)}\right]\right\vert^2-\textrm{Re}\left\{\sqrt{I(x,y,z=0)}e^{-i\phi(x,y,z=0)}(\nabla_{\perp}^2)^2\left[\sqrt{I(x,y,z=0)}e^{i\phi(x,y,z=0)}\right]\right\}\right\}.
\end{align}
\end{widetext}
The upper line is the first-order finite-difference form of the TIE [Eq.~(\ref{eq:outOfFocusConstrastVanillaForm})], which quantifies out-of-focus contrast \cite{Zernike1942,Bremmer1952,Cowley1959,CowleyBook} in a physically intuitive manner that accords with our earlier statements regarding the local convergence or divergence of the wavefield upon propagation through a small distance. The lower line of Eq.~(\ref{eq:InfinityOfTIEs--SecondOrderExpansionOfIntensity}) is a correction associated with the second-order TIE [Eq.~(\ref{eq:InfinityOfTIEs--2ndIntensityDerivative})]. 

\section{Fokker-Planck and extended Fokker-Planck generalizations of the transport-of-intensity equation}\label{sec:FPEandExtendedFPE}

For the boundary-value problem of paraxial coherent scalar wave propagation, the formal solution in Eq.~(\ref{eq:FormalSolutionToParaxialEquation}) may be written as (see, e.g., Sec.~1.4.1 of Ref.~\cite{Paganin2006})
\begin{equation}
\psi(x,y,z=\Delta)=\mathcal{D}_{\Delta}^{\textrm{(F)}}\psi(x,y,z=0). 
\label{eq:SecIII--01}
\end{equation}
Here, the Fresnel diffraction operator $\mathcal{D}_{\Delta}^{\textrm{(F)}}$ acts on the unpropagated complex wavefield $\psi(x,y,z=0)$, to yield the propagated wavefield $\psi(x,y,z=\Delta)$.  A Fourier representation of this diffraction operator is\footnote{{In writing Eq.~(\ref{eq:SecIII--02}), use has been made of the convention that cascaded operators act from right to left. Thus the Fresnel diffraction operator $\mathcal{D}_{\Delta}^{\textrm{(F)}}$ corresponds to applying the Fourier transformation $\mathscr{F}$, then multiplying by $\exp[-i\Delta(k_x^2+k_y^2)/(2k)]$, and finally applying the inverse Fourier transformation $\mathscr{F}^{-1}$ \cite{NazarathyShamir1980}.}}
\begin{equation}
\mathcal{D}_{\Delta}^{\textrm{(F)}}
= \mathscr{F}^{-1} \exp[-i\Delta(k_x^2+k_y^2)/(2k)]\mathscr{F}, 
\label{eq:SecIII--02}
\end{equation}
where $\mathscr{F}$ denotes Fourier transformation with respect to $x$ and $y$ using the convention
\begin{equation}
\mathscr{F}{g(x,y)}=\frac{1}{2\pi}\iint_{-\infty}^{\infty} g(x,y)\exp[-i(k_xx+k_yy)]dxdy
\label{eq:SecIII--03}
\end{equation}
for a well-behaved function $g(x,y)$, $\mathscr{F}^{-1}$ denotes the corresponding inverse Fourier transformation, and $(k_x,k_y)$ are Fourier-space coordinates corresponding to $(x,y)$. The exponential function in Eq.~(\ref{eq:SecIII--02}) may be spoken of as the ``Fourier representation of the Fresnel propagator'' or the ``free-space complex transfer function''.  Taken together, Eqs.~(\ref{eq:SecIII--01}) and (\ref{eq:SecIII--02}) are equivalent to the convolution formulation of Fresnel diffraction \cite{WinthropWorthington66}, thereby indicating the inverse Fourier transformation of $\exp[-i\Delta(k_x^2+k_y^2)/(2k)]$ to be proportional to the paraxial approximation for an expanding spherical wave (Huygens wavelet, outgoing Green function).\footnote{{See Eqs.~(\ref{eq:FirstPrinciples1})-(\ref{eq:FirstPrinciples2}) below.}}  

The preceding summary of basic Fresnel diffraction theory assumes exclusively coherent paraxial flow to be present, both over the plane $z=0$ and downstream of this plane (see Fig.~\ref{Fig:CoherentFlow}).  This situation may be generalized by considering both coherent and diffuse paraxial flow to be present simultaneously (see Fig.~\ref{Fig:CoherentAndDiffuseFlow}).

In particular, consider a thin sample to be positioned immediately upstream of $z=0$.  By assumption, this sample is illuminated by a normally-incident coherent scalar plane wave $\exp(ikz)$.  The spatially unresolved microstructure within the sample leads to a bifurcation of the incident paraxial flow, such that the transmitted flow---over the nominally planar exit surface $z=0$ of the sample---is a sum of both coherent and diffuse components. Restricting consideration to physical models where the intensities of the two components may be added, we may write the propagated intensity as\footnote{{See the first-principles derivation in Sec.~\ref{sec:DiffusionFieldFromFirstPrinciples}.}  Cf.~the very similar form in Eqs.~(27), (28), (33), and (42)-(44) of Ref.~\cite{PaganinMorgan2019}. }
\begin{align}
\label{eq:SecIII--04}
I(&x,y,z=\Delta)
\\ \nonumber &=\left\vert \mathcal{D}_{\Delta}^{\textrm{(F)}} \left\{\sqrt{I_c(x,y,z=0)}\exp[i\phi(x,y,z=0)]\right\}\right\vert^2
\\ \nonumber
&+ \iint_{-\infty}^{\infty}I_d(x',y',z=0)K(x',y',x,y,\Delta)\,dx'\,dy'.
\end{align}
In the middle line above, $I_c(x,y,z=0)$ denotes the coherent component of the intensity distribution over the exit surface of the sample, with $\phi(x,y,z=0)$ being the phase of the complex disturbance
\begin{equation}
\psi(x,y,z)=\sqrt{I_c(x,y,z)}  \exp[i\phi(x,y,z)]
\label{eq:SecIII--05}
\end{equation}
that is associated with this coherent flow. In the bottom line of Eq.~(\ref{eq:SecIII--04}), $I_d(x',y',z=0)$ is the diffuse component of the intensity distribution over the exit surface of the sample, and the kernel $K(x',y',x,y,\Delta)$ of the integral transform is a diffusion function [cf.~Eq.~(1) in Paper I].  The diffusion function quantifies the effect of position-dependent blur, upon propagating through a positive distance $\Delta$, with this blurring being due to the influence of unresolved random microstructure in the sample.\footnote{{Cf.~Eq.~(\ref{eq:FirstPrinciples11}). }If one extends from coherent to partially coherent illumination, blurring may also be due to the smearing effects of an extended chaotic source (see Fig.~3 of Paper I).}  

By definition, $K(x',y',x,y,\Delta)$ is the intensity distribution over the plane $z=\Delta$, as a function of transverse coordinates $(x,y)$, that arises from a pointlike unit-strength diffusing intensity distribution
\begin{equation}
    I_d(x,y,z=0)=\delta(x-x',y-y')
\label{eq:SecIII--06}
\end{equation}
at the transverse location $(x',y')$ over the plane $z=0$, where $\delta(x,y)$ is a two-dimensional Dirac delta.  Thus, $K(x',y',x,y,\Delta)$ may be viewed as the Green function associated with an anomalous diffusion equation \cite{MetzlerKlafter2000,EvangelistaLenziBook2018}, in the same sense that was given in the paragraph containing Eqs.~(10)-(12) in Paper I. In particular, ``anomalous'' rather than ``normal'' diffusion will usually be operative for the blurring of $I_d(x,y,z=\Delta)$ with increasing $\Delta$, because the local blur width will typically scale in proportion to the evolution parameter $\Delta$ (rather than the square root of this parameter, as would be the case for normal diffusion \cite{EinsteinBrownianMotion}). The diffusion function (i) is normalized to unity via  
\begin{equation}
\iint_{-\infty}^{\infty}K(x',y',x,y,\Delta)\,dx'\,dy'=1
\label{eq:SecIII--07}
\end{equation}
as a direct consequence of the conservation of energy [cf.~Eq.~(3) in Paper I], (ii) has first moments which are assumed to vanish, so that 
\begin{align}
\nonumber 0 &=\iint_{-\infty}^{\infty}(x-x')K(x',y',x,y,\Delta)\,dx'\,dy'
\\ &=\iint_{-\infty}^{\infty}(y-y')K(x',y',x,y,\Delta)\,dx'\,dy'
\label{eq:SecIII--08}
\end{align}
because we only consider models where net refraction is exclusively associated with the coherent component of the total flow [cf.~Eq.~(4) in Paper I], (iii) is assumed to be rotationally symmetric about its center, so that 
\begin{equation}
0 =\iint_{-\infty}^{\infty}
(x-x')(y-y')K(x',y',x,y,\Delta)\,dx'\,dy',
\label{eq:SecIII--09}
\end{equation}
and (iv) is of a form that implies zero blur at zero propagation distance $\Delta$, hence
\begin{equation}
K(x',y',x,y,\Delta=0)=\delta(x-x',y-y').
\label{eq:SecIII--10}
\end{equation}
Following similar logic to that given in Eqs.~(1)-(8) from Paper I, the diffusion-function properties in Eqs.~(\ref{eq:SecIII--07})-(\ref{eq:SecIII--10}) imply that the final line of Eq.~(\ref{eq:SecIII--04}) may be approximated as the energy-preserving flow
\begin{align}
\label{eq:SecIII--11}
\iint_{-\infty}^{\infty} &I_d(x',y',z=0)K(x',y',x,y,\Delta)\,dx'\,dy' 
\\ \nonumber &\approx I_d(x,y,z=0)
+\frac{1}{2}\nabla_{\perp}^2\left[ 
\sigma_{\Delta}^2(x,y) I_d(x,y,z=0)
\right],
\end{align}
provided $\Delta$ is sufficiently small. Here,
\begin{align}
\nonumber
    \sigma_{\Delta}^2(x,y)=\frac{1}{2}\iint_{-\infty}^{\infty}[(x-x')^2+(y-y')^2] \\ \times K(x',y',x,y,\Delta)\,dx'\,dy'
\end{align}
is the position-dependent squared standard deviation (half-width) of the rotationally symmetric diffusion function.

Inspired by the two-fluid model of superfluidity\footnote{See \citet{Tisza1947}, together with the textbook accounts on pp.~152-159 of \citet{YourgrauMandelstam} or pp.~311-313 of \citet{Huang_1987}, regarding the decomposition of quantum-liquid flow into a superfluid and an ordinary fluid.  This two-fluid model has some points of similarity with our decomposition of paraxial-wave intensity transport into coherent and diffuse flow channels.} we now write the total intensity as 
\begin{equation}
I(x,y,z)=I_c(x,y,z)+I_d(x,y,z).
\label{eq:SecIII--12}
\end{equation}
The position-dependent fraction of the total intensity, that is associated with the diffuse flow, is an order-parameter field \cite{SethnaBook} given by\footnote{One may also consider $F(x,y,z)$ to be a non-standard measure of the degree of partial coherence, which may be compared to commonly employed measures of the degree of coherence in standard works such as \citet{Zernike1938} and \citet{MandelWolf} (cf.~Sec.~\ref{sec:LinkWithPartialCoherence}).  Moreover, for the idealized case of a thin sample illuminated by strictly monochromatic light, $F(x,y)$ is equal to the position-dependent total scattering cross section (for diffuse scatter) per unit area.}
\begin{equation}
F(x,y,z)=\frac{I_d(x,y,z)}{I_c(x,y,z)+I_d(x,y,z)}=\frac{I_d(x,y,z)}{I(x,y,z)}.
\label{eq:SecIII--13}
\end{equation}
We may then write
\begin{align}
\nonumber I_c(x,y,z) &=[1-F(x,y,z)]\,I(x,y,z),
\\ \nonumber I_d(x,y,z)&=F(x,y,z)\,I(x,y,z), 
\\ 0 &\le F(x,y,z) \le 1,
\label{eq:SecIII--14}
\end{align}
so that Eq.~(\ref{eq:SecIII--04}) becomes\footnote{{ See the first-principles derivation in Sec.~\ref{sec:DiffusionFieldFromFirstPrinciples}.} Cf.~Eqs.~(27), (33), and (42) in \citet{PaganinMorgan2019}.}
\begin{widetext}
\begin{align}
\nonumber
I(x,y,z=\Delta)
 =&\left\vert  \mathcal{D}_{\Delta}^{\textrm{(F)}} \Big\{\sqrt{[1-F(x,y,z=0)] \,I(x,y,z=0)}\exp[i\phi(x,y,z=0)]\Big\}\right\vert^2
\\& + \iint_{-\infty}^{\infty}F(x',y',z=0)\,I(x',y',z=0) K(x',y',x,y,\Delta)\,dx'\,dy'.
\label{eq:SecIII--15}
\end{align}
\end{widetext}

As mentioned earlier, it will often be the case that the blurring-kernel half-width $\sigma_{\Delta}(x,y)$ is directly proportional to $\Delta$, when this blur is associated with rotationally symmetric paraxial diffuse-scatter cones such as those shown in Figs.~3, 5, 6, 7, and 10(b) of Paper I. With a loss of generality that is readily dropped if needed, we assume this direct proportionality and thereby write 
\begin{equation}
\tan\theta_s(x,y)\approx\theta_s(x,y)
=\frac{\sigma_{\Delta}(x,y)}{\Delta}.
\label{eq:SecIII--16}
\end{equation}
Here, $\theta_s(x,y) \ll 1$ is the paraxial apex half-angle of the diffuse-scatter cone that may be associated, e.g., with spatially unresolved random sample microstructure (see Fig.~7 of Paper I). The assumption that\footnote{The constant of proportionality, which is implicit in Eq.~(\ref{eq:SecIII--17}), is a function of $x$ and $y$. } 
\begin{equation}
\sigma_{\Delta}(x,y)\propto \Delta,
\label{eq:SecIII--17}
\end{equation}
when coupled with the approximation for the diffuse channel of the flow in Eq.~(\ref{eq:SecIII--11}), results in the bottom line of Eq.~(\ref{eq:SecIII--15}) being estimated by an expression including a term on the order of $\Delta^2$. Working to the same order in $\Delta$ for the first line of Eq.~(\ref{eq:SecIII--15}) may be achieved with the help of Eq.~(\ref{eq:InfinityOfTIEs--SecondOrderExpansionOfIntensity}). We thereby arrive at
\begin{widetext}
\begin{align}
\nonumber
I(&x,y,z=\Delta) 
= I(x,y,z=0) - \frac{\Delta}{k}\nabla_{\perp}\cdot\left\{[1-F(x,y)] \, I(x,y,z=0) \nabla_{\perp}\phi(x,y)\right\}
+   \Delta^2\nabla_{\perp}^2 \left[ D(x,y) I(x,y,z=0)\right]
\\ \nonumber &+\frac{\Delta^2}{4k^2}\Bigg(\left\vert\nabla_{\perp}^2\left[\sqrt{[1-F(x,y)]\,I(x,y,z=0)}e^{i\phi(x,y)}\right]\right\vert^2
\\ &\quad-\textrm{Re}\left\{\sqrt{[1-F(x,y)]\,I(x,y,z=0)}e^{-i\phi(x,y)}(\nabla_{\perp}^2)^2\left[\sqrt{[1-F(x,y)]\,I(x,y,z=0)}e^{i\phi(x,y)}\right]\right\}\Bigg),
\label{eq:SecIII--ExtendedFPE}
\end{align}
\end{widetext}
where we have introduced the abbreviations 
\begin{align}
\label{eq:F-abbrev} F(x,y) &\equiv F(x,y,z=0) \\
\label{eq:phi-abbrev}\phi(x,y) &\equiv \phi(x,y,z=0),    
\end{align}
and made use of the dimensionless diffusion coefficient $D(x,y)$ in Eq.~(\ref{eq:DimensionlessScalarDiffusionCoefficient}).

If rotational symmetry cannot be assumed for the diffusion function $K(x',y',x,y,\Delta)$ at every fixed value for $(x',y')$, but it can instead be approximated by an elliptical form (see Fig.~9 in Paper I), we can follow the associated development in Sec.~IV\,A of that paper by making the replacement\footnote{Cf.~pp.~5-6 of \citet{CrankBook} and p.~88 of \citet{PavliotisBook2014}, together with the form of ``directional'' (diffusive) dark-field imaging reported in \citet{jensen2010a,jensen2010b}. Note, moreover, that the tilde on the right side of Eq.~(\ref{eq:ReplacementToGetFromIsotropicToAnisotropicDiffusion}) accords with the notation in Secs.~III\,C and IV\,A of Paper I, together with Eq.~(\ref{eq:NiceNewEquationForTensorDiffusionField}) of the present paper.}
\begin{align}
    \nonumber \nabla_{\perp}^2[&D(x,y) I(x,y,z=0)] \\
    &\longrightarrow \nabla_{\perp}\cdot[\widetilde{\mathbf{D}}(x,y) \nabla_{\perp}I(x,y,z=0)]
    \label{eq:ReplacementToGetFromIsotropicToAnisotropicDiffusion}
\end{align}
in the first line of Eq.~(\ref{eq:SecIII--ExtendedFPE}).  Here, $\widetilde{\mathbf{D}}(x,y)$ is the dimensionless symmetric tensor diffusion field
\begin{align}
\label{eq:TensorDiffusionFieldRetrieval9}    
\widetilde{\mathbf{D}}&(x,y)= \frac{1}{2}F(x,y)
\\ \nonumber &\times\mathcal{R}_{\psi(x,y)} 
\begin{pmatrix}
[\theta_1(x,y)]^2 & 0\\
0 & [\theta_2(x,y)]^2
\end{pmatrix}
[\mathcal{R}_{\psi(x,y)}]^T
\end{align}
that is obtained by combining Eqs.~(103) and (106) of Paper I, the matrix field 
\begin{equation}
\label{eq:TensorDiffusionFieldRetrieval10}    
\mathcal{R}_{\psi(x,y)}=
\begin{pmatrix}
\cos\psi(x,y) & -\sin\psi(x,y)\\
\sin\psi(x,y) & \cos\psi(x,y)
\end{pmatrix}
\end{equation}
corresponds to the position-dependent principal-axis rotation angle for the approximately-elliptical smearing function (see Fig.~9 of Paper I), the transverse gradient $\nabla_{\perp}$ is taken to be a 
two-component column-vector operator $(\partial/\partial x,\partial/\partial y)^T$, a superscript $T$ denotes matrix transposition, the transverse divergence $\nabla_{\perp}\cdot$ is taken to be a row-vector operator $(\partial/\partial x,\partial/\partial y)$, and  ${\theta_1(x,y),\theta_2(x,y)}$ denote the position-dependent angular half-widths associated with the principal axes of the smearing-function ellipse (again see Fig.~9 in Paper I).   Comparison of Eqs.~(\ref{eq:NiceNewEquationForTensorDiffusionField}) and (\ref{eq:TensorDiffusionFieldRetrieval9}) shows the lower line of the latter equation to coincide with the diagonalized factorization of the angular-deflection covariance-matrix field in Eq.~(\ref{eq:CovarianceMatrix}).

We speak of Eq.~(\ref{eq:SecIII--ExtendedFPE}), together with the variant given by the replacement in Eq.~(\ref{eq:ReplacementToGetFromIsotropicToAnisotropicDiffusion}), as the {\em extended Fokker-Planck generalization of the transport-of-intensity equation for paraxial wave optics}. This terminology is chosen because, if $D(x,y)$ is sufficiently large\footnote{The dimensionless diffusion field $D(x,y)$ is always much smaller than unity, for the paraxial scenarios to which we restrict consideration.  This smallness compared to unity arises from the fact that paraxiality implies the inequality $\theta_s(x,y)\ll 1$, which may be combined with the constraint that $F(x,y)\le 1$ to observe from Eq.~(\ref{eq:DimensionlessScalarDiffusionCoefficient}) that $0\le D(x,y) \ll 1$. Thus, when writing the words ``if $D(x,y)$ is sufficiently large'' in the main text, what is meant is ``even though $D(x,y)$ is intrinsically small in the absolute sense that it is a dimensionless non-negative quantity that is much less than unity for all paraxial diffusion fields, this quantity may nevertheless be sufficiently large in the relative sense that the only term, of order $\Delta^2$, that needs to be retained in Eq.~(\ref{eq:SecIII--ExtendedFPE}) is $\Delta^2\nabla_{\perp}^2[D(x,y)\,I(x,y,z=0)]$''. From a physical perspective, this approximation corresponds to the blurring influence of the diffusion field being large in comparison to the influence of diffraction (to second order in $\Delta$).  Such an approximation is particularly likely to break down at sharp sample edges \cite{Alloo2025}.\label{footnote:Why-D-is-small}} that the second and third lines of Eq.~(\ref{eq:SecIII--ExtendedFPE}) may be neglected, then (i) Eq.~(\ref{eq:SecIII--ExtendedFPE}) reduces to Eq.~(\ref{eq:SecIII--FPE-scalar-case}) and (ii) with the replacement in Eq.~(\ref{eq:ReplacementToGetFromIsotropicToAnisotropicDiffusion}), Eq.~(\ref{eq:SecIII--ExtendedFPE}) reduces to the finite-difference form 
\begin{align}
\nonumber I(&x,y,z=\Delta)=I(x,y,z=0)
\\  \nonumber &-\frac{\Delta}{k} \nabla_{\perp}\cdot \{[1-F(x,y)] I(x,y,z=0) \nabla_{\perp}\phi(x,y)\}
\\  &+\Delta^2\nabla_{\perp}\cdot[\widetilde{\mathbf{D}}(x,y) \nabla_{\perp}I(x,y,z=0)], 
\label{eq:SecIII--FPE-tensor-case}
\end{align}
of the anisotropic-diffusion Fokker-Planck equation in Eq.~(147) of Paper I.  These special cases [Eqs.~(\ref{eq:SecIII--FPE-scalar-case}) and (\ref{eq:SecIII--FPE-tensor-case})] may both be spoken as {\em Fokker-Planck generalizations of the TIE for paraxial wave optics}.\footnote{Very similar expressions are given in Refs.~\cite{PaganinMorgan2019} and \cite{MorganPaganin2019}. For the case where $F(x,y) \ll 1$, Eqs.~(\ref{eq:SecIII--FPE-scalar-case}) and (\ref{eq:SecIII--FPE-tensor-case}) have been utilized in a number of recent papers on x-ray optics \cite{MIST,Gureyev2020, MISTdirectional,PaganinPelliccia2021,alloo2022dark, alloo2023SciRep, Leatham2023,Beltran2023,Ahlers2024,Leatham2024,Alloo2025,Ahlers2025,LiuAllooLangerPavlovPhysRevA2025} and mentioned in one paper on neutron optics \cite{PaganinNeutron2023}.} Thus Eq.~(\ref{eq:SecIII--ExtendedFPE})---together with the variant obtained via the replacement in Eq.~(\ref{eq:ReplacementToGetFromIsotropicToAnisotropicDiffusion})---is an ``extended'' form of Eqs.~(\ref{eq:SecIII--FPE-scalar-case}) and (\ref{eq:SecIII--FPE-tensor-case}), in the sense that the second and third lines of Eq.~(\ref{eq:SecIII--ExtendedFPE}) are omitted in Eqs.~(\ref{eq:SecIII--FPE-scalar-case}) and (\ref{eq:SecIII--FPE-tensor-case}).  

\section{Route from nonequilibrium statistical mechanics to paraxial beam optics, via the Fokker-Planck equation}\label{sec:FPE--GenericRemarks}

Here we establish broader context regarding the Fokker-Planck extension to the TIE. In Sec.~\ref{subsec:FPE--GenericForm} we briefly revise some salient elements regarding the Fokker-Planck equation (forward Kolmogorov equation) \cite{Risken1989} in its more usual context of nonequilibrium statistical physics. Section~\ref{subsec:FPE--RelatingToParaxialOptics} clarifies the conceptual and mathematical connection, between this master equation\footnote{``Master equations'' are ``equations of motion for various probability distributions'' (see p.~878 of \citet{MandelWolf}).} and the present context of paraxial-optics diffusion fields. 

\subsection{Fokker-Planck equation in generic form}\label{subsec:FPE--GenericForm}

The Fokker-Planck equation is a particular form of continuity equation, which governs the evolution of a probability density function as it diffuses and drifts with time.  It is often used in nonequilibrium statistical mechanics since it can allow one to study the manner in which statistical equilibrium is approached in thermodynamic systems (Ref.~\cite{KittelStatPhysBook}, p.~157). 
It is also employed in a plethora of additional settings \cite{Risken1989,Davis1994,ScullyZubairy,PavliotisBook2014,PaganinMorgan2019} that will not be reviewed here. 

Following pp.~169-170 of Ref.~\cite{BalakrishnanNonEqmStatMechBook}, this equation is
\begin{align}
    \nonumber \frac{\partial}{\partial t} p(\boldsymbol{\xi},t)
    =- &\sum_{i=1}^N \frac{\partial}{\partial \xi_i}\left[p(\boldsymbol{\xi},t)\,f_i(\boldsymbol{\xi},t)\right]
    \\ &+\sum_{i=1}^N\sum_{j=1}^N\frac{\partial^2}{\partial\xi_i\partial\xi_j}\left[\mathcal{D}_{ij}(\boldsymbol{\xi},t)\,p(\boldsymbol{\xi},t)\right].
    \label{eq:FPE--genericForm}
\end{align}
Here, $p(\boldsymbol{\xi},t)$ is a probability density function associated with the $N$-component vector 
\begin{equation}
\boldsymbol{\xi}(t)=(\xi_1(t),\xi_2(t),\cdots,\xi_N(t)) 
\end{equation}
that stochastically evolves with time $t$, the set $\{\mathcal{D}_{ij}(\boldsymbol{\xi},t)\}$ gives the elements of a symmetric $N\times N$ diffusion-matrix field $\boldsymbol{\mathcal{D}}(\boldsymbol{\xi},t)$ which governs the manner in which the probability density diffuses with time, and 
\begin{equation}
\boldsymbol{f}(\boldsymbol{\xi},t)=(f_1(\boldsymbol{\xi},t),f_2(\boldsymbol{\xi},t),\cdots,f_N(\boldsymbol{\xi},t))
\end{equation}
is an $N$-component drift-vector field which may be thought of as a local velocity vector in $\boldsymbol{\xi}$ space.  

Loosely, the diffusion matrix is associated with the manner in which the probability density smears out in $\boldsymbol{\xi}$ space with time, with the drift vector being associated with the shifting of peak values for $p(\boldsymbol{\xi},t)$ in $\boldsymbol{\xi}$ space as time evolves. Any one particular instance of $\boldsymbol{\xi}(t)$ may be viewed as a time-dependent point in $\boldsymbol{\xi}$ space whose evolution is generated by a stochastic differential equation; the drift vector is associated with deterministic generalized forces \cite{GoldsteinBook} that act on $\boldsymbol{\xi}(t)$, with the diffusion matrix being related to random generalized forces acting on $\boldsymbol{\xi}(t)$.  Under this view, there is a correspondence between the stochastic differential equation for one particular realization of $\boldsymbol{\xi}(t)$, and the associated Fokker-Planck equation for the probability density associated with an ensemble of realizations (Ref.~\cite{BalakrishnanNonEqmStatMechBook}, pp.~73, 120, and 169-170).  

 \subsection{Fokker-Planck equation in paraxial-optics form}\label{subsec:FPE--RelatingToParaxialOptics}

\begin{figure}[ht!]
\centering
\includegraphics[width=0.95\columnwidth]{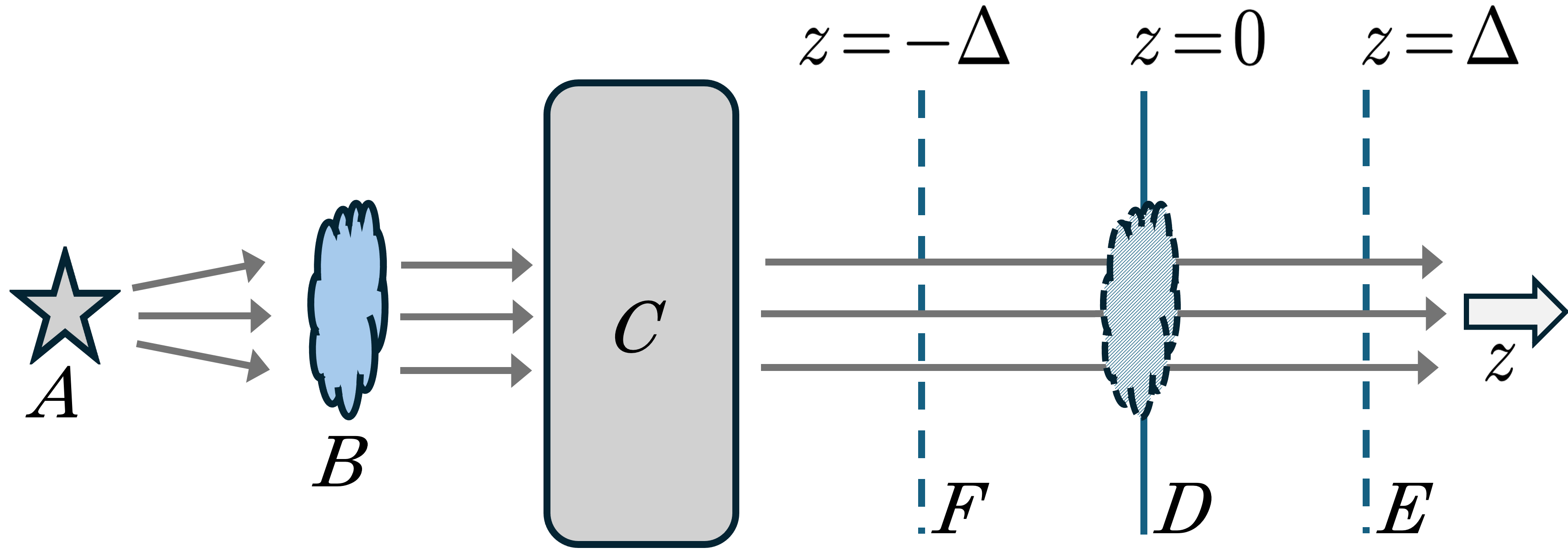}
\caption{Generic imaging setup for paraxial diffusion-field retrieval using Fokker-Planck and extended Fokker-Planck generalizations of the TIE.}
\label{Fig:Schematic}
\end{figure}

 To connect the generic Fokker-Planck equation in Eq.~(\ref{eq:FPE--genericForm}) with our context of paraxial optics, consider Fig.~\ref{Fig:Schematic}. Here, an extended chaotic source $A$ of quasimonochromatic\footnote{The source may also be broadband, in which case energy filtration (quasimonochromatization) may be performed by the detector, or by a monochromator or energy filter upstream of the detector.} radiation or matter waves, which are paraxial with respect to an optical axis $z$,  illuminates a sample $B$.  The transmitted waves pass through an imaging system $C$ that forms a real image $I(x,y,z=0)$ of $B$ over the plane $D$.  Over-focused and under-focused images may also be registered over planes $E$ and $F$, corresponding to $I(x,y,z=\Delta)$ and $I(x,y,z=-\Delta)$, respectively.\footnote{Alternatively, over-focused and under-focused images $I(x,y,z=\pm\Delta)$ may be obtained by keeping the detection plane fixed and suitably altering the state of the imaging system $C$.  Our treatment applies to either scenario, but for simplicity we refer to the free-space propagation geometry of Fig.~\ref{Fig:Schematic}, in the main text.}  The sample is considered to be sufficiently weakly scattering that the exiting beam remains paraxial. We also assume the sample to be elastically scattering, so that the energy of the incident and transmitted  fields is unchanged. The volume between the exit surface of $C$ and the detection planes $\{D,E,F\}$ is taken as vacuum.  

Conservation of energy implies that the propagating paraxial intensity distribution $I(x,y,z)$ obeys
\begin{equation}
\iint I(x,y,z)\,dx\,dy\equiv\kappa=\textrm{constant},
\label{eq:energy-conservation-over-plane}
\end{equation}
over any plane of fixed $z$ in the free space between the exit surface of $C$ and the detection planes $D$, $E$, $F$.\footnote{Cf.~the paragraph containing Eqs.~(26) and (27), in Paper I.} Over any two-dimensional plane of fixed $z$, the intensity may be related to a $z$-dependent probability density\footnote{Cf.~p.~805 of \citet{BarrettMyersBook}.}
\begin{equation}
p(x,y;z)=I(x,y,z)/\kappa.     
\label{eq:mapping-I-to-p}
\end{equation}
If, in addition, we consider $z$ rather than $t$ to be the evolution parameter, the replacements
\begin{equation}
t\longrightarrow z, \quad \boldsymbol{\xi}\longrightarrow (x,y), \quad (f_1,f_2)\longrightarrow k^{-1}\nabla_{\perp}\phi    
\label{eq:ThreeReplacemenets}\end{equation}
change the $N=2$ special case of Eq.~(\ref{eq:FPE--genericForm}) to
\begin{align}
    \nonumber \frac{\partial}{\partial z} &I(x,y,z)
    = - \frac{1}{k} \nabla_{\perp}\cdot\left[I(x,y,z)\,\nabla_{\perp}\phi (x,y,z)\right]
    \\ \nonumber &+\frac{\partial^2}{\partial x^2}\left[\mathcal{D}_{xx}(x,y,z)\,I(x,y,z)\right]
    \\ \nonumber &+\frac{\partial^2}{\partial y^2}\left[\mathcal{D}_{yy}(x,y,z)\,I(x,y,z)\right]
    \\ &+2\frac{\partial^2}{\partial x\partial y}\left[\mathcal{D}_{xy}(x,y,z)\,I(x,y,z)\right].
\label{eq:IntermediateResult01}
\end{align}

While we can already recognize the TIE [Eq.~(\ref{eq:TIE})] in the first line of Eq.~(\ref{eq:IntermediateResult01}), some care is needed regarding the subsequent lines, on account of the distinction between normal and anomalous diffusion \cite{MetzlerKlafter2000,EvangelistaLenziBook2018} that was discussed in Sec.~II of Paper I.  Accordingly, set $z=0$ in Eq.~(\ref{eq:IntermediateResult01}) and then use the asymmetric-defocus first-order finite-difference approximation in Eq.~(\ref{eq:IntermediateResult02}). Hence 
\begin{align}
    \nonumber     I(x, &y,z =\Delta)
    =  I(x,y,z=0) 
    \\ \nonumber &- \frac{\Delta}{k} \nabla_{\perp}\cdot\left[I(x,y,z=0)\,\nabla_{\perp}\phi (x,y)\right]
    \\ \nonumber &+[\Delta^{a+(1/2)}]^2\bigg\{\frac{\partial^2}{\partial x^2}\left[\mathcal{D}_{xx}(x,y)\,I(x,y,z=0)\right]
    \\ \nonumber &+\frac{\partial^2}{\partial y^2}\left[\mathcal{D}_{yy}(x,y)\,I(x,y,z=0)\right]
    \\ &+2\frac{\partial^2}{\partial x\partial y}\left[\mathcal{D}_{xy}(x,y)\,I(x,y,z=0)\right]\bigg\},
\label{eq:IntermediateResult03}
\end{align}
wherein explicit ``$z=0$'' dependence is dropped for clarity in both the phase $\phi$ and the diffusion tensor $\boldsymbol{\mathcal{D}}$, and we have introduced the anomalous diffusion coefficient $a$ ``by hand''.\footnote{Cf.~pp.~101-102 of \citet{EvangelistaLenziBook2018}. There the normal diffusion equation is converted to an anomalous diffusion equation ``by hand'', in changing one of the integer-order derivatives to a fractional-order derivative \cite{OldhamSpanier}. For similar reasoning in a Fokker-Planck context, see \citet{YANOVSKY200013}.}  When $a=0$, Eq.~(\ref{eq:IntermediateResult03}) is faithful to Eq.~(\ref{eq:IntermediateResult01}), insofar as it is a normal-diffusion Fokker-Planck equation in finite-difference form.  When $a=1/2$ [see Eq.~(12) of Paper I] we have the case of ballistic superdiffusion \cite{MetzlerKlafter2000,KellyBallisticAnomalousDiffusion} that is relevant to us.  In this case Eq.~(\ref{eq:IntermediateResult03}) is an anomalous-diffusion Fokker-Planck equation \cite{ZASLAVSKY-FractionalFPE1994,MetzlerPRL1999,YANOVSKY200013} in finite-difference form.  It may also be thought of as a crudely space-fractional form of Fokker-Planck equation \cite{SpaceFractionalFPE} that displays the required ballistic-superdiffusion behavior (here ``fractional'' refers to fractional-order differentiation \cite{OldhamSpanier,EvangelistaLenziBook2018}).  The ``$a=1/2$'' case of Eq.~(\ref{eq:IntermediateResult03}) is essentially identical to the $F(x,y) \ll 1$ special case of Eq.~(\ref{eq:SecIII--FPE-tensor-case}), albeit with a different representation for the tensorial diffusion field (cf.~the distinction between two different representations for the scalar diffusion field in Secs. III\,C and IV\,A of Paper I). Similarly, if we set $a=1/2$ and
\begin{equation}
\mathcal{D}_{xx}(x,y)=\mathcal{D}_{yy}(x,y)\equiv D(x,y), \quad \mathcal{D}_{xy}(x,y)=0,     
\label{eq:IntermediateResult04}
\end{equation}
Eq.~(\ref{eq:IntermediateResult03}) becomes the $F(x,y) \ll 1$ special case of Eq.~(\ref{eq:SecIII--FPE-scalar-case}); cf.~Eq.~(\ref{eq:TransitionFromTensorToScalarDiffusion}). Finally, recall that we chose to speak of Eq.~(\ref{eq:SecIII--ExtendedFPE}) as an ``extended'' Fokker-Planck equation, which is consistent with the fact that its second and third lines are not present in Eq.~(\ref{eq:IntermediateResult03}).    

\bigskip

We close this section by returning attention to Eq.~(\ref{eq:ThreeReplacemenets}) and briefly expanding upon the physical meaning of the three replacements that are given there.  (i) Given that the image-forming field (e.g., photons, electrons, neutrons etc.) propagates at a fixed speed for monoenergetic or quasimonoenergetic systems, paraxiality implies that $t \propto z$ for a (mathematical) point that is comoving with respect to the energy flow. In this sense the first replacement in Eq.~(\ref{eq:ThreeReplacemenets}) is merely a change of variables in which time is still the relevant evolution parameter, from a physical perspective. Moreover, for paraxial electromagnetic beams, $t \propto z$ even in the case of polyenergetic radiation. (ii) Regarding the second replacement in Eq.~(\ref{eq:ThreeReplacemenets}), $(x,y)$ is the transverse apex location of the cone of small-angle diffuse scatter arising from an ensemble of realizations of the sample microstructure (Paper I, Figs.~3, 5-7, and 10(b)). For ray models, this diffuse-scatter cone may be associated with an ensemble of continuous random walks \cite{ChernovBook1960} through the sample microstructure (Paper I, Fig.~6). For a wave model, under the first Born approximation \cite{Messiah}, see Eq.~(111) of Paper I{; cf.~Sec.~\ref{sec:DiffusionFieldFromFirstPrinciples} below}. (iii) The third replacement in Eq.~(\ref{eq:ThreeReplacemenets}) makes a physical connection between the generic concept of a Fokker-Planck drift vector, and the specific notion of deflection angles associated with the coherent flow channel, namely the paraxial angles 
\begin{equation}
(\alpha_x(x,y),\alpha_y(x,y))=k^{-1}\nabla_{\perp}\phi(x,y)    
\label{eq:ParaxialAngles}
\end{equation}
that the coherent-flow current density makes with respect to the optical axis.  While we used wave-optics language in speaking of $\phi$ as the ``phase'' of the propagating beam, if we work in geometrical-optics terms we can instead use the Hamiltonian-mechanics concept of ``action''. Under this view, the wave-optical notion of a wavefront (surface of constant phase) reduces to the corresponding geometrical-optics notion (surface of constant action) in the short-wavelength limit (Ref.~\cite{GoldsteinBook}, pp.~484-492). 

 \section{Phase and diffusion-field retrieval via the Fokker-Planck or extended Fokker-Planck equations}\label{sec:PhaseAndDiffusionFieldRetrieval}

This section has two parts. Section~\ref{sec:PhaseRetrieval} considers means for phase retrieval that are based on our Fokker-Planck TIE generalizations [e.g., Eqs.~(\ref{eq:SecIII--FPE-scalar-case}), (\ref{eq:SecIII--ExtendedFPE}), and (\ref{eq:SecIII--FPE-tensor-case})]. Here, ``phase retrieval'' refers to means by which noninterferometric intensity measurements such as $I(x,y,z=0)$ and $I(x,y,z=\pm\Delta)$ may be used to infer wavefield phase $\phi(x,y)$. Diffusion-field retrieval, namely the question of inferring $D(x,y)$ or $\widetilde{\mathbf{D}}(x,y)$ from measured intensity distributions, is treated in Sec.~\ref{sec:DiffusionFieldRetrieval}.  

\subsection{Phase retrieval}\label{sec:PhaseRetrieval}

Many papers utilize the TIE for phase retrieval.  See \citet{TIE==LongReviewArticle2020} for an extensive bibliography. Evidently, our Fokker-Planck extensions to this equation imply the need to revisit TIE-based phase retrieval. Do such analyses need to be modified when a diffuse channel to the paraxial flow is present? As we shall argue, when symmetric over-focus and under-focus intensity measurements $I(x,y,z=\pm \Delta)$ are employed, existing TIE-based approaches to phase retrieval may be used without any modification, even though a diffuse channel to the flow is present and a Fokker-Planck extension to the TIE is employed (Sec.~\ref{sec:PhaseRetrievalSymmetricDefocus}).  However when asymmetric defoci are employed, such as a focal series where all of the defocus distances $\Delta$ have the same sign, modifications to the phase-retrieval process are required when passing from the TIE to its Fokker-Planck extensions (Sec.~\ref{sec:PhaseRetrievalAsymmetricDefocus}). Structured-illumination approaches to phase retrieval are considered in Sec.~\ref{sec:PhaseRetrievalStructuredIlllumination}.

\subsubsection{Phase retrieval: Symmetric-defocus case}\label{sec:PhaseRetrievalSymmetricDefocus}

The asymmetric-defocus estimate for the longitudinal intensity derivative, as given in Eq.~(\ref{eq:IntermediateResult02}), may be improved by the symmetric-defocus estimate  
\begin{align}
    \nonumber \left[\frac{\partial}{\partial z} I(x, y, z)\right]_{z=0}\!\! &\approx \frac{I(x,y,z=\Delta)-I(x,y,z=-\Delta)}{2\Delta} \\ &\equiv -k^{-1}Q(x,y).
\label{eq:PhaseRetrieval01}
\end{align}
Approximate each term in the numerator on the right side, using the extended Fokker-Planck equation [Eq.~(\ref{eq:SecIII--ExtendedFPE})].  The second and third lines of Eq.~(\ref{eq:SecIII--ExtendedFPE}) make no contribution, with the diffuse-flow final term on the first line also dropping out since all of these terms are proportional to $\Delta^2=(-\Delta)^2$.  Hence the known data function $Q(x,y)$ acts as a source term for the elliptic second-order partial differential equation
\begin{align}
Q(x,y) 
=  \nabla_{\perp}\cdot\left\{[1-F(x,y)] \, I(x,y,z=0) \nabla_{\perp}\phi(x,y)\right\}.
\label{eq:PhaseRetrieval02}
\end{align}
Restrict consideration to the case of a weak diffusion field, so that $F(x,y)$ is everywhere much less than unity [see Eq.~(22) of Paper I].  Hence the term in square brackets in Eq.~(\ref{eq:PhaseRetrieval02}) may be approximated by unity, leaving
\begin{align}
Q(x,y) 
=  \nabla_{\perp}\cdot\left[I(x,y,z=0) \nabla_{\perp}\phi(x,y)\right],~F(x,y)\ll 1.
\label{eq:PhaseRetrieval03}
\end{align}

Given the known data function $Q(x,y)$ that is constructed using overfocus and underfocus intensities according to Eq.~(\ref{eq:PhaseRetrieval01}), the form in Eq.~(\ref{eq:PhaseRetrieval03}) is identical to that encountered when working with the ``bare'' TIE (i.e., without any Fokker-Planck extension).  Hence if we consider both $Q(x,y)$ and $I(x,y,z=0)$ to be known by direct intensity measurements in the infocus plane $z=0$ and symmetric-defocus planes $z=\pm\Delta$, {\em existing phase-retrieval methods for solving the TIE may be employed without modification} in order to solve Eq.~(\ref{eq:PhaseRetrieval03}) for the phase function $\phi(x,y)$.\footnote{For all methods of the present paper that employ two or more images obtained at different defoci, note that it is possible for such a ``focal stack'' to be acquired in a single shot.  For a recent example of such ``multifocus'' imaging systems, see \citet{MultifocusMicroscopeOptica2025}.}  

One such method, due to \citet{paganin1998}, will be employed here. This begins by following \citet{Teague1983} in introducing the auxiliary function $\Upsilon(x,y)$ via
\begin{align}
\nabla_{\perp}\Upsilon(x,y) = I(x,y,z=0)\nabla_{\perp}\phi(x,y).
\label{eq:PhaseRetrieval04}
\end{align}
We emphasize that an approximation is made in this implicit definition for $\Upsilon(x,y)$, since the Helmholtz decomposition theorem\footnote{See, e.g., pp.~52-54 of \citet{MorseFeshbach}.  For the 2D variant of Helmholtz decomposition, see, e.g., \citet{Schmalz2011}.} implies that the vector field on the right side of Eq.~(\ref{eq:PhaseRetrieval04}) should in general be written as the sum of the gradient of a scalar potential $\Upsilon$ plus the curl of a divergence-free vector potential $\boldsymbol{\eta}$.  However, omission of the vector potential is often an excellent approximation in the context of TIE-based phase retrieval for highly structured beams \cite{Schmalz2011}; the physical reason for this is that the local linear momentum density typically has a much larger magnitude than the local angular-momentum density, for highly structured paraxial beams with continuous (and therefore vortex-free) wavefronts.  If vortices are present (e.g., for fully developed speckle fields \cite{Aksenov1998})  then the vector potential cannot be neglected \cite{paganin1998}, with a similar statement applying to gently structured beams \cite{Schmalz2011} (e.g., in the context of wavefront metrology \cite{BerujonWavefrontMetrology}).

Returning to the main thread of the argument, Eq.~(\ref{eq:PhaseRetrieval04}) transforms Eq.~(\ref{eq:PhaseRetrieval03}) to the Poisson equation \cite{StraussPDEbook}
\begin{equation}
Q(x,y) 
=  \nabla_{\perp}^2\Upsilon(x,y).
\label{eq:PhaseRetrieval05}
\end{equation}
Formally solve this by applying the inverse transverse Laplacian operator $\nabla_{\perp}^{-2}$ to both sides.\footnote{Cf.~Eq.~(35) in Paper I.}  Then apply $\nabla_{\perp}$ to both sides of the resulting expression, hence
\begin{equation}
\nabla_{\perp}\nabla_{\perp}^{-2} Q(x,y) 
=  \nabla_{\perp}\Upsilon(x,y).
\label{eq:PhaseRetrieval06}
\end{equation}
Next, use Eq.~(\ref{eq:PhaseRetrieval04}) for the right side, and divide through by the infocus intensity $I(x,y,z=0)$ under the often-mild assumption that this does not vanish at any point within the field of view.  This gives
\begin{equation}
\nabla_{\perp}\phi(x,y)=\frac{\nabla_{\perp}\nabla_{\perp}^{-2} Q(x,y)}{I(x,y,z=0)}.
\label{eq:PhaseRetrieval07}
\end{equation}
We can stop at this point if the intention is to recover the transverse phase gradients $\nabla_{\perp}\phi(x,y)$, or, in what amounts to essentially the same thing, we wish to recover the position-dependent transverse deflection angles in Eq.~(\ref{eq:ParaxialAngles}). If, however, we wish to recover the phase map $\phi(x,y)$ then we can apply the transverse divergence operator $\nabla_{\perp}\cdot$ to both sides of Eq.~(\ref{eq:PhaseRetrieval07}).  The resulting estimate 
\begin{equation}
\nabla_{\perp}^2\phi(x,y)=\nabla_{\perp}\cdot\left[\frac{\nabla_{\perp}\nabla_{\perp}^{-2} Q(x,y)}{I(x,y,z=0)}\right]
\label{eq:PhaseRetrieval09}
\end{equation}
for the wavefront curvature\footnote{Cf.~the concept of ``curvature sensing'' in the context of astronomical adaptive optics \cite{Roddier1988}. Also, there are certain contexts in which the phase curvature $\nabla_{\perp}^2\phi(x,y)$ is of specific interest, e.g., in transmission electron microscopy, where the phase curvature is proportional to the projected charge density of a sufficiently thin sample (in the high-energy limit) \cite{CowleyMoodie1960}.} then gives 
\begin{widetext}
\begin{equation}
\phi(x,y)=\nabla_{\perp}^{-2}\left\{\nabla_{\perp}\!\cdot\!\left[\frac{\nabla_{\perp}\nabla_{\perp}^{-2} Q(x,y)}{I(x,y,z=0)}\right]\right\}
=-k \nabla_{\perp}^{-2}\left\{\nabla_{\perp}\!\cdot\!\left[\frac{1}{I(x,y,z=0)}\nabla_{\perp}\nabla_{\perp}^{-2} \frac{I(x,y,z=\Delta)-I(x,y,z=-\Delta)}{2\Delta}\right]\right\},
\label{eq:PhaseRetrieval10}
\end{equation}
\end{widetext}
which is identical to the Paganin-Nugent TIE-based phase-retrieval algorithm in Eq.~(21) of Ref.~\cite{paganin1998} {\em provided that the longitudinal intensity derivative $[\partial I(x,y,z)/\partial z]_{z=0}$ is estimated using the symmetric-defocus estimate in Eq.~(\ref{eq:PhaseRetrieval01}}) above.  This statement does not just apply to the particular method for phase retrieval that is used above.  More generally, {\em while the presence of a diffusion field extends the TIE into a Fokker-Planck (or extended Fokker-Planck) form, this extension leaves TIE-based algorithms for phase retrieval unchanged if symmetric defoci are used to estimate the longitudinal intensity derivative}.  

Regarding the inverse Laplacian that appears in Eq.~(\ref{eq:PhaseRetrieval10}), Ref.~\cite{paganin1998} employed the Fourier representation
\begin{equation}
\nabla_{\perp}^{-2}=-\mathscr{F}^{-1}\frac{1}{k_x^2+k_y^2+\epsilon^2}\mathscr{F}, \quad \epsilon > 0,
\label{eq:PhaseRetrieval11}
\end{equation}
where $\mathscr{F}$ denotes Fourier transformation with respect to $x$ and $y$, $\mathscr{F}^{-1}$ denotes the corresponding inverse Fourier transformation, $(k_x,k_y)$ are Fourier-space coordinates dual to $(x,y)$, and $\epsilon$ is a positive regularization parameter\footnote{The regularization parameter avoids the division-by-zero blowup that would otherwise occur in the denominator of Eq.~(\ref{eq:PhaseRetrieval11}), at the origin $(k_x,k_y)=(0,0)$ of Fourier space.} that has physical units of inverse length.\footnote{Recall that we avoid redundant brackets by considering operator products to ``act from right to left''.  Thus, for example, if $v(k_x,k_y)$ and $w(x,y)$ are arbitrary smooth functions, then the action of the operator $\mathscr{F}^{-1}v(k_x,k_y)\mathscr{F}$ on $w(x,y)$ is equal to $\mathscr{F}^{-1}(v(k_x,k_y)\{\mathscr{F}[w(x,y)]\})$.  Cf.~Eq.~(\ref{eq:SecIII--02}).}  Moreover, any Fourier-transform convention may be employed which is such that the Fourier derivative theorem \cite{Bracewellbook} maps the operators $\partial/\partial x$ and $\partial/\partial y$ to multiplication by $ik_x$ and multiplication by $ik_y$, respectively.\footnote{Equation~(\ref{eq:SecIII--03}) is an example of such a convention.}  The preceding sentence also implies that the gradient operators, appearing in Eq.~(\ref{eq:PhaseRetrieval10}), may be computed via
\begin{equation}
\nabla_{\perp}=i\mathscr{F}^{-1}(k_x,k_y)\mathscr{F}.
\label{eq:PhaseRetrieval12}
\end{equation}

From a data-analysis perspective for pixelated intensity maps [see, e.g., Eq.~(38) of Paper I], use of the Fast Fourier Transform (FFT) \cite{Press} to estimate $\mathscr{F}$ and $\mathscr{F}^{-1}$ implies an assumption of periodic boundary conditions.  For highly structured aperiodic wavefields whose number of independent degrees of freedom in $\phi(x,y)$ is on the order of several hundred or more, this assumption---which amounts, strictly speaking, to an improper treatment of boundary conditions associated with the Poisson equations whose solutions are informally denoted by the inverse Laplacian operators in Eq.~(\ref{eq:PhaseRetrieval10})---typically leads to errors that are acceptably small.  This conclusion arises from the fact that solutions to the corresponding homogeneous partial differential equation, namely the Laplace equation, will typically make a small contribution on account of their often unphysical properties that (i) every point is a saddle point, and (ii) such solutions will attain both their maximum and minimum values over the boundary of the field of view \cite{StraussPDEbook}.  See \citet{Schmalz2011} for further information on the typically mild consequences of implicitly assuming periodic boundary conditions when this is not in fact the case, in the context of TIE-based phase retrieval using an FFT implementation of the method of Ref.~\cite{paganin1998}; the conclusions of \citet{Schmalz2011} may also be applied to Eq.~(\ref{eq:PhaseRetrieval10}) above.  If boundary conditions do need to be properly taken into account, which will certainly be the case for gently structured wavefields such as those encountered in wavefront metrology, then methods are available to do so (see, e.g., Refs.~\cite{GureyevZernikePaper1995,Gureyev1996}).  Also,  if zero boundary conditions apply---e.g., in the case of plane-wave illumination of a sample that is well contained within the field of view of the intensity measurements---then the operator $\mathscr{F}$ in Eq.~(\ref{eq:PhaseRetrieval10}) may be replaced with the Fast Sine Transform, using an analogous approach to that in Sec.~III\,B of Paper I.

\subsubsection{Phase retrieval: Asymmetric-defocus case}\label{sec:PhaseRetrievalAsymmetricDefocus}

If no post-specimen imaging system $C$ is employed then negatively defocused images such as $I(x,y,z=-\Delta)$ are inaccessible. Upon deleting $C$ from Fig.~\ref{Fig:Schematic}, we return to the scenario in Fig.~3 of Paper I, where the transmitted field propagates in vacuum from the exit surface $z=0$ of the sample to the entrance surface $z=\Delta\ge 0$ of the detector. Here we solve the extended Fokker-Planck equation [Eq.~(\ref{eq:SecIII--ExtendedFPE})] for the case where $F(x,y) \ll 1$, in order to determine the phase $\phi(x,y)\equiv\phi(x,y,z=0)$ when intensity measurements are only available for $z\ge 0$.

Assume the three intensity maps 
\begin{align}
\nonumber I_0(x,y) &\equiv I(x,y,z=0)  
\\ \nonumber I_1(x,y) &\equiv I(x,y,z = \Delta_1)  
\\ I_2(x,y) &\equiv I(x,y,z = \Delta_2) 
\label{eq:PhaseRetrieval13}
\end{align}
to have been measured, where $\Delta_1$ and $\Delta_2\ne\Delta_1$ are both nonzero. Write down the corresponding ``$z=\Delta_1$'' and ``$z=\Delta_2$'' instances of Eq.~(\ref{eq:SecIII--ExtendedFPE}), which then allows the resulting system of linear equations to be solved algebraically for $\nabla_{\perp}\cdot\left\{[1-F(x,y)] \, I_0(x,y) \nabla_{\perp}\phi(x,y)\right\}$ because this term is multiplied by $\Delta$ whereas all other terms are proportional to different powers of $\Delta$ (i.e., $\Delta^2$ or $\Delta^0$).  We thereby obtain a modified form of Eq.~(\ref{eq:PhaseRetrieval02}), namely 
\begin{align}
\breve{Q}(x,y) 
=  \nabla_{\perp}\cdot\left\{[1-F(x,y)] \, I_0(x,y) \nabla_{\perp}\phi(x,y)\right\},
\label{eq:PhaseRetrieval14}
\end{align}
where we have introduced the known data function \cite{LeathamHonsThesis2019}\footnote{Cf.~Eq.~(61) in Ref.~\cite{PaganinMorgan2019} and Eq.~(10) in Ref.~\cite{Leatham2023}.} 
\begin{align}
\breve{Q}(x,y) 
=  
\frac{\Delta_2^2I_1(x,y)-\Delta_1^2I_2(x,y)+(\Delta_1^2-\Delta_2^2)I_0(x,y)}{\Delta_1\Delta_2(\Delta_1-\Delta_2)/k}.
\label{eq:PhaseRetrieval15}
\end{align}
Analogous to our passage from Eq.~(\ref{eq:PhaseRetrieval02}) to Eq.~(\ref{eq:PhaseRetrieval03}), the assumption that $F(x,y) \ll 1$ then gives
\begin{align}
\breve{Q}(x,y) 
=  \nabla_{\perp}\cdot\left[I_0(x,y) \nabla_{\perp}\phi(x,y)\right],~F(x,y)\ll 1.
\label{eq:PhaseRetrieval16}
\end{align}

We emphasize that the preceding equation holds true, whether we work with (i) the $F(x,y) \ll 1$ special case of the scalar-diffusion extended Fokker-Planck equation in the form given by Eq.~(\ref{eq:SecIII--ExtendedFPE}), or (ii) the $F(x,y) \ll 1$ special case of the tensor-diffusion extended Fokker-Planck equation that is obtained when the replacement in Eq.~(\ref{eq:ReplacementToGetFromIsotropicToAnisotropicDiffusion}) is applied to Eq.~(\ref{eq:SecIII--ExtendedFPE}).  In either case, when only nonnegative propagation distances $0<\Delta_1<\Delta_2$ are employed, the fact that the replacement $Q(x,y)\longrightarrow\breve{Q}(x,y)$ maps Eq.~(\ref{eq:PhaseRetrieval03}) to Eq.~(\ref{eq:PhaseRetrieval16}) means that the solution to the latter equation may be obtained by applying the same replacement to Eq.~(\ref{eq:PhaseRetrieval10}).  This leads to the following phase-retrieval algorithm for the case of asymmetric defocus, which is identical to that previously derived by \citet{LeathamHonsThesis2019,LeathamPhDThesis2023} in the more restricted context of the $F \ll 1$ variant of Eq.~(\ref{eq:SecIII--FPE-scalar-case}):
\begin{equation}
\phi(x,y)=\nabla_{\perp}^{-2}\left\{\nabla_{\perp}\!\cdot\!\left[\frac{\nabla_{\perp}\nabla_{\perp}^{-2} \breve{Q}(x,y)}{I_0(x,y)}\right]\right\}.
\label{eq:PhaseRetrieval17}
\end{equation}
While we have written down this result with the case $0<\Delta_1<\Delta_2$ in mind, it may also be applied when $\Delta_1<0<\Delta_2$. For the special symmetric-defocus case where $\Delta_1=-\Delta_2\equiv\Delta>0$, the three-image data function $\breve{Q}(x,y)$ in Eq.~(\ref{eq:PhaseRetrieval15}) reduces to the two-image data function $Q(x,y)$ in Eq.~(\ref{eq:PhaseRetrieval01}), with Eq.~(\ref{eq:PhaseRetrieval17}) thereby reducing to Eq.~(\ref{eq:PhaseRetrieval10}).  Also, if all three defoci $z=\Delta_1,\Delta_2,\Delta_3$ differ from zero, with $\Delta_1\ne\Delta_2\ne\Delta_3$,  then the data function in Eqs.~(\ref{eq:PhaseRetrieval15})-(\ref{eq:PhaseRetrieval17}) should be generalized via the replacement (cf.~Ref.~\cite{LeathamPhDThesis2023})
\begin{equation}
\breve{Q}\longrightarrow 
 \frac{(\Delta_2^2-\Delta_3^2)I_1+(\Delta_3^2-\Delta_1^2)I_2+(\Delta_1^2-\Delta_2^2)I_3}{(\Delta_1-\Delta_3)(\Delta_2-\Delta_3)(\Delta_1-\Delta_2)/k},
\label{eq:PhaseRetrieval18}
\end{equation}
where $I_j\equiv I(x,y,z=\Delta_j)$ for $j=1,2,3$ and functional dependence on $(x,y)$ is omitted for clarity.\footnote{Note that if the defocus values all differ from zero, then $I_0(x,y)$ in Eq.~(\ref{eq:PhaseRetrieval16}) will need to be estimated using $I_1(x,y)$, $I_2(x,y)$, and $I_3(x,y)$.  This can be done via a quadratic fit, at each transverse position $(x,y)$, to the intensity as a function of defocus $z$.}  This form may be of use, for example, when negative $\Delta$ values are inaccessible, but one has the added constraint that $\Delta>0$ cannot be made arbitrarily small (since, e.g., the detector plane $C$ for the lens-free scenario in Fig.~3 of Paper I cannot be moved to $z=0$ on account of the geometry of that detector).  If we set $\Delta_3$ to zero, Eq.~(\ref{eq:PhaseRetrieval18}) reverts to Eq.~(\ref{eq:PhaseRetrieval15}).

\subsubsection{Phase retrieval: Structured-illumination case}\label{sec:PhaseRetrievalStructuredIlllumination}

In passing from Fig.~3 to Fig.~5 in Paper I, structured illumination of a thin sample $B$ was introduced via a thin mask $\mathcal{M}$ having a position-dependent intensity transmission function $M(x,y)$. As stated in Sec.~III\,D of Paper I, the purpose of such patterned illumination is ``to introduce a spatially rapidly varying transmission function, which amplifies the measured signal''. Note that in order to produce structured illumination at the detector, masks may be designed with a position-dependent transmission function \cite{Massig1,Perciante,Massig2,wen2010,morgan2011quantitative,Medhi2022} or introduce position-dependent phase shifts \cite{MayoSexton2004,Morgan2013,Rizzi2013}. In many cases, for example when using a random speckle generator \cite{berujon2012,Morgan2012,berujon2012b,zdora2017,zdora2018,PaganinLabrietBrunBerujon2018,ZdoraSpeckleBook,Boominathan2022}, both phase and attenuation effects will play a role in creating structure in the illumination. While we previously considered such amplification in the setting of paraxial diffusion-field retrieval, here we revisit this concept for structured-illumination phase retrieval \cite{Massig1,Perciante,Massig2,MayoSexton2004,wen2010,morgan2011quantitative,berujon2012,Morgan2012,berujon2012b,zdora2017,zdora2018,PaganinLabrietBrunBerujon2018,ZdoraSpeckleBook,Boominathan2022,Medhi2022,Morgan2013,Rizzi2013}.  

We simultaneously consider two scenarios, namely (i) the configuration in Fig.~5 of Paper I, in which the lack of post-specimen optical elements implies that intensity images $I(x,y,z=\Delta)$ may only be formed for $\Delta \ge 0$, and (ii) a modified form of this figure, whereby the introduction of post-specimen optical elements allows intensity images to be formed for both positive and negative propagation distances (defoci) $\Delta\gtrless 0$ (cf.~Fig.~\ref{Fig:Schematic} above).

Let the complex transmission function of the mask be $\sqrt{M(x,y)}\exp[i\phi_m(x,y)]$, where $M(x,y)\ge 0$.  For the case of unit-intensity normally-incident coherent scalar plane-wave illumination, an infocus intensity image $I^{(m)}(x,y,z=0)$ of the mask in the absence of the sample may be used to determine $M(x,y)$ via\footnote{Cf.~the $\Delta=0$ case of Eq.~(57) in Paper I.  Note also that we are using a similar notation to that in Sec.~III\,D\,1 of Paper I.}
\begin{equation}
I^{(m)}(x,y,z=0)=M(x,y).
\label{eq:PhaseRetrieval19}
\end{equation}
Regarding the phase $\phi_m(x,y)$ of the complex transmission function of the mask, this may be obtained using either (i) the method of Eq.~(\ref{eq:PhaseRetrieval10}) for the symmetrical focal series $\{I^{(m)}(x,y,z=0),I^{(m)}(x,y,z=\widetilde{\Delta}),I^{(m)}(x,y,z=-\widetilde{\Delta})\}$, using a suitable mask-defocus distance $\widetilde{\Delta}>0$, provided that a post-specimen imaging system is present in order to register the negative-defocus image over the plane $z=-\widetilde{\Delta}$, or (ii) the method of Eq.~(\ref{eq:PhaseRetrieval17}), using the data function in Eq.~(\ref{eq:PhaseRetrieval15}) [if intensity images may be obtained in the planes $z=0$, $z=\widetilde{\Delta}_1>0$ and $z=\widetilde{\Delta}_2>\widetilde{\Delta}_1$] or the data function in Eq~(\ref{eq:PhaseRetrieval18}) [if intensity images may be obtained in the positive-defocus planes $z=\widetilde{\Delta}_1>0$, $z=\widetilde{\Delta}_2>\widetilde{\Delta}_1$ and $z=\widetilde{\Delta}_3>\widetilde{\Delta}_2$; for this case, Eq.~(\ref{eq:PhaseRetrieval19}) cannot be used to estimate $M(x,y)$, so this quantity should instead be estimated via suitable algebraic manipulation of the images measured in the planes $z=\widetilde{\Delta}_1,\widetilde{\Delta}_2,\widetilde{\Delta}_3$].

Having determined the complex transmission function $\sqrt{M(x,y)}\exp[i\phi_m(x,y)]$ of the structured-illumination mask using intensity measurements that are obtained in the absence of the sample, the sample may now be placed either immediately upstream or immediately downstream of the mask (for the latter variant, see Fig.~5 of Paper I). Assume that the speckle mask makes no contribution to the diffusion field at the exit surface $z=0$ of the sample-mask system.  Suppress functional dependence on transverse coordinates $(x,y)$ for clarity.  Moreover, assume that $F \ll 1$. For a defocus distance of $\Delta$, we may then use Eq.~(\ref{eq:SecIII--ExtendedFPE}) to generalize Eq.~(55) of Paper I.  This leads to the following expression for the intensity distribution $I_{\Delta}^{(s+m)}$ in the presence of both the sample and the mask:
\begin{align}\label{eq:BasicEquationForSpeckleTracking}
    I_{\Delta}^{(s+m)}= &TM+\Delta^2\nabla_{\perp}^2(DTM) 
    \\ \nonumber &-\frac{\Delta}{k}\nabla_{\perp}\cdot[MT\nabla_{\perp}(\phi+\phi_m)]+\Delta^2\Xi^{(s+m)}.
\end{align}
In the preceding equation, $T \equiv T(x,y)$ is the intensity transmission function of the sample, $\phi \equiv \phi(x,y)$ is the phase shift induced by the sample, and
\begin{align}\label{eq:XiForSamplePlusMaskDefined}
    \Xi^{(s+m)} = &\frac{1}{4k^2}\Big\{\left\vert\nabla_{\perp}^2\sqrt{TM}e^{i(\phi+\phi_m)}\right\vert^2
   \\ \nonumber &-\textrm{Re}\left[\sqrt{TM}e^{-i(\phi+\phi_m)}(\nabla_{\perp}^2)^2\sqrt{TM}e^{i(\phi+\phi_m)}\right]\Big\}
\end{align}
is a diffraction-induced correction associated with the second and third lines of Eq.~(\ref{eq:SecIII--ExtendedFPE}).  

Note the {\em amplification of the sample phase signal} via the divergence term appearing in Eq.~(\ref{eq:BasicEquationForSpeckleTracking}), due to the presence of the mask transmission function $M$.\footnote{If there is appreciable distance between the mask and the thin sample, which we previously illustrated as being ``almost touching'' in Fig.~5 of Paper I, the previously mentioned concept of ``out of focus contrast'' \cite{Zernike1942,Bremmer1952,Cowley1959,CowleyBook} implies that phase shifts associated with the mask can lead to intensity variations over the entrance surface of the sample.  In this case, the mask transmission function and phase-shift function refer to the complex field at the nominally planar entrance surface of the sample, rather than the exit surface of the mask.  Cf.~Fig.~\ref{Fig:HowNoiseBlursSpeckles}(a), later in this paper.}  From a physical perspective, sharply varying mask absorption leads to large values for $\nabla_{\perp}M$, amplifying the sensitivity to sample-induced phase gradients $\nabla_{\perp}\phi$ because the refractive effect of phase gradients deforms the sharp mask edge upon nonzero defocus $\Delta$.\footnote{Cf.~the remarks on mask-induced amplification of diffusion-field intensity signals, in Sec.~III\,D of Paper I.}  In view of this amplification, efficient mask choices have transmission functions $M(x,y)$ that are spatially rapidly varying, such as random patterns \cite{berujon2012,berujon2012b,Morgan2012,zdora2017,PaganinLabrietBrunBerujon2018,zdora2018,ZdoraSpeckleBook} or grid-type arrays \cite{Perciante,Massig1,wen2010,morgan2011quantitative} whose intensity varies significantly over transverse distances on the order of (say) a few pixels.      

To solve Eq.~(\ref{eq:BasicEquationForSpeckleTracking}) for the sample-induced phase shift $\phi$, one can measure ``sample plus mask'' intensities $I_{\Delta}^{(s+m)}$ for three different defocus distances, say $\Delta=0$, $\Delta=\Delta_1$, and $\Delta=\Delta_2$.  The mask phase $\phi_m$ is assumed to have been determined using the method described earlier [see the paragraph containing Eq.~(\ref{eq:PhaseRetrieval19})].  With these data the terms containing $D$ and $\Xi$ may be algebraically eliminated from the three instances of Eq.~(\ref{eq:BasicEquationForSpeckleTracking}) that correspond to the defoci $\Delta=0,\Delta_1,\Delta_2$.  This yields 
\begin{equation}\label{eq:TIE-like-form-for-speckle-tracking}
    \nabla_{\perp}\cdot(I_0^{(s+m)}\nabla_{\perp}\phi)=\widetilde{Q}
\end{equation}
where the known data function $\widetilde{Q}$ is 
\begin{align}
    \widetilde{Q} = \frac{k}{\Delta_1\Delta_2(\Delta_1-\Delta_2)}\Big[\Delta_2^2I_{\Delta_1}^{(s+m)}
    - \Delta_1^2I_{\Delta_2}^{(s+m)}
    \\ \nonumber + (\Delta_1^2-\Delta_2^2)I_0^{(s+m)} \Big]
    - \nabla_{\perp}\cdot(I_0^{(s+m)}\nabla_{\perp}\phi_m).
\end{align}
Equation~(\ref{eq:TIE-like-form-for-speckle-tracking}) is identical in form to Eq.~(\ref{eq:PhaseRetrieval03}), if $I(x,y,z=0)$ in Eq.~(\ref{eq:PhaseRetrieval03}) is replaced by $I_0^{(s+m)}\equiv I_0^{(s+m)}(x,y)$ and $Q$ is replaced by $\widetilde{Q}$.  Hence Eq.~(\ref{eq:TIE-like-form-for-speckle-tracking}) may be solved for $\phi$ in the same manner that was previously employed for solving Eq.~(\ref{eq:PhaseRetrieval03}) [see the first part of Eq.~(\ref{eq:PhaseRetrieval10})]. 

Alternative structured-illumination phase-retrieval algorithms may be constructed by transversely shifting the mask $\mathcal{M}$ to a number of different locations (see Fig.~5 in Paper I) {\cite{zdora2018, ZdoraSpeckleBook}}, but to keep this paper to a reasonable length such algorithms will not be reported here.

\subsection{Diffusion-field retrieval}\label{sec:DiffusionFieldRetrieval}

Here we consider the problem of paraxial diffusion-field retrieval, namely means to solve the Fokker-Planck or extended Fokker-Planck equations so as to obtain a scalar or tensor diffusion field, given intensity measurements $I(x,y,z=\Delta)$ that are taken for different defocus values $\Delta$. Since only sample-scattered radiation contributes to our diffusion fields,\footnote{We here neglect the influence of the term $\theta_0^2$ that appears in Eq.~(24) of Paper I [see, also, Figs.~2 and 3 of Paper I].  Alternatively, we can consider the influence of this term---which represents blurring due to an extended chaotic source---as having been subtracted from a recovered diffusion field such as $D(x,y)$.} this inverse problem may be spoken of as a form of ``dark-field imaging'' \cite{gage1920,Pfeiffer2008df}.  In particular, for a sample that is entirely contained within the imaging field of view, a map of the modulus of the sample-induced diffusion field $D(x,y)$ will appear as bright upon a dark background (see, e.g., \citet{Alloo2025}). 

The remainder of this section has three parts.  In Sec.~\ref{sec:DiffusionFieldRetrievalSymmDefocus} we consider the symmetric-defocus case where both over-focused and under-focused images are available, corresponding to defocus values that are equal in magnitude but opposite in sign.  Section \ref{sec:DiffusionFieldRetrievalAsymmDefocus} studies the case of asymmetric defoci, which is important in contexts where no imaging system is present (thereby rendering negative-defocus images inaccessible\footnote{While this case is perhaps of less importance for the field of visible-light imaging, it is of key importance in propagation-based x-ray phase contrast imaging \cite{Snigirev,Paganin2006}, namely a form of inline holography \cite{Gabor1948}  with a sample-to-detector propagation distance that is sufficiently short {such that at most one Fresnel diffraction fringe is produced by any particular sample structure}.}).  Finally, in Sec.~\ref{sec:DiffusionFieldRetrievalStructuredIllumination} we consider diffusion-field retrieval using structured illumination.

\subsubsection{Diffusion-field retrieval: Symmetric-defocus case}\label{sec:DiffusionFieldRetrievalSymmDefocus}

In this subsection we restrict consideration to scalar diffusion-field retrieval.  Two cases are considered, respectively, in the next two paragraphs: (i) the simpler case of the Fokker-Planck equation obtained by retaining only the first line of Eq.~(\ref{eq:SecIII--ExtendedFPE}) (``Case A''), and (ii) the more involved case where all three lines of Eq.~(\ref{eq:SecIII--ExtendedFPE}) are taken into account (``Case B'').    

{\em Case A:} If only the first line of Eq.~(\ref{eq:SecIII--ExtendedFPE}) is retained, we obtain the form for the Fokker-Planck extension to the TIE that was given in Eq.~(146) of Paper I (without, however, the restriction to defoci that obey $\Delta \ge 0$). Now use the symmetric overfocus and underfocus intensity images $I(x,y,z=\pm\Delta)$, together with the infocus intensity image $I(x,y,z=0)$, to form the data function $q(x,y)$ defined in Eq.~(148) of Paper I, namely
\begin{equation}\label{eq:YetAnotherDataFunction}
    q(x,y)=\frac{I(x,y,\Delta)+I(x,y,-\Delta)-2I(x,y,0)}{2\Delta^2}.
\end{equation}
Following Eq.~(149) of Paper I, the measured data function $q(x,y)$ may be related to the scalar diffusion field $D(x,y)$ and infocus intensity via the Poisson equation
\begin{equation}\label{eq:PoissonEqnForD}
    \nabla_{\perp}^2[D(x,y)I(x,y,z=0)]=q(x,y).
\end{equation}
In a similar manner to the means by which we solved Eq.~(\ref{eq:PhaseRetrieval05}), we may write a formal solution as
\begin{equation}\label{eq:FormalSolutionDiffusionFieldRetrievalSymmetricDefocus}
    D(x,y)=\frac{\nabla_{\perp}^{-2}[q(x,y)]}{I(x,y,z=0)}.
\end{equation}
The inverse Laplacian can be evaluated in a number of ways, since it merely indicates that one should ``solve the Poisson equation subject to relevant boundary conditions''.  For example, if one makes use of the Fourier representation for the inverse Laplacian in Eq.~(\ref{eq:PhaseRetrieval11}), the recovered diffusion field is
\begin{equation}\label{eq:DiffusionFieldRetrieval1of2}
    D(x,y)=\frac{-1}{I(x,y,z=0)}\mathscr{F}^{-1}\frac{1}{k_x^2+k_y^2+\epsilon^2}\mathscr{F} [q(x,y)].
\end{equation}
Other means for regularization are possible, such as
\begin{equation}\label{eq:DiffusionFieldRetrieval2of2}
    D(x,y)=\frac{-1}{I(x,y,z=0)}\mathscr{F}^{-1}\frac{k_x^2+k_y^2}{(k_x^2+k_y^2)^2+\epsilon^4}\mathscr{F} [q(x,y)].
\end{equation}
Note also that Eq.~(\ref{eq:YetAnotherDataFunction}) contains a finite-difference approximation to the second $z$ derivative of intensity, since
\begin{equation}
    \left[\frac{\partial^2}{\partial z^2}I(x,y,z)\right]_{\!z=0} \!\!\!\!\!\! \approx \frac{\frac{I(x,y,\Delta)-I(x,y,0)}{\Delta}-\frac{I(x,y,0)-I(x,y,-\Delta)}{\Delta}}{\Delta}.
\end{equation}
Hence the symmetric-defocus formula for ``Case A'' diffusion-field retrieval, in Eq.~(\ref{eq:FormalSolutionDiffusionFieldRetrievalSymmetricDefocus}), may be written as
\begin{equation}\label{eq:SimplifiedFormalSolutionDiffusionFieldRetrievalSymmetricDefocus}
    D(x,y)=\frac{\nabla_{\perp}^{-2}\left\{\left[\frac{\partial^2}{\partial z^2}I(x,y,z)\right]_{\!z=0}\right\}}{2 I(x,y,z=0)}.
\end{equation}
Thus the same focal series of three intensities employed for ``ordinary'' TIE phase retrieval in Eq.~(\ref{eq:PhaseRetrieval10})---namely an overfocus image $I(x,y,z=\Delta)$, an underfocus image $I(x,y,z=-\Delta)$, and an infocus image $I(x,y,z=0)$---may also be employed for diffusion-field retrieval using Eq.~(\ref{eq:SimplifiedFormalSolutionDiffusionFieldRetrievalSymmetricDefocus}).  {\em The infocus intensity and intensity derivative give the key information needed for phase retrieval; the infocus intensity and} second {\em intensity derivative give the key information needed for diffusion-field retrieval}.  Regarding the latter point, Eq.~(\ref{eq:SimplifiedFormalSolutionDiffusionFieldRetrievalSymmetricDefocus}) implies the second intensity derivative (with respect to defocus) to be a source term for the {\em scalar-diffusion intensity transport equation}
\begin{equation}\label{eq:ScalarDiffusionIntensityTransportEquation}
    \nabla_{\perp}^2 \left[D(x,y)I(x,y,z=0)\right]=\frac{1}{2}\left[\frac{\partial^2}{\partial z^2}I(x,y,z)\right]_{\!z=0}.
\end{equation}
Moreover, the replacement in Eq.~(\ref{eq:ReplacementToGetFromIsotropicToAnisotropicDiffusion}) leads to the {\em tensor-diffusion intensity transport equation}\footnote{We emphasize that Eqs.~(\ref{eq:ScalarDiffusionIntensityTransportEquation}) and (\ref{eq:TensorDiffusionIntensityTransportEquation}) are approximations which are valid to the extent that one can indeed  retain only the first line of Eq.~(\ref{eq:SecIII--ExtendedFPE}) (``Case A'').  This simplification is likely to become an oversimplification in the vicinity of sharp sample edges. Comparison with the edge-induced diffusion fields, as developed in Secs.~\ref{sec:DiffusionFieldModels:NegativeCaseScalar} and \ref{sec:DiffusionFieldModels:NegativeCaseTensor} below, leads to the conjecture that Eqs.~(\ref{eq:ScalarDiffusionIntensityTransportEquation}) and (\ref{eq:TensorDiffusionIntensityTransportEquation}) might have a broader domain of validity if the diffusion fields $D(x,y)$ and $\widetilde{\mathbf{D}}(x,y)$ are augmented by diffusion fields that are induced by sharp-edge scatter.}  
\begin{equation}\label{eq:TensorDiffusionIntensityTransportEquation}
 \nabla_{\perp}\cdot[\widetilde{\mathbf{D}}(x,y) \nabla_{\perp}I(x,y,z=0)]=\frac{1}{2}\left[\frac{\partial^2}{\partial z^2}I(x,y,z)\right]_{\!z=0}.
\end{equation}
If the right sides of the preceding two ``Case A'' equations were to contain the first rather than the second longitudinal derivative of intensity, they would be ``normal diffusion'' equations.  The fact that the second derivative appears on the right sides is related to the fact that they are ``anomalous diffusion'' equations \cite{EvangelistaLenziBook2018} associated with ballistic superdiffusion \cite{MetzlerKlafter2000} [see the paragraph in Paper I that contains Eqs.~(10)-(12), for further information]. 

{\em Case B:} If all three lines of Eq.~(\ref{eq:SecIII--ExtendedFPE}) are taken into account, while restricting consideration to symmetric-defocus scenarios where $F(x,y) \ll 1$, the analysis in Sec.~\ref{sec:PhaseRetrievalSymmetricDefocus} shows that the phase-retrieval algorithm of Paganin and Nugent---namely Eq.~(\ref{eq:PhaseRetrieval10}) \cite{paganin1998}---may be used to recover the phase $\phi(x,y)$ in the plane $z=0$.  Adopting a suggestion made to the authors by Mario A.~Beltran in 2022, we may then use the recovered phase and infocus intensity to calculate the second and third lines of Eq.~(\ref{eq:SecIII--ExtendedFPE}).  Denoting the calculated form for the second and third lines of this equation by $\Delta^2 \,\Xi(x,y)$, Eq.~(\ref{eq:PoissonEqnForD}) has the generalized form
\begin{equation}\label{eq:PoissonEqnForD==generalized}
    \nabla_{\perp}^2[D(x,y)I(x,y,z=0)]=q(x,y)-\Xi(x,y).
\end{equation}
Since the right side of this equation consists entirely of known functions, we can make the replacement
\begin{equation}
    q(x,y)\longrightarrow q(x,y)-\Xi(x,y)
\end{equation}
in Eqs.~(\ref{eq:DiffusionFieldRetrieval1of2}) or (\ref{eq:DiffusionFieldRetrieval2of2}), to give the diffusion field $D(x,y)$.  For later use, we write $\Xi(x,y)$ explicitly [cf.~Eq.~(\ref{eq:XiForSamplePlusMaskDefined})]:
\begin{equation}\label{eq:XiDefined}
    \Xi=\frac{1}{4k^2}\left\{\left\vert\nabla_{\perp}^2\sqrt{I_0}e^{i\phi}\right\vert^2-\textrm{Re}\left[\sqrt{I_0}e^{-i\phi}(\nabla_{\perp}^2)^2\sqrt{I_0}e^{i\phi}\right]\right\}.
\end{equation}
Above, we use the compressed notation $\Xi\equiv\Xi(x,y), I_0\equiv I(x,y,z=0)$, and $\phi\equiv\phi(x,y,z=0)$ [cf.~Eq.~(\ref{eq:PhaseRetrieval13})].

\subsubsection{Diffusion-field retrieval: Asymmetric-defocus case}\label{sec:DiffusionFieldRetrievalAsymmDefocus}

We again restrict consideration to paraxial diffusion-field retrieval when $F(x,y) \ll 1$. Three intensity images at different defoci are assumed to have been measured.  If we suitably choose the origin of optical-axis coordinates $z$, one of these images will correspond to the plane $z=0$; let the other defocus planes be denoted by $z=\Delta_1$ and $z=\Delta_2$.  Using the method outlined in Sec.~\ref{sec:PhaseRetrievalAsymmetricDefocus}, the phase $\phi(x,y)$ may be obtained in the plane $z=0$. From this point onward, we then distinguish between Cases A and B that were defined in the previous subsection.

{\em Case A:} If only the first line of Eq.~(\ref{eq:SecIII--ExtendedFPE}) is retained, and $1-F(x,y)$ is approximated by unity, then for any one of the nonzero defocus distances (e.g., $\Delta_1$), one obtains a Poisson equation of the same form as Eq.~(\ref{eq:PoissonEqnForD}), with the data function $q(x,y)$ now being obtained via the known-function replacement
\begin{equation}
    q(x,y)\longrightarrow \frac{I_1-I_0 + \Delta_1k^{-1}\nabla\cdot(I_0\nabla_{\perp}\phi)}{\Delta_1^2}.
\end{equation}
Above, we use the previously-defined compressed notation, together with $I_1\equiv I(x,y,z=\Delta_1)$.  With this replacement, $D(x,y)$ is obtained via Eqs.~(\ref{eq:DiffusionFieldRetrieval1of2}) or (\ref{eq:DiffusionFieldRetrieval2of2}).

{\em Case B:} If all three lines of Eq.~(\ref{eq:SecIII--ExtendedFPE}) are taken into account, $\Xi$ may be constructed using the measured intensity $I_0$ and retrieved phase $\phi$ [see Eq.~(\ref{eq:XiDefined})].  Equation (\ref{eq:SecIII--ExtendedFPE}) again yields the Poisson-type form in Eq.~(\ref{eq:PoissonEqnForD}), with the data function now given by
\begin{equation}\label{eq:YetYetYetAnotherDataFunction}
    q(x,y)\longrightarrow \frac{I_1-I_0 + \Delta_1k^{-1}\nabla\cdot(I_0\nabla_{\perp}\phi)}{\Delta_1^2}-\Xi.
\end{equation}
Equations (\ref{eq:DiffusionFieldRetrieval1of2}) or (\ref{eq:DiffusionFieldRetrieval2of2}) then retrieve $D(x,y)$.

\subsubsection{Diffusion-field retrieval: Structured-illumination case}\label{sec:DiffusionFieldRetrievalStructuredIllumination}

In the next four paragraphs, respectively, we study structured-illumination diffusion-field retrieval for (i) Case A scalar diffusion fields, (ii) Case B scalar diffusion fields, (iii) Case A tensor diffusion fields, and (iv) Case B tensor diffusion fields. We use the same notation that was previously employed in Sec.~\ref{sec:PhaseRetrievalStructuredIlllumination}.    

{\em Scalar diffusion field, Case A:}  Write down Eq.~(\ref{eq:BasicEquationForSpeckleTracking}) and delete $\Xi^{(s+m)}$ (as previously mentioned, this is the fundamental assumption which distinguishes Case A from Case B).  Assume that $I_{\Delta}^{(s+m)}$, $I_{\Delta=0}^{(s+m)} \equiv TM$, and $\phi+\phi_m$ have been either directly measured or retrieved; the nonzero defocus distance $\Delta$ and wavenumber $k$ are also taken to be known.  Algebraically rearrange the resulting equation to give
\begin{equation}\label{eq:YetAnotherPoissonEquation1}
\nabla_{\perp}^2(DTM)=\mathring{Q}    
\end{equation}
where the known data function is
\begin{equation}
\mathring{Q}=\frac{I_{\Delta}^{(s+m)}-TM}{\Delta^2}+\frac{1}{k\Delta}\nabla_{\perp}\cdot[MT\nabla_{\perp}(\phi+\phi_m)].    
\end{equation}
Hence the required scalar diffusion field is
\begin{equation}\label{eq:fqwrfg32}
D=(TM)^{-1}\nabla_{\perp}^{-2}\mathring{Q}.    
\end{equation}

{\em Scalar diffusion field, Case B:} Write down Eq.~(\ref{eq:BasicEquationForSpeckleTracking}).  Again assume that $I_{\Delta}^{(s+m)}$, $TM$, and $\phi + \phi_m$ have been either directly measured or retrieved; $\Delta$ and $k$ are also given. Then construct $\Xi^{(s+m)}$ using Eq.~(\ref{eq:XiForSamplePlusMaskDefined}).  Algebraically rearrange the resulting equation to give
\begin{equation}\label{eq:YetAnotherPoissonEquation2}
\nabla_{\perp}^2(DTM)=\mathring{Q} - \Xi^{(s+m)},    
\end{equation}
hence Eq.~(\ref{eq:fqwrfg32}) generalizes to
\begin{equation}\label{eq:YetAnotherPoissonEquation3}
D=(TM)^{-1}\nabla_{\perp}^{-2}[\mathring{Q}-\Xi^{(s+m)}].    
\end{equation}

{\em Tensor diffusion field, Case A:}  Approximate $\Xi_j^{(s+m)}$ by zero, in the approach that is given in the next paragraph.

{\em Tensor diffusion field, Case B:} By using a tensor diffusion field $\widetilde{\mathbf{D}}(x,y)$ [see Eqs.~(\ref{eq:NiceNewEquationForTensorDiffusionField}), (\ref{eq:ReplacementToGetFromIsotropicToAnisotropicDiffusion}), and (\ref{eq:TensorDiffusionFieldRetrieval9})] rather than a scalar diffusion field $D(x,y)$, Eq.~(\ref{eq:BasicEquationForSpeckleTracking}) generalizes to the following extension of Eq.~(125) in Paper I:
\begin{align}\label{eq:BasicEquationForSpeckleTrackingTensorForm}
    I_{j,\Delta}^{(s+m)}= &TM_j+\Delta^2\nabla_{\perp}\cdot[\widetilde{\mathbf{D}}\nabla_{\perp}(TM_j)] 
    \\ \nonumber &-\frac{\Delta}{k}\nabla_{\perp}\cdot[M_jT\nabla_{\perp}(\phi+\phi_{j,m})]+\Delta^2\Xi_j^{(s+m)}.
\end{align}
Above, the subscript $j=1,2,\cdots,j_{\textrm{max}}$ indicates a number of different masks (or transverse mask positions); similar notation was used for Eq.~(126) in Paper I. The phase shift due to the $j$th mask, namely $\phi_{j,m}$, can be determined using the same method that was given in Sec.~\ref{sec:PhaseRetrievalStructuredIlllumination}; this is because the method of that section algebraically eliminates the diffusion-field term, irrespective of whether the diffusion has a scalar or tensor character.\footnote{Alternatively, for each mask position $j$, the sample-plus-mask phase shift $\phi+\phi_{j,m}$ can be retrieved via the method in Sec.~\ref{sec:PhaseRetrievalAsymmetricDefocus}.}  Also, as was the case on several previous occasions, Eq.~(\ref{eq:XiForSamplePlusMaskDefined}) can be used to calculate $\Xi_j^{(s+m)}$, via quantities that have either been directly measured or retrieved. Thus, the only unknown quantity in Eq.~(\ref{eq:BasicEquationForSpeckleTrackingTensorForm}) is $\widetilde{\mathbf{D}}$.  Introduce the known data function
\begin{align}\label{eq:DataFunctionForSpeckleTrackingTensorForm}
\mathfrak{Q}_{j,\Delta} =& \frac{I_{j,\Delta}^{(s+m)}-TM_j}{\Delta^2} \\ \nonumber &+ \frac{1}{k\Delta}\nabla_{\perp}\cdot[M_jT\nabla_{\perp}(\phi+\phi_{j,m})] - \Xi_j^{(s+m)}
\end{align}
which corresponds to the $j$th mask and defocus $\Delta$. Hence
\begin{equation}\label{sec:TensorDiffusionFieldRetrievalWith
Masks}
\nabla_{\perp}\cdot[\widetilde{\mathbf{D}}\nabla_{\perp}(TM_j)]=\mathfrak{Q}_{j,\Delta}.    
\end{equation}
Now use Eq.~(108) from Paper I to write the symmetric diffusion tensor as a $2 \times 2$ matrix, represent $\nabla_{\perp}(TM_j)$ as a two-component column vector, then introduce the abbreviations $\partial_x\equiv\partial/\partial x$ and $\partial_y\equiv\partial/\partial y$.  Hence 
\begin{equation}\label{eq:FancyEquationForDiffusionTensor}
\nabla_{\perp}\cdot\left\{\begin{pmatrix}
\widetilde{D}_{xx} & \widetilde{D}_{xy}\\
\widetilde{D}_{xy} & \widetilde{D}_{yy}
\end{pmatrix}\begin{pmatrix}
    \partial_x(TM_j) \\ \partial_y(TM_j)
\end{pmatrix}\right\}=\mathfrak{Q}_{j,\Delta}.    
\end{equation}
Expanding the left side then gives a linear combination (with known position-dependent coefficients that will not be written here) of the same seven unknown functions contained within square brackets in Eq.~(126) from Paper I. The same method, employed in Sec.~IV\,B of Paper I, may then be used to construct the diffusion tensor, using at least seven independent data functions $\mathfrak{Q}_{j,\Delta}$.  These independent data functions may be obtained by varying either or both of the mask position $j$ and the defocus $\Delta$. {  For another reconstruction method, see  Sec.~\ref{subsec:uniquenessOfTensorDiffusionFieldRetrieval}. }

\section{Some paraxial diffusion-field models}\label{sec:DiffusionFieldModels}

Several models for paraxial scalar and tensor diffusion fields, all of which are of primary relevance to this paper, were given in Paper I.  Here we augment this list with { a first-principles model which shows how the diffusion-field concept naturally emerges from the convolution formulation of Fresnel diffraction (Sec.~\ref{sec:DiffusionFieldFromFirstPrinciples}).  Next, we develop} a model for {\em noise-induced contributions to diffusion fields}, namely an effective position-dependent blur associated with shot noise in photon-limited intensity detection scenarios (Sec.~\ref{sec:NoiseContributionToParaxialDiffusionField}).\footnote{As explained in Sec.~V\,E of Paper I, noise-induced blur can be thought of as the ``washing out of high-frequency detail by high-frequency noise''. Resolution estimation via Fourier ring correlation  \cite{SaxtonBaumeister1982} leads to the same conclusion.  Cf.~Sec.~\ref{sec:NoiseInducedVisibilityReduction}.  \label{footnote:WashingOutFineDetailByNoise}} Finally, we  study two distinct classes of model for {\em negative diffusion fields}, namely diffusion fields that locally sharpen an intensity distribution rather than locally blurring it (Sec.~\ref{sec:DiffusionFieldModels:NegativeCases}).  

{ 
\subsection{Paraxial diffusion field from first principles}\label{sec:DiffusionFieldFromFirstPrinciples}

To augment the phenomenological models for coexisting paraxial coherent-flow and diffusive-flow optical wavefields given earlier in this paper, here we study a first-principles model.  Influenced by \citet{Nugent2003} and especially \citet{Nesterets2008}, our derivation generalizes and refines the analysis in \citet{PaganinMorgan2019}.


Return to the diffraction problem for a coherent paraxial scalar field $\psi(x,y,z)$, as illustrated in Fig.~\ref{Fig:CoherentFlow}, whereby one seeks to propagate the field in vacuum from the plane $z=0$ to the plane $z=\Delta$.  The convolution theorem of Fourier analysis \cite{Bracewellbook}, applied to the Fourier representation for Fresnel diffraction in Eqs.~(\ref{eq:SecIII--01})-(\ref{eq:SecIII--02}), implies that the propagated intensity may be written as
\begin{equation}\label{eq:FirstPrinciples1}
    I(x,y,z=\Delta)=\left\vert\psi(x,y,z=0)\otimes P_{\Delta}(x,y)\right\vert^2.
\end{equation}
Here, $\otimes$ denotes two-dimensional convolution and
\begin{eqnarray}\label{eq:FirstPrinciples2}
P_{\Delta}(x,y) = -\frac{ik}{2\pi\Delta}\exp(ik\Delta)\exp\left[\frac{ik(x^2+y^2)}{2\Delta}\right]
\\ \nonumber
\end{eqnarray}
is the propagator (Green function) for the convolution formulation of Fresnel diffraction \cite{WinthropWorthington66,NazarathyShamir1980}. Writing the convolution out explicitly gives \cite{Nesterets2008}
\begin{widetext}  
\begin{equation}\label{eq:FirstPrinciples3}
    I(x,y,z=\Delta)=\iiiint_{-\infty}^{\infty}
    \psi(x',y',z=0)\psi^*(x'',y'',z=0)
    P_{\Delta}(x-x',y-y')P_{\Delta}^*(x-x'',y-y'')dx'dy'dx''dy''.
\end{equation}
\end{widetext}  

To continue, express the unpropagated field in terms of its intensity and phase via
\begin{equation}\label{eq:FirstPrinciples4}
\psi(x,y,z=0)=\sqrt{I(x,y,z=0)} \, e^{i\phi(x,y,z=0)}.
\end{equation}
Restrict attention to scenarios where the phase $\phi(x,y,z=0)$ in the plane $z=0$ can be decomposed in terms of a ``slow'' component  $\varphi_s(x,y)$ (which may have a large amplitude but is spatially slowly varying over the width $\mathcal{W}$ of an imaging system pixel) and a stochastic ``fast'' component $\varphi_f(x,y)$ (which must have a small amplitude but is spatially rapidly varying over the width $\mathcal{W}$ of an imaging system pixel) \cite{VoronovichBook,Nesterets2008,Yashiro2010,Yashiro2011}: 
\begin{equation}\label{eq:FirstPrinciples5}
\phi(x,y,z=0)=\varphi_s(x,y)+\varphi_f(x,y).
\end{equation}
As shown in Fig.~\ref{Fig:FastAndSlowPhase}, the fast phase fluctuation models ``wavefront roughness'' associated with (i) unresolved microstructure contained within a thin sample that lies immediately upstream of the plane $z=0$, (ii) the effect of partially coherent illumination, and (iii) noise-related effects.  Note also the strong ``separation of length scales'' assumption implicit in the resolution-dependent decomposition given by Eq.~(\ref{eq:FirstPrinciples5}), whereby the influence of wavefield phase variations over length scales intermediate between the ``fast'' and ``slow'' scales is neglected.   

\begin{figure}[ht!]
\centering
\includegraphics[width=0.8\columnwidth]{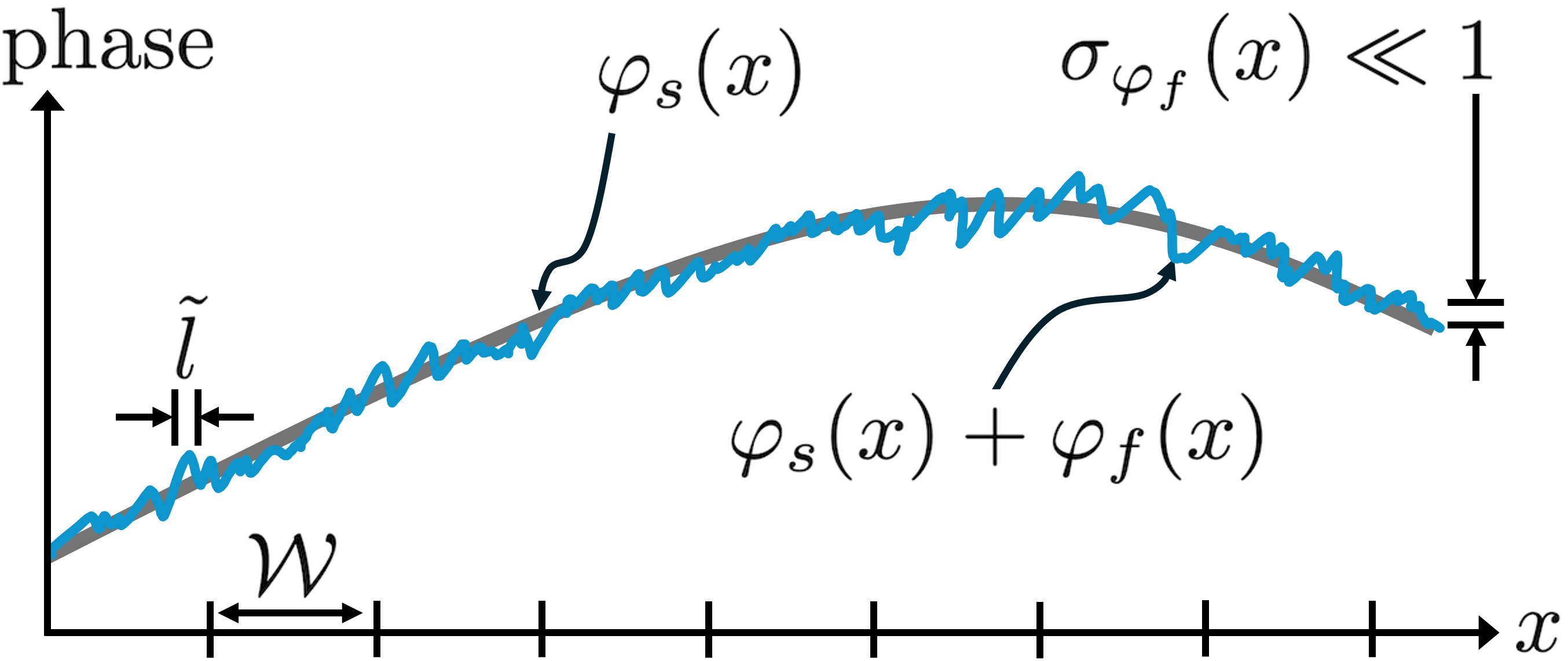}
\caption{{Illustrating the notation of Eq.~(\ref{eq:FirstPrinciples5}), for the case of ``wavefront roughening'' in one transverse dimension $x$.}}
\label{Fig:FastAndSlowPhase}
\end{figure}

The detector resolution is no smaller than $\mathcal{W}$, while the characteristic length scale $\tilde{\ell}$ associated with $\varphi_f(x,y)$ is by assumption significantly smaller than $\mathcal{W}$.  Now follow the lead of \citet{Nugent2003}, which states that $\tilde{\ell}$ ``is sufficiently small that the detection process forms a {\em de facto}
ensemble average.'' Under this view, the intensity average performed over the surface of each pixel may be modeled via an ensemble of statistical instances (realizations) of the stochastic process $\varphi_f(x,y)$.  
Denoting the associated ensemble average with an overline, we require
\begin{equation}\label{eq:FirstPrinciples6}
\overline{\varphi_f(x,y)}=0
\end{equation}
at every position coordinate $(x,y)$. 

Substitute Eq.~(\ref{eq:FirstPrinciples5}) into Eq.~(\ref{eq:FirstPrinciples4}), which may then be substituted into Eq.~(\ref{eq:FirstPrinciples3}). Ensemble averaging gives
\begin{widetext}  
\begin{align}
    \nonumber I(x,y,z=\Delta)=\iiiint_{-\infty}^{\infty}&
    \sqrt{I(x',y',z=0)}\sqrt{I(x'',y'',z=0)} \exp\{i[\varphi_s(x',y') -\varphi_s(x'',y'')]\}\\
    &\times \overline{\exp\{i[\varphi_f(x',y') -\varphi_f(x'',y'')]\}}
    P_{\Delta}(x-x',y-y')P_{\Delta}^*(x-x'',y-y'') 
     dx'dy'dx''dy''
     \label{eq:FirstPrinciples7}
\end{align}
for the propagated intensity.  To proceed further we work henceforth to second order in $\varphi_f(x,y)$, so that
\begin{align}
\nonumber \overline{\exp\{i[\varphi_f(x',y') -\varphi_f(x'',y'')]\}} 
&\approx
1-\tfrac{1}{2}\sigma_{\varphi_f}^2(x',y')-\tfrac{1}{2}\sigma_{\varphi_f}^2(x'',y'')+\overline{\varphi_f(x',y')\varphi_f(x'',y'')}
\\  &\approx \sqrt{1-\sigma_{\varphi_f}^2(x',y')}\sqrt{1-\sigma_{\varphi_f}^2(x'',y'')}+\overline{\varphi_f(x',y')\varphi_f(x'',y'')},
\label{eq:FirstPrinciples8}
\end{align}
where $\sigma_{\varphi_f}^2(x,y)\equiv\overline{[\varphi_f(x,y)]^2}\ll 1$ is the variance of the ensemble $\{\varphi_f(x,y)\}$ at each spatial position.  Also, in passing from the first to the second line of Eq.~(\ref{eq:FirstPrinciples8}), we make use of the reverse binomial approximation $1-\tfrac{1}{2}a\approx\sqrt{1-a}$ if $\vert a\vert\ll 1$; thus both right hand sides in Eq.~(\ref{eq:FirstPrinciples8}) are equivalent to second order in $\varphi_f(x,y)$. Hence Eq.~(\ref{eq:FirstPrinciples7}) becomes
\begin{align}
    \nonumber I(x,y,z=\Delta)
    = &\left\vert\mathcal{D}_{\Delta}^{\textrm{(F)}}\left\{\sqrt{[1-\sigma_{\varphi_f}^2(x,y)]I(x,y,z=0)}\exp[i\varphi_s(x,y)]\right\}\right\vert^2  
    \\ \nonumber &\quad\quad+\iiiint_{-\infty}^{\infty}
    \sqrt{I(x',y',z=0)}\sqrt{I(x'',y'',z=0)} 
    \exp\{i[\varphi_s(x',y') -\varphi_s(x'',y'')]\}
    \\
 &\quad\quad\quad\quad\quad\quad\quad\times \overline{\varphi_f(x',y')\varphi_f(x'',y'')}
    P_{\Delta}(x-x',y-y')P_{\Delta}^*(x-x'',y-y'') 
     dx'dy'dx''dy'',
     \label{eq:FirstPrinciples9}
\end{align}
where $\mathcal{D}_{\Delta}^{\textrm{(F)}}\equiv P_{\Delta}(x,y)\otimes$ denotes the Fresnel diffraction operator in the notation of Eq.~(\ref{eq:SecIII--02}) [cf.~Eq.~(\ref{eq:FirstPrinciples1})].  

Now make the key approximation, used in a very closely related context by \citet{Nesterets2008}, that the fast-phase covariance  $\overline{\varphi_f(x',y')\varphi_f(x'',y'')}$ will only be appreciably different from zero when $(x',y')$ and $(x'',y'')$ are separated by a distance that is on the order of (or less than on the order of) the spatial scale $\tilde{\ell}$ over which $\varphi_f(x,y)$ fluctuates appreciably; this ``fast'' length scale is small compared to the lower bound $\mathcal{W}$ of the scale over which $I(x,y,z=0)$ and $\varphi_s(x,y)$ fluctuate appreciably.  Hence, negligible error is invoked upon replacing $\sqrt{I(x'',y'',z=0)}$ by $\sqrt{I(x',y',z=0)}$ and $\varphi_s(x'',y'')$ by $\varphi_s(x',y')$ in the integrand of Eq.~(\ref{eq:FirstPrinciples9}).  This approximation immediately leads to
\begin{align}
    \nonumber I(x,y,z=\Delta)
    = &\left\vert\mathcal{D}_{\Delta}^{\textrm{(F)}}\left\{\sqrt{[1-\sigma_{\varphi_f}^2(x,y)]I(x,y,z=0)}\exp[i\varphi_s(x,y)]\right\}\right\vert^2  
    \\ &+\iint_{-\infty}^{\infty}
    I(x',y',z=0) \overset{\circ}{K}(x',y',x,y,\Delta)dx'dy',
     \label{eq:FirstPrinciples10}
\end{align}
\end{widetext} 
wherein the real diffuse-flow smearing kernel is given by
\begin{align} 
\nonumber &\overset{\circ}{K}(x',y',x,y,\Delta)=\iint_{-\infty}^{\infty}
\overline{\varphi_f(x',y')\varphi_f(x'',y'')}
\\ &\times P_{\Delta}(x-x',y-y')P_{\Delta}^*(x-x'',y-y'') 
dx''dy''.
\label{eq:FirstPrinciples11}
\end{align}

Equation (\ref{eq:FirstPrinciples10}) exhibits the first-principles bifurcation of a coherent paraxial flow into coexisting coherent and diffuse flows, which is a key theme of the present paper (recall, for example, the passage from Fig.~\ref{Fig:CoherentFlow} to Fig.~\ref{Fig:CoherentAndDiffuseFlow}).  With this in mind, identify the diffuse-flow fraction (cf.~Eq.~(29) in Ref.~\cite{PaganinMorgan2019})
\begin{equation}\label{eq:FirstPrinciples12}
F(x,y,z=0) = \sigma_{\varphi_f}^2(x,y) \equiv \overline{[\varphi_f(x,y)]^2}, 
\end{equation}
so that the first term in square brackets in Eq.~(\ref{eq:FirstPrinciples10}) represents a reduction in the coherently-flowing intensity associated with the ``wavefront roughness'' transferring energy from the coherent to the diffuse energy-flow channels.  The resulting effective complex wavefield, contained within the braces in Eq.~(\ref{eq:FirstPrinciples10}), has a phase
\begin{equation}\label{eq:FirstPrinciples13}
     \phi(x,y,z=0) = \varphi_s(x,y) 
\end{equation}
which has the ``fast'' fluctuations filtered away. Moreover, the unitarity of the Fresnel diffraction operator \cite{Montgomery1981} (and the associated conservation of energy) implies that for nonzero $F(x',y',z=0)$ the diffuse-flow kernel
\begin{equation}\label{eq:FirstPrinciples14}
    K(x',y',x,y,\Delta) = \frac{\overset{\circ}{K}(x',y',x,y,\Delta)}{F(x',y',z=0)}
\end{equation}
is normalized to unity [cf.~Eq.~(3) in Paper I].  

The identifications in Eqs.~(\ref{eq:FirstPrinciples12})-(\ref{eq:FirstPrinciples14}) transform Eq.~(\ref{eq:FirstPrinciples10}) into the coherent-plus-diffuse propagation model written down in Eq.~(\ref{eq:SecIII--15}).  Since Eq.~(\ref{eq:SecIII--15}) was the key point of departure for the main Fokker-Planck models of the present paper, the preceding formalism shows how those results may be obtained via a connected chain of argument which commences with the Fresnel diffraction integral. Moreover, the present first-principles derivation relates the fast-phase covariance  $\overline{\varphi_f(x',y')\varphi_f(x'',y'')}$ to the form for the diffuse-flow smearing kernel, via Eqs.~(\ref{eq:FirstPrinciples11}), (\ref{eq:FirstPrinciples12}) and (\ref{eq:FirstPrinciples14}). If $\Delta$ is sufficiently large to be in the far field of the ``fast'' phase fluctuations, the form of the propagator in Eq.~(\ref{eq:FirstPrinciples2}) implies that we recover the standard result from small-angle scattering which states that the diffuse-scatter angular distribution is proportional to the Fourier transform of the autocorrelation of the random microstructure \cite{GlatterKratky1982, SiviaScatteringTheoryBook2011}.\footnote{{Cf.~Eq.~(111), Fig.~10, and Sec.~V\,A in Paper I. See also \citet{Magnin2026} and \citet{Strobl2014}.}}  However, if $\Delta$ is insufficiently large for this approximation to be made, the preceding formalism still allows $K(x',y',x,y,\Delta)$ to be computed (and thence $\widetilde{\mathbf{D}}(x,y;\Delta)$ to be approximated, if the latter quantity is required).  In this context it will often be useful to adopt a particular model for the fast-phase covariance, such as that of \citet{Sinha1988}.  Also, if we are in the far field of the fast-phase fluctuations then we can consider $K(x',y',x,y,\Delta)$ to be a local differential scattering cross section (in the language of collision theory \cite{Messiah}).  Finally, while we have focused particular attention on the contribution to the covariance $\overline{\varphi_f(x',y')\varphi_f(x'',y'')}$---and by implication its associated smearing kernel $K$ and resulting diffusion field $\widetilde{\mathbf{D}}$---that is due to unresolved random microstructure within the sample,\footnote{{Cf.~Figs.~3, 6, and 10(b) in Paper I.}} this covariance will more generally be a function of both the sample microstructure and the state of coherence of the field which illuminates it.  In this sense, the scalar and tensor paraxial diffusion fields which have been our central concern constitute an effective parameter field which aggregates several different effects such as (i) local scattering cross sections driven by characteristic correlation lengths of unresolved sample microstructure, (ii) the effects of using an extended source which is spatially and temporally partially coherent, (iii) finite detector-pixel size, and (iv) the influence of noise.

}

\subsection{Noise-induced contribution to diffusion field}\label{sec:NoiseContributionToParaxialDiffusionField}

Paper I { and Sec.~\ref{sec:DiffusionFieldFromFirstPrinciples}} give models for the contribution to a paraxial diffusion field that arises from position-dependent blur due to diffuse scatter from spatially random microstructure in a sample.  Here we study a physically distinct contribution due to photon shot noise.  The analysis, which complements Sec.~V\,E of Paper I, has three parts. Section \Ref{sec:NoiseResolutionTradeoff} develops a simple form of noise-resolution tradeoff.  This is used, in Sec.~\ref{sec:NoiseInducedFokkerPlanckD}, to obtain a noise-induced contribution to the Fokker-Planck diffusion field.  Finally, Sec.~\ref{sec:NoiseInducedVisibilityReduction} develops a variant of these ideas that is applicable to the case of randomly structured illumination (cf.~Fig.~5 in Paper I, for the special case where a spatially random mask $\mathcal{M}$ is employed).

\subsubsection{Noise-resolution tradeoff}\label{sec:NoiseResolutionTradeoff}

\begin{figure}[ht!]
\centering
\includegraphics[width=0.75\columnwidth]{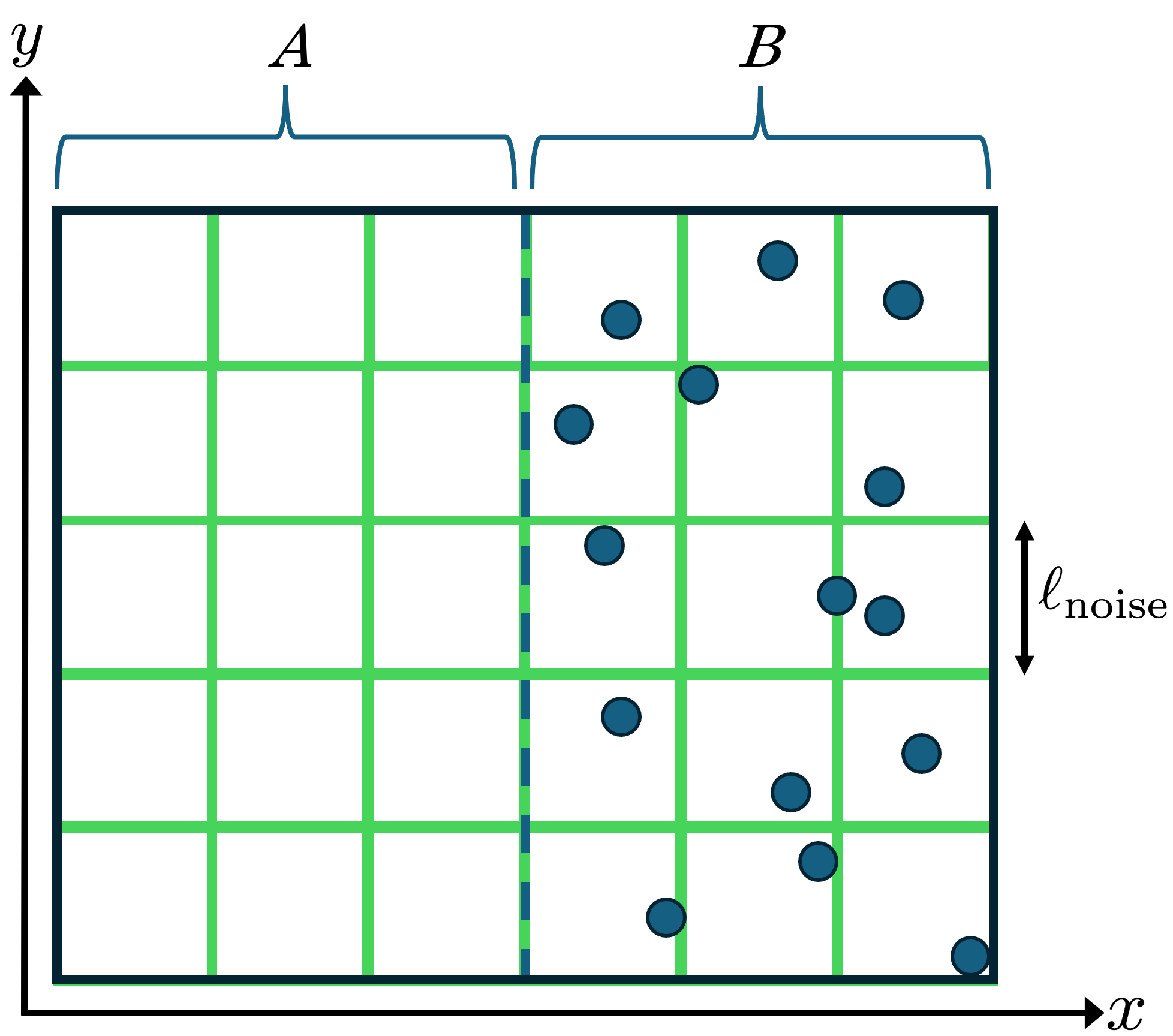}
\caption{Geometry for derivation of noise-resolution tradeoff.}
\label{Fig:NRU}
\end{figure}

Consider the detector-surface region in Fig.~\ref{Fig:NRU}.  By assumption the rectangular region $A$ is not illuminated, while the rectangular region $B$ is uniformly illuminated with a photon fluence $f$ (stated more precisely, the mean number of detected photons per unit area is $f$, assuming perfect detector efficiency).\footnote{This definition for $f$ follows p.~652 of \citet{BarrettMyersBook}.} Now imagine a fictitious square-pixel grid to be imposed over the illuminated region, as shown in green. By definition, let $\ell_{\textrm{noise}}$ be the width of the smallest ``fictitious pixel'' that can resolve the ``shadow edge'' indicated by the dashed line in Fig.~\ref{Fig:NRU}.  A rough estimate for $\ell_{\textrm{noise}}$ corresponds to a mean of one photon per pixel, hence $f \ell_{\textrm{noise}}^2=1$.  If we now consider the more general case where fluence is a function of detector-plane Cartesian coordinates $(x,y)$, $\ell_{\textrm{noise}}$ will also become position dependent.  Thus \cite{NRU==1998,NRU=D}
\begin{equation}
    \ell_{\textrm{noise}}(x,y)=\frac{1}{\sqrt{f(x,y)}}
    \label{eq:NRU1}
\end{equation}
is a position-dependent lower bound on the spatial resolution associated with a given fluence $f(x,y)$.   This result is an example of the well-known tradeoff between noise and spatial resolution, namely the fact that for fixed fluence, improvement in per-pixel signal-to-noise ratio implies coarsened spatial resolution \cite{NRU=1997,NRU==1998,NRU==2001,NRU=A,NRU=B,NRU=C,NRU=D}. From a dimensional-analysis perspective \cite{BridgmanBook,LemonsBook}, Eq.~(\ref{eq:NRU1}) is the only length scale that can be formed from the fluence alone. Moreover, Eq.~(\ref{eq:NRU1}) bears some conceptual similarity to the position-momentum uncertainty principle \cite{Messiah}, since $\ell_{\textrm{noise}}$ is a lower-bound measure of positional uncertainty and $\sqrt{f}$ is proportional to the uncertainty in transverse momentum associated with ``salt and pepper'' shot noise \cite{BarrettMyersBook}.  In particular, for two contiguous square detector areas that correspond to the same fluence $f$ [see, e.g., two neighboring green boxes in region $B$ of Fig.~\ref{Fig:NRU}], Poisson statistics implies the detected number of photons in the contiguous areas to differ by a zero-mean random number whose standard deviation is proportional to $\sqrt{f}$; this difference may be conceptualized in terms of a fluctuating transverse momentum, or in terms of a spatially random phase distribution associated with unresolved spatiotemporal speckle.

\subsubsection{Noise-induced Fokker-Planck diffusion field}\label{sec:NoiseInducedFokkerPlanckD}

We now seek an expression for the Fokker-Planck paraxial diffusion field induced by Eq.~(\ref{eq:NRU1}).  While it is somewhat artificial to do so, the noise-induced blur width $\ell_{\textrm{noise}}$ may be associated with the deflection half-angle $\theta_{\textrm{noise}}$ given by truncating Eq.~(142) in Paper I to
\begin{equation}
    \theta_{\textrm{noise}} = \tan^{-1}\left(\frac{\ell_{\textrm{noise}}}{\Delta}\right)\approx\frac{\ell_{\textrm{noise}}}{\Delta},
    \label{eq:NRU2}
\end{equation}
where $\Delta$ is the sample-to-detector distance.  Then take Eq.~(\ref{eq:DimensionlessScalarDiffusionCoefficient}), replace $\theta_s$ with $\theta_{\textrm{noise}}$, set $F$ to unity, and use Eq.~(\ref{eq:NRU2}).  Now employ Eq.~(\ref{eq:NRU1}) and then discard a factor of one half on account of the crudeness of the present analysis, to give the following expression for the noise-induced contribution to the Fokker-Planck diffusion field:
\begin{equation}
    D_{\textrm{noise}}(x,y)=\frac{1}{\Delta^2 f(x,y)}.
    \label{eq:NRU3}
\end{equation}
This is the simplest dimensionless function that can be created using the only available quantities (namely $f$ and $\Delta$), which is such that the noise-induced diffusion field is a decreasing function of increasing fluence.  

Having obtained an expression for the noise-induced diffusion field, we now examine conditions under which it is ignorable. Since the angle subtended by one physical (rather than fictitious) pixel with respect to a point on an imaged sample is paraxially equal to the width $\mathcal{W}$ of that pixel divided by the sample-to-detector distance (defocus distance) $\Delta$, the condition for a negligible contribution due to the noise-induced diffusion field is
\begin{equation}
    \theta_{\textrm{noise}}(x,y) \ll \frac{\mathcal{W}}{\Delta}.
    \label{eq:NRU4}
\end{equation}
Using Eq.~(\ref{eq:NRU2}) reduces this to the natural requirement
\begin{equation}
    \ell_{\textrm{noise}}(x,y) \ll \mathcal{W}.
    \label{eq:NRU5}
\end{equation}
Now use Eq.~(\ref{eq:NRU1}), hence
\begin{equation}
    \mathcal{W}^2 f(x,y) \gg 1.
    \label{eq:NRU6}
\end{equation}
Since the left side is equal to the mean number of photons per pixel, we conclude that {\em the noise-induced contribution to a paraxial diffusion field is negligible when the mean number of photons per pixel is much greater than unity}.  Conversely, in weak-field (low sample-dose) situations where the mean number of photons per pixel is on the order of (say) 10 or less, the noise-induced contribution to the diffusion field should be taken into account. 

In the common situation where a sample with intensity transmission function $T(x,y)$ is uniformly illuminated at normal incidence with spatially uniform fluence $f_0$, we may write the absorption-contrast relation
\begin{equation}
    f(x,y) \approx f_0 \, T(x,y), \quad 0 < T(x,y) \le 1.
    \label{eq:NRU7}
\end{equation}
Hence Eq.~(\ref{eq:NRU3}) becomes
\begin{equation}
    D_{\textrm{noise}}(x,y)=\frac{1}{\Delta^2 f_0 \, T(x,y)}.
    \label{eq:NRU8}
\end{equation}
The more attenuating the sample, the stronger the diffusion field induced by shot noise.  Lower incident fluence $f_0$ also increases the strength of this diffusion field.

\subsubsection{Noise-induced visibility-reduction field}\label{sec:NoiseInducedVisibilityReduction}

\begin{figure}[ht!]
\centering
\includegraphics[width=0.9\columnwidth]{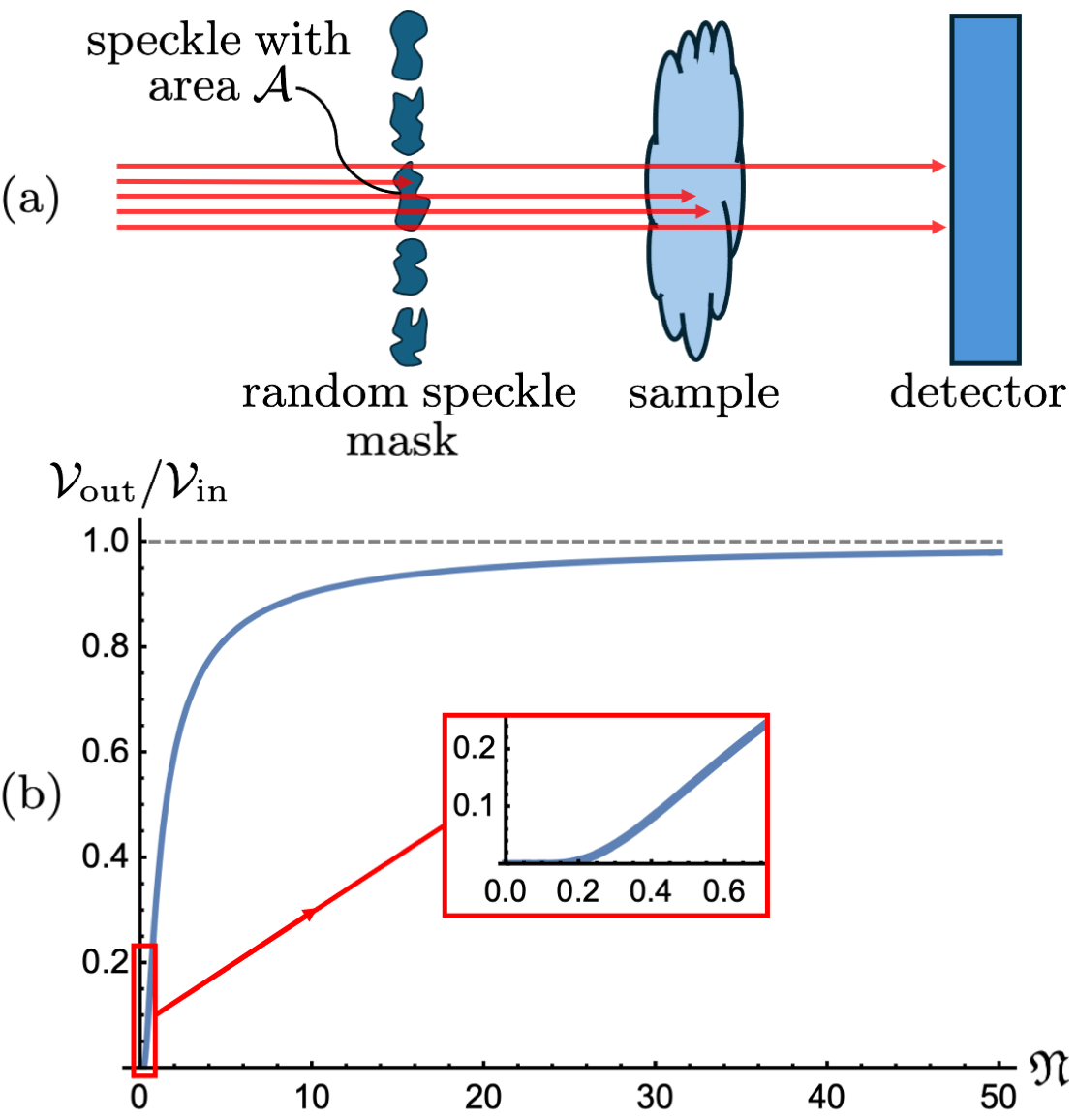}
\caption{(a) Basic scenario for ``speckle tracking'' approach to structured-illumination imaging \cite{berujon2012,Morgan2012,zdora2018,ZdoraSpeckleBook}.   (b) Plot of the noise-induced relative reduction $\mathcal{V}_{\textrm{out}}/\mathcal{V}_{\textrm{in}}$ in the visibility of a random-mask speckle, according to Eq.~(\ref{eq:NR18}), where $\mathfrak{N}$ is the number of photons that pass through both the particular speckle and the sample. The inset zooms into the region in the vicinity of the essential singularity at $\mathfrak{N}=0$ (see, e.g., p.~380 of \citet{MorseFeshbach}).}
\label{Fig:HowNoiseBlursSpeckles}
\end{figure}

In the context of random-illumination (speckle tracking \cite{berujon2012,Morgan2012}) approaches to paraxial diffusion-field retrieval \cite{zdora2018,ZdoraSpeckleBook}, such as that sketched in Fig.~\ref{Fig:HowNoiseBlursSpeckles}(a),\footnote{The basic idea underpinning speckle tracking is as follows.  A uniformly illuminated random mask creates speckles over a detector plane, which are (i) attenuated in accord with the absorption profile $T(x,y)$ of the sample, (ii) transversely displaced via the refractive properties $\phi(x,y)$ of the sample, and (iii) blurred by the presence of a diffusion field $D(x,y)$ [or $\widetilde{\mathbf{D}}(x,y)$] induced by unresolved random structure in the sample.  Speckle tracking seeks to recover $T(x,y)$ from the sample-induced attenuation of the mask speckles, in addition to recovering $\phi(x,y)$ from the sample-induced displacement of the speckles, and the diffusion field (or visibility-reduction field) from the position-dependent blur of the speckles.  The book by \citet{ZdoraSpeckleBook} reviews several methods.} let an illuminating speckle be roughly approximated as ``one bump'' of a sinusoid.\footnote{(a) While the intensity distribution of the illuminating random speckles---namely the intensity distribution immediately downstream of the random mask in Fig.~\ref{Fig:HowNoiseBlursSpeckles}(a)---is not periodic, each individual speckle is here being modeled as having an intensity distribution that is locally sinusoidal. (b) Note also that the argument that follows can be applied to other methods of phase and dark-field imaging, for example, simply replacing the speckle by a beamlet used in single-grid imaging, edge illumination, or grating interferometry \cite{endrizzi2018,MorganPaganin2019}.} Working in one transverse spatial dimension $x$, the intensity of the illuminating speckle may therefore be written as \cite{MorganPaganin2019}
\begin{equation}
I_{\textrm{in}}(x)=a \sin(x/p)+b,
    \label{eq:NRU9}
\end{equation}
where $a > 0$ is the speckle-intensity amplitude, $b \ge a$ is an additive offset, and $p$ is a measure of the speckle width.  The Michelson visibility \cite{Michelson95} of the illuminating speckle is obtained from the maximum intensity $I_{\textrm{max}}$ and the minimum intensity $I_{\textrm{min}}$ via   
\begin{equation}
\mathcal{V}_{\textrm{in}}=\frac{I_{\textrm{max}}-I_{\textrm{min}}}{I_{\textrm{max}}+I_{\textrm{min}}}=\frac{(a+b)-(b-a)}{(a+b)+(b-a)}=\frac{a}{b} \le 1.
\label{eq:NR10}
\end{equation}
The noise-induced smearing of this speckle, over a transverse length scale of $\ell_{\textrm{noise}}$, may be modeled via the intensity distribution
\begin{equation}
I_{\textrm{out}}(x)=[a \sin(x/p)+b] \otimes G(\mu=0,\sigma=\ell_{\textrm{noise}},x),
    \label{eq:NR11}
\end{equation}
where $\otimes$ denotes convolution and $G(\mu,\sigma,x)$ is a unit-area Gaussian function of $x$ with mean $\mu$ and standard deviation $\sigma$.  Evaluate the convolution integral to give
\begin{equation}
I_{\textrm{out}}(x)=a \exp\left\{-\frac{[\ell_{\textrm{noise}}(x)]^2}{2p^2}\right\}\sin\left(\frac{x}{p}\right)+b,
    \label{eq:NR12}
\end{equation}
with the speckle visibility now being reduced to
\begin{equation}
\mathcal{V}_{\textrm{out}}(x)=\exp\left\{-\frac{[\ell_{\textrm{noise}}(x)]^2}{2p^2}\right\}\frac{a}{b}.
    \label{eq:NR13}
\end{equation}
Note that the speckle width $p$ is independent of $x$, by assumption; this corresponds to the random mask in Fig.~\ref{Fig:HowNoiseBlursSpeckles}(a) being statistically spatially stationary. Take the ratio of Eqs.~(\ref{eq:NR13}) and (\ref{eq:NR10}), revert to two spatial dimensions, then use Eq.~(\ref{eq:NRU1}).  Hence\footnote{Cf.~Eq.~(132) in Paper I, whose functional form is consistent with the expression obtained by eliminating $f(x,y)$ from Eqs.~(\ref{eq:NRU3}) and (\ref{eq:NR14}) in the present paper.}
\begin{align}
\nonumber \frac{\mathcal{V}_{\textrm{out}}(x,y)}{\mathcal{V}_{\textrm{in}}(x,y)} &=\exp\left\{-\frac{[\ell_{\textrm{noise}}(x,y)]^2}{2p^2}\right\} \\ &=\exp\left[-\frac{1}{2p^2 f(x,y)}\right].
    \label{eq:NR14}
\end{align}
Now assume uniform-fluence illumination of a sample with transmission function $T(x,y)$, hence Eq.~(\ref{eq:NRU7}) gives 
\begin{equation}
\frac{\mathcal{V}_{\textrm{out}}(x,y)}{\mathcal{V}_{\textrm{in}}(x,y)}=\exp\left[-\frac{1}{2p^2 f_0 \, T(x,y)}\right].
    \label{eq:NR15}
\end{equation}
Approximate the area $\mathcal{A}$ of one random-illumination speckle as $p^2$, then discard a factor of two on account of the crudeness of our calculation, thus

\begin{equation}
\frac{\mathcal{V}_{\textrm{out}}(x,y)}{\mathcal{V}_{\textrm{in}}(x,y)}=\exp\left[-\frac{1}{\mathcal{A} f_0 \, T(x,y)}\right].
    \label{eq:NR16}
\end{equation}
Now let
\begin{equation}
n= \mathcal{A} f_0
\label{eq:NR17}
\end{equation}
be the mean number of photons that illuminate each random-mask speckle, during the exposure time needed to record the intensity image.  Hence
\begin{equation}
\frac{\mathcal{V}_{\textrm{out}}(x,y)}{\mathcal{V}_{\textrm{in}}(x,y)}=\exp\left[-\frac{1}{n \, T(x,y)}\right]=\exp\left[\frac{-1}{\mathfrak{N}(x,y)
}\right],
    \label{eq:NR18}
\end{equation}
where the dimensionless quantity
\begin{equation}
\mathfrak{N}\equiv n T     
\end{equation}
is the total number of photons transmitted through both a given speckle and the sample.  For example, as shown in Fig.~\ref{Fig:HowNoiseBlursSpeckles}(a), if $f_0$ photons per unit area illuminate a random-mask speckle with area $\mathcal{A}$, on the order of $n=\mathcal{A} f_0$ photons will impinge on the corresponding area of the sample; upon subsequently passing through a region of the sample with local intensity transmission coefficient $T$, a total of $n T \equiv\mathfrak{N}$ photons will strike the detector.  This value for $\mathfrak{N}$ is unchanged if the mask-sample sequence is reversed.

From Eq.~(\ref{eq:NR18}), which is plotted in Fig.~\ref{Fig:HowNoiseBlursSpeckles}(b), we conclude that the noise-induced visibility reduction of an illuminating speckle is only non-negligible when at most a handful of photons $\mathfrak{N}$ is transmitted through both that particular speckle and the sample that is being imaged.  For example, a noise-induced blur that alters speckle visibility by 1\% or more implies the left side of Eq.~(\ref{eq:NR18}) to be no greater than 0.99.  In this case, $nT\lesssim 100$, implying no more than 100 photons to be transmitted through any one random-mask speckle. If $\mathfrak{N} \gtrsim 5$, as will typically be the case if the illumination is not too weak or the random-mask speckles are not too absorbing, then we may make the Taylor-series approximation $\exp(
-1/\mathfrak{N})\approx 1-(1/\mathfrak{N})$.  This leads to the simple rule-of-thumb
\begin{equation}
\frac{\Delta\mathcal{V}}{\mathcal{V}_{\textrm{in}}}\times 100\%\approx\frac{100}{\mathfrak{N}} \quad\textrm{if}~~\mathfrak{N} \gtrsim 5
\label{eq:NR19}
\end{equation}
for the relative percentage reduction in mask-speckle visibility that is induced by photon shot noise, where 
\begin{equation}
\Delta\mathcal{V}\equiv \mathcal{V}_{\textrm{in}}-\mathcal{V}_{\textrm{out}}
\label{eq:NR20}
\end{equation}
is the absolute change in mask-speckle visibility.

We close this subsection with a point of clarification.  The idea that noise leads to local blurring, as explained in footnote~\ref{footnote:WashingOutFineDetailByNoise}, is consistent with the concept of a noise-induced visibility-{\em reduction} field. However, differing methods for measuring visibility from an experimental image can change how noise influences the outcome.  For example, if one takes the absolute maximum and minimum pixel values in a local region then noise will induce a local {\em increase} in visibility \cite{Chabior2011}; conversely, if one were to fit a sinusoid (or some other smooth function such as a Gaussian) to the intensity in the local region then noise will induce a local {\em decrease} in visibility \cite{GratingInterferometryDFbiasDueToNoise2017}.  Our analysis is consistent with the latter approach, but distinct from the former.

\subsection{Negative diffusion-field models}\label{sec:DiffusionFieldModels:NegativeCases}

For the visibility-reduction structured-illumination approach to diffusive dark-field imaging, in the sense explained in Secs.~III\,D and V\,B of Paper I, several articles have reported the measurement of paraxial diffusion fields whose origin is due at least in part to scattering from sharp edges \cite{Pfeiffer2008df, Strobl2008, pfeiffer2009,zanette2014,Yashiro2015} (cf.~Refs.~\cite{MorrisonBrowne1992,SuzukiUchida1995}). Similar remarks apply to edge-scatter contributions for both the scalar \cite{MIST, alloo2023SciRep,Alloo2025} and tensor \cite{MISTdirectional} Fokker-Planck diffusion fields. The visibility-reduction and Fokker-Planck viewpoints, respectively associated with the previous two sentences, may be related to one another through Eq.~(132) of Paper I. For a recent direct experimental example, of the local visibility reduction of structured-illumination intensity due to edge scatter, see the borders of the dielectric spheres in Fig.~2(b) of \citet{Groenendijk2020}. As explained by \citet{Yashiro2015}, the physical origin---of such edge-scatter local visibility reduction---is different from the contribution due to diffuse scatter from spatially unresolved sample microstructure. Indeed, this edge scatter may be variously conceptualized as due to (i) Young boundary waves in scalar-wave diffraction theory \cite{YoungOnTheBoundaryWave,Maggi,Rubinowicz,MiyamotoWolf1,MiyamotoWolf2,HannayParaxialBoundaryWave,BorghiParaxialBoundaryWave}, (ii) edge scatter associated with critical points of the second kind in the asymptotic expansion of the Fresnel diffraction integral in the presence of sharp absorptive edges \cite{MandelWolf}, or (iii) Keller diffracted rays associated with the geometrical theory of diffraction \cite{Keller}.  

The idea of a {\em negative diffusion coefficient associated with scattering from sharp sample edges in the context of coherent imaging}  has several precedents in the literature (see, e.g., Refs.~\cite{Esposito2022, Croughan2023, Esposito2023, Alloo2025}).  Following \citet{Croughan2023}, local {\em sharpening of visibility due to diffuse scatter} may be modeled as  a diffuse-scatter cone with complex apex angle.  In cases where the cone-angle $\theta_s(x,y)$ is purely imaginary, the expression for the dimensionless diffusion field in Eq.~(\ref{eq:DimensionlessScalarDiffusionCoefficient}) becomes negative.  In this context, the following terms may be taken as synonymous: ``locally negative diffusion field'', ``local deblur (negative blur)'', ``local deconvolution'', and ``local diffusion-induced sharpening ''.  Regarding this final term in particular, which might superficially appear to contain a logical contradiction, we refer to the argument centered about Fig.~4 in Ref.~\cite{MorganPaganin2019}.  There, it is shown that when a diffusion-field is spatially rapidly varying---for example, in the vicinity of a sharp sample edge---it can lead to local sharpening rather than local smearing (see also Supplementary Sec.~1 of ~\citet{Ahlers2024}).  

Additional context provided by previous work on negative diffusion in a Fokker-Planck setting \cite{NegativeDiffusion1986,NegativeDiffusion1987} encourages exploration of the possibility of negative diffusion fields associated with Fokker-Planck extensions to the TIE.  Below we consider scalar (Sec.~\ref{sec:DiffusionFieldModels:NegativeCaseScalar}) and then tensor (Sec.~\ref{sec:DiffusionFieldModels:NegativeCaseTensor}) negative diffusion fields associated with scattering from sharp sample edges, followed by an outline of how out-of-focus contrast may also lead to a negative diffusion field (Sec.~\ref{sec:DiffusionFieldModels:SingleMaterialSampleCase}).  In the context of the presented models, note that it can be challenging to achieve quantitative agreement between experimental data and models of the signal observed at a sample edge. This is due to the multiple possible contributions to the edge signal, the concentration of contrast into only a few pixels, and the potential diffusion of that very local contrast due to source size or detector blurring. 

\subsubsection{Negative diffusion field induced by edge scatter: Scalar case}\label{sec:DiffusionFieldModels:NegativeCaseScalar}

The dimensionless scalar diffusion field $D(x,y)\ge 0$ in Eq.~(24) of Paper I, resulting from both unresolved sample microstructure and source-size blur, may be augmented by a negative diffusion field associated with scattering from sharp edges.  Hence we make the replacement 
\begin{equation}\label{eq:NegativeDiffusion01}
D(x,y)\longrightarrow D(x,y)-  D_{\textrm{edge}}(x,y;\Delta,\cdots),
\end{equation}
where $D_{\textrm{edge}}(x,y;\Delta,\cdots)\ge 0$, $\Delta$ is the defocus, and the ellipsis indicates additional relevant parameters. In particular, the functional form of the negative diffusion field will in general depend upon the sharpness of the transverse intensity gradients, and therefore upon the appropriate model for the associated edges. 
If more than one edge model is needed in a particular context, then more than one negative-diffusion field may be required.  For example, if two diffusion fields are needed to model two different classes of sharp edge, then we may write
\begin{equation}\label{eq:NegativeDiffusion02}
D(x,y)\longrightarrow D(x,y)- \sum_{j=1}^{2}D^{(j)}_{\textrm{edge}}(x,y;\Delta,\cdots),
\end{equation}
where $D(x,y)\ge 0$ and $D^{(j)}_{\textrm{edge}}(x,y;\Delta,\cdots)\ge 0$.

\begin{figure}[ht!]
\centering
\includegraphics[width=0.9\columnwidth]{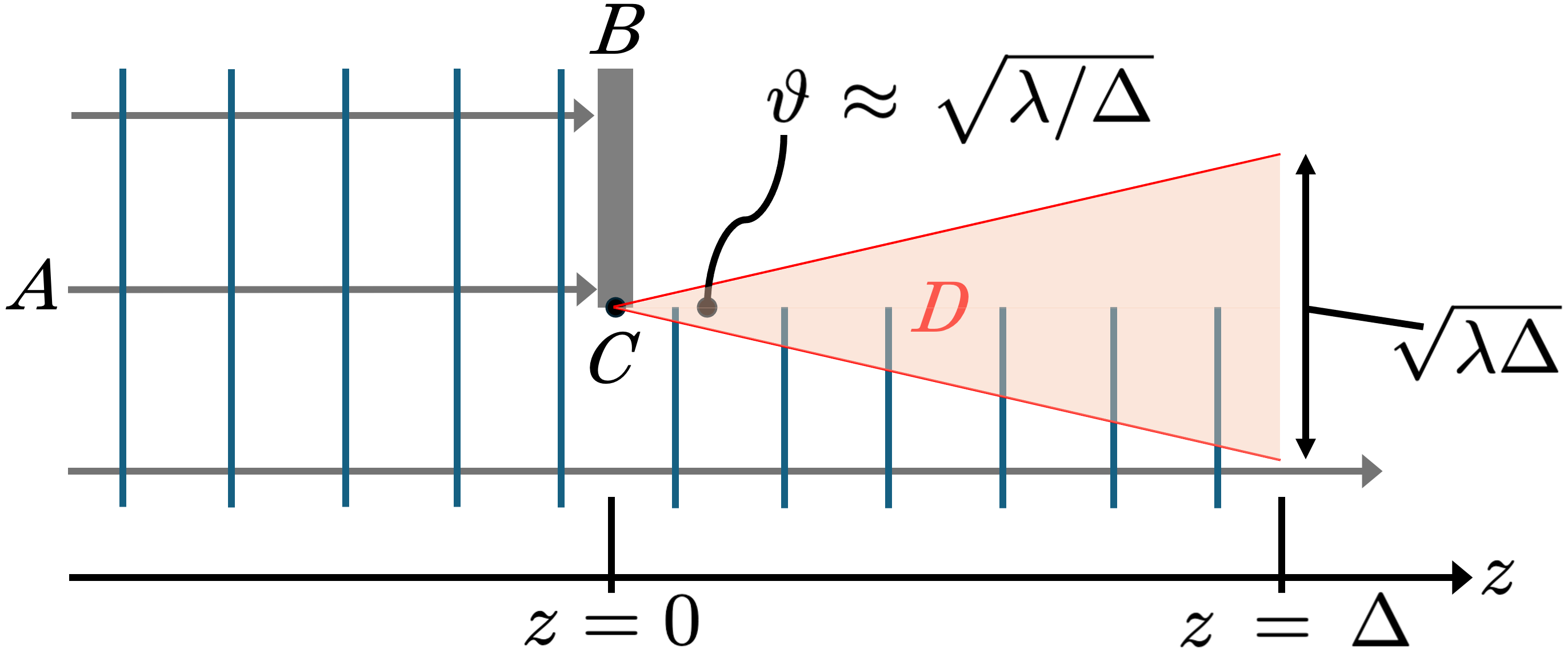}
\caption{Fresnel diffraction of normally-incident monochromatic plane waves $A$ by a thin opaque screen $B$ (side view).  The edge $C$ of the screen gives a scattered wave $D$ whose width is on the order of $\sqrt{\lambda\Delta}$, upon propagating a distance $z = \Delta \ge 0$ downstream of $C$. The corresponding angular width of $D$ is $\vartheta \approx \sqrt{\lambda/\Delta}$.   If $\sqrt{\lambda\Delta}$ is small relative to the width of the field of view of the ``defocused'' intensity distribution over the plane $z=\Delta$, the narrow primary Fresnel diffraction fringe may be viewed as locally sharpening the image of the screen's edge.}
\label{Fig:FirstFresnelZone}
\end{figure}

A simple negative-diffusion model recalls the theory of Fresnel diffraction of normally incident plane waves with wavelength $\lambda$, from a perfectly sharp straight edge of an opaque screen: see Fig.~\ref{Fig:FirstFresnelZone}.  The width of the Fresnel diffraction peak, in the vicinity of the sharp edge, is on the order of the width $\sqrt{\lambda\Delta}$ of the first Fresnel zone \cite{CosslettNixonBook, GureyevRulesOfThumb2008}.\footnote{Importantly, this corresponds to normal rather than anomalous diffusion, in the sense explained in Sec.~II of Paper I. As shall be seen in a moment, this is related to the introduction of the $\Delta$ dependence of $D_{\textrm{edge}}(x,y;\Delta,\cdots)$ in Eq.~(\ref{eq:NegativeDiffusion01}).}  The associated angular width for the fan of edge-scatter radiation is $\sqrt{\lambda/\Delta}$.  Now set $F$ to roughly unity in Eq.~(\ref{eq:DimensionlessScalarDiffusionCoefficient}), on account of the fact that, for small $\Delta$, edge scatter dominates the defocused intensity in the vicinity of the sharp scattering edge.  This gives the rough estimate\footnote{Since we here model edges as perfectly sharp, there is no length scale associated with the width of such edges.  From a dimensional-analysis perspective \cite{BridgmanBook,LemonsBook}, the only available nondimensionless parameters are $\lambda$ and $\Delta$; since both have units of length, and the left side of Eq.~(\ref{eq:NegativeDiffusion03}) is dimensionless, only the ratio $\lambda/\Delta$ (or some power $(\lambda/\Delta)^p$ or function $q(\lambda/\Delta)$ thereof) may appear on the right side of this equation. Also, since $\lambda$ must explicitly influence this expression since its physical underpinnings pertain to diffraction, $p$ cannot be zero and $q$ cannot be constant. Hence $\Delta$ must appear on the right side of our equation, which is why functional $\Delta$ dependence was included on the left side.  From a physical perspective, and for the particular negative-diffusion model considered here, the $\Delta$ dependence on the right side of Eq.~(\ref{eq:NegativeDiffusion03}) relates to the transition from ballistic superdiffusion to normal diffusion (see Sec.~II of Paper I); cf.~footnote \ref{footnote:SoftEdgeNegativeDiffusionField}. \label{footnote:HardEdgeNegativeDiffusionField}}
\begin{align}
\nonumber
D^{(1)}_{\textrm{edge}}(x,y;\Delta)
&\approx \frac{1}{2}\times 1 \times \left(\sqrt{\frac{\lambda}{\Delta}}\right)^2 \mathscr{I}_1(x,y)
\\ &\approx\frac{\lambda}{\Delta} \mathscr{I}_1(x,y),
\label{eq:NegativeDiffusion03}
\end{align}
where $\mathscr{I}_1(x,y) \ge 0$ is a dimensionless indicator function that is (i) defined to be on the order of unity in the vicinity of sharp edges where the infocus intensity $I(x,y,z=0)$ changes abruptly from zero to unity, and (ii) equal to zero elsewhere; also, a factor of one half has been discarded in the final line, by absorbing it into the indicator function. Note that this unusual means of framing small-distance Fresnel diffraction, in terms of a negative diffusion coefficient, may be justified by the fact that boundary-wave scatter sharpens rather than blurs the intensity distribution of the edge (in the immediate transverse vicinity of that edge, for propagation distances $\Delta$ that are sufficiently small), with the corresponding transverse length scale corresponding to the width $\sqrt{\lambda\Delta}$ of the first Fresnel zone.  For similar reasons, the transverse thickness of the indicator function will also be on the order of $\sqrt{\lambda\Delta}$ (cf.~the notion of a Fresnel volume, on p.~14 of \citet{KravtsovOrlovBook}).  Also, via a slight extension in the preceding chain of logic---for example in the presence of sharp transverse phase gradients, or sharp transverse intensity gradients where the intensity rapidly changes from one nonzero value to a different nonzero value---$\mathscr{I}_1(x,y)$ may be viewed as a dimensionless sharp-edge negative-diffusion field whose magnitude varies between zero and a number on the order of unity.  

A different class of negative-diffusion field may be considered, by returning to the second and third lines of the extended Fokker-Planck equation [Eq.~(\ref{eq:SecIII--ExtendedFPE})].  Let $\ell_{\textrm{edge}} \gg \lambda$ be the width over which a soft absorptive edge has its associated intensity transmission function vary from its minimum to its maximum value.  Dimensional analysis then gives the associated soft-edge negative-diffusion field 
\begin{equation}
\label{eq:NegativeDiffusion04}
D^{(2)}_{\textrm{edge}}(x,y;\ell_{\textrm{edge}})
\approx \left(\frac{\lambda}{\ell_{\textrm{edge}}}\right)^2 \mathscr{I}_2(x,y),
\end{equation}
where $\mathscr{I}_2(x,y)$ is a dimensionless soft-edge negative-diffusion indicator function that is not necessarily on the order of unity [unlike $\mathscr{I}_1(x,y)$, which is typically on the order of unity in regions where it is not negligible].\footnote{The lack of $\Delta$ dependence in Eq.~(\ref{eq:NegativeDiffusion04}) is consistent with ballistic superdiffusion (Paper I, Sec.~II). 
 Cf.~the end of footnote \ref{footnote:HardEdgeNegativeDiffusionField}.\label{footnote:SoftEdgeNegativeDiffusionField}}

When the sharp-edge and soft-edge negative-diffusion fields of the preceding two paragraphs are employed for the replacement in Eq.~(\ref{eq:NegativeDiffusion02}), additional edge-related terms are generated in the Fokker-Planck and extended Fokker-Planck generalizations of the TIE.  Restricting ourselves to one example, the $F \ll 1$ special case of Eq.~(\ref{eq:SecIII--FPE-scalar-case}) leads to
\begin{align}
\nonumber I(x,y, &z=\Delta) =I(x,y,z=0) 
\\  \nonumber &-\frac{\lambda\Delta}{2\pi} \nabla_{\perp}\cdot [I(x,y,z=0) \nabla_{\perp}\phi(x,y)]
\\ &+\Delta^2\nabla_{\perp}^2[D(x,y) I(x,y,z=0)]
\label{eq:NegativeDiffusion05a}
\\ \nonumber &-\lambda\Delta\nabla_{\perp}^2[ \mathscr{I}_1(x,y) I(x,y,z=0)]
\\ \nonumber &-\left(\frac{\lambda\Delta}{\ell_{\textrm{edge}}}\right)^2 \nabla_{\perp}^2[\mathscr{I}_2(x,y) I(x,y,z=0)].
\end{align}

The preceding rough calculations are broadly consistent with the Fokker-Planck analysis in \citet{Beltran2023}, where the discarding of negative components of a recovered paraxial diffusion field was seen to suppress contributions due to edge scatter and thereby emphasize the contribution due to unresolved sample microstructure. Such clipping of a recovered paraxial diffusion field to its nonnegative component, in order to separate bulk-microstructure scatter from edge scatter, has also been clearly observed in separate investigations by \citet{Croughan2023} and \citet{Alloo2025}.  More generally, the final two terms on the right side of Eq.~(\ref{eq:NegativeDiffusion05a}) can be viewed as modeling ``left over'' contrast that is not adequately modeled by the first three terms on the right side.  Under this view, the indicator functions $\mathscr{I}_1(x,y)$ and $\mathscr{I}_2(x,y)$ are generic corrections, of order $\Delta$ and $\Delta^2$ respectively, associated with the failure of the other terms [on the right side of Eq.~(\ref{eq:NegativeDiffusion05a})] to model $I(x,y,z=\Delta)$ in an accurate manner. Such indicator functions are of particular relevance to sharp sample edges since it is difficult to correctly model a very high contrast fringe that is sampled over only a few pixels and with some detector point spread function.  Note that the nature of these correction terms will in general differ depending on whether one uses a forward-evolution Fokker-Planck approach (as employed in the present paper) or a reverse-evolution approach (see \citet{Beltran2023} and \citet{Alloo2025}).

We close this section by pointing out that a more systematic means for obtaining a series of negative-diffusion terms associated with edge-induced scatter---each of which depend on a different power of $\Delta$---is to consider the role of critical points of the second kind in the asymptotic expansion of double integrals for Fresnel diffraction (see, e.g., Sec.~3.3.3 of \citet{MandelWolf}).

\subsubsection{Negative diffusion field induced by edge scatter: Tensor case}\label{sec:DiffusionFieldModels:NegativeCaseTensor}

In structured-illumination experiments that recover the Fokker-Planck diffusion tensor \cite{MISTdirectional}, it is evident that the principal axis of the corresponding diffusion ellipse (see Fig.~9 in Paper I) has an inclination angle $\psi(x,y)$ that makes its long axis perpendicular to sharp sample edges.\footnote{Recall footnote \ref{footnote:Disambiguate-psi-notation}, regarding our dual use of the symbol $\psi$.} This phenomenon is evident in the experimental reconstruction from Fig.~3(c) of \citet{MISTdirectional}; see, also, the very closely related earlier examples in Fig.~2(d) of \citet{jensen2010a} and Fig.~4(f) of \citet{jensen2010b}.

\begin{figure}[ht!]
\centering
\includegraphics[width=0.9\columnwidth]{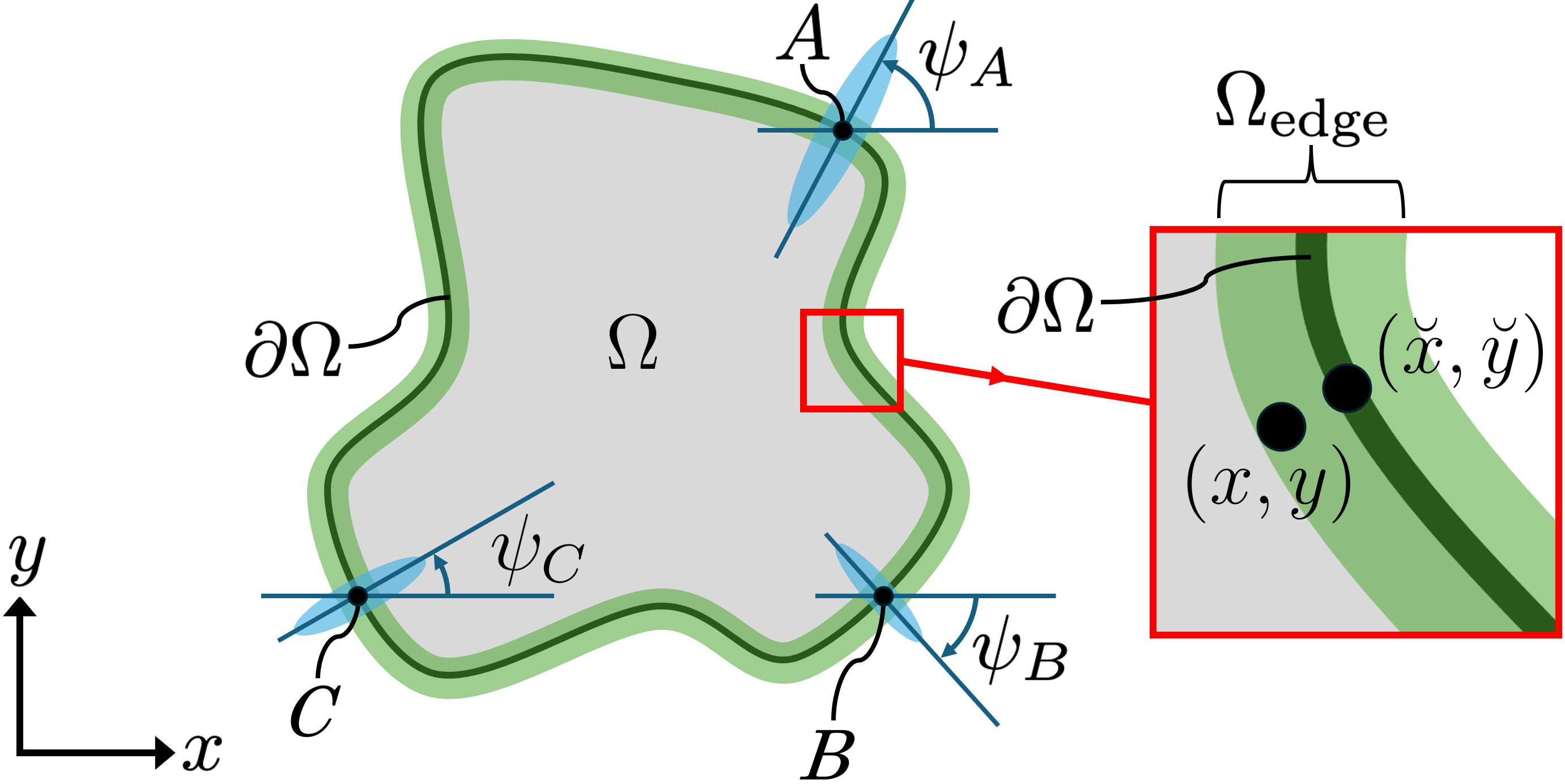}
\caption{Diffusion-field ellipses associated with edge scatter.}
\label{Fig:TangentDiffusionEllipses}
\end{figure}

To help develop a preliminary model for diffusion-field ellipses induced by sharp-edge scatter, consider Fig.~\ref{Fig:TangentDiffusionEllipses}.  Here, a solid sample with a smooth surface is shown projected onto the $(x,y)$ plane which is transverse to the optical axis of $z$-directed illumination (not shown).  The shaded region $\Omega$ denotes the interior of the $z$-projected sample, with smooth boundary $\partial\Omega$.  Diffusion-field ellipses are shown in blue, for three locations $A$, $B$, and $C$ on $\partial\Omega$; the size of these ellipses is exaggerated for clarity, since they are in fact constrained to lie within the region $\Omega_{\textrm{edge}}$ shaded in green, which by definition corresponds to the set of points where the diffusion field associated with sharp-edge scatter is non-negligible.

It is beyond the scope of this paper to derive particular models for edge-scatter diffusion tensor fields, e.g., using the geometrical theory of diffraction \cite{Keller} or the method of stationary phase for two-dimensional diffraction integrals \cite{MandelWolf}.  In lieu of such a formal development, take Eq.~(\ref{eq:TensorDiffusionFieldRetrieval9}), make the approximation $\theta_1(x,y) \approx \sqrt{\lambda/\Delta}$ (see Fig.~\ref{Fig:FirstFresnelZone} from this paper and Fig.~9 from Paper I) and then ignore the influence of $\theta_2(x,y)$. Next, replace $\tfrac{1}{2}F(x,y)$ with a dimensionless indicator function $-\mathscr{I}_3(x,y)$ that is only significantly different from zero in the vicinity of a scattering edge (as shaded in green in Fig.~\ref{Fig:TangentDiffusionEllipses}).  This estimates the sharp-edge negative-diffusion tensor field as the symmetric matrix function
\begin{equation}
\label{eq:NegativeDiffusion05}
    \widetilde{\mathbf{D}}_{\textrm{edge}}(x,y) \approx 
    -\frac{\lambda}{\Delta} \mathscr{I}_3(x,y)
\mathcal{R}_{\psi(\breve{x},\breve{y})} 
\begin{pmatrix}
1 & 0\\
0 & 0
\end{pmatrix}
[\mathcal{R}_{\psi(\breve{x},\breve{y})}]^T,
\end{equation}
where $\psi(\breve{x},\breve{y}) \in (-\pi/2,\pi/2]$ is the angle that the major axis of each edge-scatter ellipse makes with the positive $x$ axis, $\mathcal{R}_{\psi(x,y)}$ is the rotation-matrix field defined in Eq.~(\ref{eq:TensorDiffusionFieldRetrieval10}), $(\breve{x},\breve{y})$ denotes spatial coordinates of the point on the scattering edge $\partial\Omega$ that is closest to $(x,y)$ (see Fig.~\ref{Fig:TangentDiffusionEllipses}), and a superscript $T$ means matrix transposition.

\subsubsection{Negative diffusion field: Case for a single-material sample }\label{sec:DiffusionFieldModels:SingleMaterialSampleCase}

Our preceding discussions, relating to negative diffusion fields in the context of Fokker-Planck extensions to the TIE, all consider scattering from sharp sample edges.  Here we consider a distinct mechanism for negative diffusion, which may be associated with out-of-focus contrast \cite{Zernike1942,Bremmer1952,Cowley1959,CowleyBook} of a single-material sample \cite{paganin2002}.

Consider the simple special case of a thin sample that is {\em composed of a single material} described by a fixed nonzero linear attenuation coefficient $\mu$ and fixed refractive index 
\begin{equation}
\label{eq:RefractiveIndexDecrement}
n=1+\delta     
\end{equation}
that is not equal to unity.\footnote{The following formalism is also applicable if the single material is of variable density.  In this more general case, both $\delta$ and $\mu$ will be functions of position at any point within the sample, but the ratio of these quantities will be constant. Note, further, that Eq.~(\ref{eq:RefractiveIndexDecrement}) in the present paper repeats the first half of Eq.~(74) in Paper I, without the ``$\vert\delta\vert\ll 1$'' restriction in the second half of the latter equation.  Cf.~footnote 11 in Paper I.}  Let the projected thickness of the sample be denoted by $T_{\textrm{proj}}(x,y)$, with this projection being taken along the direction of the optical axis of a paraxial imaging system.  Assuming normally-incident near-monochromatic illumination by a plane wave with uniform intensity $I_0$, with the thin sample being placed immediately upstream of the plane $z=0$, the intensity at the exit surface of the sample is given by the Beer--Lambert law as
\begin{equation}
\label{eq:BeerLawSingleMaterialSample}
    I(x,y,z=0)=I_0\exp[-\mu T_{\textrm{proj}}(x,y)].
\end{equation}
Under the projection approximation \cite{Paganin2006} whereby local ray paths within the sample are well approximated by straight-line trajectories that are parallel to the optical axis $z$, the phase $\phi(x,y)$ of the field---over the sample exit surface in the plane $z=0$---is approximated by \cite{paganin2002}
\begin{equation}
\label{eq:ProjectionApproximationForSingleMaterialSample}
    \phi(x,y)= k \delta T_{\textrm{proj}}(x,y).
\end{equation}
Substitute the preceding two equations  into Eq.~(\ref{eq:SecIII--FPE-scalar-case}) and then make a weak-diffuse-scatter approximation by replacing $1-F(x,y)$ with unity.  Hence, without any further approximation, we obtain [Ref.~\cite{PaganinMorgan2019}, Eq.~(59)]\footnote{Note that a non-dimensionless form for the diffusion field is employed in Ref.~\cite{PaganinMorgan2019}.  Moreover, the quantity $\delta$ in that x-ray paper is the negative of the $\delta$ employed in the present paper.  This difference in notation arises from the fact that refractive indices for hard x rays are typically (slightly) less than unity, whereas for visible light refractive indices are typically greater than unity.}
\begin{align}
\nonumber I(&x,y,z=\Delta) =I(x,y,z=0) 
\\ &+\Delta^2\nabla_{\perp}^2[\breve{D}(x,y;\Delta) I(x,y,z=0)], \quad F(x,y)\ll 1,
\label{eq:NegativeDiffusionFieldsingleMaterialA}
\end{align}
wherein we have introduced the dimensionless effective diffusion field
\begin{equation}
    \breve{D}(x,y;\Delta)=D(x,y)+\frac{\delta}{\mu \Delta}.
\label{eq:NegativeDiffusionFieldsingleMaterialB}
\end{equation}

Equation~(\ref{eq:NegativeDiffusionFieldsingleMaterialA}) has the mathematical form of a finite-difference approximation to the ordinary diffusion equation, with the important difference that {\em the associated effective diffusion coefficient may be either positive or negative}.  In particular, if the blurring influence of the diffusion field $D(x,y)$ is negligible, this term may be dropped to leave the constant {\em defocus-induced diffusion field}
\begin{equation}
    \breve{D}(x,y; \Delta)\longrightarrow\frac{\delta}{\mu \Delta}.
\label{eq:NegativeDiffusionFieldsingleMaterialC}
\end{equation}
If we exclude the case of optically-pumped active media under conditions of population inversion \cite{LoudonBook}, the linear attenuation coefficient $\mu$ will be positive.  However, $\delta/\Delta$ may be either positive or negative.  For example, for visible-light imaging, $\delta$ will typically be positive but $\Delta$ can be made negative via an imaging system that can access negative defoci $\Delta$.  As another example, for hard-x-ray imaging, $\delta$ will typically be negative (and much less than unity in magnitude),\footnote{See, e.g., p.~114 of Ref.~\cite{Paganin2006}.} with positive $\Delta$ being accessible via free-space propagation (from the exit surface of the sample, to the entrance surface of a position-sensitive detector).  Under such conditions, the defocus-induced diffusion field in Eq.~(\ref{eq:NegativeDiffusionFieldsingleMaterialC}) will be negative.    

The uniform negative diffusion field corresponding to the ``negative $\delta/\Delta$'' case of Eq.~(\ref{eq:NegativeDiffusionFieldsingleMaterialC}) is possible because the single-material assumption [as quantified in Eqs.~(\ref{eq:BeerLawSingleMaterialSample}) and (\ref{eq:ProjectionApproximationForSingleMaterialSample})] enables the phase $\phi(x,y)$ over the sample's exit surface to be written in terms of the intensity $I(x,y,z=0)$ over the same surface.  This formally eliminates phase from the finite-difference approximation to the Fokker-Planck generalization of the TIE in the form given by Eq.~(\ref{eq:NegativeDiffusionFieldsingleMaterialA}). The associated conceptual adjustment---from (i) viewing transverse phase gradients as associated with defocus-induced transverse intensity flow to (ii) instead viewing transverse intensity gradients as associated with such flow---recasts out-of-focus phase contrast in entirely diffusive terms. For indicative experimental examples, of the sharpening effects of the negative diffusion field associated with out-of-focus contrast for single-material samples, see, e.g., the visible-light images in \citet{Poola2017}, the electron-microscopy image in \citet{GaborOldTEMBook}, the x-ray images in \citet{Snigirev}, and the neutron images in \citet{Allman2000}.

It is also interesting to consider how the image-blurring effects of an extended chaotic source may be balanced by the image-sharpening effects due to out-of-focus contrast and its associated negative diffusion field, for the case of a thin single-material sample. Such a balance has previously been studied by \citet{DeblurByDefocus,Gureyev2026} and \citet{Beltran2018}. By assumption, let us here ignore the blurring influence of a nonzero sample-induced diffuse-flow fraction by setting $F$ to zero in Eq.~(24) from Paper I.  The associated source-blur uniform diffusion field is given by $D=\tfrac{1}{2}\theta_0^2$, where $\theta_0$ is the half-angle subtended by the illuminating source at the entrance surface of the sample (see Fig.~2 in Paper I).  Hence Eq.~(\ref{eq:NegativeDiffusionFieldsingleMaterialB}) becomes
\begin{equation}
\breve{D}(\Delta)= \frac{1}{2}\theta_0^2+\frac{\delta}{\mu \Delta}.
\label{eq:NegativeDiffusionFieldsingleMaterialD}
\end{equation}
Depending on the defocus $\Delta$, $\breve{D}$ can assume negative or positive values.  Negative values for $\breve{D}$ correspond to defoci $\Delta$ which are such that the image-sharpening effects of out-of-focus contrast are stronger than the image-blurring effects of a nonzero source size.  Conversely, positive values for $\breve{D}$ correspond to defoci where the image-sharpening effects of out-of-focus contrast are weaker than the image-blurring effects of a nonzero source size.  A zero value for $\breve{D}$ is attained for $\Delta$ equal to $-2\delta/(\mu\theta_0^2)$; this reproduces a result reported in \citet{DeblurByDefocus}, wherein the previously mentioned effects---namely out-of-focus sharpening and source-size blur (or, more generally, the blurring effects of an imaging system's point spread function (PSF))---cancel one another.\footnote{This cancellation when $\Delta=-2\delta/(\mu\theta_0^2)$ may be reframed by following Fig.~2 in Paper I to set $\sigma_{\Delta}=\theta_0\Delta$, where $\sigma_{\Delta}$ is the standard deviation associated with {\em any} position-invariant PSF (i.e., a PSF which is not necessarily associated with source-size blur). Hence the cancellation of blur and out-of-focus contrast occurs when $\Delta=-2\delta/[\mu(\sigma_{\Delta}/\Delta)^2]$, namely when $\Delta=-\mu\sigma_{\Delta}^2/(2\delta)$.  Now let $\mu=2k\beta$ (see p.~74 of Ref.~\cite{Paganin2006}), where $\beta$ is the imaginary part of the complexified refractive index $n=1+\delta+i\beta$ of the single-material sample [cf.~Eq.~(\ref{eq:RefractiveIndexDecrement})], and $k=2\pi/\lambda$ is the wavenumber corresponding to wavelength $\lambda$.  Hence the cancellation occurs when $\Delta=-k\beta\sigma_{\Delta}^2/\delta$, in agreement with the result of \citet{DeblurByDefocus,Gureyev2026}.  Note that the sign convention for $\delta$, in the just-cited papers, is the negative of that employed in Eq.~(\ref{eq:RefractiveIndexDecrement}).}

{ 

\section{Practical aspects of phase retrieval and diffusion-field retrieval}\label{secs:PracticalAspects}

As pointed out earlier in the paper, dimensionless scalar diffusion fields $D(x,y)$ can be recovered by employing the same focal-series datasets previously employed for TIE-based phase retrieval. Here we discuss several practical aspects regarding this inverse problem.  These are reconstruction uniqueness (Sec.~\ref{subsec:Practicalities1}), reconstruction stability (Sec.~\ref{subsec:Practicalities2}), noise sensitivity (Sec.~\ref{subsec:Practicalities3}), the required number of focal-series intensity measurements (Sec.~\ref{subsec:Practicalities4}), 
conditions under which the influence of higher-order TIE terms in our equations is ignorable (Sec.~\ref{subsec:Practicalities6}), and conditions under which the influence of the diffusion-field term is ignorable (Sec.~\ref{subsec:Practicalities7}). 

\subsection{Uniqueness of phase and diffusion-field reconstructions}\label{subsec:Practicalities1} 

Our preceding formulae for phase retrieval and diffusion-field retrieval give {\em particular} solutions to the associated inverse problems based on our Fokker-Planck and extended Fokker-Planck formalisms, but are these solutions {\em unique}?  In the following two subsections we consider, respectively, reconstruction uniqueness for (i) phase retrieval in the presence of a nonzero diffusion field $D(x,y)$ or $\mathbf{D}(x,y)$ and (ii) diffusion-field retrieval in the presence of a nonzero phase field $\phi(x,y)$.

\subsubsection{Uniqueness of phase retrieval in the presence of a diffusion field}

As a special case of the general formalism for elliptic second-order partial differential equations given on pp.~177-180 of the book by \citet{GilbargTrudingerBook}, \citet{Gureyev1995} showed that the TIE [Eq.~(\ref{eq:TIE})] may be uniquely solved for the smooth phase $\phi(x,y,z)$ of a monochromatic paraxial scalar wave field over some plane of constant $z$, when (i) the intensity distribution is strictly positive over some simply-connected field of view\footnote{{As noted by \citet{Gureyev1995}, strictly positive intensity maps are a key assumption since they exclude the possibility of screw-type tears in the wave front associated with phase vortices \cite{Dirac1931, Nye1999}.  Under some rather special circumstances---namely, certain intensity distributions with a high degree of symmetry---vortices can lead to nonuniqueness in TIE-based phase recovery \cite{Gori1993,Gureyev1995}.}} and (ii) Dirichlet boundary conditions are specified.  For normally-incident plane waves that impinge upon a compact object that is entirely contained within a simply connected field of view (FOV), zero Dirichlet boundary conditions---namely the fact that the phase can be set to zero at each point on the boundary of the FOV---can be employed.  Also, for a rectangular field of view, zero Dirichlet boundary conditions are a special case of the periodic boundary conditions implicit in FFT implementations of the method of \citet{paganin1998}.  Since Eq.~(\ref{eq:PhaseRetrieval10}) of the present paper shows the Paganin-Nugent phase retrieval method to be {\em unchanged by the presence of a diffusion field when symmetrically overfocused and underfocused images are employed} for the purposes of TIE-based phase retrieval, we can invoke the previously published uniqueness results and thereby conclude that the recovered phase is unique provided that the previously specified conditions (strictly positive intensity, compact object, use of symmetrically defocused images, etc.) are met.  If the symmetric defocus condition is dropped, we may instead consider the elliptic partial differential equation for the phase in Eq.~(\ref{eq:PhaseRetrieval16}); since this is mathematically identical in form to the TIE, albeit with a different known data function (that is derived from the Fokker-Planck extension to the TIE rather than from the TIE itself), we can again leverage the findings of \citet{Gureyev1995} to conclude that the recovered phase is unique.  If structured illumination is employed, similar reasoning applied to Eq.~(\ref{eq:TIE-like-form-for-speckle-tracking}) again shows the reconstructed phase to be unique.

\subsubsection{Uniqueness of diffusion-field retrieval in the presence of a phase field}\label{subsec:uniquenessOfTensorDiffusionFieldRetrieval}

{\em Scalar diffusion-field retrieval using symmetrically defocused images}. This case is studied in Sec.~\ref{sec:DiffusionFieldRetrievalSymmDefocus}. The relevant expressions are Eqs.~(\ref{eq:PoissonEqnForD}) and (\ref{eq:PoissonEqnForD==generalized}), wherein all functions with the exception of the scalar diffusion field $D(x,y)$ are known.  For a compact object that is entirely contained within the FOV $\Omega$ over which intensity measurements are taken, Eq.~(24) of Paper I shows that the values taken by $D(x,y)$ over the boundary $\partial\Omega$ of the FOV are equal to $\tfrac{1}{2}\theta_0^2$, where $\theta_0$ is the half-angle subtended by an on-axis chaotic source at each point on the entrance surface of the sample (see Figs.~3 and 4 in Paper I). Since the infocus intensity distribution $I(x,y,z=0)$ is known, the term in square brackets in Eqs.~(\ref{eq:PoissonEqnForD}) or (\ref{eq:PoissonEqnForD==generalized}) have known values of  $\tfrac{1}{2}\theta_0^2 I(x_{\partial\Omega},y_{\partial\Omega},z=0)$ at every point $(x_{\partial\Omega},y_{\partial\Omega})$ on the FOV boundary. Hence solving Eqs.~(\ref{eq:PoissonEqnForD}) or (\ref{eq:PoissonEqnForD==generalized}) for the term $D(x,y)I(x,y,z=0)$ in square brackets amounts to solving a Poisson equation with known Dirichet boundary conditions.  This solution is unique \cite{JacksonElectrodynamicsBook}.  Provided that the intensity is again taken to be strictly positive at every point in the FOV, we may then divide this intensity out to give a unique reconstruction for the scalar diffusion field $D(x,y)$ [see Eq.~(\ref{eq:FormalSolutionDiffusionFieldRetrievalSymmetricDefocus})].  

{\em Scalar diffusion-field retrieval using asymmetrically defocused images}. This case is studied in Sec.~\ref{sec:DiffusionFieldRetrievalAsymmDefocus}.  Only the known data functions change (in comparison to the data functions associated with symmetric defoci).  Hence similar considerations, to those given above, demonstrate the uniqueness of the reconstructed scalar diffusion field. 

{\em Scalar diffusion-field retrieval using structured illumination}. This case is studied in the first half of Sec.~\ref{sec:DiffusionFieldRetrievalStructuredIllumination}.  We may trivially adapt the preceding arguments to Eqs.~(\ref{eq:YetAnotherPoissonEquation1})-(\ref{eq:YetAnotherPoissonEquation3}), to see once again that the reconstructed scalar diffusion field $D(x,y)$ is unique. 

{\em Tensor diffusion-field retrieval using structured illumination}. This case is studied in the second half of Sec.~\ref{sec:DiffusionFieldRetrievalStructuredIllumination}. As explained there, varying either or both of the mask transverse position $j$ and the defocus $\Delta$ will lead to known linear combinations (with known position-dependent coefficients) of the seven unknown functions 
\begin{subequations}\label{eq:SevenDiffusionRelatedFunctions}
\begin{align}
Q_1(x,y) &\equiv \frac{\partial}{\partial x} \widetilde{D}_{xx}(x,y), \\
Q_2(x,y) &\equiv \frac{\partial}{\partial x}\widetilde{D}_{xy}(x,y), \\
Q_3(x,y) &\equiv \frac{\partial}{\partial y}\widetilde{D}_{yy}(x,y), \\
Q_4(x,y) &\equiv \frac{\partial}{\partial y}\widetilde{D}_{xy}(x,y), \\
Q_5(x,y) &\equiv \widetilde{D}_{xx}(x,y), \\
Q_6(x,y) &\equiv \widetilde{D}_{yy}(x,y), \\
Q_7(x,y) &\equiv \widetilde{D}_{xy}(x,y)    
\end{align}
\end{subequations}
corresponding to the three independent components of the second-rank symmetric diffusion tensor field $\widetilde{\mathbf{D}}(x,y)$ [see Eq.~(\ref{eq:FancyEquationForDiffusionTensor})]. For the purposes of demonstrating reconstruction uniqueness, it is convenient to consider a different algorithm to that given in Sec.~\ref{sec:DiffusionFieldRetrievalStructuredIllumination}. Suppose seven different states of the mask-plus-sample imaging system have been constructed, corresponding (say) to seven different mask positions $j$ that all have the same defocus $\Delta$.  Write the associated special cases of Eq.~(\ref{sec:TensorDiffusionFieldRetrievalWith
Masks}) as
\begin{equation}\label{eq:sevenLinearEquationsA}
\mathbf{M}(x,y)\mathbf{Q}(x,y)=\mathbf{G}(x,y) 
\end{equation}
where $\mathbf{M}(x,y)$ is a known $7 \times 7$ matrix function, 
\begin{equation}
\mathbf{Q}(x,y)\equiv(Q_1(x,y),Q_2(x,y),\cdots,Q_7(x,y))^T
\end{equation}
is a 7-component column vector, a superscript $T$ denotes matrix (or vector) transposition, and the 7-component column vector $\mathbf{G}(x,y)$ is a known data function.  At every transverse location $(x,y)$, the determinant of the mask-dependent matrix $\mathbf{M}(x,y)$ will either be zero or nonzero.  If its determinant is nonzero then the matrix inverse $\mathbf{M}^{-1}(x,y)$ exists and one may write
\begin{equation}\label{eq:sevenLinearEquationsB}
\mathbf{Q}(x,y)=\mathbf{M}^{-1}(x,y)\mathbf{G}(x,y). 
\end{equation}
If $\mathbf{M}(x,y)$ has zero determinant at any location\footnote{\label{footnote:conditionNumberDefinition} This requirement may be more precisely formulated in terms of the condition number $\mathscr{C}$ of the square matrix $\mathbf{M}(x,y)$, namely the magnitude of the ratio of its maximum-modulus singular value and minimum-modulus singular value \cite{Press}. If $\mathscr{C}$ is much greater than unity then the matrix is ill conditioned, with zero determinant corresponding to the limit of infinitely large $\mathscr{C}$.  If $\mathscr{C}$ is too large at a given transverse location $(x,y)$, different mask transverse displacements should be chosen such that $\mathbf{M}(x,y)$ is not ill conditioned at or near that location.} then a different combination of masks can be chosen for that location, such that $\mathbf{M}(x,y)$ has nonzero determinant there. Having thereby uniquely reconstructed the seven right-hand sides of Eq.~(\ref{eq:SevenDiffusionRelatedFunctions}), these may be aggregated into three screened Poisson equations \cite{GPM2020}
\begin{subequations}\label{eq:threeScreenedPoissonEquations}
\begin{align}
(\nabla_{\perp}^2-\mathscr{W})\widetilde{D}_{xx}(x,y)&=g_1(x,y) \\
(\nabla_{\perp}^2-\mathscr{W})\widetilde{D}_{yy}(x,y)&=g_2(x,y) \\
(\nabla_{\perp}^2-\mathscr{W})\widetilde{D}_{xy}(x,y)&=g_3(x,y)
\end{align}
\end{subequations}
where
\begin{subequations}\label{eq:threeKnownDataFunctions}
\begin{align}
g_1(x,y) &= \frac{\partial}{\partial x} Q_1(x,y) 
+ \left( \frac{\partial^2}{\partial y^2} - \mathscr{W}\right) Q_5(x,y)\\
g_2(x,y) &= \frac{\partial}{\partial y} Q_3(x,y) 
+ \left( \frac{\partial^2}{\partial x^2} - \mathscr{W}\right) Q_6(x,y)\\
g_3(x,y) &= \frac{\partial}{\partial x} Q_2(x,y) \!+\! \frac{\partial}{\partial y} Q_4(x,y) \!-\! \mathscr{W} Q_7(x,y)
\end{align}
\end{subequations}
are known data functions and $\mathscr{W}$ is any positive number. In the notation of Eqs.~(\ref{eq:SecIII--02})-(\ref{eq:SecIII--03}), the screened Poisson equations may be solved via \cite{paganin2002}
\begin{subequations}\label{eq:SolnToThreeScreenedPoissonEquations}
\begin{align}
\widetilde{D}_{xx}(x,y)&=-\mathscr{F}^{-1}\left\{\frac{\mathscr{F}[g_1(x,y)]}{k_x^2+k_y^2+\mathscr{W}}\right\} \\
\widetilde{D}_{yy}(x,y)&=-\mathscr{F}^{-1}\left\{\frac{\mathscr{F}[g_2(x,y)]}{k_x^2+k_y^2+\mathscr{W}}\right\} \\
\widetilde{D}_{xy}(x,y)&=-\mathscr{F}^{-1}\left\{\frac{\mathscr{F}[g_3(x,y)]}{k_x^2+k_y^2+\mathscr{W}}\right\}.
\end{align}
\end{subequations}
To proceed further, we adapt an argument from the first paragraph of this subsection.  For a compact object entirely contained within the FOV over which intensity measurements are taken, Eq.~(24) of Paper I---together with Eq.~(\ref{eq:TransitionFromTensorToScalarDiffusion}) of the present paper---shows that the values taken by $\widetilde{\mathbf{D}}(x,y)$ at points on the FOV boundary $\partial\Omega$ are equal to $\tfrac{1}{2}\theta_0^2 \mathbf{I}_2$.  This gives the following Dirichlet boundary conditions, respectively, for each line of Eq.~(\ref{eq:threeScreenedPoissonEquations}):
\begin{subequations}\label{eq:threeScreenedPoissonEquations==BoundaryConditions}
\begin{align}
[\widetilde{D}_{xx}(x,y)]_{(x,y)\in\partial\Omega}&=\tfrac{1}{2}\theta_0^2 \\
[\widetilde{D}_{yy}(x,y)]_{(x,y)\in\partial\Omega}&=\tfrac{1}{2}\theta_0^2 \\
[\widetilde{D}_{xy}(x,y)]_{(x,y)\in\partial\Omega}&=0.
\end{align}
\end{subequations}
Thus, to determine each of the three independent scalar functions that constitute the tensorial diffusion field, a decoupled screened Poisson equation of the form
\begin{equation}
    (\nabla_{\perp}^2-\mathscr{W})\varpi(x,y)=\mathfrak{g}(x,y)
\end{equation}
needs to be solved for $\varpi(x,y)\equiv \widetilde{D}_{xx}(x,y)$, $\widetilde{D}_{yy}(x,y)$, or $\widetilde{D}_{xy}(x,y)$; both the data function $\mathfrak{g}(x,y)$ and constant Dirichlet boundary conditions are known. For any one of these boundary-value problems, assume two distinct single-valued twice-differentiable solutions $\varpi_A(x,y)$ and $\varpi_B(x,y)$ exist. The difference
\begin{equation}
    \widetilde{\varpi}(x,y)\equiv\varpi_A(x,y)-\varpi_B(x,y)
\end{equation}
obeys the homogeneous equation 
\begin{equation}
    (\nabla_{\perp}^2-\mathscr{W})\widetilde{\varpi}(x,y)=0
\end{equation}
with zero Dirichlet boundary conditions.  Now note that
\begin{align}
    \nonumber\nabla_{\perp}\cdot&[\widetilde{\varpi}(x,y)\nabla_{\perp}\widetilde{\varpi}(x,y)] \\ \nonumber
    &=\vert\nabla_{\perp}\widetilde{\varpi}(x,y)\vert^2
    +\widetilde{\varpi}(x,y)\nabla_{\perp}^2\widetilde{\varpi}(x,y)
    \\
    &=\vert\nabla_{\perp}\widetilde{\varpi}(x,y)\vert^2
    +\mathscr{W}[\widetilde{\varpi}(x,y)]^2.
\end{align}
Integrate over the FOV $\Omega$, apply the Gauss divergence theorem to the left side of the resulting expression, then invoke the zero Dirichet boundary conditions for $\widetilde{\varpi}(x,y)$ [cf.~Eq.~(27) in Paper I].  The left side vanishes, leaving 
\begin{align}
    0=\iint_{\Omega}\left\{
\mathscr{W}[\widetilde{\varpi}(x,y)]^2
+\vert\nabla_{\perp}\widetilde{\varpi}(x,y)\vert^2    
    \right\}dx\,dy. 
\end{align}
Since each term in the integrand is never negative, each must vanish individually at every location within the FOV.  Hence $\widetilde{\varpi}(x,y)$ vanishes at every point within the FOV, thereby establishing uniqueness for tensor diffusion-field retrieval using structured illumination.

\subsection{Stability of phase and diffusion-field reconstructions}\label{subsec:Practicalities2}   

Inverse problems \cite{Sabatier2000} such as phase and diffusion-field retrieval are said to be {\em well posed in the sense of Hadamard} if there exists a unique solution depending continuously on the specified data.\footnote{{See, e.g., pp.~297-298 of \citet{KressBook}.}}  Uniqueness (and by implication existence) was considered in Sec.~\ref{subsec:Practicalities1}, so here we examine the question of reconstruction stability in the particular mathematical sense of a solution which depends continuously on the input data.\footnote{{See, e.g., p.~301 of \citet{KressBook}.}}

As emphasized in the previous subsection, all of our phase and diffusion-field retrieval schemes may be reduced to solution of either a Poisson equation or a screened Poisson equation.  Accordingly, the stability of our phase and diffusion-field reconstructions reduces to studying the continuity with respect to input data of smooth single-valued solutions to the Poisson or screened Poisson equations, subject to specified constant Dirichlet boundary conditions.  The stability of both classes of boundary-value problem, for H\"{o}lder continuous data functions and smooth-boundary simply-connected regions with smooth Dirichlet boundary conditions, is proven in the book by \citet{GilbargTrudingerBook}.

\subsection{Sensitivity of reconstructions to noise}\label{subsec:Practicalities3}

The formal mathematical stability of our reconstructions, as studied in the previous subsection, does not necessarily imply an acceptable level of numerical stability.  In particular, we must examine the possibility that our formally stable solutions may nevertheless be ill-conditioned \cite{Press} in the sense that they display an unacceptably large degree of sensitivity with respect to the noise inevitably present in measured data functions.  Such noise arises, for example, due to the Poisson statistics associated with the photon detection process \cite{MandelWolf,LoudonBook}. 

Numerical stability, or lack thereof, may be studied via the notion of a condition number defined in footnote \ref{footnote:conditionNumberDefinition}. As pointed out there, a linear inverse problem is {\em ill conditioned} when the (dimensionless) condition number is too large compared to unity \cite{Press}. Given that all of our methods for phase retrieval and diffusion-field retrieval have key steps which solve either a Poisson equation or a screened Poisson equation, the numerical stability of each is separately considered in the following two paragraphs.

{\em Numerical stability of solutions to the Poisson equation.} When working with intensity data measured using pixellated two-dimensional detectors, corresponding to the coordinate lattice in Eq.~(38) from Paper I, the inverse Laplacian associated with solving Poisson-type equations may be calculated via its Fourier representation.  In the continuum limit (zero pixel size) the Fourier representation of the inverse Laplacian is given by the special case of Eq.~(\ref{eq:PhaseRetrieval11}) where the regularization parameter $\epsilon$ has a vanishingly small positive value; for the spatially discrete form, see Eq.~(43) from Paper I. For pixellated images, the smallest nonzero eigenvalues of the Laplacian---which are crucial in the context of numerical stability since they correspond to the imaging degrees of freedom \cite{Gabor1961} which are most strongly suppressed in the forward problem---scale relative to the largest eigenvalues as the inverse square of the spatial frequency (in a Fourier representation).  For pixellated two-dimensional images whose number of pixels in each transverse dimension is approximately equal, we immediately conclude that {\em the condition number $\mathscr{C}$ associated with an unregularized inverse Laplacian is on the order of the number of pixels in the image}.  Thus $\mathscr{C}$ is much greater than unity for realistic images.  Hence, notwithstanding the formal mathematical stability of the inverse Laplacian discussed in Sec.~\ref{subsec:Practicalities2}, this operator is numerically unstable.  It is for this reason that regularized forms of the inverse Laplacian are given in Eqs.~(\ref{eq:PhaseRetrieval11}) for phase retrieval and (\ref{eq:DiffusionFieldRetrieval1of2})-(\ref{eq:DiffusionFieldRetrieval2of2}) for diffusion-field retrieval.  These regularized forms promote numerical stability by reducing the condition number of the inverse problem, with this increased stability coming at the cost of some suppression of the lowest Fourier frequencies in the reconstruction.  The higher the degree of noise present in the measured intensity data, the larger the value for the regularization parameter needed to establish numerical stability. We can also improve the numerical stability of the inverse Laplacian, at the cost of measuring additional images, by adapting the ``multiple passband'' approach of Ref.~\cite{QPAMIII} to the present context.  Regarding the noise-induced artifacts, associated with the Fourier representation of the inverse Laplacian, they will often have a cloud-like appearance \cite{QPAMIII,PaganinPhDthesis1999} consistent with the random-fractal character \cite{SethnaBook} which results from filtration of white noise with an envelope having power-law decay.

{\em Numerical stability of solutions to the screened Poisson equation.} A short calculation gives the rough estimate
\begin{equation}
    \mathscr{C}(\mathscr{W} > 0) \approx \frac{N_{\textrm{pixels}}+\mathscr{W}\mathscr{A}}{1+\mathscr{W}\mathscr{A}}
\end{equation}
for the condition number of the screened Poisson equation as a function of the positive screening parameter $\mathscr{W}$, where $N_{\textrm{pixels}}$ is the number of pixels in each image and $\mathscr{A}$ is the image FOV area.  Evidently, the arbitrary positive value for $\mathscr{W}$ can be chosen to very significantly reduce the condition number in comparison to the upper bound of $N_{\textrm{pixels}}$ attained in the limit of zero screening. Accordingly, the screened Poisson equation has dramatically improved numerical stability in comparison to the unscreened Poisson equation.  Interestingly, the structured-illumination tensorial diffusion-field retrieval formulae in Eq.~(\ref{eq:SolnToThreeScreenedPoissonEquations}) have the same mathematical form as the so-called ``Paganin filter'' \cite{paganin2002} which has been very widely employed for x-ray TIE-based phase retrieval due to its extreme stability with respect to input-data noise (see, e.g., Refs.~\cite{Nesterets2014,Gureyev2014,Gureyev2017Unreasonable,Kitchen2017}).  Note also that since any positive value for $\mathscr{W}$ will give a unique reconstruction for $\widetilde{\mathbf{D}}(x,y)$, in practice one can choose a value for $\mathscr{W}$ which maximizes the match between the measured data and the data that would be measured should the reconstruction be correct.

We close this subsection by noting that, while our emphasis in studying numerical stability has been on the key steps in our phase retrieval and diffusion-field retrieval formalism which require either an unscreened or screened Poisson equation to be solved, there are other steps in the analysis chain---such as the division by intensity in Eq.~(\ref{eq:SimplifiedFormalSolutionDiffusionFieldRetrievalSymmetricDefocus}) or the numerical differentiations in Eq.~(\ref{eq:threeKnownDataFunctions})---which would also need to be examined in order to give a complete picture of the numerical stability.  To keep the paper to a reasonable length, these additional stability-analysis steps will not be studied here.

\subsection{Number of measurements required}\label{subsec:Practicalities4}

Regarding the number of intensity measurements required for the phase-retrieval and diffusion-field-retrieval methods considered in this paper, the next two paragraphs consider the mask-free and structured-illumination cases, respectively.

All of the mask-free schemes considered in this paper require three intensity measurements, corresponding to a focal series with three different values for the propagation distance (or defocus) $\Delta$. Examples include Eqs.~(\ref{eq:PhaseRetrieval10}), (\ref{eq:PhaseRetrieval14})-(\ref{eq:PhaseRetrieval15}), (\ref{eq:YetAnotherDataFunction})-(\ref{eq:FormalSolutionDiffusionFieldRetrievalSymmetricDefocus}), and (\ref{eq:YetYetYetAnotherDataFunction}) of the present paper; see, also, the three-intensity data function in Eq.~(148) of Paper I.  While these examples all employ three intensity measurements, this is in fact a minimum number.  In particular, focal series with more than three intensity measurements may be employed, for the purposes of trading off the disadvantage of taking more than the minimum number of measurements against the advantage of increased reconstruction accuracy, precision, or signal-to-noise ratio. For an example of such a strategy in the context of TIE-based phase retrieval, which (as previously mentioned) may be adapted to the more general context of the present paper, see Ref.~\cite{QPAMIII}.  

Regarding mask-based (i.e., structured illumination) schemes for diffusion-tensor retrieval, the methods in this paper rely on measurements obtained using seven different transverse mask positions (see Fig.~\ref{Fig:HowNoiseBlursSpeckles}(a); cf.~Fig.~5 from Paper I).  For each transverse position of the mask, Eqs.~(\ref{eq:DataFunctionForSpeckleTrackingTensorForm}) and (\ref{sec:TensorDiffusionFieldRetrievalWith
Masks}) show that three different measurements are required to construct any one row of the data vector on the right side of Eq.~(\ref{eq:sevenLinearEquationsA}).  Hence a minimum of 21 images is needed for diffusion-tensor retrieval.  This minimum number can be increased on account of the tradeoff mentioned in the previous paragraph. For example, more than seven mask positions will increase the number of elements in the matrix $\mathbf{M}(x,y)$ appearing in Eq.~(\ref{eq:sevenLinearEquationsA}), in which case a Moore-Penrose pseudoinverse \cite{BarrettMyersBook} may be employed to obtain a numerically robust solution to the overdetermined system of linear equations.

\subsection{Condition for ignorability of higher-order-TIE effects}\label{subsec:Practicalities6} 

Define the dimensionless defocus-induced contrast
\begin{equation}\label{eq:DimensionlessContrastMeasure}
\mathscr{K}(x,y;\Delta) \equiv \frac{I(x,y,z=\Delta)-I(x,y,z=0)}{I(x,y,z=0)}.  \end{equation}
For simplicity, assume the local diffuse-scatter fraction obeys $F(x,y) \ll 1$, in addition to considering the dimensionless diffusion field $D(x,y)$ to be a sufficiently slowly varying function such that it approximately commutes with the transverse Laplacian $\nabla_{\perp}^2$. With these assumptions in place, a crude order-of-magnitude analysis of Eq.~(\ref{eq:SecIII--ExtendedFPE})  gives---with the help of Eq.~(\ref{eq:InfinityOfTIEs--MainResult})---the estimate\footnote{{In writing Eq.~(\ref{eq:RoughFormulaForContrast}), the phase of the unpropagated complex wavefield has been assumed to vary by on the order of one radian over a transverse distance on the order of $\mathscr{L}$.  This excludes the case of a weak phase object from the present rough analysis.  We have also assumed for simplicity that the phase and intensity, of the unpropagated field, vary appreciably over the same characteristic transverse length scale $\mathscr{L}$.  Both of these assumptions are readily dropped, if necessary.}} 
\begin{equation}\label{eq:RoughFormulaForContrast}
    \mathscr{K} \approx \mathcal{K}_1 \mathcal{N}_{\textrm{F}}^{-1}
    + \mathcal{K}_2 \mathcal{N}_{\textrm{F}}^{-2}
    + \mathcal{K}_3 \Delta^2D\mathscr{L}^{-2}.
\end{equation}
Here functional dependencies have been dropped for clarity, $\mathcal{K}_j\equiv\mathcal{K}_j(x,y;\Delta)$ for $j=1,2,3$ are dimensionless weights that are all on the order of unity, $\mathscr{L}\equiv\mathscr{L}(x,y)$ is the characteristic transverse length scale over which the unpropagated complex amplitude $\psi(x,y,z=0)$ changes appreciably in either phase or amplitude, and
\begin{equation}\label{eq:coherentFresnelNumber}
    \mathcal{N}_{\textrm{F}} \equiv\frac{\mathscr{L}^2 k}{\Delta}
\end{equation}
is the coherent-flow Fresnel number \cite{SalehTeichBook}.\footnote{Cf.~the Fresnel number $N_{\textrm{F}}$ in Eq.~(17) of Paper I, which considers a different transverse length scale in the numerator ($\ell$ rather than $\mathscr{L}$) as well as in-sample propagation distance $T_s$ rather than sample-to-detector propagation distance $\Delta$ in the denominator.}  

The term in Eq.~(\ref{eq:RoughFormulaForContrast}) that is proportional to $\mathcal{K}_1$, which originates from the second term on the right side of Eq.~(\ref{eq:SecIII--ExtendedFPE}), estimates the contrast due to the first-order TIE as being on the order of the inverse Fresnel number $\mathcal{N}_{\textrm{F}}^{-1}$ \cite{GureyevCompositeTechniques2003}. The term proportional to $\mathcal{K}_2$, which originates from the second and third lines of Eq.~(\ref{eq:SecIII--ExtendedFPE}), estimates the contrast due to the second-order TIE as being on the order of the squared inverse Fresnel number $\mathcal{N}_{\textrm{F}}^{-2}$. The term proportional to $\mathcal{K}_3$ originates from the final term on the first line of Eq.~(\ref{eq:SecIII--ExtendedFPE}), namely the Fokker-Planck diffusion term; note also that we have assumed $I(x,y,z=0)$ to appreciably vary over the same transverse length scale $\mathscr{L}$ as the complex amplitude $\psi(x,y,z=0)$ -- this simplifying assumption is ready dropped, if needed.

If $\Delta$ is sufficiently small to make the inverse Fresnel number much less than unity, then the second term on the right side of our rough contrast formula [Eq.~(\ref{eq:RoughFormulaForContrast})] is ignorably small in comparison to the first term on the right side.  This condition, namely \cite{GureyevCompositeTechniques2003}
\begin{equation}
    \mathcal{N}_{\textrm{F}}^{-1} \ll 1,
\end{equation}
specifies the domain of validity for ignoring higher-order TIE effects in the context of the present paper.  If $\mathcal{N}_{\textrm{F}}^{-1}$ is less than {\em but not significantly less than} unity, the contribution due to the second-order TIE cannot be ignored.  As a very rough guide, if the out-of-focus contrast is on the order of $10\%$ or less then the second-order TIE effects are ignorable, but if the contrast is between (say) $10\%$ and $30\%$ then the effects of the second-order TIE cannot be ignored; if the contrast is stronger still, the formalism of the present pair of papers becomes inapplicable. 

\subsection{Condition for ignorability of diffusion-field effects}\label{subsec:Practicalities7}

Consistent with the domain of validity stated at the end of the previous paragraph, assume the inverse Fresnel number to be less than (say) $1/3$.  In this case, the first term on the right side of Eq.~(\ref{eq:RoughFormulaForContrast}) will always be bigger than the second term.  Therefore the condition for the ignorability of diffusion-field effects is that the third term be much smaller than the first term.  Hence if the dimensionless diffusion field is sufficiently weak such that
\begin{equation}\label{eq:WhenDCanBeIgnored}
    D(x,y) \ll  (k\Delta)^{-1},
\end{equation}
then its influence may be ignored.  

This inequality is equivalent to the requirement that the influence of the diffusion field is negligible if its local blur width (at the image plane $z=\Delta$) is much smaller than the angular width
\begin{equation}
    \vartheta \approx \sqrt{\lambda/\Delta}
\end{equation}
associated with the first Fresnel zone (see Fig.~\ref{Fig:FirstFresnelZone}). Indeed, if we recall the concept of an ``effective blur angle''
$\theta_{\textrm{eff}}$
associated with the diffusion field that was given in Eq.~(62) of Paper I, and then make use of Eq.~(\ref{eq:DimensionlessScalarDiffusionCoefficient}) from the present paper, then the ``ignorability condition'' in Eq.~(\ref{eq:WhenDCanBeIgnored}) reduces to the natural condition that
\begin{equation}
    \theta_{\textrm{eff}}(x,y) \ll \vartheta(\lambda,\Delta).
\end{equation}
Intuitively, this states that the diffusion-field blur is negligible in comparison to the out-of-focus contrast because it does not smear over a sufficiently large length scale.  

The above inequalities delineate experimental conditions under which diffusion fields are so feeble that they cannot realistically be reconstructed from intensity measurements via the methods of this paper.  If the diffusion field is sufficiently strong to violate these inequalities, the experimental conditions are such that the diffusion field may be realistically reconstructed using our approaches.

}

\section{Discussion}\label{sec:Discussion}

This discussion has five parts. Section~\ref{sec:RunningCoupling} examines how paraxial diffusion fields are a ``running coupling'' associated with the statistical-physics idea of a renormalization group.  Section \ref{sec:HowDiffusionFieldVariesWithResolution} treats the resolution dependence of a paraxial-flow partition into coherent and diffuse channels, emphasizing that {\em progressively finer spatiotemporal resolution leads to a progressively smaller fraction of the flow being assigned to the diffuse channel}.  This is closely related to Sec.~\ref{sec:Liouville&UnresolvedSpeckle}, which considers Liouville's theorem and the role of unresolved speckle.  Connections between the diffusion-field concept and the notion of partial coherence are discussed in Sec.~\ref{sec:LinkWithPartialCoherence}.  We close with Sec.~\ref{sec:FPEforElectronsNeutronsPhotonsEtc}, which suggests how some of the ideas developed in the present pair of papers might be applied to imaging using a variety of radiation and matter-wave fields, such as visible-light microscopy, transmission electron microscopy, and neutron imaging.\footnote{As mentioned on several occasions, Fokker-Planck diffusion-field retrieval has already been applied in a number of publications employing hard-x-ray photons.  However, to the best of our knowledge, it has yet to be applied to visible-light microscopy, transmission electron microscopy, or neutron imaging.}

\subsection{Paraxial-optics diffusion fields, renormalization group, and running couplings}\label{sec:RunningCoupling}

From a fundamental perspective, there is an important relation between the Fokker--Planck diffusion field and the notion of a renormalization group.  The renormalization group is a concept often employed in statistical physics and quantum field theory \cite{Huang_1987,PeskinSchroeder1995}.  The basic idea is that, upon suitable coarse graining (``blurring'') of a physical system, the governing equations---which may be differential equations, as is the case in our work---are unchanged in mathematical form.  However, the physical quantities that appear in the equations {\em are} changed by the process of coarse graining.  In our case, the coarse graining is in the spatial dimension, with characteristic length scale $\mathcal{W}$ specified by the pixel size of the optical imaging system.  Features smaller than $\mathcal{W}$ are not spatially resolved, by definition, with such fine features contributing to the Fokker--Planck diffusion field in either its scalar form $D(x,y)$ or tensor form $\widetilde{\mathbf{D}}(x,y)$.  Hence such paraxial {\em diffusion fields will be a function of} $\mathcal{W}$.\footnote{The diffusion field will also depend on other parameters such as energy and detector acquisition time.  Regarding the dependence on detector acquisition time,  cf.~\citet{MagyarMandel1963} and \citet{paganinsanchezdelrio2019}.}  More generally, the bifurcation of energy flows into both the diffuse and coherent channels is dependent on $\mathcal{W}$.  In this context, we can view the Fokker--Planck diffusion scalar $D(x,y;\mathcal{W},\cdots)$ or tensor $\widetilde{\mathbf{D}}(x,y;\mathcal{W},\cdots)$ as a {\em running coupling} \cite{PeskinSchroeder1995} that quantifies the degree to which an illuminated object transfers incident photons (or electrons, neutrons, etc.) from the coherent-flow to the diffuse-flow channel.  The invariance in mathematical form of the Fokker--Planck equation upon a change of $\mathcal{W}$, albeit with a change in the running couplings such as that encoded in the diffusion scalar or tensor, is typical of the renormalization-group concept.  This statement holds for a particular range of the sliding scale $\mathcal{W}$, which should be (i) large enough that higher-order diffusion-field terms\footnote{We here refer to the Kramers-Moyal extension of the Fokker-Planck equation (see the final paragraph of Sec.~\ref{sec:Second
RoughDerivation}).} may be neglected to a good approximation, and (ii) small enough that one has an image, i.e., $\mathcal{W}$ must be significantly smaller than the transverse extent of the sample at the image plane.  Moreover, the manner in which the diffusion tensor varies with $\mathcal{W}$ (or with energy) may give further information regarding the nature of the unresolved microstructure. For example, if the unresolved structure is a random fractal \cite{SethnaBook,BaleSchmidt1984,Sinha1988,Teixeira1988,JakemanRidleyBook} with a given fractal dimension, this will influence the manner in which the Fokker-Planck diffusion field scales with changing $\mathcal{W}$.  

\subsection{Resolution dependence of paraxial-flow partition into coherent and diffuse channels}\label{sec:HowDiffusionFieldVariesWithResolution}

The considerations of Sec.~\ref{sec:RunningCoupling} have implications for the use of pixelated image-intensity measurements.  Consider an imaging detector that uses square pixels of width $\mathcal{W}$, which measures intensity maps in the context of paraxial-wave diffusion-field retrieval. Changing the pixel size will change the decomposition of the paraxial optical flow, into its coherent-flow and diffuse-flow components. In particular, the scalar and tensor diffusion fields will carry a dependence upon $\mathcal{W}$.  In the idealized limit of an elastically-scattering sample transmitting paraxial scalar monochromatic radiation whose intensity is measured using a detector having pixels that are arbitrarily small, all of the detected energy will correspond to the coherent-flow channel, with none being assigned to the diffuse-flow channel. More realistically, for an elastically-scattering sample which is illuminated by paraxial quasimonochromatic radiation that remains paraxial after passing through the sample, the scalar and tensor diffusion fields will in general reduce in magnitude as $\mathcal{W}$ decreases. Moreover, if the sample-to-detector defocus distance $\Delta$ may be altered via free-space propagation\footnote{This situation is distinct from scenarios where defocus is induced using a suitable imaging system, in which case the text of the remainder of this paragraph is inapplicable.} through a vacuum-filled region between the exit surface of a fixed sample $B$ and a movable entrance-surface of a detector $C$ (see Fig.~3 of Paper I), then increasing $\Delta$ will also alter the scalar and tensor diffusion fields associated with the sample, because this increase in $\Delta$ will alter the angle that any pixel subtends with respect to any particular point in the sample.

\begin{figure}[ht!]
\centering
\includegraphics[width=0.85\columnwidth]{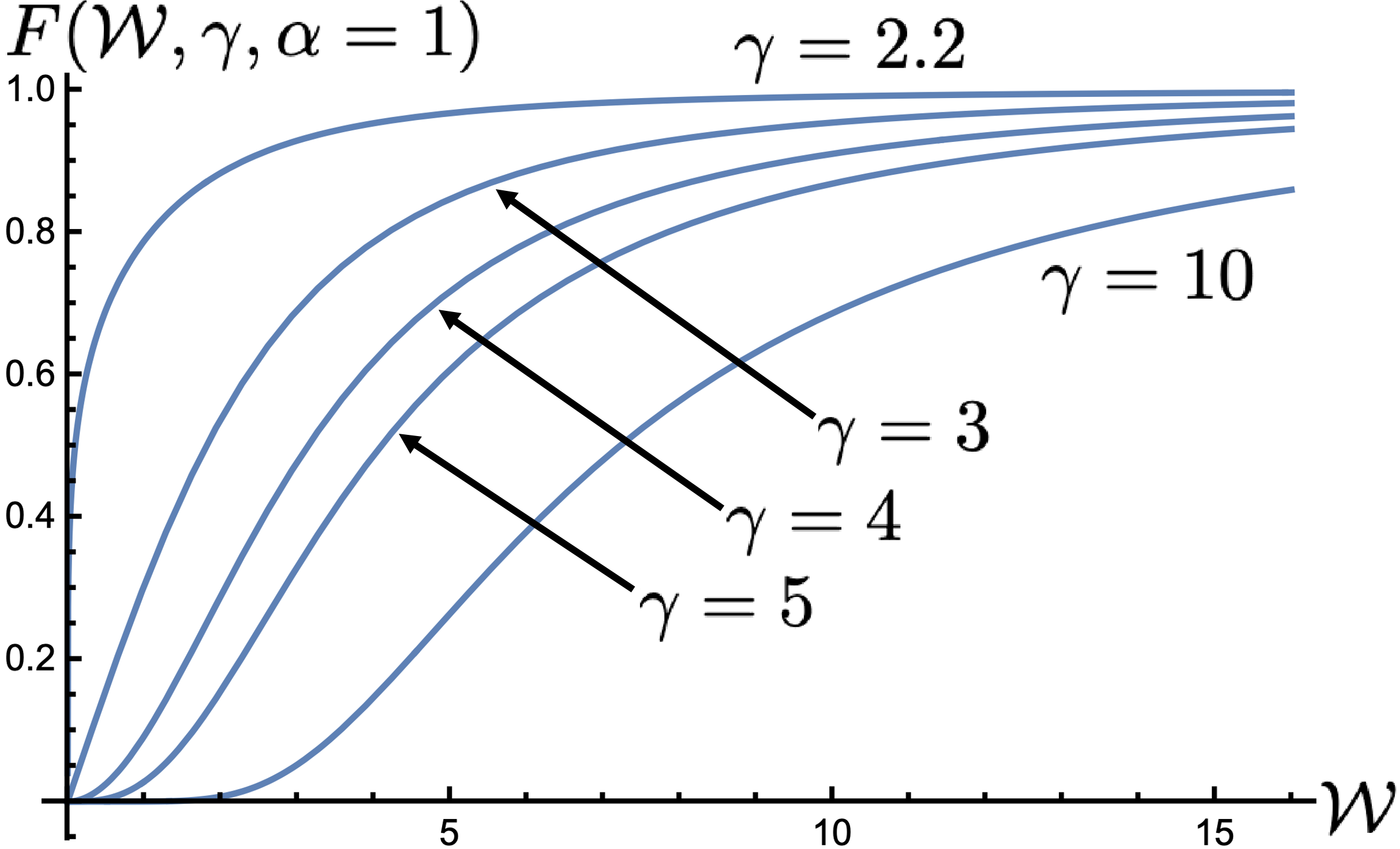}
\caption{Variation of the diffuse-flow fraction $F$ as a function of detector spatial resolution $\mathcal{W}$, according to Eq.~(\ref{eq:RunningCoupling4}).}
\label{Fig:RunningCoupling}
\end{figure}

In light of the preceding paragraph, it is clarifying to consider the following simple model for the dependence of a diffusion field upon pixel size $\mathcal{W}$. Adopting the notation in Sec.~V\,E of Paper I, let $P(k_R)$ denote the intensity power spectrum illuminating a planar detector surface, where $k_R$ denotes radial spatial frequency. Let 
\begin{equation}
\breve{k}_R \equiv \pi/\mathcal{W}
\label{eq:RunningCoupling1}
\end{equation}
be the radial Nyquist spatial frequency \cite{Press}, namely the highest radial spatial frequency that is properly sampled by the detector.  At any one particular pixel location, the fraction $F$ of the field that is associated with diffusive flow (rather than coherent flow) is  \cite{LeathamPhDThesis2023}
\begin{equation}
F(\mathcal{W}) = \frac{\int_{\breve{k}_R}^{\infty} k_R P(k_R) d k_R}{\int_0^{\infty} k_R P(k_R) d k_R}.
\label{eq:RunningCoupling2}
\end{equation}
As a model power spectrum, use the ``Pearson VII'' form 
\begin{equation}
P(k_R)=\frac{\mathcal{N}}{(1+\alpha k_R^2)^{\gamma/2}},
\label{eq:RunningCoupling3}
\end{equation}
where $\alpha > 0$ and $\gamma > 2$ are positive constants\footnote{The condition on $\gamma$ ensures that the integrals in Eq.~(\ref{eq:RunningCoupling2}) are finite.  Also, $\alpha$ has units of squared length.} and $\mathcal{N}$ is a normalization factor \cite{PearsonVII}.  Note that the case where $\gamma \longrightarrow 2$ corresponds to a Lorentzian distribution (Breit-Wigner distribution), with $\gamma \longrightarrow \infty$ corresponding to a Gaussian distribution provided that $\alpha\propto\gamma^{-1}$ in the same limit. Also note that for large radial spatial frequency, the asymptotic behavior 
\begin{equation}
P(k_R) \sim k_R^{-\gamma} 
\label{eq:RunningCoupling3.5}
\end{equation}
of the power spectrum agrees with the form in Eq.~(137) of Paper I.  Substitute Eq.~(\ref{eq:RunningCoupling3}) into Eq.~(\ref{eq:RunningCoupling2}) and then evaluate the resulting integrals to give \cite{LeathamPhDThesis2023}
\begin{equation}
F(\mathcal{W},\gamma,\alpha)=\mathcal{W}^{\gamma-2}(\alpha \pi^2+\mathcal{W}^2)^{1-(\gamma/2)}.
\label{eq:RunningCoupling4}
\end{equation}
%
With trivial loss of generality we choose units of length such that $\alpha=1$.  In this case $F$ may be plotted versus $\mathcal{W}$ for various values of $\gamma > 2$.  As shown in Fig.~\ref{Fig:RunningCoupling}, we observe the expected behaviors that (i) $F$ is a nonconstant function of $\mathcal{W}$, (ii) $F$ tends to zero in the infinite-resolution limit where $\mathcal{W}$ tends to zero, and (iii) $F$ becomes progressively closer to unity as the resolution becomes sufficiently coarser, namely as $\mathcal{W}$ becomes progressively larger.  The ``broader the tails'' of the power spectrum, namely the smaller the value of $\gamma$, the greater the fraction $F$ of the paraxial flow that is classified as diffusive rather than coherent.  Since the diffusion field is directly proportional to the scattering fraction $F$, via Eq.~(\ref{eq:DimensionlessScalarDiffusionCoefficient}),\footnote{For a more general expression, which takes the effects of source-size blur into account, see Eq.~(24) in Paper I.} we reach the interesting conclusion that measuring the resolution dependence of the diffusion field may enable a measurement of $\gamma$, which governs the asymptotic rate of decay of the local power spectrum.

{ The particular model considered here illustrates  the more general point that the decomposition of paraxial quasimonochromatic optical flow into coherent and diffuse components is not unique, in the sense that it depends on the spatial resolution of the intensity detection.  More generally, this decomposition also depends on temporal resolution \cite{MagyarMandel1963}.  For fixed spatial and temporal resolution of the intensity detection, the paraxial-flow
partition into coherent and diffuse channels corresponds to a physically meaningful and experimentally measurable separation of coherent and diffuse energy flows.}

\subsection{Liouville's theorem, spatial coarse graining, and the role of unresolved speckle}\label{sec:Liouville&UnresolvedSpeckle}

Consider the following two models for beamlike flow through an infinitely extended two-dimensional planar mathematical surface $\mathcal{S}$ that is perpendicular to the optical axis associated with a partially coherent \cite{Zernike1938} quasimonochromatic \cite{BornWolf} paraxial complex scalar wave field.  (i) Using the Fokker-Planck formalism of the present pair of papers, we may describe this propagating beam using four fields, namely the (scalar) intensity $I(x,y)$, the (scalar) phase $\phi(x,y)$, the (scalar) fraction $F(x,y)$ of the flow that is associated with the diffuse channel, and the diffusion field [which may be a scalar field $D(x,y)$ or a rank-two symmetric matrix field $\mathbf{D}(x,y)$].  (ii) More generally, under the space-frequency description of partially coherent fields \cite{Wolf1982,ThinWolfBook}, we may instead model the beam over $\mathcal{S}$ via a statistical ensemble $\mathcal{E}$ of strictly monochromatic fields, all of which have the same temporal frequency (i.e., the same energy). Each member of $\mathcal{E}$ is a complex scalar wavefunction that may be thought of a single point in a high-dimensional state space (phase space) \cite{Huang_1987,GoldsteinBook}; hence $\mathcal{E}$ corresponds to a cloud $\mathcal{C}$ of state-space points.  Upon propagating along the optical axis downstream of $\mathcal{S}$, the state-space cloud will flow. The unitary nature of the paraxial wavefield diffraction operator \cite{Montgomery1981} implies that Liouville's theorem is operative, i.e., $\mathcal{C}$ flows in a volume-preserving manner that corresponds to an incompressible phase-space fluid \cite{Huang_1987,GoldsteinBook}.  

The volume preserving (and therefore entropy preserving) phase-space flow in the second model might appear to contradict the entropy increase implied by the diffusion field in the first model.  However, there is no contradiction, for the following reason.  Following the explanation on pp.~183-184 of \citet{ThorneBlandfordBook}, the unresolved speckles implicit in the ensemble $\mathcal{E}$ will cause the cloud $\mathcal{C}$ to evolve in a ``mixing'' \cite{McCauleyChaosBook} manner whereby the volume of the cloud is folded (in a highly convoluted way) through a progressively larger volume of the phase space; while the volume of $\mathcal{C}$ does not change upon free-space propagation, it quickly becomes so broadly dispersed (filamented) that {\em upon performing a spatial coarse graining that is necessitated by the finite spatial resolution of pixelated imaging detectors} the effective phase-space volume increases with increased propagation distance downstream of $\mathcal{S}$.  Stated differently, the phase-space volume does not grow, but the coarse-grained phase-space volume {\em does} grow; this latter statement is consistent with the entropy increase implied by nonzero Fokker-Planck diffusion fields.  

The preceding discussion on spatial coarse graining leads us to the closely related notion of unresolved speckle. It is typical, in imaging scenarios employing partially coherent radiation, for unresolved speckle to play a key albeit often implicit role \cite{Nesterets2008,Nugent2003, paganinsanchezdelrio2019,PaganinMorgan2019,PaganinPelliccia2021}.  For the work outlined here, two distinct foundational roles for unresolved speckle should be highlighted.  The first relates to spatiotemporal speckle \cite{Alperin2019} associated with partially coherent illumination and its subsequent transport through an imaging system.  The time integration in an intensity measurement averages over the temporal speckles, with spatial averages over finite-area detector elements coarse graining over the spatial speckles {(cf.~Fig.~\ref{Fig:FastAndSlowPhase})}.  These time averages may be replaced by ensemble averages, if ergodicity can be assumed \cite{GoodmanStatisticalOpticsBook}.  Under this view, the time-averaged energy-flow vector (or ensemble-averaged flow) exhibits the bifurcation into coherent-flow and diffuse-flow channels as described by the Fokker-Planck generalization \cite{MorganPaganin2019,PaganinMorgan2019} of the TIE \cite{Teague1983}.  A second role for unresolved speckle lies with spatially-unresolved sample microstructure (``internal roughness''), which may itself be viewed as a form of speckle insofar as it is associated with spatially random variations in the scattering potential \cite{GoodmanStatisticalOpticsBook,GoodmanSpeckleBook,ThinWolfBook}.

\subsection{Link between paraxial diffusion fields and partial coherence}\label{sec:LinkWithPartialCoherence}

Recall Eq.~(24) from Paper I, which models two distinct contributions to a paraxial scalar diffusion field $D(x,y)$. The first contribution is the approximately position-independent diffusive influence of the nonzero angular radius $\theta_0$ of an extended chaotic source (see Fig.~3 from Paper I), with the second being the position-dependent diffusive contribution due to scattering from unresolved sample microstructure.  Notwithstanding this partition of the influence of source size (``spatial coherence'') and sample microstructure, the resulting {\em diffusion field is a property of the paraxial optical beam} at the exit surface of the sample (again see Fig.~3 from Paper I).  While the diffusion field may be  influenced by both sample and source, together with the spatial and temporal resolution of the detector, the concept of a diffusion field may be decoupled from such particular influences. 

This observation invites comparison with formalisms for partially coherent optical fields \cite{BornWolf,MandelWolf}.  Indeed, the concept of partial optical coherence is often couched in terms of the ability of an optical field to form interference fringes, with high fringe visibility associated with a high degree of coherence, and low fringe visibility associated with a low degree of coherence \cite{Zernike1938}.  If we now think of lower interference-fringe visibility as being a diffused form of higher visibility,\footnote{Cf.~Eqs.~(\ref{eq:NRU9})-(\ref{eq:NR12}).} we are immediately led to inquire into the connection between the diffusion-field concept and standard notions of partial coherence.  

In an imaging context, partial coherence may exist at two distinct levels which are related to the imaging system and sample (if present), respectively.  Both mechanisms may lead to a diffuse component in energy flow through the system, which augments the coherent energy flow \cite{PaganinMorgan2019}.  We discuss each mechanism in turn.  (i) The optical system in Fig.~\ref{Fig:Schematic}---which consists of a source $A$, an imaging system $C$, and a detector (not shown) with finite spatial and temporal resolution---will in general lead to a partially coherent field over the imaging plane $D$.  If it is not laser-like, the source itself will typically generate illumination that is partially coherent to a non-negligible degree.  This state of partial coherence may be described using a variety of formalisms such as the cross-spectral density \cite{MandelWolf}, density matrix \cite{Messiah}, Wigner function \cite{Alonso2011}, generalized radiance \cite{MarchandWolf1974}, ambiguity function \cite{Papoulis1974}, etc.  Regardless of the adopted formalism, the corresponding coherence function will evolve as the field propagates through free space.\footnote{The van Cittert-Zernike theorem \cite{GoodmanStatisticalOpticsBook} is a key example.}  The imaging system $C$ also influences coherence-function transport.  Moreover, finite spatial and temporal detector resolution are key parameters.  For example, finite pixel size  will smear spatial features or speckles that are smaller than the pixel size (or point-spread-function width if this is larger than the pixel size);\footnote{See, e.g., Sec.~3.6.3 of \citet{Paganin2006}.} finite temporal resolution will smear over temporal fluctuations occurring at timescales smaller than the temporal resolution \cite{MagyarMandel1963,FrozenPhononModel}. More generally, a complete description of the state of coherence of the field involves an ensemble of correlation functions of progressively higher order.\footnote{See, e.g., pp.~422-428 of \citet{MandelWolf}.} In comparison, the diffusion-field concept only gives a partial description of the state of a partially coherent field.  (ii) Spatially unresolved microstructure within the sample alters the degree of coherence of transmitted radiation.  Stated differently, we may consider any of the previously-mentioned coherence functions as being scattered by the microstructure that is present in the sample.\footnote{See, e.g., Sec.~6.3 of \citet{ThinWolfBook}.}        

In light of the preceding comments, both the diffusion field $D(x,y,z)$ and the diffuse-flow fraction $F(x,y,z)$ may be viewed as single-point measures---or order parameters \cite{Huang_1987}---that are influenced by the partially coherent nature of the underlying field. These single-point functions, which are only nonvanishing for partially coherent fields, contain partial rather than complete information regarding the state of coherence of the field. 

\subsection{Paraxial diffusion-field retrieval using a variety of radiation and matter wave fields}\label{sec:FPEforElectronsNeutronsPhotonsEtc}

At the time of writing, and to the best of our knowledge, no Fokker-Planck TIE experiments outside of the domain of x-ray optics have been reported.  Accordingly, it would be interesting to apply the results of the present pair of papers, to Fokker-Planck paraxial diffusion-field retrieval using visible light, electrons, and neutrons. While the same principles apply to all of these paraxial-beam scenarios, some important points of difference should be mentioned.  These are discussed below.

In the case of both visible-light and electron-optical implementations of the method, relatively aberration-free coherent imaging systems are readily available \cite{BornWolf,Erni2015}.  These enable both over-focus and under-focus images to be obtained for the purposes of phase and diffusion-field retrieval; as seen earlier in this paper, the ability to work with symmetrically defocused images considerably simplifies the analysis.  Conversely, for the case of both x-ray imaging \cite{Paganin2006} and neutron imaging \cite{NeutronImagingAndItsApplications}, it is often (but not always) the case that negatively defocused images are not available since a relatively aberration-free imaging system is not incorporated into the experiment; in this case, positively-defocused images may be formed (in the absence of an imaging system) via free-space propagation \cite{Snigirev,Cloetens,KleinOpat1976,Allman2000} but negatively-defocused images are not available.  It is for this reason that some of the approaches in this paper have been restricted to positively defocused images.  

The influence of partial spatial coherence \cite{ThinWolfBook}, namely the image-blurring effects of nonzero source size, may also differ across different implementations.  For experiments using visible-light microscopes \cite{BornWolf}, synchrotron x rays \cite{PellicciaPaganin2025}, and transmission electron microscopes \cite{ReimerKohlBook}, it will often be the case that the image-blurring effects of an extended incoherent source may be rendered negligible, on account of the sufficiently high degree of brightness of the respective sources.  In this small-source case, the contribution of angular source-size blur to a paraxial diffusion field---as quantified by nonzero $\theta_0$ in Eq.~(24) of Paper I---may often be ignored.  However, for nuclear-reactor \cite{KleinWerner1983} and spallation \cite{SpallationSources} neutron sources, microfocus x-ray sources \cite{WilkinsFish,Wilkins2014}, and liquid metal-jet x-ray sources \cite{LiquidMetalJetSources}, it will often be the case that the source-blur contribution to the paraxial diffusion field cannot be neglected.

Finally, regarding appropriate scattering-theory diffusion-field models, these will often differ depending on whether the Fokker-Planck extension of the TIE is being applied to visible-light photons, x-ray photons, electrons, neutrons, etc. { Moreover, working with particular models may allow one to investigate the sensitivity of diffusion-field retrieval to modeling assumptions.} We now give two examples { regarding context-specific diffusion-field models}. (i) The projection approximation, whereby both the phase shift and the diffusion field at some point $P$ on the exit surface of a sample may be obtained via a suitable integral taken over a straight line passing through the sample, will often be applicable for hard x rays \cite{Paganin2006} and neutrons \cite{NeutronImagingAndItsApplications} passing through a macroscopic-sized non-crystalline object [see, e.g., Eqs.~(88) and (95) in Paper I].  However, the projection approximation will fail in visible-light scattering through optically thick media, such as thick turbid media where the transverse scatter within the volume of the sample cannot be neglected; in such cases, a more sophisticated scattering formalism is needed to calculate the intensity, phase, and diffusion field over the exit surface of the sample \cite{JakemanRidleyBook,CarminatiSchotlandBook}. 
 (ii) Since hard x rays primarily interact with the electron density in a sample \cite{Paganin2006} and neutrons primarily interact with the atomic nuclei \cite{SiviaScatteringTheoryBook2011}, the corresponding diffusion fields will in general contain complementary information regarding unresolved microstructure.

\section{Conclusion}\label{sec:Conclusion}

In this direct continuation of Paper I \cite{PaganinPellicciaMorgan2023} the concept of a paraxial diffusion field was augmented to take into account the presence of a coherent component to the flow.  This gave Fokker-Planck extensions to the transport-of-intensity equation of paraxial wave optics, whereby coherent flow associated with phase gradients is augmented by diffuse flow associated with intensity gradients.  Several derivations of this extension were given, together with several different means for phase retrieval and diffusion-field retrieval.  Existing methods for transport-of-intensity phase retrieval need no modification in light of any of the modifications of this equation discussed in the present paper, provided that symmetrically overfocused and underfocused images are employed.  Perhaps more importantly, {\em diffusion-field retrieval enables a complementary additional channel of information to be obtained from paraxial focal series, beyond the phase and intensity fields that are reconstructed in transport-of-intensity analyses}.  While the Fokker-Planck extension of the transport-of-intensity equation has been applied to experiments in x-ray optics, to the best of our knowledge the method has yet to be applied to experiments employing visible light, electrons, and neutrons.      

\section*{Acknowledgments}

K.S.M. acknowledges funding via Australian Research Council Future Fellowship FT180100374. K.S.M. and D.M.P. acknowledge funding from Australian Research Council Discovery Project DP230101327.  We acknowledge useful discussions with Jannis Ahlers, Samantha Alloo, Alaleh Aminzadeh, Alex Aumann, Mario Beltran, {Victorien Bouffetier,} Emmanuel Brun, Michelle Croughan, Tim Davis, Carsten Detlefs, Christian Dwyer, Jayan Gunasekera, Timur Gureyev, Ying Ying How, Dominik John, Samy Kefs, Andrew Kingston, Marcus Kitchen, Christoph Koch, Kieran Larkin, {Alejandro Laso Garcia,} Thomas Leatham, Amelia Liu, Clara Magnin, Eduardo Miqueles, Konstantin Pavlov, Daniele Pelliccia, Tim Petersen, James Pollock, {Caitlyn Smith,} Rahil Valani, and Marie-Christine Zdora.


\begin{thebibliography}{214}%
\makeatletter
\providecommand \@ifxundefined [1]{%
 \@ifx{#1\undefined}
}%
\providecommand \@ifnum [1]{%
 \ifnum #1\expandafter \@firstoftwo
 \else \expandafter \@secondoftwo
 \fi
}%
\providecommand \@ifx [1]{%
 \ifx #1\expandafter \@firstoftwo
 \else \expandafter \@secondoftwo
 \fi
}%
\providecommand \natexlab [1]{#1}%
\providecommand \enquote  [1]{``#1''}%
\providecommand \bibnamefont  [1]{#1}%
\providecommand \bibfnamefont [1]{#1}%
\providecommand \citenamefont [1]{#1}%
\providecommand \href@noop [0]{\@secondoftwo}%
\providecommand \href [0]{\begingroup \@sanitize@url \@href}%
\providecommand \@href[1]{\@@startlink{#1}\@@href}%
\providecommand \@@href[1]{\endgroup#1\@@endlink}%
\providecommand \@sanitize@url [0]{\catcode `\\12\catcode `\$12\catcode
  `\&12\catcode `\#12\catcode `\^12\catcode `\_12\catcode `\%12\relax}%
\providecommand \@@startlink[1]{}%
\providecommand \@@endlink[0]{}%
\providecommand \url  [0]{\begingroup\@sanitize@url \@url }%
\providecommand \@url [1]{\endgroup\@href {#1}{\urlprefix }}%
\providecommand \urlprefix  [0]{URL }%
\providecommand \Eprint [0]{\href }%
\providecommand \doibase [0]{https://doi.org/}%
\providecommand \selectlanguage [0]{\@gobble}%
\providecommand \bibinfo  [0]{\@secondoftwo}%
\providecommand \bibfield  [0]{\@secondoftwo}%
\providecommand \translation [1]{[#1]}%
\providecommand \BibitemOpen [0]{}%
\providecommand \bibitemStop [0]{}%
\providecommand \bibitemNoStop [0]{.\EOS\space}%
\providecommand \EOS [0]{\spacefactor3000\relax}%
\providecommand \BibitemShut  [1]{\csname bibitem#1\endcsname}%
\let\auto@bib@innerbib\@empty
\bibitem [{\citenamefont {Paganin}\ \emph
  {et~al.}(2023{\natexlab{a}})\citenamefont {Paganin}, \citenamefont
  {Pelliccia},\ and\ \citenamefont {Morgan}}]{PaganinPellicciaMorgan2023}%
  \BibitemOpen
  \bibfield  {author} {\bibinfo {author} {\bibfnamefont {D.~M.}\ \bibnamefont
  {Paganin}}, \bibinfo {author} {\bibfnamefont {D.}~\bibnamefont {Pelliccia}},\
  and\ \bibinfo {author} {\bibfnamefont {K.~S.}\ \bibnamefont {Morgan}},\
  }\bibfield  {title} {\bibinfo {title} {Paraxial diffusion-field retrieval},\
  }\href {https://doi.org/10.1103/PhysRevA.108.013517} {\bibfield  {journal}
  {\bibinfo  {journal} {Phys. Rev. A}\ }\textbf {\bibinfo {volume} {108}},\
  \bibinfo {pages} {013517} (\bibinfo {year} {2023}{\natexlab{a}})}\BibitemShut
  {NoStop}%
\bibitem [{\citenamefont {Sabatier}(2000)}]{Sabatier2000}%
  \BibitemOpen
  \bibfield  {author} {\bibinfo {author} {\bibfnamefont {P.~C.}\ \bibnamefont
  {Sabatier}},\ }\bibfield  {title} {\bibinfo {title} {Past and future of
  inverse problems},\ }\href@noop {} {\bibfield  {journal} {\bibinfo  {journal}
  {J. Math. Phys.}\ }\textbf {\bibinfo {volume} {41}},\ \bibinfo {pages} {4082}
  (\bibinfo {year} {2000})}\BibitemShut {NoStop}%
\bibitem [{\citenamefont {Risken}(1989)}]{Risken1989}%
  \BibitemOpen
  \bibfield  {author} {\bibinfo {author} {\bibfnamefont {H.}~\bibnamefont
  {Risken}},\ }\href@noop {} {\emph {\bibinfo {title} {The Fokker-Planck
  Equation}}},\ \bibinfo {edition} {2nd}\ ed.\ (\bibinfo  {publisher} {Springer
  Verlag, Berlin},\ \bibinfo {year} {1989})\BibitemShut {NoStop}%
\bibitem [{\citenamefont {Teague}(1983)}]{Teague1983}%
  \BibitemOpen
  \bibfield  {author} {\bibinfo {author} {\bibfnamefont {M.~R.}\ \bibnamefont
  {Teague}},\ }\bibfield  {title} {\bibinfo {title} {Deterministic phase
  retrieval: a {G}reen's function solution},\ }\href
  {http://www.opticsinfobase.org/abstract.cfm?URI=josa-73-11-1434} {\bibfield
  {journal} {\bibinfo  {journal} {J. Opt. Soc. Am.}\ }\textbf {\bibinfo
  {volume} {73}},\ \bibinfo {pages} {1434} (\bibinfo {year}
  {1983})}\BibitemShut {NoStop}%
\bibitem [{\citenamefont {Gerchberg}\ and\ \citenamefont
  {Saxton}(1972)}]{gerchberg1972practical}%
  \BibitemOpen
  \bibfield  {author} {\bibinfo {author} {\bibfnamefont {R.~W.}\ \bibnamefont
  {Gerchberg}}\ and\ \bibinfo {author} {\bibfnamefont {W.~O.}\ \bibnamefont
  {Saxton}},\ }\bibfield  {title} {\bibinfo {title} {A practical algorithm for
  the determination of phase from image and diffraction plane pictures},\
  }\href@noop {} {\bibfield  {journal} {\bibinfo  {journal} {Optik}\ }\textbf
  {\bibinfo {volume} {35}},\ \bibinfo {pages} {237} (\bibinfo {year}
  {1972})}\BibitemShut {NoStop}%
\bibitem [{\citenamefont {Paganin}\ and\ \citenamefont
  {Nugent}(1998)}]{paganin1998}%
  \BibitemOpen
  \bibfield  {author} {\bibinfo {author} {\bibfnamefont {D.}~\bibnamefont
  {Paganin}}\ and\ \bibinfo {author} {\bibfnamefont {K.~A.}\ \bibnamefont
  {Nugent}},\ }\bibfield  {title} {\bibinfo {title} {Noninterferometric phase
  imaging with partially coherent light},\ }\href
  {https://link.aps.org/doi/10.1103/PhysRevLett.80.2586} {\bibfield  {journal}
  {\bibinfo  {journal} {Phys. Rev. Lett.}\ }\textbf {\bibinfo {volume} {80}},\
  \bibinfo {pages} {2586} (\bibinfo {year} {1998})}\BibitemShut {NoStop}%
\bibitem [{\citenamefont {Zuo}\ \emph {et~al.}(2020)\citenamefont {Zuo},
  \citenamefont {Li}, \citenamefont {Sun}, \citenamefont {Fan}, \citenamefont
  {Zhang}, \citenamefont {Lu}, \citenamefont {Zhang}, \citenamefont {Wang},
  \citenamefont {Huang},\ and\ \citenamefont
  {Chen}}]{TIE==LongReviewArticle2020}%
  \BibitemOpen
  \bibfield  {author} {\bibinfo {author} {\bibfnamefont {C.}~\bibnamefont
  {Zuo}}, \bibinfo {author} {\bibfnamefont {J.}~\bibnamefont {Li}}, \bibinfo
  {author} {\bibfnamefont {J.}~\bibnamefont {Sun}}, \bibinfo {author}
  {\bibfnamefont {Y.}~\bibnamefont {Fan}}, \bibinfo {author} {\bibfnamefont
  {J.}~\bibnamefont {Zhang}}, \bibinfo {author} {\bibfnamefont
  {L.}~\bibnamefont {Lu}}, \bibinfo {author} {\bibfnamefont {R.}~\bibnamefont
  {Zhang}}, \bibinfo {author} {\bibfnamefont {B.}~\bibnamefont {Wang}},
  \bibinfo {author} {\bibfnamefont {L.}~\bibnamefont {Huang}},\ and\ \bibinfo
  {author} {\bibfnamefont {Q.}~\bibnamefont {Chen}},\ }\bibfield  {title}
  {\bibinfo {title} {Transport of intensity equation: a tutorial},\ }\href
  {https://doi.org/https://doi.org/10.1016/j.optlaseng.2020.106187} {\bibfield
  {journal} {\bibinfo  {journal} {Opt. Lasers Eng.}\ }\textbf {\bibinfo
  {volume} {135}},\ \bibinfo {pages} {106187} (\bibinfo {year}
  {2020})}\BibitemShut {NoStop}%
\bibitem [{\citenamefont {Paganin}\ and\ \citenamefont
  {Morgan}(2019)}]{PaganinMorgan2019}%
  \BibitemOpen
  \bibfield  {author} {\bibinfo {author} {\bibfnamefont {D.~M.}\ \bibnamefont
  {Paganin}}\ and\ \bibinfo {author} {\bibfnamefont {K.~S.}\ \bibnamefont
  {Morgan}},\ }\bibfield  {title} {\bibinfo {title} {X-ray {F}okker-{P}lanck
  equation for paraxial imaging},\ }\href@noop {} {\bibfield  {journal}
  {\bibinfo  {journal} {Sci. Rep.}\ }\textbf {\bibinfo {volume} {9}},\ \bibinfo
  {pages} {17537} (\bibinfo {year} {2019})}\BibitemShut {NoStop}%
\bibitem [{\citenamefont {Morgan}\ and\ \citenamefont
  {Paganin}(2019)}]{MorganPaganin2019}%
  \BibitemOpen
  \bibfield  {author} {\bibinfo {author} {\bibfnamefont {K.~S.}\ \bibnamefont
  {Morgan}}\ and\ \bibinfo {author} {\bibfnamefont {D.~M.}\ \bibnamefont
  {Paganin}},\ }\bibfield  {title} {\bibinfo {title} {Applying the
  {F}okker-{P}lanck equation to grating-based x-ray phase and dark-field
  imaging},\ }\href@noop {} {\bibfield  {journal} {\bibinfo  {journal} {Sci.
  Rep.}\ }\textbf {\bibinfo {volume} {9}},\ \bibinfo {pages} {17465} (\bibinfo
  {year} {2019})}\BibitemShut {NoStop}%
\bibitem [{\citenamefont {Pavlov}\ \emph {et~al.}(2020)\citenamefont {Pavlov},
  \citenamefont {Paganin}, \citenamefont {Li}, \citenamefont {Berujon},
  \citenamefont {Roug{\'e}-Labriet},\ and\ \citenamefont {Brun}}]{MIST}%
  \BibitemOpen
  \bibfield  {author} {\bibinfo {author} {\bibfnamefont {K.~M.}\ \bibnamefont
  {Pavlov}}, \bibinfo {author} {\bibfnamefont {D.~M.}\ \bibnamefont {Paganin}},
  \bibinfo {author} {\bibfnamefont {H.~T.}\ \bibnamefont {Li}}, \bibinfo
  {author} {\bibfnamefont {S.}~\bibnamefont {Berujon}}, \bibinfo {author}
  {\bibfnamefont {H.}~\bibnamefont {Roug{\'e}-Labriet}},\ and\ \bibinfo
  {author} {\bibfnamefont {E.}~\bibnamefont {Brun}},\ }\bibfield  {title}
  {\bibinfo {title} {X-ray multi-modal intrinsic-speckle-tracking},\
  }\href@noop {} {\bibfield  {journal} {\bibinfo  {journal} {J. Opt.}\ }\textbf
  {\bibinfo {volume} {22}},\ \bibinfo {pages} {125604} (\bibinfo {year}
  {2020})}\BibitemShut {NoStop}%
\bibitem [{\citenamefont {Gureyev}\ \emph {et~al.}(2020)\citenamefont
  {Gureyev}, \citenamefont {Paganin}, \citenamefont {Arhatari}, \citenamefont
  {Taba}, \citenamefont {Lewis}, \citenamefont {Brennan},\ and\ \citenamefont
  {Quiney}}]{Gureyev2020}%
  \BibitemOpen
  \bibfield  {author} {\bibinfo {author} {\bibfnamefont {T.~E.}\ \bibnamefont
  {Gureyev}}, \bibinfo {author} {\bibfnamefont {D.~M.}\ \bibnamefont
  {Paganin}}, \bibinfo {author} {\bibfnamefont {B.~D.}\ \bibnamefont
  {Arhatari}}, \bibinfo {author} {\bibfnamefont {S.~T.}\ \bibnamefont {Taba}},
  \bibinfo {author} {\bibfnamefont {S.}~\bibnamefont {Lewis}}, \bibinfo
  {author} {\bibfnamefont {P.~C.}\ \bibnamefont {Brennan}},\ and\ \bibinfo
  {author} {\bibfnamefont {H.~M.}\ \bibnamefont {Quiney}},\ }\bibfield  {title}
  {\bibinfo {title} {Dark-field signal extraction in propagation-based
  phase-contrast imaging},\ }\href@noop {} {\bibfield  {journal} {\bibinfo
  {journal} {Phys. Med. Biol.}\ }\textbf {\bibinfo {volume} {65}},\ \bibinfo
  {pages} {215029} (\bibinfo {year} {2020})}\BibitemShut {NoStop}%
\bibitem [{\citenamefont {Pavlov}\ \emph {et~al.}(2021)\citenamefont {Pavlov},
  \citenamefont {Paganin}, \citenamefont {Morgan}, \citenamefont {Li},
  \citenamefont {Berujon}, \citenamefont {Qu\'enot},\ and\ \citenamefont
  {Brun}}]{MISTdirectional}%
  \BibitemOpen
  \bibfield  {author} {\bibinfo {author} {\bibfnamefont {K.~M.}\ \bibnamefont
  {Pavlov}}, \bibinfo {author} {\bibfnamefont {D.~M.}\ \bibnamefont {Paganin}},
  \bibinfo {author} {\bibfnamefont {K.~S.}\ \bibnamefont {Morgan}}, \bibinfo
  {author} {\bibfnamefont {H.~T.}\ \bibnamefont {Li}}, \bibinfo {author}
  {\bibfnamefont {S.}~\bibnamefont {Berujon}}, \bibinfo {author} {\bibfnamefont
  {L.}~\bibnamefont {Qu\'enot}},\ and\ \bibinfo {author} {\bibfnamefont
  {E.}~\bibnamefont {Brun}},\ }\bibfield  {title} {\bibinfo {title}
  {Directional dark-field implicit x-ray speckle tracking using an
  anisotropic-diffusion {F}okker-{P}lanck equation},\ }\href@noop {} {\bibfield
   {journal} {\bibinfo  {journal} {Phys. Rev. A}\ }\textbf {\bibinfo {volume}
  {104}},\ \bibinfo {pages} {053505} (\bibinfo {year} {2021})}\BibitemShut
  {NoStop}%
\bibitem [{\citenamefont {Paganin}\ and\ \citenamefont
  {Pelliccia}(2021)}]{PaganinPelliccia2021}%
  \BibitemOpen
  \bibfield  {author} {\bibinfo {author} {\bibfnamefont {D.~M.}\ \bibnamefont
  {Paganin}}\ and\ \bibinfo {author} {\bibfnamefont {D.}~\bibnamefont
  {Pelliccia}},\ }\bibfield  {title} {\bibinfo {title} {{X}-ray phase-contrast
  imaging: a broad overview of some fundamentals},\ }\href@noop {} {\bibfield
  {journal} {\bibinfo  {journal} {Adv. Imaging Electron Phys.}\ }\textbf
  {\bibinfo {volume} {218}},\ \bibinfo {pages} {63} (\bibinfo {year}
  {2021})}\BibitemShut {NoStop}%
\bibitem [{\citenamefont {Alloo}\ \emph {et~al.}(2022)\citenamefont {Alloo},
  \citenamefont {Paganin}, \citenamefont {Morgan}, \citenamefont {Kitchen},
  \citenamefont {Stevenson}, \citenamefont {Mayo}, \citenamefont {Li},
  \citenamefont {Kennedy}, \citenamefont {Maksimenko}, \citenamefont {Bowden},\
  and\ \citenamefont {Pavlov}}]{alloo2022dark}%
  \BibitemOpen
  \bibfield  {author} {\bibinfo {author} {\bibfnamefont {S.~J.}\ \bibnamefont
  {Alloo}}, \bibinfo {author} {\bibfnamefont {D.~M.}\ \bibnamefont {Paganin}},
  \bibinfo {author} {\bibfnamefont {K.~S.}\ \bibnamefont {Morgan}}, \bibinfo
  {author} {\bibfnamefont {M.~J.}\ \bibnamefont {Kitchen}}, \bibinfo {author}
  {\bibfnamefont {A.~W.}\ \bibnamefont {Stevenson}}, \bibinfo {author}
  {\bibfnamefont {S.~C.}\ \bibnamefont {Mayo}}, \bibinfo {author}
  {\bibfnamefont {H.~T.}\ \bibnamefont {Li}}, \bibinfo {author} {\bibfnamefont
  {B.~M.}\ \bibnamefont {Kennedy}}, \bibinfo {author} {\bibfnamefont
  {A.}~\bibnamefont {Maksimenko}}, \bibinfo {author} {\bibfnamefont {J.~C.}\
  \bibnamefont {Bowden}},\ and\ \bibinfo {author} {\bibfnamefont {K.~M.}\
  \bibnamefont {Pavlov}},\ }\bibfield  {title} {\bibinfo {title} {Dark-field
  tomography of an attenuating object using intrinsic x-ray speckle tracking},\
  }\href@noop {} {\bibfield  {journal} {\bibinfo  {journal} {J. Med. Imaging}\
  }\textbf {\bibinfo {volume} {9}},\ \bibinfo {pages} {031502} (\bibinfo {year}
  {2022})}\BibitemShut {NoStop}%
\bibitem [{\citenamefont {Alloo}\ \emph {et~al.}(2023)\citenamefont {Alloo},
  \citenamefont {Morgan}, \citenamefont {Paganin},\ and\ \citenamefont
  {Pavlov}}]{alloo2023SciRep}%
  \BibitemOpen
  \bibfield  {author} {\bibinfo {author} {\bibfnamefont {S.~J.}\ \bibnamefont
  {Alloo}}, \bibinfo {author} {\bibfnamefont {K.~S.}\ \bibnamefont {Morgan}},
  \bibinfo {author} {\bibfnamefont {D.~M.}\ \bibnamefont {Paganin}},\ and\
  \bibinfo {author} {\bibfnamefont {K.~M.}\ \bibnamefont {Pavlov}},\ }\bibfield
   {title} {\bibinfo {title} {Multimodal intrinsic speckle‐tracking ({MIST})
  to extract images of rapidly‐varying diffuse {X}‐ray dark‐field},\
  }\href@noop {} {\bibfield  {journal} {\bibinfo  {journal} {Sci. Rep.}\
  }\textbf {\bibinfo {volume} {13}},\ \bibinfo {pages} {5424} (\bibinfo {year}
  {2023})}\BibitemShut {NoStop}%
\bibitem [{\citenamefont {Leatham}\ \emph {et~al.}(2023)\citenamefont
  {Leatham}, \citenamefont {Paganin},\ and\ \citenamefont
  {Morgan}}]{Leatham2023}%
  \BibitemOpen
  \bibfield  {author} {\bibinfo {author} {\bibfnamefont {T.~A.}\ \bibnamefont
  {Leatham}}, \bibinfo {author} {\bibfnamefont {D.~M.}\ \bibnamefont
  {Paganin}},\ and\ \bibinfo {author} {\bibfnamefont {K.~S.}\ \bibnamefont
  {Morgan}},\ }\bibfield  {title} {\bibinfo {title} {X-ray dark-field and phase
  retrieval without optics, via the {F}okker-{P}lanck equation},\ }\href
  {https://doi.org/10.1109/TMI.2023.3234901} {\bibfield  {journal} {\bibinfo
  {journal} {IEEE Trans. Med. Imaging}\ }\textbf {\bibinfo {volume} {42}},\
  \bibinfo {pages} {1681} (\bibinfo {year} {2023})}\BibitemShut {NoStop}%
\bibitem [{\citenamefont {Beltran}\ \emph {et~al.}(2023)\citenamefont
  {Beltran}, \citenamefont {Paganin}, \citenamefont {Croughan},\ and\
  \citenamefont {Morgan}}]{Beltran2023}%
  \BibitemOpen
  \bibfield  {author} {\bibinfo {author} {\bibfnamefont {M.~A.}\ \bibnamefont
  {Beltran}}, \bibinfo {author} {\bibfnamefont {D.~M.}\ \bibnamefont
  {Paganin}}, \bibinfo {author} {\bibfnamefont {M.~K.}\ \bibnamefont
  {Croughan}},\ and\ \bibinfo {author} {\bibfnamefont {K.~S.}\ \bibnamefont
  {Morgan}},\ }\bibfield  {title} {\bibinfo {title} {Fast implicit diffusive
  dark-field retrieval for single-exposure, single-mask x-ray imaging},\ }\href
  {https://doi.org/10.1364/OPTICA.480489} {\bibfield  {journal} {\bibinfo
  {journal} {Optica}\ }\textbf {\bibinfo {volume} {10}},\ \bibinfo {pages}
  {422} (\bibinfo {year} {2023})}\BibitemShut {NoStop}%
\bibitem [{\citenamefont {Esposito}\ \emph {et~al.}(2023)\citenamefont
  {Esposito}, \citenamefont {Buchanan}, \citenamefont {Massimi}, \citenamefont
  {Ferrara}, \citenamefont {Shearing}, \citenamefont {Olivo},\ and\
  \citenamefont {Endrizzi}}]{Esposito2023}%
  \BibitemOpen
  \bibfield  {author} {\bibinfo {author} {\bibfnamefont {M.}~\bibnamefont
  {Esposito}}, \bibinfo {author} {\bibfnamefont {I.}~\bibnamefont {Buchanan}},
  \bibinfo {author} {\bibfnamefont {L.}~\bibnamefont {Massimi}}, \bibinfo
  {author} {\bibfnamefont {J.~D.}\ \bibnamefont {Ferrara}}, \bibinfo {author}
  {\bibfnamefont {P.~R.}\ \bibnamefont {Shearing}}, \bibinfo {author}
  {\bibfnamefont {A.}~\bibnamefont {Olivo}},\ and\ \bibinfo {author}
  {\bibfnamefont {M.}~\bibnamefont {Endrizzi}},\ }\bibfield  {title} {\bibinfo
  {title} {Laboratory-based x-ray dark-field microscopy},\ }\href
  {https://doi.org/10.1103/PhysRevApplied.20.064039} {\bibfield  {journal}
  {\bibinfo  {journal} {Phys. Rev. Appl.}\ }\textbf {\bibinfo {volume} {20}},\
  \bibinfo {pages} {064039} (\bibinfo {year} {2023})}\BibitemShut {NoStop}%
\bibitem [{\citenamefont {Magnin}\ \emph {et~al.}(2023)\citenamefont {Magnin},
  \citenamefont {Qu\'{e}not}, \citenamefont {Bohic}, \citenamefont {Cenda},
  \citenamefont {{Fern\'{a}ndez Mart\'{i}nez}}, \citenamefont {Lantz},
  \citenamefont {Faure},\ and\ \citenamefont {Brun}}]{Magnin2023}%
  \BibitemOpen
  \bibfield  {author} {\bibinfo {author} {\bibfnamefont {C.}~\bibnamefont
  {Magnin}}, \bibinfo {author} {\bibfnamefont {L.}~\bibnamefont {Qu\'{e}not}},
  \bibinfo {author} {\bibfnamefont {S.}~\bibnamefont {Bohic}}, \bibinfo
  {author} {\bibfnamefont {D.~M.}\ \bibnamefont {Cenda}}, \bibinfo {author}
  {\bibfnamefont {M.}~\bibnamefont {{Fern\'{a}ndez Mart\'{i}nez}}}, \bibinfo
  {author} {\bibfnamefont {B.}~\bibnamefont {Lantz}}, \bibinfo {author}
  {\bibfnamefont {B.}~\bibnamefont {Faure}},\ and\ \bibinfo {author}
  {\bibfnamefont {E.}~\bibnamefont {Brun}},\ }\bibfield  {title} {\bibinfo
  {title} {Dark-field and directional dark-field on low-coherence x ray sources
  with random mask modulations: validation with {SAXS} anisotropy
  measurements},\ }\href {https://doi.org/10.1364/OL.501716} {\bibfield
  {journal} {\bibinfo  {journal} {Opt. Lett.}\ }\textbf {\bibinfo {volume}
  {48}},\ \bibinfo {pages} {5839} (\bibinfo {year} {2023})}\BibitemShut
  {NoStop}%
\bibitem [{\citenamefont {Ahlers}\ \emph {et~al.}(2024)\citenamefont {Ahlers},
  \citenamefont {Pavlov}, \citenamefont {Kitchen},\ and\ \citenamefont
  {Morgan}}]{Ahlers2024}%
  \BibitemOpen
  \bibfield  {author} {\bibinfo {author} {\bibfnamefont {J.~N.}\ \bibnamefont
  {Ahlers}}, \bibinfo {author} {\bibfnamefont {K.~M.}\ \bibnamefont {Pavlov}},
  \bibinfo {author} {\bibfnamefont {M.~J.}\ \bibnamefont {Kitchen}},\ and\
  \bibinfo {author} {\bibfnamefont {K.~S.}\ \bibnamefont {Morgan}},\ }\bibfield
   {title} {\bibinfo {title} {X-ray dark-field via spectral propagation-based
  imaging},\ }\href {https://doi.org/10.1364/OPTICA.506742} {\bibfield
  {journal} {\bibinfo  {journal} {Optica}\ }\textbf {\bibinfo {volume} {11}},\
  \bibinfo {pages} {1182} (\bibinfo {year} {2024})}\BibitemShut {NoStop}%
\bibitem [{\citenamefont {Leatham}\ \emph {et~al.}(2024)\citenamefont
  {Leatham}, \citenamefont {Paganin},\ and\ \citenamefont
  {Morgan}}]{Leatham2024}%
  \BibitemOpen
  \bibfield  {author} {\bibinfo {author} {\bibfnamefont {T.~A.}\ \bibnamefont
  {Leatham}}, \bibinfo {author} {\bibfnamefont {D.~M.}\ \bibnamefont
  {Paganin}},\ and\ \bibinfo {author} {\bibfnamefont {K.~S.}\ \bibnamefont
  {Morgan}},\ }\bibfield  {title} {\bibinfo {title} {X-ray phase and dark-field
  computed tomography without optical elements},\ }\href
  {https://doi.org/10.1364/OE.509604} {\bibfield  {journal} {\bibinfo
  {journal} {Opt. Express}\ }\textbf {\bibinfo {volume} {32}},\ \bibinfo
  {pages} {4588} (\bibinfo {year} {2024})}\BibitemShut {NoStop}%
\bibitem [{\citenamefont {Celestre}\ \emph {et~al.}(2025)\citenamefont
  {Celestre}, \citenamefont {Qu{\'{e}}not}, \citenamefont {Ninham},
  \citenamefont {Brun},\ and\ \citenamefont {Fardin}}]{CelestreReview2025}%
  \BibitemOpen
  \bibfield  {author} {\bibinfo {author} {\bibfnamefont {R.}~\bibnamefont
  {Celestre}}, \bibinfo {author} {\bibfnamefont {L.}~\bibnamefont
  {Qu{\'{e}}not}}, \bibinfo {author} {\bibfnamefont {C.}~\bibnamefont
  {Ninham}}, \bibinfo {author} {\bibfnamefont {E.}~\bibnamefont {Brun}},\ and\
  \bibinfo {author} {\bibfnamefont {L.}~\bibnamefont {Fardin}},\ }\bibfield
  {title} {\bibinfo {title} {{Review and experimental comparison of
  speckle-tracking algorithms for {X}-ray phase contrast imaging}},\ }\href
  {https://doi.org/10.1107/S1600577524010117} {\bibfield  {journal} {\bibinfo
  {journal} {J. Synchrotron Rad.}\ }\textbf {\bibinfo {volume} {32}},\ \bibinfo
  {pages} {180} (\bibinfo {year} {2025})}\BibitemShut {NoStop}%
\bibitem [{\citenamefont {Alloo}\ \emph {et~al.}(2025)\citenamefont {Alloo},
  \citenamefont {Paganin}, \citenamefont {Croughan}, \citenamefont {Ahlers},
  \citenamefont {Pavlov},\ and\ \citenamefont {Morgan}}]{Alloo2025}%
  \BibitemOpen
  \bibfield  {author} {\bibinfo {author} {\bibfnamefont {S.~J.}\ \bibnamefont
  {Alloo}}, \bibinfo {author} {\bibfnamefont {D.~M.}\ \bibnamefont {Paganin}},
  \bibinfo {author} {\bibfnamefont {M.~K.}\ \bibnamefont {Croughan}}, \bibinfo
  {author} {\bibfnamefont {J.~N.}\ \bibnamefont {Ahlers}}, \bibinfo {author}
  {\bibfnamefont {K.~M.}\ \bibnamefont {Pavlov}},\ and\ \bibinfo {author}
  {\bibfnamefont {K.~S.}\ \bibnamefont {Morgan}},\ }\bibfield  {title}
  {\bibinfo {title} {Separating edges from microstructure in x-ray dark-field
  imaging: evolving and devolving perspectives via the {X}-ray
  {F}okker-{P}lanck equation},\ }\href {https://doi.org/10.1364/OE.545960}
  {\bibfield  {journal} {\bibinfo  {journal} {Opt. Express}\ }\textbf {\bibinfo
  {volume} {33}},\ \bibinfo {pages} {3577} (\bibinfo {year}
  {2025})}\BibitemShut {NoStop}%
\bibitem [{\citenamefont {Ahlers}\ \emph {et~al.}(2025)\citenamefont {Ahlers},
  \citenamefont {Pavlov}, \citenamefont {Kitchen}, \citenamefont {Harker},
  \citenamefont {Pryor}, \citenamefont {Pollock}, \citenamefont {Croughan},
  \citenamefont {How}, \citenamefont {Zdora}, \citenamefont {Costello},
  \citenamefont {O'Connell}, \citenamefont {Hall},\ and\ \citenamefont
  {Morgan}}]{Ahlers2025}%
  \BibitemOpen
  \bibfield  {author} {\bibinfo {author} {\bibfnamefont {J.~N.}\ \bibnamefont
  {Ahlers}}, \bibinfo {author} {\bibfnamefont {K.~M.}\ \bibnamefont {Pavlov}},
  \bibinfo {author} {\bibfnamefont {M.~J.}\ \bibnamefont {Kitchen}}, \bibinfo
  {author} {\bibfnamefont {S.~A.}\ \bibnamefont {Harker}}, \bibinfo {author}
  {\bibfnamefont {E.~J.}\ \bibnamefont {Pryor}}, \bibinfo {author}
  {\bibfnamefont {J.~A.}\ \bibnamefont {Pollock}}, \bibinfo {author}
  {\bibfnamefont {M.~K.}\ \bibnamefont {Croughan}}, \bibinfo {author}
  {\bibfnamefont {Y.~Y.}\ \bibnamefont {How}}, \bibinfo {author} {\bibfnamefont
  {M.-C.}\ \bibnamefont {Zdora}}, \bibinfo {author} {\bibfnamefont {L.~F.}\
  \bibnamefont {Costello}}, \bibinfo {author} {\bibfnamefont {D.~W.}\
  \bibnamefont {O'Connell}}, \bibinfo {author} {\bibfnamefont {C.}~\bibnamefont
  {Hall}},\ and\ \bibinfo {author} {\bibfnamefont {K.~S.}\ \bibnamefont
  {Morgan}},\ }\bibfield  {title} {\bibinfo {title} {Single-exposure x-ray
  dark-field imaging via a dual-energy propagation-based setup},\ }\href
  {https://doi.org/10.1364/OL.553310} {\bibfield  {journal} {\bibinfo
  {journal} {Opt. Lett.}\ }\textbf {\bibinfo {volume} {50}},\ \bibinfo {pages}
  {2171} (\bibinfo {year} {2025})}\BibitemShut {NoStop}%
\bibitem [{\citenamefont {Liu}\ \emph {et~al.}(2025)\citenamefont {Liu},
  \citenamefont {Alloo}, \citenamefont {Langer},\ and\ \citenamefont
  {Pavlov}}]{LiuAllooLangerPavlovPhysRevA2025}%
  \BibitemOpen
  \bibfield  {author} {\bibinfo {author} {\bibfnamefont {J.}~\bibnamefont
  {Liu}}, \bibinfo {author} {\bibfnamefont {S.~J.}\ \bibnamefont {Alloo}},
  \bibinfo {author} {\bibfnamefont {M.}~\bibnamefont {Langer}},\ and\ \bibinfo
  {author} {\bibfnamefont {K.~M.}\ \bibnamefont {Pavlov}},\ }\bibfield  {title}
  {\bibinfo {title} {Low-exposure high-quality multimodal speckle x-ray imaging
  via an intrinsic gradient-flow approach},\ }\href
  {https://doi.org/10.1103/scnz-pqy3} {\bibfield  {journal} {\bibinfo
  {journal} {Phys. Rev. A}\ }\textbf {\bibinfo {volume} {112}},\ \bibinfo
  {pages} {063530} (\bibinfo {year} {2025})}\BibitemShut {NoStop}%
\bibitem [{\citenamefont {Alloo}\ and\ \citenamefont
  {Morgan}(2025)}]{AllooPhysicaScripta2025}%
  \BibitemOpen
  \bibfield  {author} {\bibinfo {author} {\bibfnamefont {S.~J.}\ \bibnamefont
  {Alloo}}\ and\ \bibinfo {author} {\bibfnamefont {K.~S.}\ \bibnamefont
  {Morgan}},\ }\bibfield  {title} {\bibinfo {title} {Stabilizing {L}aplacian
  inversion in {F}okker–{P}lanck image retrieval using the
  transport-of-intensity equation},\ }\href
  {https://doi.org/10.1088/1402-4896/ade834} {\bibfield  {journal} {\bibinfo
  {journal} {Phys. Scr.}\ }\textbf {\bibinfo {volume} {100}},\ \bibinfo {pages}
  {075566} (\bibinfo {year} {2025})}\BibitemShut {NoStop}%
\bibitem [{\citenamefont {Yuan}\ and\ \citenamefont {Das}(2025)}]{YuanDas25}%
  \BibitemOpen
  \bibfield  {author} {\bibinfo {author} {\bibfnamefont {J.}~\bibnamefont
  {Yuan}}\ and\ \bibinfo {author} {\bibfnamefont {M.}~\bibnamefont {Das}},\
  }\bibfield  {title} {\bibinfo {title} {Single-shot, single-mask {X}-ray
  dark-field, and phase-contrast imaging},\ }\href@noop {} {\bibfield
  {journal} {\bibinfo  {journal} {Optica}\ }\textbf {\bibinfo {volume} {12}},\
  \bibinfo {pages} {1895} (\bibinfo {year} {2025})}\BibitemShut {NoStop}%
\bibitem [{\citenamefont {Alloo}\ \emph
  {et~al.}(2026{\natexlab{a}})\citenamefont {Alloo}, \citenamefont {How},
  \citenamefont {Ahlers}, \citenamefont {Paganin}, \citenamefont {Croughan},\
  and\ \citenamefont {Morgan}}]{Alloo2026JSynchRad}%
  \BibitemOpen
  \bibfield  {author} {\bibinfo {author} {\bibfnamefont {S.~J.}\ \bibnamefont
  {Alloo}}, \bibinfo {author} {\bibfnamefont {Y.~Y.}\ \bibnamefont {How}},
  \bibinfo {author} {\bibfnamefont {J.~N.}\ \bibnamefont {Ahlers}}, \bibinfo
  {author} {\bibfnamefont {D.~M.}\ \bibnamefont {Paganin}}, \bibinfo {author}
  {\bibfnamefont {M.~K.}\ \bibnamefont {Croughan}},\ and\ \bibinfo {author}
  {\bibfnamefont {K.~S.}\ \bibnamefont {Morgan}},\ }\bibfield  {title}
  {\bibinfo {title} {{Demonstration of a family of {X}-ray dark-field retrieval
  approaches on a common set of samples}},\ }\href@noop {} {\bibfield
  {journal} {\bibinfo  {journal} {J. Synchrotron Rad.}\ }\textbf {\bibinfo
  {volume} {33}},\ \bibinfo {pages} {437} (\bibinfo {year}
  {2026}{\natexlab{a}})}\BibitemShut {NoStop}%
\bibitem [{\citenamefont {Alloo}\ \emph
  {et~al.}(2026{\natexlab{b}})\citenamefont {Alloo}, \citenamefont {Schaff},
  \citenamefont {Gradl}, \citenamefont {Gunther}, \citenamefont {Pfeiffer},\
  and\ \citenamefont {Morgan}}]{alloo2026arXiv}%
  \BibitemOpen
  \bibfield  {author} {\bibinfo {author} {\bibfnamefont {S.~J.}\ \bibnamefont
  {Alloo}}, \bibinfo {author} {\bibfnamefont {F.}~\bibnamefont {Schaff}},
  \bibinfo {author} {\bibfnamefont {R.}~\bibnamefont {Gradl}}, \bibinfo
  {author} {\bibfnamefont {B.}~\bibnamefont {Gunther}}, \bibinfo {author}
  {\bibfnamefont {F.}~\bibnamefont {Pfeiffer}},\ and\ \bibinfo {author}
  {\bibfnamefont {K.~S.}\ \bibnamefont {Morgan}},\ }\href
  {https://arxiv.org/abs/2605.01265} {\bibinfo {title} {X-ray dark-field
  imaging from intensity flow: A {F}okker-{P}lanck approach to grating
  interferometry}} (\bibinfo {year} {2026}{\natexlab{b}}),\ \Eprint
  {https://arxiv.org/abs/2605.01265} {arXiv:2605.01265} \BibitemShut {NoStop}%
\bibitem [{\citenamefont {Allan}\ \emph {et~al.}(2026)\citenamefont {Allan},
  \citenamefont {Navarrete-León}, \citenamefont {Doherty}, \citenamefont
  {Marathe}, \citenamefont {Wanelik},\ and\ \citenamefont
  {Endrizzi}}]{allan2026arXiv}%
  \BibitemOpen
  \bibfield  {author} {\bibinfo {author} {\bibfnamefont {H.}~\bibnamefont
  {Allan}}, \bibinfo {author} {\bibfnamefont {C.}~\bibnamefont
  {Navarrete-León}}, \bibinfo {author} {\bibfnamefont {A.}~\bibnamefont
  {Doherty}}, \bibinfo {author} {\bibfnamefont {S.}~\bibnamefont {Marathe}},
  \bibinfo {author} {\bibfnamefont {K.}~\bibnamefont {Wanelik}},\ and\ \bibinfo
  {author} {\bibfnamefont {M.}~\bibnamefont {Endrizzi}},\ }\href
  {https://arxiv.org/abs/2606.05269} {\bibinfo {title} {Spatial resolution of
  {X}-ray beam-tracking microscopy}} (\bibinfo {year} {2026}),\ \Eprint
  {https://arxiv.org/abs/2606.05269} {arXiv:2606.05269} \BibitemShut {NoStop}%
\bibitem [{\citenamefont {Magnin}\ \emph {et~al.}(2026)\citenamefont {Magnin},
  \citenamefont {Paganin}, \citenamefont {Pavlov}, \citenamefont {Morgan},
  \citenamefont {Faure}, \citenamefont {{Fern\'{a}ndez Mart\'{i}nez}},\ and\
  \citenamefont {Brun}}]{Magnin2026}%
  \BibitemOpen
  \bibfield  {author} {\bibinfo {author} {\bibfnamefont {C.}~\bibnamefont
  {Magnin}}, \bibinfo {author} {\bibfnamefont {D.~M.}\ \bibnamefont {Paganin}},
  \bibinfo {author} {\bibfnamefont {K.~M.}\ \bibnamefont {Pavlov}}, \bibinfo
  {author} {\bibfnamefont {K.~S.}\ \bibnamefont {Morgan}}, \bibinfo {author}
  {\bibfnamefont {B.}~\bibnamefont {Faure}}, \bibinfo {author} {\bibfnamefont
  {M.}~\bibnamefont {{Fern\'{a}ndez Mart\'{i}nez}}},\ and\ \bibinfo {author}
  {\bibfnamefont {E.}~\bibnamefont {Brun}},\ }\bibfield  {title} {\bibinfo
  {title} {X-ray dark-field contrast as a function of sample-to-detector
  propagation distance},\ }\href@noop {} {\bibfield  {journal} {\bibinfo
  {journal} {Optica}\ } (\bibinfo {year} {2026})},\ \bibinfo {note} {in
  press}\BibitemShut {NoStop}%
\bibitem [{\citenamefont {Lindberg}\ \emph {et~al.}(2026)\citenamefont
  {Lindberg}, \citenamefont {Rossi},\ and\ \citenamefont
  {Zboray}}]{Lindberg2026PMB}%
  \BibitemOpen
  \bibfield  {author} {\bibinfo {author} {\bibfnamefont {W.~A.}\ \bibnamefont
  {Lindberg}}, \bibinfo {author} {\bibfnamefont {R.~M.}\ \bibnamefont
  {Rossi}},\ and\ \bibinfo {author} {\bibfnamefont {R.}~\bibnamefont
  {Zboray}},\ }\bibfield  {title} {\bibinfo {title} {Performance and
  feasibility of single-grid dark-field radiography with a clinical x-ray
  source},\ }\href@noop {} {\bibfield  {journal} {\bibinfo  {journal} {Phys.
  Med. Biol.}\ } (\bibinfo {year} {2026})},\ \bibinfo {note} {in
  press}\BibitemShut {NoStop}%
\bibitem [{\citenamefont {Zboray}\ \emph {et~al.}(2026)\citenamefont {Zboray},
  \citenamefont {Lindberg},\ and\ \citenamefont
  {Rossi}}]{Zboray2026OpticaOpenPreprint}%
  \BibitemOpen
  \bibfield  {author} {\bibinfo {author} {\bibfnamefont {R.}~\bibnamefont
  {Zboray}}, \bibinfo {author} {\bibfnamefont {W.~A.}\ \bibnamefont
  {Lindberg}},\ and\ \bibinfo {author} {\bibfnamefont {R.~M.}\ \bibnamefont
  {Rossi}},\ }\href {https://doi.org/10.1364/opticaopen.33039593} {\bibinfo
  {title} {Revisiting the {F}okker-{P}lanck approach and its relation to the
  sample’s real space correlation function in single-grid {X}-ray dark-field
  imaging}} (\bibinfo {year} {2026}),\ \Eprint {https://arxiv.org/abs/134485}
  {Optica Open preprint:134485} \BibitemShut {NoStop}%
\bibitem [{\citenamefont {Paganin}\ \emph
  {et~al.}(2023{\natexlab{b}})\citenamefont {Paganin}, \citenamefont {Sales},
  \citenamefont {Kadletz}, \citenamefont {Kockelmann}, \citenamefont {Beltran},
  \citenamefont {Poulsen},\ and\ \citenamefont {Schmidt}}]{PaganinNeutron2023}%
  \BibitemOpen
  \bibfield  {author} {\bibinfo {author} {\bibfnamefont {D.~M.}\ \bibnamefont
  {Paganin}}, \bibinfo {author} {\bibfnamefont {M.}~\bibnamefont {Sales}},
  \bibinfo {author} {\bibfnamefont {P.~M.}\ \bibnamefont {Kadletz}}, \bibinfo
  {author} {\bibfnamefont {W.}~\bibnamefont {Kockelmann}}, \bibinfo {author}
  {\bibfnamefont {M.~A.}\ \bibnamefont {Beltran}}, \bibinfo {author}
  {\bibfnamefont {H.~F.}\ \bibnamefont {Poulsen}},\ and\ \bibinfo {author}
  {\bibfnamefont {S.}~\bibnamefont {Schmidt}},\ }\bibfield  {title} {\bibinfo
  {title} {Revisiting neutron propagation-based phase-contrast imaging and
  tomography: Use of phase retrieval to amplify the effective degree of
  brilliance},\ }\href {https://doi.org/10.1103/PhysRevApplied.19.034005}
  {\bibfield  {journal} {\bibinfo  {journal} {Phys. Rev. Appl.}\ }\textbf
  {\bibinfo {volume} {19}},\ \bibinfo {pages} {034005} (\bibinfo {year}
  {2023}{\natexlab{b}})}\BibitemShut {NoStop}%
\bibitem [{\citenamefont {Lanczos}(1970)}]{Lanczos1970}%
  \BibitemOpen
  \bibfield  {author} {\bibinfo {author} {\bibfnamefont {C.}~\bibnamefont
  {Lanczos}},\ }\href@noop {} {\emph {\bibinfo {title} {The Variational
  Principles of Mechanics}}},\ \bibinfo {edition} {4th}\ ed.\ (\bibinfo
  {publisher} {University of Toronto Press, Toronto},\ \bibinfo {year}
  {1970})\BibitemShut {NoStop}%
\bibitem [{\citenamefont {Born}\ and\ \citenamefont {Wolf}(1999)}]{BornWolf}%
  \BibitemOpen
  \bibfield  {author} {\bibinfo {author} {\bibfnamefont {M.}~\bibnamefont
  {Born}}\ and\ \bibinfo {author} {\bibfnamefont {E.}~\bibnamefont {Wolf}},\
  }\href@noop {} {\emph {\bibinfo {title} {Principles of Optics}}},\ \bibinfo
  {edition} {7th}\ ed.\ (\bibinfo  {publisher} {Cambridge University Press,
  Cambridge},\ \bibinfo {year} {1999})\BibitemShut {NoStop}%
\bibitem [{\citenamefont {Jackson}(1999)}]{JacksonElectrodynamicsBook}%
  \BibitemOpen
  \bibfield  {author} {\bibinfo {author} {\bibfnamefont {J.~D.}\ \bibnamefont
  {Jackson}},\ }\href@noop {} {\emph {\bibinfo {title} {Classical
  Electrodynamics}}},\ \bibinfo {edition} {3rd}\ ed.\ (\bibinfo  {publisher}
  {John Wiley \& Sons, New York},\ \bibinfo {year} {1999})\BibitemShut
  {NoStop}%
\bibitem [{\citenamefont {Berry}(2013)}]{BerryFiveMomenta2013}%
  \BibitemOpen
  \bibfield  {author} {\bibinfo {author} {\bibfnamefont {M.~V.}\ \bibnamefont
  {Berry}},\ }\bibfield  {title} {\bibinfo {title} {Five momenta},\ }\href
  {https://doi.org/10.1088/0143-0807/34/6/1337} {\bibfield  {journal} {\bibinfo
   {journal} {Eur. J. Phys.}\ }\textbf {\bibinfo {volume} {34}},\ \bibinfo
  {pages} {1337} (\bibinfo {year} {2013})}\BibitemShut {NoStop}%
\bibitem [{\citenamefont {Zernike}(1942)}]{Zernike1942}%
  \BibitemOpen
  \bibfield  {author} {\bibinfo {author} {\bibfnamefont {F.}~\bibnamefont
  {Zernike}},\ }\bibfield  {title} {\bibinfo {title} {Phase contrast, a new
  method for the microscopic observation of transparent objects},\ }\href@noop
  {} {\bibfield  {journal} {\bibinfo  {journal} {Physica}\ }\textbf {\bibinfo
  {volume} {9}},\ \bibinfo {pages} {686} (\bibinfo {year} {1942})}\BibitemShut
  {NoStop}%
\bibitem [{\citenamefont {Bremmer}(1952)}]{Bremmer1952}%
  \BibitemOpen
  \bibfield  {author} {\bibinfo {author} {\bibfnamefont {H.}~\bibnamefont
  {Bremmer}},\ }\bibfield  {title} {\bibinfo {title} {On the asymptotic
  evaluation of diffraction integrals with a special view to the theory of
  defocusing and optical contrast},\ }\href
  {https://doi.org/https://doi.org/10.1016/S0031-8914(52)80079-5} {\bibfield
  {journal} {\bibinfo  {journal} {Physica}\ }\textbf {\bibinfo {volume} {18}},\
  \bibinfo {pages} {469} (\bibinfo {year} {1952})}\BibitemShut {NoStop}%
\bibitem [{\citenamefont {Cowley}(1959)}]{Cowley1959}%
  \BibitemOpen
  \bibfield  {author} {\bibinfo {author} {\bibfnamefont {J.~M.}\ \bibnamefont
  {Cowley}},\ }\bibfield  {title} {\bibinfo {title} {The electron-optical
  imaging of crystal lattices},\ }\href@noop {} {\bibfield  {journal} {\bibinfo
   {journal} {Acta Cryst.}\ }\textbf {\bibinfo {volume} {12}},\ \bibinfo
  {pages} {367} (\bibinfo {year} {1959})}\BibitemShut {NoStop}%
\bibitem [{\citenamefont {Cowley}(1995)}]{CowleyBook}%
  \BibitemOpen
  \bibfield  {author} {\bibinfo {author} {\bibfnamefont {J.~M.}\ \bibnamefont
  {Cowley}},\ }\href@noop {} {\emph {\bibinfo {title} {Diffraction Physics}}},\
  \bibinfo {edition} {3rd}\ ed.\ (\bibinfo  {publisher} {North Holland},\
  \bibinfo {address} {Amsterdam},\ \bibinfo {year} {1995})\BibitemShut
  {NoStop}%
\bibitem [{\citenamefont {Gilbarg}\ and\ \citenamefont
  {Trudinger}(1998)}]{GilbargTrudingerBook}%
  \BibitemOpen
  \bibfield  {author} {\bibinfo {author} {\bibfnamefont {D.}~\bibnamefont
  {Gilbarg}}\ and\ \bibinfo {author} {\bibfnamefont {N.~S.}\ \bibnamefont
  {Trudinger}},\ }\href@noop {} {\emph {\bibinfo {title} {Elliptic Partial
  Differential Equations of Second Order}}},\ \bibinfo {edition} {2nd}\ ed.\
  (\bibinfo  {publisher} {Springer, Berlin},\ \bibinfo {year}
  {1998})\BibitemShut {NoStop}%
\bibitem [{\citenamefont {Gureyev}\ \emph
  {et~al.}(1995{\natexlab{a}})\citenamefont {Gureyev}, \citenamefont
  {Roberts},\ and\ \citenamefont {Nugent}}]{Gureyev1995}%
  \BibitemOpen
  \bibfield  {author} {\bibinfo {author} {\bibfnamefont {T.~E.}\ \bibnamefont
  {Gureyev}}, \bibinfo {author} {\bibfnamefont {A.}~\bibnamefont {Roberts}},\
  and\ \bibinfo {author} {\bibfnamefont {K.~A.}\ \bibnamefont {Nugent}},\
  }\bibfield  {title} {\bibinfo {title} {Partially coherent fields, the
  transport-of-intensity equation, and phase uniqueness},\ }\href
  {https://doi.org/10.1364/JOSAA.12.001942} {\bibfield  {journal} {\bibinfo
  {journal} {J. Opt. Soc. Am. A}\ }\textbf {\bibinfo {volume} {12}},\ \bibinfo
  {pages} {1942} (\bibinfo {year} {1995}{\natexlab{a}})}\BibitemShut {NoStop}%
\bibitem [{\citenamefont {Hecht}(2017)}]{HechtOpticsBook}%
  \BibitemOpen
  \bibfield  {author} {\bibinfo {author} {\bibfnamefont {E.}~\bibnamefont
  {Hecht}},\ }\href@noop {} {\emph {\bibinfo {title} {Optics}}},\ \bibinfo
  {edition} {5th}\ ed.\ (\bibinfo  {publisher} {Pearson, Boston},\ \bibinfo
  {year} {2017})\BibitemShut {NoStop}%
\bibitem [{\citenamefont {Paganin}\ \emph {et~al.}(2020)\citenamefont
  {Paganin}, \citenamefont {Favre-Nicolin}, \citenamefont {Mirone},
  \citenamefont {Rack}, \citenamefont {Villanova}, \citenamefont {Olbinado},
  \citenamefont {Fernandez}, \citenamefont {da~Silva},\ and\ \citenamefont
  {Pelliccia}}]{GPM2020}%
  \BibitemOpen
  \bibfield  {author} {\bibinfo {author} {\bibfnamefont {D.~M.}\ \bibnamefont
  {Paganin}}, \bibinfo {author} {\bibfnamefont {V.}~\bibnamefont
  {Favre-Nicolin}}, \bibinfo {author} {\bibfnamefont {A.}~\bibnamefont
  {Mirone}}, \bibinfo {author} {\bibfnamefont {A.}~\bibnamefont {Rack}},
  \bibinfo {author} {\bibfnamefont {J.}~\bibnamefont {Villanova}}, \bibinfo
  {author} {\bibfnamefont {M.~P.}\ \bibnamefont {Olbinado}}, \bibinfo {author}
  {\bibfnamefont {V.}~\bibnamefont {Fernandez}}, \bibinfo {author}
  {\bibfnamefont {J.~C.}\ \bibnamefont {da~Silva}},\ and\ \bibinfo {author}
  {\bibfnamefont {D.}~\bibnamefont {Pelliccia}},\ }\bibfield  {title} {\bibinfo
  {title} {Boosting spatial resolution by incorporating periodic boundary
  conditions into single-distance hard-x-ray phase retrieval},\ }\href@noop {}
  {\bibfield  {journal} {\bibinfo  {journal} {J. Opt.}\ }\textbf {\bibinfo
  {volume} {22}},\ \bibinfo {pages} {115607} (\bibinfo {year}
  {2020})}\BibitemShut {NoStop}%
\bibitem [{\citenamefont {Davis}(1991)}]{Davis1991}%
  \BibitemOpen
  \bibfield  {author} {\bibinfo {author} {\bibfnamefont {T.~J.}\ \bibnamefont
  {Davis}},\ }\bibfield  {title} {\bibinfo {title} {Imperfect crystals and
  dynamical {X}-ray diffraction in the complex reflectance plane},\ }\href@noop
  {} {\bibfield  {journal} {\bibinfo  {journal} {Aust. J. Phys.}\ }\textbf
  {\bibinfo {volume} {44}},\ \bibinfo {pages} {693} (\bibinfo {year}
  {1991})}\BibitemShut {NoStop}%
\bibitem [{\citenamefont {Davis}(1992)}]{Davis1992}%
  \BibitemOpen
  \bibfield  {author} {\bibinfo {author} {\bibfnamefont {T.~J.}\ \bibnamefont
  {Davis}},\ }\bibfield  {title} {\bibinfo {title} {A stochastic model for
  {X}-ray diffraction from imperfect crystals},\ }\href@noop {} {\bibfield
  {journal} {\bibinfo  {journal} {Acta Cryst. A}\ }\textbf {\bibinfo {volume}
  {48}},\ \bibinfo {pages} {872} (\bibinfo {year} {1992})}\BibitemShut
  {NoStop}%
\bibitem [{\citenamefont {Davis}(1993)}]{Davis1993}%
  \BibitemOpen
  \bibfield  {author} {\bibinfo {author} {\bibfnamefont {T.~J.}\ \bibnamefont
  {Davis}},\ }\bibfield  {title} {\bibinfo {title} {The measurement of defect
  parameters in imperfect crystals using {X}-ray diffraction},\ }\href@noop {}
  {\bibfield  {journal} {\bibinfo  {journal} {Acta Cryst. A}\ }\textbf
  {\bibinfo {volume} {49}},\ \bibinfo {pages} {755} (\bibinfo {year}
  {1993})}\BibitemShut {NoStop}%
\bibitem [{\citenamefont {Davis}(1994)}]{Davis1994}%
  \BibitemOpen
  \bibfield  {author} {\bibinfo {author} {\bibfnamefont {T.~J.}\ \bibnamefont
  {Davis}},\ }\bibfield  {title} {\bibinfo {title} {Dynamical {X}-ray
  diffraction from imperfect crystals: a solution based on the
  {F}okker-{P}lanck equation},\ }\href@noop {} {\bibfield  {journal} {\bibinfo
  {journal} {Acta Cryst. A}\ }\textbf {\bibinfo {volume} {50}},\ \bibinfo
  {pages} {224} (\bibinfo {year} {1994})}\BibitemShut {NoStop}%
\bibitem [{\citenamefont {Sakurai}(1967)}]{SakuraiAdvancedQuantumMechanics}%
  \BibitemOpen
  \bibfield  {author} {\bibinfo {author} {\bibfnamefont {J.~J.}\ \bibnamefont
  {Sakurai}},\ }\href@noop {} {\emph {\bibinfo {title} {Advanced Quantum
  Mechanics}}}\ (\bibinfo  {publisher} {Addison-Wesley},\ \bibinfo {address}
  {Reading},\ \bibinfo {year} {1967})\BibitemShut {NoStop}%
\bibitem [{\citenamefont {Wolf}(2007)}]{ThinWolfBook}%
  \BibitemOpen
  \bibfield  {author} {\bibinfo {author} {\bibfnamefont {E.}~\bibnamefont
  {Wolf}},\ }\href@noop {} {\emph {\bibinfo {title} {Introduction to the Theory
  of Coherence and Polarization of Light}}}\ (\bibinfo  {publisher} {Cambridge
  University Press},\ \bibinfo {address} {Cambridge},\ \bibinfo {year}
  {2007})\BibitemShut {NoStop}%
\bibitem [{\citenamefont {Loudon}(2000)}]{LoudonBook}%
  \BibitemOpen
  \bibfield  {author} {\bibinfo {author} {\bibfnamefont {R.}~\bibnamefont
  {Loudon}},\ }\href@noop {} {\emph {\bibinfo {title} {The Quantum Theory of
  Light}}},\ \bibinfo {edition} {3rd}\ ed.\ (\bibinfo  {publisher} {Oxford
  University Press},\ \bibinfo {address} {Oxford},\ \bibinfo {year}
  {2000})\BibitemShut {NoStop}%
\bibitem [{\citenamefont {Loane}\ \emph {et~al.}(1991)\citenamefont {Loane},
  \citenamefont {Xu},\ and\ \citenamefont {Silcox}}]{FrozenPhononModel}%
  \BibitemOpen
  \bibfield  {author} {\bibinfo {author} {\bibfnamefont {R.~F.}\ \bibnamefont
  {Loane}}, \bibinfo {author} {\bibfnamefont {P.}~\bibnamefont {Xu}},\ and\
  \bibinfo {author} {\bibfnamefont {J.}~\bibnamefont {Silcox}},\ }\bibfield
  {title} {\bibinfo {title} {Thermal vibrations in convergent-beam electron
  diffraction},\ }\href
  {https://doi.org/https://doi.org/10.1107/S0108767391000375} {\bibfield
  {journal} {\bibinfo  {journal} {Acta Cryst. A}\ }\textbf {\bibinfo {volume}
  {47}},\ \bibinfo {pages} {267} (\bibinfo {year} {1991})}\BibitemShut
  {NoStop}%
\bibitem [{\citenamefont {Nesterets}(2008)}]{Nesterets2008}%
  \BibitemOpen
  \bibfield  {author} {\bibinfo {author} {\bibfnamefont {{\relax Ya}.~I.}\
  \bibnamefont {Nesterets}},\ }\bibfield  {title} {\bibinfo {title} {On the
  origins of decoherence and extinction contrast in phase-contrast imaging},\
  }\href@noop {} {\bibfield  {journal} {\bibinfo  {journal} {Opt. Commun.}\
  }\textbf {\bibinfo {volume} {281}},\ \bibinfo {pages} {533} (\bibinfo {year}
  {2008})}\BibitemShut {NoStop}%
\bibitem [{\citenamefont {Crank}(1975)}]{CrankBook}%
  \BibitemOpen
  \bibfield  {author} {\bibinfo {author} {\bibfnamefont {J.}~\bibnamefont
  {Crank}},\ }\href@noop {} {\emph {\bibinfo {title} {The Mathematics of
  Diffusion}}},\ \bibinfo {edition} {2nd}\ ed.\ (\bibinfo  {publisher} {Oxford
  University Press, Oxford},\ \bibinfo {year} {1975})\BibitemShut {NoStop}%
\bibitem [{\citenamefont {Mandl}\ and\ \citenamefont
  {Shaw}(2010)}]{MandlShawBook}%
  \BibitemOpen
  \bibfield  {author} {\bibinfo {author} {\bibfnamefont {F.}~\bibnamefont
  {Mandl}}\ and\ \bibinfo {author} {\bibfnamefont {G.}~\bibnamefont {Shaw}},\
  }\href@noop {} {\emph {\bibinfo {title} {Quantum Field Theory}}},\ \bibinfo
  {edition} {2nd}\ ed.\ (\bibinfo  {publisher} {Wiley},\ \bibinfo {address}
  {Chichester},\ \bibinfo {year} {2010})\BibitemShut {NoStop}%
\bibitem [{\citenamefont {Goldberg}\ \emph {et~al.}(1967)\citenamefont
  {Goldberg}, \citenamefont {Schey},\ and\ \citenamefont
  {Schwartz}}]{Goldberg1967}%
  \BibitemOpen
  \bibfield  {author} {\bibinfo {author} {\bibfnamefont {A.}~\bibnamefont
  {Goldberg}}, \bibinfo {author} {\bibfnamefont {H.~M.}\ \bibnamefont
  {Schey}},\ and\ \bibinfo {author} {\bibfnamefont {J.~L.}\ \bibnamefont
  {Schwartz}},\ }\bibfield  {title} {\bibinfo {title} {Computer-generated
  motion pictures of one-dimensional quantum-mechanical transmission and
  reﬂection phenomena},\ }\href@noop {} {\bibfield  {journal} {\bibinfo
  {journal} {Am. J. Phys.}\ }\textbf {\bibinfo {volume} {35}},\ \bibinfo
  {pages} {177} (\bibinfo {year} {1967})}\BibitemShut {NoStop}%
\bibitem [{\citenamefont {Paganin}(1999)}]{PaganinPhDthesis1999}%
  \BibitemOpen
  \bibfield  {author} {\bibinfo {author} {\bibfnamefont {D.~M.}\ \bibnamefont
  {Paganin}},\ }\emph {\bibinfo {title} {Studies in Phase Retrieval}},\
  \href@noop {} {Ph.D. thesis},\ \bibinfo  {school} {University of Melbourne}
  (\bibinfo {year} {1999})\BibitemShut {NoStop}%
\bibitem [{\citenamefont {Paganin}(2006)}]{Paganin2006}%
  \BibitemOpen
  \bibfield  {author} {\bibinfo {author} {\bibfnamefont {D.~M.}\ \bibnamefont
  {Paganin}},\ }\href@noop {} {\emph {\bibinfo {title} {Coherent X-{R}ay
  Optics}}}\ (\bibinfo  {publisher} {Oxford University Press, Oxford},\
  \bibinfo {year} {2006})\BibitemShut {NoStop}%
\bibitem [{\citenamefont {Nazarathy}\ and\ \citenamefont
  {Shamir}(1980)}]{NazarathyShamir1980}%
  \BibitemOpen
  \bibfield  {author} {\bibinfo {author} {\bibfnamefont {M.}~\bibnamefont
  {Nazarathy}}\ and\ \bibinfo {author} {\bibfnamefont {J.}~\bibnamefont
  {Shamir}},\ }\bibfield  {title} {\bibinfo {title} {Fourier optics described
  by operator algebra},\ }\href@noop {} {\bibfield  {journal} {\bibinfo
  {journal} {J. Opt. Soc. Am.}\ }\textbf {\bibinfo {volume} {70}},\ \bibinfo
  {pages} {150} (\bibinfo {year} {1980})}\BibitemShut {NoStop}%
\bibitem [{\citenamefont {Winthrop}\ and\ \citenamefont
  {Worthington}(1966)}]{WinthropWorthington66}%
  \BibitemOpen
  \bibfield  {author} {\bibinfo {author} {\bibfnamefont {J.~T.}\ \bibnamefont
  {Winthrop}}\ and\ \bibinfo {author} {\bibfnamefont {C.~R.}\ \bibnamefont
  {Worthington}},\ }\bibfield  {title} {\bibinfo {title} {Convolution
  formulation of {F}resnel diffraction},\ }\href
  {https://doi.org/10.1364/JOSA.56.000588} {\bibfield  {journal} {\bibinfo
  {journal} {J. Opt. Soc. Am.}\ }\textbf {\bibinfo {volume} {56}},\ \bibinfo
  {pages} {588} (\bibinfo {year} {1966})}\BibitemShut {NoStop}%
\bibitem [{\citenamefont {Metzler}\ and\ \citenamefont
  {Klafter}(2000)}]{MetzlerKlafter2000}%
  \BibitemOpen
  \bibfield  {author} {\bibinfo {author} {\bibfnamefont {R.}~\bibnamefont
  {Metzler}}\ and\ \bibinfo {author} {\bibfnamefont {J.}~\bibnamefont
  {Klafter}},\ }\bibfield  {title} {\bibinfo {title} {The random walk's guide
  to anomalous diffusion: a fractional dynamics approach},\ }\href
  {https://doi.org/https://doi.org/10.1016/S0370-1573(00)00070-3} {\bibfield
  {journal} {\bibinfo  {journal} {Phys. Rep.}\ }\textbf {\bibinfo {volume}
  {339}},\ \bibinfo {pages} {1} (\bibinfo {year} {2000})}\BibitemShut {NoStop}%
\bibitem [{\citenamefont {Evangelista}\ and\ \citenamefont
  {Lenzi}(2018)}]{EvangelistaLenziBook2018}%
  \BibitemOpen
  \bibfield  {author} {\bibinfo {author} {\bibfnamefont {L.~R.}\ \bibnamefont
  {Evangelista}}\ and\ \bibinfo {author} {\bibfnamefont {E.~K.}\ \bibnamefont
  {Lenzi}},\ }\href@noop {} {\emph {\bibinfo {title} {Fractional Diffusion
  Equations and Anomalous Diffusion}}}\ (\bibinfo  {publisher} {Cambridge
  University Press},\ \bibinfo {address} {Cambridge},\ \bibinfo {year}
  {2018})\BibitemShut {NoStop}%
\bibitem [{\citenamefont {Einstein}(1956)}]{EinsteinBrownianMotion}%
  \BibitemOpen
  \bibfield  {author} {\bibinfo {author} {\bibfnamefont {A.}~\bibnamefont
  {Einstein}},\ }\href@noop {} {\emph {\bibinfo {title} {Investigations on the
  Theory of the Brownian Movement}}}\ (\bibinfo  {publisher} {Dover
  Publications, New York},\ \bibinfo {year} {1956})\BibitemShut {NoStop}%
\bibitem [{\citenamefont {Tisza}(1947)}]{Tisza1947}%
  \BibitemOpen
  \bibfield  {author} {\bibinfo {author} {\bibfnamefont {L.}~\bibnamefont
  {Tisza}},\ }\bibfield  {title} {\bibinfo {title} {The theory of liquid
  helium},\ }\href {https://doi.org/10.1103/PhysRev.72.838} {\bibfield
  {journal} {\bibinfo  {journal} {Phys. Rev.}\ }\textbf {\bibinfo {volume}
  {72}},\ \bibinfo {pages} {838} (\bibinfo {year} {1947})}\BibitemShut
  {NoStop}%
\bibitem [{\citenamefont {Yourgrau}\ and\ \citenamefont
  {Mandelstam}(1979)}]{YourgrauMandelstam}%
  \BibitemOpen
  \bibfield  {author} {\bibinfo {author} {\bibfnamefont {W.}~\bibnamefont
  {Yourgrau}}\ and\ \bibinfo {author} {\bibfnamefont {S.}~\bibnamefont
  {Mandelstam}},\ }\href@noop {} {\emph {\bibinfo {title} {Variational
  Principles in Dynamics and Quantum Theory}}},\ \bibinfo {edition} {3rd}\ ed.\
  (\bibinfo  {publisher} {Dover Publications, New York},\ \bibinfo {year}
  {1979})\BibitemShut {NoStop}%
\bibitem [{\citenamefont {Huang}(1987)}]{Huang_1987}%
  \BibitemOpen
  \bibfield  {author} {\bibinfo {author} {\bibfnamefont {K.}~\bibnamefont
  {Huang}},\ }\href@noop {} {\emph {\bibinfo {title} {{S}tatistical
  {M}echanics}}},\ \bibinfo {edition} {2nd}\ ed.\ (\bibinfo  {publisher}
  {Wiley},\ \bibinfo {address} {New York, NY, USA},\ \bibinfo {year}
  {1987})\BibitemShut {NoStop}%
\bibitem [{\citenamefont {Sethna}(2006)}]{SethnaBook}%
  \BibitemOpen
  \bibfield  {author} {\bibinfo {author} {\bibfnamefont {J.~P.}\ \bibnamefont
  {Sethna}},\ }\href@noop {} {\emph {\bibinfo {title} {Statistical Mechanics:
  Entropy, Order Parameters and Complexity}}}\ (\bibinfo  {publisher} {Oxford
  University Press, Oxford},\ \bibinfo {year} {2006})\BibitemShut {NoStop}%
\bibitem [{\citenamefont {Zernike}(1938)}]{Zernike1938}%
  \BibitemOpen
  \bibfield  {author} {\bibinfo {author} {\bibfnamefont {F.}~\bibnamefont
  {Zernike}},\ }\bibfield  {title} {\bibinfo {title} {The concept of degree of
  coherence and its application to optical problems},\ }\href
  {https://doi.org/https://doi.org/10.1016/S0031-8914(38)80203-2} {\bibfield
  {journal} {\bibinfo  {journal} {Physica}\ }\textbf {\bibinfo {volume} {5}},\
  \bibinfo {pages} {785} (\bibinfo {year} {1938})}\BibitemShut {NoStop}%
\bibitem [{\citenamefont {Mandel}\ and\ \citenamefont
  {Wolf}(1995)}]{MandelWolf}%
  \BibitemOpen
  \bibfield  {author} {\bibinfo {author} {\bibfnamefont {L.}~\bibnamefont
  {Mandel}}\ and\ \bibinfo {author} {\bibfnamefont {E.}~\bibnamefont {Wolf}},\
  }\href@noop {} {\emph {\bibinfo {title} {Optical Coherence and Quantum
  Optics}}}\ (\bibinfo  {publisher} {Cambridge University Press, Cambridge},\
  \bibinfo {year} {1995})\BibitemShut {NoStop}%
\bibitem [{\citenamefont {Pavliotis}(2014)}]{PavliotisBook2014}%
  \BibitemOpen
  \bibfield  {author} {\bibinfo {author} {\bibfnamefont {G.~A.}\ \bibnamefont
  {Pavliotis}},\ }\href@noop {} {\emph {\bibinfo {title} {Stochastic Processes
  and Applications: {D}iffusion Processes, the {F}okker-{P}lanck and {L}angevin
  Equations}}}\ (\bibinfo  {publisher} {Springer, New York},\ \bibinfo {year}
  {2014})\BibitemShut {NoStop}%
\bibitem [{\citenamefont {Jensen}\ \emph
  {et~al.}(2010{\natexlab{a}})\citenamefont {Jensen}, \citenamefont {Bech},
  \citenamefont {Bunk}, \citenamefont {Donath}, \citenamefont {David},
  \citenamefont {Feidenhans'l},\ and\ \citenamefont {Pfeiffer}}]{jensen2010a}%
  \BibitemOpen
  \bibfield  {author} {\bibinfo {author} {\bibfnamefont {T.~H.}\ \bibnamefont
  {Jensen}}, \bibinfo {author} {\bibfnamefont {M.}~\bibnamefont {Bech}},
  \bibinfo {author} {\bibfnamefont {O.}~\bibnamefont {Bunk}}, \bibinfo {author}
  {\bibfnamefont {T.}~\bibnamefont {Donath}}, \bibinfo {author} {\bibfnamefont
  {C.}~\bibnamefont {David}}, \bibinfo {author} {\bibfnamefont
  {R.}~\bibnamefont {Feidenhans'l}},\ and\ \bibinfo {author} {\bibfnamefont
  {F.}~\bibnamefont {Pfeiffer}},\ }\bibfield  {title} {\bibinfo {title}
  {Directional x-ray dark-field imaging},\ }\href@noop {} {\bibfield  {journal}
  {\bibinfo  {journal} {Phys. Med. Biol.}\ }\textbf {\bibinfo {volume} {55}},\
  \bibinfo {pages} {3317} (\bibinfo {year} {2010}{\natexlab{a}})}\BibitemShut
  {NoStop}%
\bibitem [{\citenamefont {Jensen}\ \emph
  {et~al.}(2010{\natexlab{b}})\citenamefont {Jensen}, \citenamefont {Bech},
  \citenamefont {Zanette}, \citenamefont {Weitkamp}, \citenamefont {David},
  \citenamefont {Deyhle}, \citenamefont {Rutishauser}, \citenamefont
  {Reznikova}, \citenamefont {Mohr}, \citenamefont {Feidenhans'l},\ and\
  \citenamefont {Pfeiffer}}]{jensen2010b}%
  \BibitemOpen
  \bibfield  {author} {\bibinfo {author} {\bibfnamefont {T.~H.}\ \bibnamefont
  {Jensen}}, \bibinfo {author} {\bibfnamefont {M.}~\bibnamefont {Bech}},
  \bibinfo {author} {\bibfnamefont {I.}~\bibnamefont {Zanette}}, \bibinfo
  {author} {\bibfnamefont {T.}~\bibnamefont {Weitkamp}}, \bibinfo {author}
  {\bibfnamefont {C.}~\bibnamefont {David}}, \bibinfo {author} {\bibfnamefont
  {H.}~\bibnamefont {Deyhle}}, \bibinfo {author} {\bibfnamefont
  {S.}~\bibnamefont {Rutishauser}}, \bibinfo {author} {\bibfnamefont
  {E.}~\bibnamefont {Reznikova}}, \bibinfo {author} {\bibfnamefont
  {J.}~\bibnamefont {Mohr}}, \bibinfo {author} {\bibfnamefont {R.}~\bibnamefont
  {Feidenhans'l}},\ and\ \bibinfo {author} {\bibfnamefont {F.}~\bibnamefont
  {Pfeiffer}},\ }\bibfield  {title} {\bibinfo {title} {Directional x-ray
  dark-field imaging of strongly ordered systems},\ }\href
  {https://doi.org/10.1103/PhysRevB.82.214103} {\bibfield  {journal} {\bibinfo
  {journal} {Phys. Rev. B}\ }\textbf {\bibinfo {volume} {82}},\ \bibinfo
  {pages} {214103} (\bibinfo {year} {2010}{\natexlab{b}})}\BibitemShut
  {NoStop}%
\bibitem [{\citenamefont {Kittel}(1958)}]{KittelStatPhysBook}%
  \BibitemOpen
  \bibfield  {author} {\bibinfo {author} {\bibfnamefont {C.}~\bibnamefont
  {Kittel}},\ }\href@noop {} {\emph {\bibinfo {title} {Elementary Statistical
  Physics}}}\ (\bibinfo  {publisher} {John Wiley \& Sons, New York},\ \bibinfo
  {year} {1958})\BibitemShut {NoStop}%
\bibitem [{\citenamefont {Scully}\ and\ \citenamefont
  {Zubairy}(1997)}]{ScullyZubairy}%
  \BibitemOpen
  \bibfield  {author} {\bibinfo {author} {\bibfnamefont {M.~O.}\ \bibnamefont
  {Scully}}\ and\ \bibinfo {author} {\bibfnamefont {M.~S.}\ \bibnamefont
  {Zubairy}},\ }\href@noop {} {\emph {\bibinfo {title} {Quantum Optics}}}\
  (\bibinfo  {publisher} {Cambridge University Press, Cambridge},\ \bibinfo
  {year} {1997})\BibitemShut {NoStop}%
\bibitem [{\citenamefont
  {Balakrishnan}(2021)}]{BalakrishnanNonEqmStatMechBook}%
  \BibitemOpen
  \bibfield  {author} {\bibinfo {author} {\bibfnamefont {V.}~\bibnamefont
  {Balakrishnan}},\ }\href@noop {} {\emph {\bibinfo {title} {Elements of
  Nonequilibrium Statistical Mechanics}}}\ (\bibinfo  {publisher} {Springer,
  Cham},\ \bibinfo {year} {2021})\BibitemShut {NoStop}%
\bibitem [{\citenamefont {Goldstein}(1980)}]{GoldsteinBook}%
  \BibitemOpen
  \bibfield  {author} {\bibinfo {author} {\bibfnamefont {H.}~\bibnamefont
  {Goldstein}},\ }\href@noop {} {\emph {\bibinfo {title} {Classical
  Mechanics}}},\ \bibinfo {edition} {2nd}\ ed.\ (\bibinfo  {publisher}
  {Addison-Wesley Publishing Company, Reading MA},\ \bibinfo {year}
  {1980})\BibitemShut {NoStop}%
\bibitem [{\citenamefont {Barrett}\ and\ \citenamefont
  {Myers}(2004)}]{BarrettMyersBook}%
  \BibitemOpen
  \bibfield  {author} {\bibinfo {author} {\bibfnamefont {H.~H.}\ \bibnamefont
  {Barrett}}\ and\ \bibinfo {author} {\bibfnamefont {K.~J.}\ \bibnamefont
  {Myers}},\ }\href@noop {} {\emph {\bibinfo {title} {Foundations of Image
  Science}}}\ (\bibinfo  {publisher} {John Wiley \& Sons, Hoboken NJ},\
  \bibinfo {year} {2004})\BibitemShut {NoStop}%
\bibitem [{\citenamefont {Oldham}\ and\ \citenamefont
  {Spanier}(2006)}]{OldhamSpanier}%
  \BibitemOpen
  \bibfield  {author} {\bibinfo {author} {\bibfnamefont {K.~B.}\ \bibnamefont
  {Oldham}}\ and\ \bibinfo {author} {\bibfnamefont {J.}~\bibnamefont
  {Spanier}},\ }\href@noop {} {\emph {\bibinfo {title} {The Fractional
  Calculus: Theory and Applications of Differentiation and Integration to
  Arbitrary Order}}}\ (\bibinfo  {publisher} {Dover Publications, New York},\
  \bibinfo {year} {2006})\BibitemShut {NoStop}%
\bibitem [{\citenamefont {Yanovsky}\ \emph {et~al.}(2000)\citenamefont
  {Yanovsky}, \citenamefont {Chechkin}, \citenamefont {Schertzer},\ and\
  \citenamefont {Tur}}]{YANOVSKY200013}%
  \BibitemOpen
  \bibfield  {author} {\bibinfo {author} {\bibfnamefont {V.~V.}\ \bibnamefont
  {Yanovsky}}, \bibinfo {author} {\bibfnamefont {A.~V.}\ \bibnamefont
  {Chechkin}}, \bibinfo {author} {\bibfnamefont {D.}~\bibnamefont
  {Schertzer}},\ and\ \bibinfo {author} {\bibfnamefont {A.~V.}\ \bibnamefont
  {Tur}},\ }\bibfield  {title} {\bibinfo {title} {L\'{e}vy anomalous diffusion
  and fractional {F}okker–{P}lanck equation},\ }\href
  {https://doi.org/https://doi.org/10.1016/S0378-4371(99)00565-8} {\bibfield
  {journal} {\bibinfo  {journal} {Physica A}\ }\textbf {\bibinfo {volume}
  {282}},\ \bibinfo {pages} {13} (\bibinfo {year} {2000})}\BibitemShut
  {NoStop}%
\bibitem [{\citenamefont {Kelly}\ \emph {et~al.}(2018)\citenamefont {Kelly},
  \citenamefont {Li},\ and\ \citenamefont
  {Meerschaert}}]{KellyBallisticAnomalousDiffusion}%
  \BibitemOpen
  \bibfield  {author} {\bibinfo {author} {\bibfnamefont {J.~F.}\ \bibnamefont
  {Kelly}}, \bibinfo {author} {\bibfnamefont {C.-G.}\ \bibnamefont {Li}},\ and\
  \bibinfo {author} {\bibfnamefont {M.~M.}\ \bibnamefont {Meerschaert}},\
  }\bibfield  {title} {\bibinfo {title} {Anomalous diffusion with ballistic
  scaling: A new fractional derivative},\ }\href
  {https://doi.org/https://doi.org/10.1016/j.cam.2017.11.012} {\bibfield
  {journal} {\bibinfo  {journal} {J. Comput. Appl. Math.}\ }\textbf {\bibinfo
  {volume} {339}},\ \bibinfo {pages} {161} (\bibinfo {year}
  {2018})}\BibitemShut {NoStop}%
\bibitem [{\citenamefont {Zaslavsky}(1994)}]{ZASLAVSKY-FractionalFPE1994}%
  \BibitemOpen
  \bibfield  {author} {\bibinfo {author} {\bibfnamefont {G.}~\bibnamefont
  {Zaslavsky}},\ }\bibfield  {title} {\bibinfo {title} {Fractional kinetic
  equation for {H}amiltonian chaos},\ }\href
  {https://doi.org/https://doi.org/10.1016/0167-2789(94)90254-2} {\bibfield
  {journal} {\bibinfo  {journal} {Physica D}\ }\textbf {\bibinfo {volume}
  {76}},\ \bibinfo {pages} {110} (\bibinfo {year} {1994})}\BibitemShut
  {NoStop}%
\bibitem [{\citenamefont {Metzler}\ \emph {et~al.}(1999)\citenamefont
  {Metzler}, \citenamefont {Barkai},\ and\ \citenamefont
  {Klafter}}]{MetzlerPRL1999}%
  \BibitemOpen
  \bibfield  {author} {\bibinfo {author} {\bibfnamefont {R.}~\bibnamefont
  {Metzler}}, \bibinfo {author} {\bibfnamefont {E.}~\bibnamefont {Barkai}},\
  and\ \bibinfo {author} {\bibfnamefont {J.}~\bibnamefont {Klafter}},\
  }\bibfield  {title} {\bibinfo {title} {Anomalous diffusion and relaxation
  close to thermal equilibrium: A fractional {F}okker-{P}lanck equation
  approach},\ }\href {https://doi.org/10.1103/PhysRevLett.82.3563} {\bibfield
  {journal} {\bibinfo  {journal} {Phys. Rev. Lett.}\ }\textbf {\bibinfo
  {volume} {82}},\ \bibinfo {pages} {3563} (\bibinfo {year}
  {1999})}\BibitemShut {NoStop}%
\bibitem [{\citenamefont {Liu}\ \emph {et~al.}(2004)\citenamefont {Liu},
  \citenamefont {Anh},\ and\ \citenamefont {Turner}}]{SpaceFractionalFPE}%
  \BibitemOpen
  \bibfield  {author} {\bibinfo {author} {\bibfnamefont {F.}~\bibnamefont
  {Liu}}, \bibinfo {author} {\bibfnamefont {V.}~\bibnamefont {Anh}},\ and\
  \bibinfo {author} {\bibfnamefont {I.}~\bibnamefont {Turner}},\ }\bibfield
  {title} {\bibinfo {title} {Numerical solution of the space fractional
  {F}okker-{P}lanck equation},\ }\href
  {https://doi.org/https://doi.org/10.1016/j.cam.2003.09.028} {\bibfield
  {journal} {\bibinfo  {journal} {J. Comput. Appl. Math.}\ }\textbf {\bibinfo
  {volume} {166}},\ \bibinfo {pages} {209} (\bibinfo {year}
  {2004})}\BibitemShut {NoStop}%
\bibitem [{\citenamefont {Chernov}(1960)}]{ChernovBook1960}%
  \BibitemOpen
  \bibfield  {author} {\bibinfo {author} {\bibfnamefont {L.~A.}\ \bibnamefont
  {Chernov}},\ }\href@noop {} {\emph {\bibinfo {title} {Wave Propagation in a
  Random Medium}}}\ (\bibinfo  {publisher} {McGraw-Hill Book Company},\
  \bibinfo {address} {New York},\ \bibinfo {year} {1960})\BibitemShut {NoStop}%
\bibitem [{\citenamefont {Messiah}(1961)}]{Messiah}%
  \BibitemOpen
  \bibfield  {author} {\bibinfo {author} {\bibfnamefont {A.}~\bibnamefont
  {Messiah}},\ }\href@noop {} {\emph {\bibinfo {title} {Quantum Mechanics}}}\
  (\bibinfo  {publisher} {North-Holland, Amsterdam},\ \bibinfo {year} {1961})\
  \bibinfo {note} {2 vols.}\BibitemShut {Stop}%
\bibitem [{\citenamefont {Miyasaki}\ \emph {et~al.}(2025)\citenamefont
  {Miyasaki}, \citenamefont {Bajor}, \citenamefont {Pettersson}, \citenamefont
  {Senftleben}, \citenamefont {Fouke}, \citenamefont {Graham}, \citenamefont
  {John}, \citenamefont {Morgan}, \citenamefont {Haspel},\ and\ \citenamefont
  {Abrahamsson}}]{MultifocusMicroscopeOptica2025}%
  \BibitemOpen
  \bibfield  {author} {\bibinfo {author} {\bibfnamefont {E.~H.}\ \bibnamefont
  {Miyasaki}}, \bibinfo {author} {\bibfnamefont {A.~A.}\ \bibnamefont {Bajor}},
  \bibinfo {author} {\bibfnamefont {G.~M.}\ \bibnamefont {Pettersson}},
  \bibinfo {author} {\bibfnamefont {M.~L.}\ \bibnamefont {Senftleben}},
  \bibinfo {author} {\bibfnamefont {K.~E.}\ \bibnamefont {Fouke}}, \bibinfo
  {author} {\bibfnamefont {T.~G.~W.}\ \bibnamefont {Graham}}, \bibinfo {author}
  {\bibfnamefont {D.~D.}\ \bibnamefont {John}}, \bibinfo {author}
  {\bibfnamefont {J.~R.}\ \bibnamefont {Morgan}}, \bibinfo {author}
  {\bibfnamefont {G.}~\bibnamefont {Haspel}},\ and\ \bibinfo {author}
  {\bibfnamefont {S.}~\bibnamefont {Abrahamsson}},\ }\bibfield  {title}
  {\bibinfo {title} {High-speed 3{D} imaging with a 25-camera multifocus
  microscope},\ }\href {https://doi.org/10.1364/OPTICA.563617} {\bibfield
  {journal} {\bibinfo  {journal} {Optica}\ }\textbf {\bibinfo {volume} {12}},\
  \bibinfo {pages} {1230} (\bibinfo {year} {2025})}\BibitemShut {NoStop}%
\bibitem [{\citenamefont {{Morse}}\ and\ \citenamefont
  {{Feshbach}}(1953)}]{MorseFeshbach}%
  \BibitemOpen
  \bibfield  {author} {\bibinfo {author} {\bibfnamefont {P.~M.}\ \bibnamefont
  {{Morse}}}\ and\ \bibinfo {author} {\bibfnamefont {H.}~\bibnamefont
  {{Feshbach}}},\ }\href@noop {} {\emph {\bibinfo {title} {Methods of
  Theoretical Physics}}}\ (\bibinfo  {publisher} {McGraw-Hill, New York},\
  \bibinfo {year} {1953})\ \bibinfo {note} {2 vols.}\BibitemShut {Stop}%
\bibitem [{\citenamefont {Schmalz}\ \emph {et~al.}(2011)\citenamefont
  {Schmalz}, \citenamefont {Gureyev}, \citenamefont {Paganin},\ and\
  \citenamefont {Pavlov}}]{Schmalz2011}%
  \BibitemOpen
  \bibfield  {author} {\bibinfo {author} {\bibfnamefont {J.~A.}\ \bibnamefont
  {Schmalz}}, \bibinfo {author} {\bibfnamefont {T.~E.}\ \bibnamefont
  {Gureyev}}, \bibinfo {author} {\bibfnamefont {D.~M.}\ \bibnamefont
  {Paganin}},\ and\ \bibinfo {author} {\bibfnamefont {K.~M.}\ \bibnamefont
  {Pavlov}},\ }\bibfield  {title} {\bibinfo {title} {Phase retrieval using
  radiation and matter-wave fields: Validity of {T}eague's method for solution
  of the transport-of-intensity equation},\ }\href
  {https://doi.org/10.1103/PhysRevA.84.023808} {\bibfield  {journal} {\bibinfo
  {journal} {Phys. Rev. A}\ }\textbf {\bibinfo {volume} {84}},\ \bibinfo
  {pages} {023808} (\bibinfo {year} {2011})}\BibitemShut {NoStop}%
\bibitem [{\citenamefont {Aksenov}\ \emph {et~al.}(1998)\citenamefont
  {Aksenov}, \citenamefont {Banakh},\ and\ \citenamefont
  {Tikhomirova}}]{Aksenov1998}%
  \BibitemOpen
  \bibfield  {author} {\bibinfo {author} {\bibfnamefont {V.}~\bibnamefont
  {Aksenov}}, \bibinfo {author} {\bibfnamefont {V.}~\bibnamefont {Banakh}},\
  and\ \bibinfo {author} {\bibfnamefont {O.}~\bibnamefont {Tikhomirova}},\
  }\bibfield  {title} {\bibinfo {title} {Potential and vortex features of
  optical speckle fields and visualization of wave-front singularities},\
  }\href@noop {} {\bibfield  {journal} {\bibinfo  {journal} {Appl. Opt.}\
  }\textbf {\bibinfo {volume} {37}},\ \bibinfo {pages} {4536} (\bibinfo {year}
  {1998})}\BibitemShut {NoStop}%
\bibitem [{\citenamefont {Berujon}\ \emph {et~al.}(2015)\citenamefont
  {Berujon}, \citenamefont {Ziegler},\ and\ \citenamefont
  {Cloetens}}]{BerujonWavefrontMetrology}%
  \BibitemOpen
  \bibfield  {author} {\bibinfo {author} {\bibfnamefont {S.}~\bibnamefont
  {Berujon}}, \bibinfo {author} {\bibfnamefont {E.}~\bibnamefont {Ziegler}},\
  and\ \bibinfo {author} {\bibfnamefont {P.}~\bibnamefont {Cloetens}},\
  }\bibfield  {title} {\bibinfo {title} {{X-ray pulse wavefront metrology using
  speckle tracking}},\ }\href {https://doi.org/10.1107/S1600577515005433}
  {\bibfield  {journal} {\bibinfo  {journal} {J. Synchrotron Rad.}\ }\textbf
  {\bibinfo {volume} {22}},\ \bibinfo {pages} {886} (\bibinfo {year}
  {2015})}\BibitemShut {NoStop}%
\bibitem [{\citenamefont {Strauss}(1992)}]{StraussPDEbook}%
  \BibitemOpen
  \bibfield  {author} {\bibinfo {author} {\bibfnamefont {W.~A.}\ \bibnamefont
  {Strauss}},\ }\href@noop {} {\emph {\bibinfo {title} {Partial Differential
  Equations: An Introduction}}}\ (\bibinfo  {publisher} {John Wiley \& Sons,
  New York},\ \bibinfo {year} {1992})\BibitemShut {NoStop}%
\bibitem [{\citenamefont {Roddier}(1988)}]{Roddier1988}%
  \BibitemOpen
  \bibfield  {author} {\bibinfo {author} {\bibfnamefont {F.}~\bibnamefont
  {Roddier}},\ }\bibfield  {title} {\bibinfo {title} {Curvature sensing and
  compensation: a new concept in adaptive optics},\ }\href
  {https://doi.org/10.1364/AO.27.001223} {\bibfield  {journal} {\bibinfo
  {journal} {Appl. Opt.}\ }\textbf {\bibinfo {volume} {27}},\ \bibinfo {pages}
  {1223} (\bibinfo {year} {1988})}\BibitemShut {NoStop}%
\bibitem [{\citenamefont {Cowley}\ and\ \citenamefont
  {Moodie}(1960)}]{CowleyMoodie1960}%
  \BibitemOpen
  \bibfield  {author} {\bibinfo {author} {\bibfnamefont {J.~M.}\ \bibnamefont
  {Cowley}}\ and\ \bibinfo {author} {\bibfnamefont {A.~F.}\ \bibnamefont
  {Moodie}},\ }\bibfield  {title} {\bibinfo {title} {Fourier images {IV}: The
  phase grating},\ }\href {https://doi.org/10.1088/0370-1328/76/3/308}
  {\bibfield  {journal} {\bibinfo  {journal} {Proc. Phys. Soc.}\ }\textbf
  {\bibinfo {volume} {76}},\ \bibinfo {pages} {378} (\bibinfo {year}
  {1960})}\BibitemShut {NoStop}%
\bibitem [{\citenamefont {Bracewell}(1986)}]{Bracewellbook}%
  \BibitemOpen
  \bibfield  {author} {\bibinfo {author} {\bibfnamefont {R.~N.}\ \bibnamefont
  {Bracewell}},\ }\href@noop {} {\emph {\bibinfo {title} {The Fourier Transform
  and its Applications}}},\ \bibinfo {edition} {2nd}\ ed.\ (\bibinfo
  {publisher} {McGraw-Hill Book Company, New York},\ \bibinfo {year}
  {1986})\BibitemShut {NoStop}%
\bibitem [{\citenamefont {Press}\ \emph {et~al.}(2007)\citenamefont {Press},
  \citenamefont {Teukolsky}, \citenamefont {Vetterling},\ and\ \citenamefont
  {Flannery}}]{Press}%
  \BibitemOpen
  \bibfield  {author} {\bibinfo {author} {\bibfnamefont {W.~H.}\ \bibnamefont
  {Press}}, \bibinfo {author} {\bibfnamefont {S.~A.}\ \bibnamefont
  {Teukolsky}}, \bibinfo {author} {\bibfnamefont {W.~T.}\ \bibnamefont
  {Vetterling}},\ and\ \bibinfo {author} {\bibfnamefont {B.~P.}\ \bibnamefont
  {Flannery}},\ }\href@noop {} {\emph {\bibinfo {title} {Numerical {R}ecipes:
  {T}he {A}rt of {S}cientific {C}omputing}}},\ \bibinfo {edition} {3rd}\ ed.\
  (\bibinfo  {publisher} {Cambridge University Press},\ \bibinfo {address}
  {Cambridge},\ \bibinfo {year} {2007})\BibitemShut {NoStop}%
\bibitem [{\citenamefont {Gureyev}\ \emph
  {et~al.}(1995{\natexlab{b}})\citenamefont {Gureyev}, \citenamefont
  {Roberts},\ and\ \citenamefont {Nugent}}]{GureyevZernikePaper1995}%
  \BibitemOpen
  \bibfield  {author} {\bibinfo {author} {\bibfnamefont {T.~E.}\ \bibnamefont
  {Gureyev}}, \bibinfo {author} {\bibfnamefont {A.}~\bibnamefont {Roberts}},\
  and\ \bibinfo {author} {\bibfnamefont {K.~A.}\ \bibnamefont {Nugent}},\
  }\bibfield  {title} {\bibinfo {title} {Phase retrieval with the
  transport-of-intensity equation: matrix solution with use of {Z}ernike
  polynomials},\ }\href {https://doi.org/10.1364/JOSAA.12.001932} {\bibfield
  {journal} {\bibinfo  {journal} {J. Opt. Soc. Am. A}\ }\textbf {\bibinfo
  {volume} {12}},\ \bibinfo {pages} {1932} (\bibinfo {year}
  {1995}{\natexlab{b}})}\BibitemShut {NoStop}%
\bibitem [{\citenamefont {Gureyev}\ and\ \citenamefont
  {Nugent}(1996)}]{Gureyev1996}%
  \BibitemOpen
  \bibfield  {author} {\bibinfo {author} {\bibfnamefont {T.~E.}\ \bibnamefont
  {Gureyev}}\ and\ \bibinfo {author} {\bibfnamefont {K.~A.}\ \bibnamefont
  {Nugent}},\ }\bibfield  {title} {\bibinfo {title} {Phase retrieval with the
  transport-of-intensity equation. {II}. {O}rthogonal series solution for
  nonuniform illumination},\ }\href {https://doi.org/10.1364/JOSAA.13.001670}
  {\bibfield  {journal} {\bibinfo  {journal} {J. Opt. Soc. Am. A}\ }\textbf
  {\bibinfo {volume} {13}},\ \bibinfo {pages} {1670} (\bibinfo {year}
  {1996})}\BibitemShut {NoStop}%
\bibitem [{\citenamefont {Leatham}(2019)}]{LeathamHonsThesis2019}%
  \BibitemOpen
  \bibfield  {author} {\bibinfo {author} {\bibfnamefont {T.~A.}\ \bibnamefont
  {Leatham}},\ }\emph {\bibinfo {title} {Retrieving {X}-ray {P}araxial
  {I}maging {M}odalities via the {F}okker-{P}lanck equation}},\ \href@noop {}
  {\bibinfo {type} {Honours thesis}},\ \bibinfo  {school} {Monash University}
  (\bibinfo {year} {2019})\BibitemShut {NoStop}%
\bibitem [{\citenamefont {Leatham}(2023)}]{LeathamPhDThesis2023}%
  \BibitemOpen
  \bibfield  {author} {\bibinfo {author} {\bibfnamefont {T.~A.}\ \bibnamefont
  {Leatham}},\ }\emph {\bibinfo {title} {Applying the {F}okker-{P}lanck
  {E}quation to {X}-ray {P}hase and {D}ark-{F}ield {R}etrieval}},\ \href@noop
  {} {Ph.D. thesis},\ \bibinfo  {school} {Monash University} (\bibinfo {year}
  {2023})\BibitemShut {NoStop}%
\bibitem [{\citenamefont {Massig}(1999)}]{Massig1}%
  \BibitemOpen
  \bibfield  {author} {\bibinfo {author} {\bibfnamefont {J.~H.}\ \bibnamefont
  {Massig}},\ }\bibfield  {title} {\bibinfo {title} {Measurement of phase
  objects by simple means},\ }\href@noop {} {\bibfield  {journal} {\bibinfo
  {journal} {Appl. Opt.}\ }\textbf {\bibinfo {volume} {38}},\ \bibinfo {pages}
  {4103} (\bibinfo {year} {1999})}\BibitemShut {NoStop}%
\bibitem [{\citenamefont {Perciante}\ and\ \citenamefont
  {Ferrari}(2000)}]{Perciante}%
  \BibitemOpen
  \bibfield  {author} {\bibinfo {author} {\bibfnamefont {C.~D.}\ \bibnamefont
  {Perciante}}\ and\ \bibinfo {author} {\bibfnamefont {J.~A.}\ \bibnamefont
  {Ferrari}},\ }\bibfield  {title} {\bibinfo {title} {Visualization of
  two-dimensional phase gradients by subtraction of a reference periodic
  pattern},\ }\href@noop {} {\bibfield  {journal} {\bibinfo  {journal} {Appl.
  Opt.}\ }\textbf {\bibinfo {volume} {39}},\ \bibinfo {pages} {2081} (\bibinfo
  {year} {2000})}\BibitemShut {NoStop}%
\bibitem [{\citenamefont {Massig}(2001)}]{Massig2}%
  \BibitemOpen
  \bibfield  {author} {\bibinfo {author} {\bibfnamefont {J.~H.}\ \bibnamefont
  {Massig}},\ }\bibfield  {title} {\bibinfo {title} {Deformation measurement on
  specular surfaces by simple means},\ }\href@noop {} {\bibfield  {journal}
  {\bibinfo  {journal} {Opt. Eng.}\ }\textbf {\bibinfo {volume} {40}},\
  \bibinfo {pages} {2315} (\bibinfo {year} {2001})}\BibitemShut {NoStop}%
\bibitem [{\citenamefont {Wen}\ \emph {et~al.}(2010)\citenamefont {Wen},
  \citenamefont {Bennett}, \citenamefont {Kopace}, \citenamefont {Stein},\ and\
  \citenamefont {Pai}}]{wen2010}%
  \BibitemOpen
  \bibfield  {author} {\bibinfo {author} {\bibfnamefont {H.~H.}\ \bibnamefont
  {Wen}}, \bibinfo {author} {\bibfnamefont {E.~E.}\ \bibnamefont {Bennett}},
  \bibinfo {author} {\bibfnamefont {R.}~\bibnamefont {Kopace}}, \bibinfo
  {author} {\bibfnamefont {A.~F.}\ \bibnamefont {Stein}},\ and\ \bibinfo
  {author} {\bibfnamefont {V.}~\bibnamefont {Pai}},\ }\bibfield  {title}
  {\bibinfo {title} {Single-shot x-ray differential phase-contrast and
  diffraction imaging using two-dimensional transmission gratings},\
  }\href@noop {} {\bibfield  {journal} {\bibinfo  {journal} {Opt. Lett.}\
  }\textbf {\bibinfo {volume} {35}},\ \bibinfo {pages} {1932} (\bibinfo {year}
  {2010})}\BibitemShut {NoStop}%
\bibitem [{\citenamefont {Morgan}\ \emph {et~al.}(2011)\citenamefont {Morgan},
  \citenamefont {Paganin},\ and\ \citenamefont {Siu}}]{morgan2011quantitative}%
  \BibitemOpen
  \bibfield  {author} {\bibinfo {author} {\bibfnamefont {K.~S.}\ \bibnamefont
  {Morgan}}, \bibinfo {author} {\bibfnamefont {D.~M.}\ \bibnamefont
  {Paganin}},\ and\ \bibinfo {author} {\bibfnamefont {K.~K.}\ \bibnamefont
  {Siu}},\ }\bibfield  {title} {\bibinfo {title} {Quantitative single-exposure
  x-ray phase contrast imaging using a single attenuation grid},\ }\href@noop
  {} {\bibfield  {journal} {\bibinfo  {journal} {Opt. Express}\ }\textbf
  {\bibinfo {volume} {19}},\ \bibinfo {pages} {19781} (\bibinfo {year}
  {2011})}\BibitemShut {NoStop}%
\bibitem [{\citenamefont {Medhi}(2022)}]{Medhi2022}%
  \BibitemOpen
  \bibfield  {author} {\bibinfo {author} {\bibfnamefont {B.}~\bibnamefont
  {Medhi}},\ }\bibfield  {title} {\bibinfo {title} {Structured
  illumination-based transport-of-intensity phase imaging for quantitative
  diagnostics of high-speed flows},\ }\href {https://doi.org/10.1364/AO.440213}
  {\bibfield  {journal} {\bibinfo  {journal} {Appl. Opt.}\ }\textbf {\bibinfo
  {volume} {61}},\ \bibinfo {pages} {945} (\bibinfo {year} {2022})}\BibitemShut
  {NoStop}%
\bibitem [{\citenamefont {Mayo}\ and\ \citenamefont
  {Sexton}(2004)}]{MayoSexton2004}%
  \BibitemOpen
  \bibfield  {author} {\bibinfo {author} {\bibfnamefont {S.~C.}\ \bibnamefont
  {Mayo}}\ and\ \bibinfo {author} {\bibfnamefont {B.}~\bibnamefont {Sexton}},\
  }\bibfield  {title} {\bibinfo {title} {Refractive microlens array for
  wave-front analysis in the medium to hard x-ray range},\ }\href@noop {}
  {\bibfield  {journal} {\bibinfo  {journal} {Opt. Lett.}\ }\textbf {\bibinfo
  {volume} {29}},\ \bibinfo {pages} {866} (\bibinfo {year} {2004})}\BibitemShut
  {NoStop}%
\bibitem [{\citenamefont {Morgan}\ \emph {et~al.}(2013)\citenamefont {Morgan},
  \citenamefont {Modregger}, \citenamefont {Irvine}, \citenamefont
  {Rutishauser}, \citenamefont {Guzenko}, \citenamefont {Stampanoni},\ and\
  \citenamefont {David}}]{Morgan2013}%
  \BibitemOpen
  \bibfield  {author} {\bibinfo {author} {\bibfnamefont {K.~S.}\ \bibnamefont
  {Morgan}}, \bibinfo {author} {\bibfnamefont {P.}~\bibnamefont {Modregger}},
  \bibinfo {author} {\bibfnamefont {S.~C.}\ \bibnamefont {Irvine}}, \bibinfo
  {author} {\bibfnamefont {S.}~\bibnamefont {Rutishauser}}, \bibinfo {author}
  {\bibfnamefont {V.~A.}\ \bibnamefont {Guzenko}}, \bibinfo {author}
  {\bibfnamefont {M.}~\bibnamefont {Stampanoni}},\ and\ \bibinfo {author}
  {\bibfnamefont {C.}~\bibnamefont {David}},\ }\bibfield  {title} {\bibinfo
  {title} {A sensitive x-ray phase contrast technique for rapid imaging using a
  single phase grid analyzer},\ }\href@noop {} {\bibfield  {journal} {\bibinfo
  {journal} {Opt. Lett.}\ }\textbf {\bibinfo {volume} {38}},\ \bibinfo {pages}
  {4605} (\bibinfo {year} {2013})}\BibitemShut {NoStop}%
\bibitem [{\citenamefont {Rizzi}\ \emph {et~al.}(2013)\citenamefont {Rizzi},
  \citenamefont {Mercere}, \citenamefont {Idir}, \citenamefont {Silva},
  \citenamefont {Vincent},\ and\ \citenamefont {Primot}}]{Rizzi2013}%
  \BibitemOpen
  \bibfield  {author} {\bibinfo {author} {\bibfnamefont {J.}~\bibnamefont
  {Rizzi}}, \bibinfo {author} {\bibfnamefont {P.}~\bibnamefont {Mercere}},
  \bibinfo {author} {\bibfnamefont {M.}~\bibnamefont {Idir}}, \bibinfo {author}
  {\bibfnamefont {P.~D.}\ \bibnamefont {Silva}}, \bibinfo {author}
  {\bibfnamefont {G.}~\bibnamefont {Vincent}},\ and\ \bibinfo {author}
  {\bibfnamefont {J.}~\bibnamefont {Primot}},\ }\bibfield  {title} {\bibinfo
  {title} {X-ray phase contrast imaging and noise evaluation using a single
  phase grating interferometer},\ }\href@noop {} {\bibfield  {journal}
  {\bibinfo  {journal} {Opt. Express}\ }\textbf {\bibinfo {volume} {21}},\
  \bibinfo {pages} {17340} (\bibinfo {year} {2013})}\BibitemShut {NoStop}%
\bibitem [{\citenamefont {B{\'e}rujon}\ \emph {et~al.}(2012)\citenamefont
  {B{\'e}rujon}, \citenamefont {Ziegler}, \citenamefont {Cerbino},\ and\
  \citenamefont {Peverini}}]{berujon2012}%
  \BibitemOpen
  \bibfield  {author} {\bibinfo {author} {\bibfnamefont {S.}~\bibnamefont
  {B{\'e}rujon}}, \bibinfo {author} {\bibfnamefont {E.}~\bibnamefont
  {Ziegler}}, \bibinfo {author} {\bibfnamefont {R.}~\bibnamefont {Cerbino}},\
  and\ \bibinfo {author} {\bibfnamefont {L.}~\bibnamefont {Peverini}},\
  }\bibfield  {title} {\bibinfo {title} {Two-dimensional x-ray beam phase
  sensing},\ }\href@noop {} {\bibfield  {journal} {\bibinfo  {journal} {Phys.
  Rev. Lett.}\ }\textbf {\bibinfo {volume} {108}},\ \bibinfo {pages} {158102}
  (\bibinfo {year} {2012})}\BibitemShut {NoStop}%
\bibitem [{\citenamefont {Morgan}\ \emph {et~al.}(2012)\citenamefont {Morgan},
  \citenamefont {Paganin},\ and\ \citenamefont {Siu}}]{Morgan2012}%
  \BibitemOpen
  \bibfield  {author} {\bibinfo {author} {\bibfnamefont {K.~S.}\ \bibnamefont
  {Morgan}}, \bibinfo {author} {\bibfnamefont {D.~M.}\ \bibnamefont
  {Paganin}},\ and\ \bibinfo {author} {\bibfnamefont {K.~K.}\ \bibnamefont
  {Siu}},\ }\bibfield  {title} {\bibinfo {title} {X-ray phase imaging with a
  paper analyzer},\ }\href@noop {} {\bibfield  {journal} {\bibinfo  {journal}
  {Appl. Phys. Lett.}\ }\textbf {\bibinfo {volume} {100}},\ \bibinfo {pages}
  {124102} (\bibinfo {year} {2012})}\BibitemShut {NoStop}%
\bibitem [{\citenamefont {Berujon}\ \emph {et~al.}(2012)\citenamefont
  {Berujon}, \citenamefont {Wang},\ and\ \citenamefont
  {Sawhney}}]{berujon2012b}%
  \BibitemOpen
  \bibfield  {author} {\bibinfo {author} {\bibfnamefont {S.}~\bibnamefont
  {Berujon}}, \bibinfo {author} {\bibfnamefont {H.}~\bibnamefont {Wang}},\ and\
  \bibinfo {author} {\bibfnamefont {K.}~\bibnamefont {Sawhney}},\ }\bibfield
  {title} {\bibinfo {title} {X-ray multimodal imaging using a random-phase
  object},\ }\href@noop {} {\bibfield  {journal} {\bibinfo  {journal} {Phys.
  Rev. A}\ }\textbf {\bibinfo {volume} {86}},\ \bibinfo {pages} {063813}
  (\bibinfo {year} {2012})}\BibitemShut {NoStop}%
\bibitem [{\citenamefont {Zdora}\ \emph {et~al.}(2017)\citenamefont {Zdora},
  \citenamefont {Thibault}, \citenamefont {Zhou}, \citenamefont {Koch},
  \citenamefont {Romell}, \citenamefont {Sala}, \citenamefont {Last},
  \citenamefont {Rau},\ and\ \citenamefont {Zanette}}]{zdora2017}%
  \BibitemOpen
  \bibfield  {author} {\bibinfo {author} {\bibfnamefont {M.-C.}\ \bibnamefont
  {Zdora}}, \bibinfo {author} {\bibfnamefont {P.}~\bibnamefont {Thibault}},
  \bibinfo {author} {\bibfnamefont {T.}~\bibnamefont {Zhou}}, \bibinfo {author}
  {\bibfnamefont {F.~J.}\ \bibnamefont {Koch}}, \bibinfo {author}
  {\bibfnamefont {J.}~\bibnamefont {Romell}}, \bibinfo {author} {\bibfnamefont
  {S.}~\bibnamefont {Sala}}, \bibinfo {author} {\bibfnamefont {A.}~\bibnamefont
  {Last}}, \bibinfo {author} {\bibfnamefont {C.}~\bibnamefont {Rau}},\ and\
  \bibinfo {author} {\bibfnamefont {I.}~\bibnamefont {Zanette}},\ }\bibfield
  {title} {\bibinfo {title} {X-ray phase-contrast imaging and metrology through
  unified modulated pattern analysis},\ }\href@noop {} {\bibfield  {journal}
  {\bibinfo  {journal} {Phys. Rev. Lett.}\ }\textbf {\bibinfo {volume} {118}},\
  \bibinfo {pages} {203903} (\bibinfo {year} {2017})}\BibitemShut {NoStop}%
\bibitem [{\citenamefont {Zdora}(2018)}]{zdora2018}%
  \BibitemOpen
  \bibfield  {author} {\bibinfo {author} {\bibfnamefont {M.-C.}\ \bibnamefont
  {Zdora}},\ }\bibfield  {title} {\bibinfo {title} {State of the art of {X}-ray
  speckle-based phase-contrast and dark-field imaging},\ }\href@noop {}
  {\bibfield  {journal} {\bibinfo  {journal} {J. Imaging}\ }\textbf {\bibinfo
  {volume} {4}},\ \bibinfo {pages} {60} (\bibinfo {year} {2018})}\BibitemShut
  {NoStop}%
\bibitem [{\citenamefont {Paganin}\ \emph {et~al.}(2018)\citenamefont
  {Paganin}, \citenamefont {Labriet}, \citenamefont {Brun},\ and\ \citenamefont
  {Berujon}}]{PaganinLabrietBrunBerujon2018}%
  \BibitemOpen
  \bibfield  {author} {\bibinfo {author} {\bibfnamefont {D.~M.}\ \bibnamefont
  {Paganin}}, \bibinfo {author} {\bibfnamefont {H.}~\bibnamefont {Labriet}},
  \bibinfo {author} {\bibfnamefont {E.}~\bibnamefont {Brun}},\ and\ \bibinfo
  {author} {\bibfnamefont {S.}~\bibnamefont {Berujon}},\ }\bibfield  {title}
  {\bibinfo {title} {Single-image geometric-flow x-ray speckle tracking},\
  }\href {https://journals.aps.org/pra/abstract/10.1103/PhysRevA.98.053813}
  {\bibfield  {journal} {\bibinfo  {journal} {Phys. Rev. A}\ }\textbf {\bibinfo
  {volume} {98}},\ \bibinfo {pages} {053813} (\bibinfo {year}
  {2018})}\BibitemShut {NoStop}%
\bibitem [{\citenamefont {Zdora}(2021)}]{ZdoraSpeckleBook}%
  \BibitemOpen
  \bibfield  {author} {\bibinfo {author} {\bibfnamefont {M.-C.}\ \bibnamefont
  {Zdora}},\ }\href@noop {} {\emph {\bibinfo {title} {X-ray Phase-Contrast
  Imaging Using Near-Field Speckles}}}\ (\bibinfo  {publisher} {Springer,
  Cham},\ \bibinfo {year} {2021})\BibitemShut {NoStop}%
\bibitem [{\citenamefont {Boominathan}\ \emph {et~al.}(2022)\citenamefont
  {Boominathan}, \citenamefont {Robinson}, \citenamefont {Waller},\ and\
  \citenamefont {Veeraraghavan}}]{Boominathan2022}%
  \BibitemOpen
  \bibfield  {author} {\bibinfo {author} {\bibfnamefont {V.}~\bibnamefont
  {Boominathan}}, \bibinfo {author} {\bibfnamefont {J.~T.}\ \bibnamefont
  {Robinson}}, \bibinfo {author} {\bibfnamefont {L.}~\bibnamefont {Waller}},\
  and\ \bibinfo {author} {\bibfnamefont {A.}~\bibnamefont {Veeraraghavan}},\
  }\bibfield  {title} {\bibinfo {title} {Recent advances in lensless imaging},\
  }\href {https://doi.org/10.1364/OPTICA.431361} {\bibfield  {journal}
  {\bibinfo  {journal} {Optica}\ }\textbf {\bibinfo {volume} {9}},\ \bibinfo
  {pages} {1} (\bibinfo {year} {2022})}\BibitemShut {NoStop}%
\bibitem [{\citenamefont {Gage}(1920)}]{gage1920}%
  \BibitemOpen
  \bibfield  {author} {\bibinfo {author} {\bibfnamefont {S.~H.}\ \bibnamefont
  {Gage}},\ }\bibfield  {title} {\bibinfo {title} {Modern dark-field microscopy
  and the history of its development},\ }\href
  {https://doi.org/10.2307/3221838} {\bibfield  {journal} {\bibinfo  {journal}
  {Trans. Am. Microsc. Soc.}\ }\textbf {\bibinfo {volume} {39}},\ \bibinfo
  {pages} {95} (\bibinfo {year} {1920})}\BibitemShut {NoStop}%
\bibitem [{\citenamefont {Pfeiffer}\ \emph {et~al.}(2008)\citenamefont
  {Pfeiffer}, \citenamefont {Bech}, \citenamefont {Bunk}, \citenamefont
  {Kraft}, \citenamefont {Eikenberry}, \citenamefont {Br{\"o}nnimann},
  \citenamefont {Gr{\"u}nzweig},\ and\ \citenamefont {David}}]{Pfeiffer2008df}%
  \BibitemOpen
  \bibfield  {author} {\bibinfo {author} {\bibfnamefont {F.}~\bibnamefont
  {Pfeiffer}}, \bibinfo {author} {\bibfnamefont {M.}~\bibnamefont {Bech}},
  \bibinfo {author} {\bibfnamefont {O.}~\bibnamefont {Bunk}}, \bibinfo {author}
  {\bibfnamefont {P.}~\bibnamefont {Kraft}}, \bibinfo {author} {\bibfnamefont
  {E.~F.}\ \bibnamefont {Eikenberry}}, \bibinfo {author} {\bibfnamefont
  {{\relax{Ch}}.}~\bibnamefont {Br{\"o}nnimann}}, \bibinfo {author}
  {\bibfnamefont {C.}~\bibnamefont {Gr{\"u}nzweig}},\ and\ \bibinfo {author}
  {\bibfnamefont {C.}~\bibnamefont {David}},\ }\bibfield  {title} {\bibinfo
  {title} {Hard-{X}-ray dark-field imaging using a grating interferometer},\
  }\href@noop {} {\bibfield  {journal} {\bibinfo  {journal} {Nat. Mater.}\
  }\textbf {\bibinfo {volume} {7}},\ \bibinfo {pages} {134} (\bibinfo {year}
  {2008})}\BibitemShut {NoStop}%
\bibitem [{\citenamefont {Snigirev}\ \emph {et~al.}(1995)\citenamefont
  {Snigirev}, \citenamefont {Snigireva}, \citenamefont {Kohn}, \citenamefont
  {Kuznetsov},\ and\ \citenamefont {Schelokov}}]{Snigirev}%
  \BibitemOpen
  \bibfield  {author} {\bibinfo {author} {\bibfnamefont {A.}~\bibnamefont
  {Snigirev}}, \bibinfo {author} {\bibfnamefont {I.}~\bibnamefont {Snigireva}},
  \bibinfo {author} {\bibfnamefont {V.}~\bibnamefont {Kohn}}, \bibinfo {author}
  {\bibfnamefont {S.}~\bibnamefont {Kuznetsov}},\ and\ \bibinfo {author}
  {\bibfnamefont {I.}~\bibnamefont {Schelokov}},\ }\bibfield  {title} {\bibinfo
  {title} {On the possibilities of x-ray phase contrast microimaging by
  coherent high-energy synchrotron radiation},\ }\href@noop {} {\bibfield
  {journal} {\bibinfo  {journal} {Rev. Sci. Instrum.}\ }\textbf {\bibinfo
  {volume} {66}},\ \bibinfo {pages} {5486} (\bibinfo {year}
  {1995})}\BibitemShut {NoStop}%
\bibitem [{\citenamefont {Gabor}(1948{\natexlab{a}})}]{Gabor1948}%
  \BibitemOpen
  \bibfield  {author} {\bibinfo {author} {\bibfnamefont {D.}~\bibnamefont
  {Gabor}},\ }\bibfield  {title} {\bibinfo {title} {A new microscopic
  principle},\ }\href@noop {} {\bibfield  {journal} {\bibinfo  {journal}
  {Nature}\ }\textbf {\bibinfo {volume} {161}},\ \bibinfo {pages} {777}
  (\bibinfo {year} {1948}{\natexlab{a}})}\BibitemShut {NoStop}%
\bibitem [{\citenamefont {Saxton}\ and\ \citenamefont
  {Baumeister}(1982)}]{SaxtonBaumeister1982}%
  \BibitemOpen
  \bibfield  {author} {\bibinfo {author} {\bibfnamefont {W.~O.}\ \bibnamefont
  {Saxton}}\ and\ \bibinfo {author} {\bibfnamefont {W.}~\bibnamefont
  {Baumeister}},\ }\bibfield  {title} {\bibinfo {title} {The correlation
  averaging of a regularly arranged bacterial cell envelope protein},\
  }\href@noop {} {\bibfield  {journal} {\bibinfo  {journal} {J. Microsc.}\
  }\textbf {\bibinfo {volume} {127}},\ \bibinfo {pages} {127} (\bibinfo {year}
  {1982})}\BibitemShut {NoStop}%
\bibitem [{\citenamefont {Nugent}\ \emph {et~al.}(2003)\citenamefont {Nugent},
  \citenamefont {Tran},\ and\ \citenamefont {Roberts}}]{Nugent2003}%
  \BibitemOpen
  \bibfield  {author} {\bibinfo {author} {\bibfnamefont {K.~A.}\ \bibnamefont
  {Nugent}}, \bibinfo {author} {\bibfnamefont {C.~Q.}\ \bibnamefont {Tran}},\
  and\ \bibinfo {author} {\bibfnamefont {A.}~\bibnamefont {Roberts}},\
  }\bibfield  {title} {\bibinfo {title} {Coherence transport through imperfect
  x-ray optical systems},\ }\href@noop {} {\bibfield  {journal} {\bibinfo
  {journal} {Opt. Express}\ }\textbf {\bibinfo {volume} {11}},\ \bibinfo
  {pages} {2323} (\bibinfo {year} {2003})}\BibitemShut {NoStop}%
\bibitem [{\citenamefont {Voronovich}(1999)}]{VoronovichBook}%
  \BibitemOpen
  \bibfield  {author} {\bibinfo {author} {\bibfnamefont {A.~G.}\ \bibnamefont
  {Voronovich}},\ }\href@noop {} {\emph {\bibinfo {title} {Wave Scattering from
  Rough Surfaces}}},\ \bibinfo {edition} {2nd}\ ed.\ (\bibinfo  {publisher}
  {Springer, Berlin},\ \bibinfo {year} {1999})\BibitemShut {NoStop}%
\bibitem [{\citenamefont {Yashiro}\ \emph {et~al.}(2010)\citenamefont
  {Yashiro}, \citenamefont {Terui}, \citenamefont {Kawabata},\ and\
  \citenamefont {Momose}}]{Yashiro2010}%
  \BibitemOpen
  \bibfield  {author} {\bibinfo {author} {\bibfnamefont {W.}~\bibnamefont
  {Yashiro}}, \bibinfo {author} {\bibfnamefont {Y.}~\bibnamefont {Terui}},
  \bibinfo {author} {\bibfnamefont {K.}~\bibnamefont {Kawabata}},\ and\
  \bibinfo {author} {\bibfnamefont {A.}~\bibnamefont {Momose}},\ }\bibfield
  {title} {\bibinfo {title} {On the origin of visibility contrast in x-ray
  {T}albot interferometry},\ }\href@noop {} {\bibfield  {journal} {\bibinfo
  {journal} {Opt. Express}\ }\textbf {\bibinfo {volume} {18}},\ \bibinfo
  {pages} {16890} (\bibinfo {year} {2010})}\BibitemShut {NoStop}%
\bibitem [{\citenamefont {Yashiro}\ \emph {et~al.}(2011)\citenamefont
  {Yashiro}, \citenamefont {Harasse}, \citenamefont {Kawabata}, \citenamefont
  {Kuwabara}, \citenamefont {Yamazaki},\ and\ \citenamefont
  {Momose}}]{Yashiro2011}%
  \BibitemOpen
  \bibfield  {author} {\bibinfo {author} {\bibfnamefont {W.}~\bibnamefont
  {Yashiro}}, \bibinfo {author} {\bibfnamefont {S.}~\bibnamefont {Harasse}},
  \bibinfo {author} {\bibfnamefont {K.}~\bibnamefont {Kawabata}}, \bibinfo
  {author} {\bibfnamefont {H.}~\bibnamefont {Kuwabara}}, \bibinfo {author}
  {\bibfnamefont {T.}~\bibnamefont {Yamazaki}},\ and\ \bibinfo {author}
  {\bibfnamefont {A.}~\bibnamefont {Momose}},\ }\bibfield  {title} {\bibinfo
  {title} {Distribution of unresolvable anisotropic microstructures revealed in
  visibility-contrast images using x-ray {T}albot interferometry},\ }\href@noop
  {} {\bibfield  {journal} {\bibinfo  {journal} {Phys. Rev. B}\ }\textbf
  {\bibinfo {volume} {84}},\ \bibinfo {pages} {094106} (\bibinfo {year}
  {2011})}\BibitemShut {NoStop}%
\bibitem [{\citenamefont {Montgomery}(1981)}]{Montgomery1981}%
  \BibitemOpen
  \bibfield  {author} {\bibinfo {author} {\bibfnamefont {W.~D.}\ \bibnamefont
  {Montgomery}},\ }\bibfield  {title} {\bibinfo {title} {Unitary operators in
  the homogeneous wave field},\ }\href {https://doi.org/10.1364/OL.6.000314}
  {\bibfield  {journal} {\bibinfo  {journal} {Opt. Lett.}\ }\textbf {\bibinfo
  {volume} {6}},\ \bibinfo {pages} {314} (\bibinfo {year} {1981})}\BibitemShut
  {NoStop}%
\bibitem [{\citenamefont {Glatter}\ and\ \citenamefont
  {Kratky}(1982)}]{GlatterKratky1982}%
  \BibitemOpen
  \bibinfo {editor} {\bibfnamefont {O.}~\bibnamefont {Glatter}}\ and\ \bibinfo
  {editor} {\bibfnamefont {O.}~\bibnamefont {Kratky}},\ eds.,\ \href@noop {}
  {\emph {\bibinfo {title} {Small Angle X-Ray Scattering}}}\ (\bibinfo
  {publisher} {Academic Press},\ \bibinfo {address} {London},\ \bibinfo {year}
  {1982})\BibitemShut {NoStop}%
\bibitem [{\citenamefont {Sivia}(2011)}]{SiviaScatteringTheoryBook2011}%
  \BibitemOpen
  \bibfield  {author} {\bibinfo {author} {\bibfnamefont {D.~S.}\ \bibnamefont
  {Sivia}},\ }\href@noop {} {\emph {\bibinfo {title} {Elementary Scattering
  Theory}}}\ (\bibinfo  {publisher} {Oxford University Press},\ \bibinfo
  {address} {Oxford},\ \bibinfo {year} {2011})\BibitemShut {NoStop}%
\bibitem [{\citenamefont {Strobl}(2014)}]{Strobl2014}%
  \BibitemOpen
  \bibfield  {author} {\bibinfo {author} {\bibfnamefont {M.}~\bibnamefont
  {Strobl}},\ }\bibfield  {title} {\bibinfo {title} {General solution for
  quantitative dark-field contrast imaging with grating interferometers},\
  }\href@noop {} {\bibfield  {journal} {\bibinfo  {journal} {Sci. Rep.}\
  }\textbf {\bibinfo {volume} {4}},\ \bibinfo {pages} {7243} (\bibinfo {year}
  {2014})}\BibitemShut {NoStop}%
\bibitem [{\citenamefont {Sinha}\ \emph {et~al.}(1988)\citenamefont {Sinha},
  \citenamefont {Sirota}, \citenamefont {Garoff},\ and\ \citenamefont
  {Stanley}}]{Sinha1988}%
  \BibitemOpen
  \bibfield  {author} {\bibinfo {author} {\bibfnamefont {S.~K.}\ \bibnamefont
  {Sinha}}, \bibinfo {author} {\bibfnamefont {E.~B.}\ \bibnamefont {Sirota}},
  \bibinfo {author} {\bibfnamefont {S.}~\bibnamefont {Garoff}},\ and\ \bibinfo
  {author} {\bibfnamefont {H.~B.}\ \bibnamefont {Stanley}},\ }\bibfield
  {title} {\bibinfo {title} {X-ray and neutron scattering from rough
  surfaces},\ }\href@noop {} {\bibfield  {journal} {\bibinfo  {journal} {Phys.
  Rev. B}\ }\textbf {\bibinfo {volume} {38}},\ \bibinfo {pages} {2297}
  (\bibinfo {year} {1988})}\BibitemShut {NoStop}%
\bibitem [{\citenamefont {Neifeld}(1998)}]{NRU==1998}%
  \BibitemOpen
  \bibfield  {author} {\bibinfo {author} {\bibfnamefont {M.~A.}\ \bibnamefont
  {Neifeld}},\ }\bibfield  {title} {\bibinfo {title} {Information, resolution,
  and space-bandwidth product},\ }\href {https://doi.org/10.1364/OL.23.001477}
  {\bibfield  {journal} {\bibinfo  {journal} {Opt. Lett.}\ }\textbf {\bibinfo
  {volume} {23}},\ \bibinfo {pages} {1477} (\bibinfo {year}
  {1998})}\BibitemShut {NoStop}%
\bibitem [{\citenamefont {Gureyev}\ \emph {et~al.}(2015)\citenamefont
  {Gureyev}, \citenamefont {de~Hoog}, \citenamefont {Nesterets},\ and\
  \citenamefont {Paganin}}]{NRU=D}%
  \BibitemOpen
  \bibfield  {author} {\bibinfo {author} {\bibfnamefont {T.~E.}\ \bibnamefont
  {Gureyev}}, \bibinfo {author} {\bibfnamefont {F.}~\bibnamefont {de~Hoog}},
  \bibinfo {author} {\bibfnamefont {{\relax{Ya}}.~I.}\ \bibnamefont
  {Nesterets}},\ and\ \bibinfo {author} {\bibfnamefont {D.~M.}\ \bibnamefont
  {Paganin}},\ }\bibfield  {title} {\bibinfo {title} {On the noise-resolution
  duality, {H}eisenberg uncertainty and {S}hannon's information},\ }\href@noop
  {} {\bibfield  {journal} {\bibinfo  {journal} {ANZIAM J.}\ }\textbf {\bibinfo
  {volume} {56}},\ \bibinfo {pages} {C1} (\bibinfo {year} {2015})}\BibitemShut
  {NoStop}%
\bibitem [{\citenamefont {den Dekker}\ and\ \citenamefont {van~den
  Bos}(1997)}]{NRU=1997}%
  \BibitemOpen
  \bibfield  {author} {\bibinfo {author} {\bibfnamefont {A.~J.}\ \bibnamefont
  {den Dekker}}\ and\ \bibinfo {author} {\bibfnamefont {A.}~\bibnamefont
  {van~den Bos}},\ }\bibfield  {title} {\bibinfo {title} {Resolution: a
  survey},\ }\href {https://doi.org/10.1364/JOSAA.14.000547} {\bibfield
  {journal} {\bibinfo  {journal} {J. Opt. Soc. Am. A}\ }\textbf {\bibinfo
  {volume} {14}},\ \bibinfo {pages} {547} (\bibinfo {year} {1997})}\BibitemShut
  {NoStop}%
\bibitem [{\citenamefont {{van den Bos}}\ and\ \citenamefont {{den
  Dekker}}(2001)}]{NRU==2001}%
  \BibitemOpen
  \bibfield  {author} {\bibinfo {author} {\bibfnamefont {A.}~\bibnamefont {{van
  den Bos}}}\ and\ \bibinfo {author} {\bibfnamefont {A.}~\bibnamefont {{den
  Dekker}}},\ }\bibfield  {title} {\bibinfo {title} {Resolution
  reconsidered--{C}onventional approaches and an alternative},\ }\href
  {https://doi.org/https://doi.org/10.1016/S1076-5670(01)80114-2} {\bibfield
  {journal} {\bibinfo  {journal} {Adv. Imaging Electron Phys.}\ }\textbf
  {\bibinfo {volume} {117}},\ \bibinfo {pages} {241} (\bibinfo {year}
  {2001})}\BibitemShut {NoStop}%
\bibitem [{\citenamefont {Gureyev}\ \emph
  {et~al.}(2014{\natexlab{a}})\citenamefont {Gureyev}, \citenamefont
  {Nesterets}, \citenamefont {de~Hoog}, \citenamefont {Schmalz}, \citenamefont
  {Mayo}, \citenamefont {Mohammadi},\ and\ \citenamefont {Tromba}}]{NRU=A}%
  \BibitemOpen
  \bibfield  {author} {\bibinfo {author} {\bibfnamefont {T.~E.}\ \bibnamefont
  {Gureyev}}, \bibinfo {author} {\bibfnamefont {{\relax{Ya}}.~I.}\ \bibnamefont
  {Nesterets}}, \bibinfo {author} {\bibfnamefont {F.}~\bibnamefont {de~Hoog}},
  \bibinfo {author} {\bibfnamefont {G.}~\bibnamefont {Schmalz}}, \bibinfo
  {author} {\bibfnamefont {S.~C.}\ \bibnamefont {Mayo}}, \bibinfo {author}
  {\bibfnamefont {S.}~\bibnamefont {Mohammadi}},\ and\ \bibinfo {author}
  {\bibfnamefont {G.}~\bibnamefont {Tromba}},\ }\bibfield  {title} {\bibinfo
  {title} {Duality between noise and spatial resolution in linear systems},\
  }\href@noop {} {\bibfield  {journal} {\bibinfo  {journal} {Opt. Express}\
  }\textbf {\bibinfo {volume} {22}},\ \bibinfo {pages} {9087} (\bibinfo {year}
  {2014}{\natexlab{a}})}\BibitemShut {NoStop}%
\bibitem [{\citenamefont {de~Hoog}\ \emph {et~al.}(2014)\citenamefont
  {de~Hoog}, \citenamefont {Schmalz},\ and\ \citenamefont {Gureyev}}]{NRU=B}%
  \BibitemOpen
  \bibfield  {author} {\bibinfo {author} {\bibfnamefont {F.}~\bibnamefont
  {de~Hoog}}, \bibinfo {author} {\bibfnamefont {G.}~\bibnamefont {Schmalz}},\
  and\ \bibinfo {author} {\bibfnamefont {T.~E.}\ \bibnamefont {Gureyev}},\
  }\bibfield  {title} {\bibinfo {title} {An uncertainty inequality},\
  }\href@noop {} {\bibfield  {journal} {\bibinfo  {journal} {Appl. Math.
  Lett.}\ }\textbf {\bibinfo {volume} {38}},\ \bibinfo {pages} {84} (\bibinfo
  {year} {2014})}\BibitemShut {NoStop}%
\bibitem [{\citenamefont {Gureyev}\ \emph {et~al.}(2016)\citenamefont
  {Gureyev}, \citenamefont {Nesterets},\ and\ \citenamefont {de~Hoog}}]{NRU=C}%
  \BibitemOpen
  \bibfield  {author} {\bibinfo {author} {\bibfnamefont {T.~E.}\ \bibnamefont
  {Gureyev}}, \bibinfo {author} {\bibfnamefont {{\relax{Ya}}.~I.}\ \bibnamefont
  {Nesterets}},\ and\ \bibinfo {author} {\bibfnamefont {F.}~\bibnamefont
  {de~Hoog}},\ }\bibfield  {title} {\bibinfo {title} {Spatial resolution,
  signal-to-noise and information capacity of linear imaging systems},\
  }\href@noop {} {\bibfield  {journal} {\bibinfo  {journal} {Opt. Express}\
  }\textbf {\bibinfo {volume} {24}},\ \bibinfo {pages} {17168} (\bibinfo {year}
  {2016})}\BibitemShut {NoStop}%
\bibitem [{\citenamefont {Bridgman}(1922)}]{BridgmanBook}%
  \BibitemOpen
  \bibfield  {author} {\bibinfo {author} {\bibfnamefont {P.~W.}\ \bibnamefont
  {Bridgman}},\ }\href@noop {} {\emph {\bibinfo {title} {Dimensional
  Analysis}}}\ (\bibinfo  {publisher} {Yale University Press, New Haven CT},\
  \bibinfo {year} {1922})\BibitemShut {NoStop}%
\bibitem [{\citenamefont {Lemons}(2017)}]{LemonsBook}%
  \BibitemOpen
  \bibfield  {author} {\bibinfo {author} {\bibfnamefont {D.~S.}\ \bibnamefont
  {Lemons}},\ }\href@noop {} {\emph {\bibinfo {title} {A Student’s Guide to
  Dimensional Analysis}}}\ (\bibinfo  {publisher} {Cambridge University Press,
  Cambridge},\ \bibinfo {year} {2017})\BibitemShut {NoStop}%
\bibitem [{\citenamefont {Endrizzi}(2018)}]{endrizzi2018}%
  \BibitemOpen
  \bibfield  {author} {\bibinfo {author} {\bibfnamefont {M.}~\bibnamefont
  {Endrizzi}},\ }\bibfield  {title} {\bibinfo {title} {X-ray phase-contrast
  imaging},\ }\href@noop {} {\bibfield  {journal} {\bibinfo  {journal} {Nucl.
  Instrum. Methods Phys. Res. A}\ }\textbf {\bibinfo {volume} {878}},\ \bibinfo
  {pages} {88} (\bibinfo {year} {2018})}\BibitemShut {NoStop}%
\bibitem [{\citenamefont {Michelson}(1927)}]{Michelson95}%
  \BibitemOpen
  \bibfield  {author} {\bibinfo {author} {\bibfnamefont {A.~A.}\ \bibnamefont
  {Michelson}},\ }\href@noop {} {\emph {\bibinfo {title} {Studies in Optics}}}\
  (\bibinfo  {publisher} {University of Chicago Press, Chicago},\ \bibinfo
  {year} {1927})\BibitemShut {NoStop}%
\bibitem [{\citenamefont {Chabior}\ \emph {et~al.}(2011)\citenamefont
  {Chabior}, \citenamefont {Donath}, \citenamefont {David}, \citenamefont
  {Schuster}, \citenamefont {Schroer},\ and\ \citenamefont
  {Pfeiffer}}]{Chabior2011}%
  \BibitemOpen
  \bibfield  {author} {\bibinfo {author} {\bibfnamefont {M.}~\bibnamefont
  {Chabior}}, \bibinfo {author} {\bibfnamefont {T.}~\bibnamefont {Donath}},
  \bibinfo {author} {\bibfnamefont {C.}~\bibnamefont {David}}, \bibinfo
  {author} {\bibfnamefont {M.}~\bibnamefont {Schuster}}, \bibinfo {author}
  {\bibfnamefont {C.}~\bibnamefont {Schroer}},\ and\ \bibinfo {author}
  {\bibfnamefont {F.}~\bibnamefont {Pfeiffer}},\ }\bibfield  {title} {\bibinfo
  {title} {Signal-to-noise ratio in x ray dark-field imaging using a grating
  interferometer},\ }\href@noop {} {\bibfield  {journal} {\bibinfo  {journal}
  {J. Appl. Phys.}\ }\textbf {\bibinfo {volume} {110}},\ \bibinfo {pages}
  {053105} (\bibinfo {year} {2011})}\BibitemShut {NoStop}%
\bibitem [{\citenamefont {Ji}\ \emph {et~al.}(2017)\citenamefont {Ji},
  \citenamefont {Ge}, \citenamefont {Zhang}, \citenamefont {Li},\ and\
  \citenamefont {Chen}}]{GratingInterferometryDFbiasDueToNoise2017}%
  \BibitemOpen
  \bibfield  {author} {\bibinfo {author} {\bibfnamefont {X.}~\bibnamefont
  {Ji}}, \bibinfo {author} {\bibfnamefont {Y.}~\bibnamefont {Ge}}, \bibinfo
  {author} {\bibfnamefont {R.}~\bibnamefont {Zhang}}, \bibinfo {author}
  {\bibfnamefont {K.}~\bibnamefont {Li}},\ and\ \bibinfo {author}
  {\bibfnamefont {G.-H.}\ \bibnamefont {Chen}},\ }\bibfield  {title} {\bibinfo
  {title} {Studies of signal estimation bias in grating-based x-ray
  multicontrast imaging},\ }\href@noop {} {\bibfield  {journal} {\bibinfo
  {journal} {Med. Phys.}\ }\textbf {\bibinfo {volume} {44}},\ \bibinfo {pages}
  {2453} (\bibinfo {year} {2017})}\BibitemShut {NoStop}%
\bibitem [{\citenamefont {Strobl}\ \emph {et~al.}(2008)\citenamefont {Strobl},
  \citenamefont {Gr\"unzweig}, \citenamefont {Hilger}, \citenamefont {Manke},
  \citenamefont {Kardjilov}, \citenamefont {David},\ and\ \citenamefont
  {Pfeiffer}}]{Strobl2008}%
  \BibitemOpen
  \bibfield  {author} {\bibinfo {author} {\bibfnamefont {M.}~\bibnamefont
  {Strobl}}, \bibinfo {author} {\bibfnamefont {C.}~\bibnamefont {Gr\"unzweig}},
  \bibinfo {author} {\bibfnamefont {A.}~\bibnamefont {Hilger}}, \bibinfo
  {author} {\bibfnamefont {I.}~\bibnamefont {Manke}}, \bibinfo {author}
  {\bibfnamefont {N.}~\bibnamefont {Kardjilov}}, \bibinfo {author}
  {\bibfnamefont {C.}~\bibnamefont {David}},\ and\ \bibinfo {author}
  {\bibfnamefont {F.}~\bibnamefont {Pfeiffer}},\ }\bibfield  {title} {\bibinfo
  {title} {Neutron dark-field tomography},\ }\href
  {https://doi.org/10.1103/PhysRevLett.101.123902} {\bibfield  {journal}
  {\bibinfo  {journal} {Phys. Rev. Lett.}\ }\textbf {\bibinfo {volume} {101}},\
  \bibinfo {pages} {123902} (\bibinfo {year} {2008})}\BibitemShut {NoStop}%
\bibitem [{\citenamefont {Pfeiffer}\ \emph {et~al.}(2009)\citenamefont
  {Pfeiffer}, \citenamefont {Bech}, \citenamefont {Bunk}, \citenamefont
  {Donath}, \citenamefont {Henrich}, \citenamefont {Kraft},\ and\ \citenamefont
  {David}}]{pfeiffer2009}%
  \BibitemOpen
  \bibfield  {author} {\bibinfo {author} {\bibfnamefont {F.}~\bibnamefont
  {Pfeiffer}}, \bibinfo {author} {\bibfnamefont {M.}~\bibnamefont {Bech}},
  \bibinfo {author} {\bibfnamefont {O.}~\bibnamefont {Bunk}}, \bibinfo {author}
  {\bibfnamefont {T.}~\bibnamefont {Donath}}, \bibinfo {author} {\bibfnamefont
  {B.}~\bibnamefont {Henrich}}, \bibinfo {author} {\bibfnamefont
  {P.}~\bibnamefont {Kraft}},\ and\ \bibinfo {author} {\bibfnamefont
  {C.}~\bibnamefont {David}},\ }\bibfield  {title} {\bibinfo {title} {X-ray
  dark-field and phase-contrast imaging using a grating interferometer},\
  }\href@noop {} {\bibfield  {journal} {\bibinfo  {journal} {J. Appl. Phys.}\
  }\textbf {\bibinfo {volume} {105}},\ \bibinfo {pages} {102006} (\bibinfo
  {year} {2009})}\BibitemShut {NoStop}%
\bibitem [{\citenamefont {Zanette}\ \emph {et~al.}(2014)\citenamefont
  {Zanette}, \citenamefont {Zhou}, \citenamefont {Burvall}, \citenamefont
  {Lundstr{\"o}m}, \citenamefont {Larsson}, \citenamefont {Zdora},
  \citenamefont {Thibault}, \citenamefont {Pfeiffer},\ and\ \citenamefont
  {Hertz}}]{zanette2014}%
  \BibitemOpen
  \bibfield  {author} {\bibinfo {author} {\bibfnamefont {I.}~\bibnamefont
  {Zanette}}, \bibinfo {author} {\bibfnamefont {T.}~\bibnamefont {Zhou}},
  \bibinfo {author} {\bibfnamefont {A.}~\bibnamefont {Burvall}}, \bibinfo
  {author} {\bibfnamefont {U.}~\bibnamefont {Lundstr{\"o}m}}, \bibinfo {author}
  {\bibfnamefont {D.~H.}\ \bibnamefont {Larsson}}, \bibinfo {author}
  {\bibfnamefont {M.}~\bibnamefont {Zdora}}, \bibinfo {author} {\bibfnamefont
  {P.}~\bibnamefont {Thibault}}, \bibinfo {author} {\bibfnamefont
  {F.}~\bibnamefont {Pfeiffer}},\ and\ \bibinfo {author} {\bibfnamefont
  {H.~M.}\ \bibnamefont {Hertz}},\ }\bibfield  {title} {\bibinfo {title}
  {Speckle-based x-ray phase-contrast and dark-field imaging with a laboratory
  source},\ }\href@noop {} {\bibfield  {journal} {\bibinfo  {journal} {Phys.
  Rev. Lett.}\ }\textbf {\bibinfo {volume} {112}},\ \bibinfo {pages} {253903}
  (\bibinfo {year} {2014})}\BibitemShut {NoStop}%
\bibitem [{\citenamefont {Yashiro}\ and\ \citenamefont
  {Momose}(2015)}]{Yashiro2015}%
  \BibitemOpen
  \bibfield  {author} {\bibinfo {author} {\bibfnamefont {W.}~\bibnamefont
  {Yashiro}}\ and\ \bibinfo {author} {\bibfnamefont {A.}~\bibnamefont
  {Momose}},\ }\bibfield  {title} {\bibinfo {title} {Effects of unresolvable
  edges in grating-based {X}-ray differential phase imaging},\ }\href@noop {}
  {\bibfield  {journal} {\bibinfo  {journal} {Opt. Express}\ }\textbf {\bibinfo
  {volume} {23}},\ \bibinfo {pages} {9233} (\bibinfo {year}
  {2015})}\BibitemShut {NoStop}%
\bibitem [{\citenamefont {Morrison}\ and\ \citenamefont
  {Browne}(1992)}]{MorrisonBrowne1992}%
  \BibitemOpen
  \bibfield  {author} {\bibinfo {author} {\bibfnamefont {G.~R.}\ \bibnamefont
  {Morrison}}\ and\ \bibinfo {author} {\bibfnamefont {M.~T.}\ \bibnamefont
  {Browne}},\ }\bibfield  {title} {\bibinfo {title} {Dark-field imaging with
  the scanning transmission x-ray microscope},\ }\href@noop {} {\bibfield
  {journal} {\bibinfo  {journal} {Rev. Sci. Instrum.}\ }\textbf {\bibinfo
  {volume} {63}},\ \bibinfo {pages} {611} (\bibinfo {year} {1992})}\BibitemShut
  {NoStop}%
\bibitem [{\citenamefont {Suzuki}\ and\ \citenamefont
  {Uchida}(1995)}]{SuzukiUchida1995}%
  \BibitemOpen
  \bibfield  {author} {\bibinfo {author} {\bibfnamefont {Y.}~\bibnamefont
  {Suzuki}}\ and\ \bibinfo {author} {\bibfnamefont {F.}~\bibnamefont
  {Uchida}},\ }\bibfield  {title} {\bibinfo {title} {Dark-field imaging in hard
  x-ray scanning microscopy},\ }\href@noop {} {\bibfield  {journal} {\bibinfo
  {journal} {Rev. Sci. Instrum.}\ }\textbf {\bibinfo {volume} {66}},\ \bibinfo
  {pages} {1468} (\bibinfo {year} {1995})}\BibitemShut {NoStop}%
\bibitem [{\citenamefont {Groenendijk}\ \emph {et~al.}(2020)\citenamefont
  {Groenendijk}, \citenamefont {Schaff}, \citenamefont {Croton}, \citenamefont
  {Kitchen},\ and\ \citenamefont {Morgan}}]{Groenendijk2020}%
  \BibitemOpen
  \bibfield  {author} {\bibinfo {author} {\bibfnamefont {C.~F.}\ \bibnamefont
  {Groenendijk}}, \bibinfo {author} {\bibfnamefont {F.}~\bibnamefont {Schaff}},
  \bibinfo {author} {\bibfnamefont {L.~C.~P.}\ \bibnamefont {Croton}}, \bibinfo
  {author} {\bibfnamefont {M.~J.}\ \bibnamefont {Kitchen}},\ and\ \bibinfo
  {author} {\bibfnamefont {K.~S.}\ \bibnamefont {Morgan}},\ }\bibfield  {title}
  {\bibinfo {title} {Material decomposition from a single x-ray projection via
  single-grid phase contrast imaging},\ }\href
  {https://doi.org/10.1364/OL.389770} {\bibfield  {journal} {\bibinfo
  {journal} {Opt. Lett.}\ }\textbf {\bibinfo {volume} {45}},\ \bibinfo {pages}
  {4076} (\bibinfo {year} {2020})}\BibitemShut {NoStop}%
\bibitem [{\citenamefont {Young}(1802)}]{YoungOnTheBoundaryWave}%
  \BibitemOpen
  \bibfield  {author} {\bibinfo {author} {\bibfnamefont {T.}~\bibnamefont
  {Young}},\ }\bibfield  {title} {\bibinfo {title} {The {B}akerian lecture: On
  the theory of light and colours},\ }\href@noop {} {\bibfield  {journal}
  {\bibinfo  {journal} {Phil. Trans. R. Soc. Lond.}\ }\textbf {\bibinfo
  {volume} {92}},\ \bibinfo {pages} {12} (\bibinfo {year} {1802})}\BibitemShut
  {NoStop}%
\bibitem [{\citenamefont {Maggi}(1888)}]{Maggi}%
  \BibitemOpen
  \bibfield  {author} {\bibinfo {author} {\bibfnamefont {G.~A.}\ \bibnamefont
  {Maggi}},\ }\bibfield  {title} {\bibinfo {title} {Sulla propagazione libera e
  perturbata delle onde luminose in un mezzo isotropo},\ }\href@noop {}
  {\bibfield  {journal} {\bibinfo  {journal} {Annali di Mat. {II}}\ }\textbf
  {\bibinfo {volume} {16}},\ \bibinfo {pages} {21} (\bibinfo {year}
  {1888})}\BibitemShut {NoStop}%
\bibitem [{\citenamefont {Rubinowicz}(1917)}]{Rubinowicz}%
  \BibitemOpen
  \bibfield  {author} {\bibinfo {author} {\bibfnamefont {A.}~\bibnamefont
  {Rubinowicz}},\ }\bibfield  {title} {\bibinfo {title} {Die {B}eugungswelle in
  der {K}irchhoffschen {T}heorie der {B}eugungserscheinungen},\ }\href@noop {}
  {\bibfield  {journal} {\bibinfo  {journal} {Ann. Physik}\ }\textbf {\bibinfo
  {volume} {53}},\ \bibinfo {pages} {257} (\bibinfo {year} {1917})}\BibitemShut
  {NoStop}%
\bibitem [{\citenamefont {Miyamoto}\ and\ \citenamefont
  {Wolf}(1962{\natexlab{a}})}]{MiyamotoWolf1}%
  \BibitemOpen
  \bibfield  {author} {\bibinfo {author} {\bibfnamefont {K.}~\bibnamefont
  {Miyamoto}}\ and\ \bibinfo {author} {\bibfnamefont {E.}~\bibnamefont
  {Wolf}},\ }\bibfield  {title} {\bibinfo {title} {Generalization of the
  {M}aggi-{R}ubinowicz theory of the boundary diffraction wave--{P}art {I}},\
  }\href@noop {} {\bibfield  {journal} {\bibinfo  {journal} {J. Opt. Soc. Am.}\
  }\textbf {\bibinfo {volume} {52}},\ \bibinfo {pages} {615} (\bibinfo {year}
  {1962}{\natexlab{a}})}\BibitemShut {NoStop}%
\bibitem [{\citenamefont {Miyamoto}\ and\ \citenamefont
  {Wolf}(1962{\natexlab{b}})}]{MiyamotoWolf2}%
  \BibitemOpen
  \bibfield  {author} {\bibinfo {author} {\bibfnamefont {K.}~\bibnamefont
  {Miyamoto}}\ and\ \bibinfo {author} {\bibfnamefont {E.}~\bibnamefont
  {Wolf}},\ }\bibfield  {title} {\bibinfo {title} {Generalization of the
  {M}aggi-{R}ubinowicz theory of the boundary diffraction wave--{P}art {II}},\
  }\href@noop {} {\bibfield  {journal} {\bibinfo  {journal} {J. Opt. Soc. Am.}\
  }\textbf {\bibinfo {volume} {52}},\ \bibinfo {pages} {626} (\bibinfo {year}
  {1962}{\natexlab{b}})}\BibitemShut {NoStop}%
\bibitem [{\citenamefont {Hannay}(2000)}]{HannayParaxialBoundaryWave}%
  \BibitemOpen
  \bibfield  {author} {\bibinfo {author} {\bibfnamefont {J.~H.}\ \bibnamefont
  {Hannay}},\ }\bibfield  {title} {\bibinfo {title} {Fresnel diffraction as an
  aperture edge integral},\ }\href {https://doi.org/10.1080/09500340008231410}
  {\bibfield  {journal} {\bibinfo  {journal} {J. Mod. Opt.}\ }\textbf {\bibinfo
  {volume} {47}},\ \bibinfo {pages} {121} (\bibinfo {year} {2000})}\BibitemShut
  {NoStop}%
\bibitem [{\citenamefont {Borghi}(2015)}]{BorghiParaxialBoundaryWave}%
  \BibitemOpen
  \bibfield  {author} {\bibinfo {author} {\bibfnamefont {R.}~\bibnamefont
  {Borghi}},\ }\bibfield  {title} {\bibinfo {title} {Uniform asymptotics of
  paraxial boundary diffraction waves},\ }\href
  {https://doi.org/10.1364/JOSAA.32.000685} {\bibfield  {journal} {\bibinfo
  {journal} {J. Opt. Soc. Am. A}\ }\textbf {\bibinfo {volume} {32}},\ \bibinfo
  {pages} {685} (\bibinfo {year} {2015})}\BibitemShut {NoStop}%
\bibitem [{\citenamefont {Keller}(1962)}]{Keller}%
  \BibitemOpen
  \bibfield  {author} {\bibinfo {author} {\bibfnamefont {J.~B.}\ \bibnamefont
  {Keller}},\ }\bibfield  {title} {\bibinfo {title} {Geometrical theory of
  diffraction},\ }\href@noop {} {\bibfield  {journal} {\bibinfo  {journal} {J.
  Opt. Soc. Am.}\ }\textbf {\bibinfo {volume} {52}},\ \bibinfo {pages} {116}
  (\bibinfo {year} {1962})}\BibitemShut {NoStop}%
\bibitem [{\citenamefont {Esposito}\ \emph {et~al.}(2022)\citenamefont
  {Esposito}, \citenamefont {Massimi}, \citenamefont {Buchanan}, \citenamefont
  {Ferrara}, \citenamefont {Endrizzi},\ and\ \citenamefont
  {Olivo}}]{Esposito2022}%
  \BibitemOpen
  \bibfield  {author} {\bibinfo {author} {\bibfnamefont {M.}~\bibnamefont
  {Esposito}}, \bibinfo {author} {\bibfnamefont {L.}~\bibnamefont {Massimi}},
  \bibinfo {author} {\bibfnamefont {I.}~\bibnamefont {Buchanan}}, \bibinfo
  {author} {\bibfnamefont {J.~D.}\ \bibnamefont {Ferrara}}, \bibinfo {author}
  {\bibfnamefont {M.}~\bibnamefont {Endrizzi}},\ and\ \bibinfo {author}
  {\bibfnamefont {A.}~\bibnamefont {Olivo}},\ }\bibfield  {title} {\bibinfo
  {title} {A laboratory-based, low-energy, multi-modal x-ray microscope with
  user-defined resolution},\ }\href {https://doi.org/10.1063/5.0082968}
  {\bibfield  {journal} {\bibinfo  {journal} {Appl. Phys. Lett.}\ }\textbf
  {\bibinfo {volume} {120}},\ \bibinfo {pages} {234101} (\bibinfo {year}
  {2022})}\BibitemShut {NoStop}%
\bibitem [{\citenamefont {Croughan}\ \emph {et~al.}(2023)\citenamefont
  {Croughan}, \citenamefont {How}, \citenamefont {Pennings},\ and\
  \citenamefont {Morgan}}]{Croughan2023}%
  \BibitemOpen
  \bibfield  {author} {\bibinfo {author} {\bibfnamefont {M.~K.}\ \bibnamefont
  {Croughan}}, \bibinfo {author} {\bibfnamefont {Y.~Y.}\ \bibnamefont {How}},
  \bibinfo {author} {\bibfnamefont {A.}~\bibnamefont {Pennings}},\ and\
  \bibinfo {author} {\bibfnamefont {K.~S.}\ \bibnamefont {Morgan}},\ }\bibfield
   {title} {\bibinfo {title} {Directional dark-field retrieval with single-grid
  x-ray imaging},\ }\href {https://doi.org/10.1364/OE.480031} {\bibfield
  {journal} {\bibinfo  {journal} {Opt. Express}\ }\textbf {\bibinfo {volume}
  {31}},\ \bibinfo {pages} {11578} (\bibinfo {year} {2023})}\BibitemShut
  {NoStop}%
\bibitem [{\citenamefont {Yuen}\ and\ \citenamefont
  {Tombesi}(1986)}]{NegativeDiffusion1986}%
  \BibitemOpen
  \bibfield  {author} {\bibinfo {author} {\bibfnamefont {H.~P.}\ \bibnamefont
  {Yuen}}\ and\ \bibinfo {author} {\bibfnamefont {P.}~\bibnamefont {Tombesi}},\
  }\bibfield  {title} {\bibinfo {title} {Langevin equations with negative
  diffusion coefficients. {A} new approach to quantum optics},\ }\href
  {https://doi.org/https://doi.org/10.1016/0030-4018(86)90469-4} {\bibfield
  {journal} {\bibinfo  {journal} {Opt. Commun.}\ }\textbf {\bibinfo {volume}
  {59}},\ \bibinfo {pages} {155} (\bibinfo {year} {1986})}\BibitemShut
  {NoStop}%
\bibitem [{\citenamefont {Tan}\ \emph {et~al.}(1987)\citenamefont {Tan},
  \citenamefont {Li},\ and\ \citenamefont {Zhang}}]{NegativeDiffusion1987}%
  \BibitemOpen
  \bibfield  {author} {\bibinfo {author} {\bibfnamefont {W.}~\bibnamefont
  {Tan}}, \bibinfo {author} {\bibfnamefont {Y.}~\bibnamefont {Li}},\ and\
  \bibinfo {author} {\bibfnamefont {W.}~\bibnamefont {Zhang}},\ }\bibfield
  {title} {\bibinfo {title} {The solution of the {F}okker-{P}lanck equation
  with zero or negative diffusion coefficients in quantum optics},\ }\href
  {https://doi.org/https://doi.org/10.1016/0030-4018(87)90052-6} {\bibfield
  {journal} {\bibinfo  {journal} {Opt. Commun.}\ }\textbf {\bibinfo {volume}
  {64}},\ \bibinfo {pages} {195} (\bibinfo {year} {1987})}\BibitemShut
  {NoStop}%
\bibitem [{\citenamefont {Cosslett}\ and\ \citenamefont
  {Nixon}(1960)}]{CosslettNixonBook}%
  \BibitemOpen
  \bibfield  {author} {\bibinfo {author} {\bibfnamefont {V.~E.}\ \bibnamefont
  {Cosslett}}\ and\ \bibinfo {author} {\bibfnamefont {W.~C.}\ \bibnamefont
  {Nixon}},\ }\href@noop {} {\emph {\bibinfo {title} {X-ray Microscopy}}}\
  (\bibinfo  {publisher} {Cambridge University Press},\ \bibinfo {address}
  {Cambridge},\ \bibinfo {year} {1960})\BibitemShut {NoStop}%
\bibitem [{\citenamefont {Gureyev}\ \emph {et~al.}(2008)\citenamefont
  {Gureyev}, \citenamefont {Nesterets}, \citenamefont {Stevenson},
  \citenamefont {Miller}, \citenamefont {Pogany},\ and\ \citenamefont
  {Wilkins}}]{GureyevRulesOfThumb2008}%
  \BibitemOpen
  \bibfield  {author} {\bibinfo {author} {\bibfnamefont {T.~E.}\ \bibnamefont
  {Gureyev}}, \bibinfo {author} {\bibfnamefont {{\relax Ya}.~I.}\ \bibnamefont
  {Nesterets}}, \bibinfo {author} {\bibfnamefont {A.~W.}\ \bibnamefont
  {Stevenson}}, \bibinfo {author} {\bibfnamefont {P.~R.}\ \bibnamefont
  {Miller}}, \bibinfo {author} {\bibfnamefont {A.}~\bibnamefont {Pogany}},\
  and\ \bibinfo {author} {\bibfnamefont {S.~W.}\ \bibnamefont {Wilkins}},\
  }\bibfield  {title} {\bibinfo {title} {Some simple rules for contrast,
  signal-to-noise and resolution in in-line x-ray phase-contrast imaging},\
  }\href {https://doi.org/10.1364/OE.16.003223} {\bibfield  {journal} {\bibinfo
   {journal} {Opt. Express}\ }\textbf {\bibinfo {volume} {16}},\ \bibinfo
  {pages} {3223} (\bibinfo {year} {2008})}\BibitemShut {NoStop}%
\bibitem [{\citenamefont {Kravtsov}\ and\ \citenamefont
  {Orlov}(1998)}]{KravtsovOrlovBook}%
  \BibitemOpen
  \bibfield  {author} {\bibinfo {author} {\bibfnamefont {{\relax{Yu}}.~A.}\
  \bibnamefont {Kravtsov}}\ and\ \bibinfo {author} {\bibfnamefont
  {{\relax{Yu}}.~I.}\ \bibnamefont {Orlov}},\ }\href@noop {} {\emph {\bibinfo
  {title} {Caustics, Catastrophes and Wave Fields}}},\ \bibinfo {edition}
  {2nd}\ ed.\ (\bibinfo  {publisher} {Springer, Berlin},\ \bibinfo {year}
  {1998})\BibitemShut {NoStop}%
\bibitem [{\citenamefont {Paganin}\ \emph {et~al.}(2002)\citenamefont
  {Paganin}, \citenamefont {Mayo}, \citenamefont {Gureyev}, \citenamefont
  {Miller},\ and\ \citenamefont {Wilkins}}]{paganin2002}%
  \BibitemOpen
  \bibfield  {author} {\bibinfo {author} {\bibfnamefont {D.}~\bibnamefont
  {Paganin}}, \bibinfo {author} {\bibfnamefont {S.~C.}\ \bibnamefont {Mayo}},
  \bibinfo {author} {\bibfnamefont {T.~E.}\ \bibnamefont {Gureyev}}, \bibinfo
  {author} {\bibfnamefont {P.~R.}\ \bibnamefont {Miller}},\ and\ \bibinfo
  {author} {\bibfnamefont {S.~W.}\ \bibnamefont {Wilkins}},\ }\bibfield
  {title} {\bibinfo {title} {Simultaneous phase and amplitude extraction from a
  single defocused image of a homogeneous object},\ }\href
  {http://dx.doi.org/10.1046/j.1365-2818.2002.01010.x} {\bibfield  {journal}
  {\bibinfo  {journal} {J. Microsc.}\ }\textbf {\bibinfo {volume} {206}},\
  \bibinfo {pages} {33} (\bibinfo {year} {2002})}\BibitemShut {NoStop}%
\bibitem [{\citenamefont {Poola}\ and\ \citenamefont {John}(2017)}]{Poola2017}%
  \BibitemOpen
  \bibfield  {author} {\bibinfo {author} {\bibfnamefont {P.~K.}\ \bibnamefont
  {Poola}}\ and\ \bibinfo {author} {\bibfnamefont {R.}~\bibnamefont {John}},\
  }\bibfield  {title} {\bibinfo {title} {{Label-free nanoscale characterization
  of red blood cell structure and dynamics using single-shot transport of
  intensity equation}},\ }\href@noop {} {\bibfield  {journal} {\bibinfo
  {journal} {J. Biomed. Opt.}\ }\textbf {\bibinfo {volume} {22}},\ \bibinfo
  {pages} {106001} (\bibinfo {year} {2017})}\BibitemShut {NoStop}%
\bibitem [{\citenamefont {Gabor}(1948{\natexlab{b}})}]{GaborOldTEMBook}%
  \BibitemOpen
  \bibfield  {author} {\bibinfo {author} {\bibfnamefont {D.}~\bibnamefont
  {Gabor}},\ }\href@noop {} {\emph {\bibinfo {title} {The Electron
  Microscope:~Its Development, Present Performance and Future Possibilities}}}\
  (\bibinfo  {publisher} {Chemical Publishing Co., Brooklyn and New York},\
  \bibinfo {year} {1948})\ p.~\bibinfo {pages} {48}\BibitemShut {NoStop}%
\bibitem [{\citenamefont {Allman}\ \emph {et~al.}(2000)\citenamefont {Allman},
  \citenamefont {McMahon}, \citenamefont {Nugent}, \citenamefont {Paganin},
  \citenamefont {Jacobson}, \citenamefont {Arif},\ and\ \citenamefont
  {Werner}}]{Allman2000}%
  \BibitemOpen
  \bibfield  {author} {\bibinfo {author} {\bibfnamefont {B.~E.}\ \bibnamefont
  {Allman}}, \bibinfo {author} {\bibfnamefont {P.~J.}\ \bibnamefont {McMahon}},
  \bibinfo {author} {\bibfnamefont {K.~A.}\ \bibnamefont {Nugent}}, \bibinfo
  {author} {\bibfnamefont {D.}~\bibnamefont {Paganin}}, \bibinfo {author}
  {\bibfnamefont {D.~L.}\ \bibnamefont {Jacobson}}, \bibinfo {author}
  {\bibfnamefont {M.}~\bibnamefont {Arif}},\ and\ \bibinfo {author}
  {\bibfnamefont {S.~A.}\ \bibnamefont {Werner}},\ }\bibfield  {title}
  {\bibinfo {title} {Phase radiography with neutrons},\ }\href@noop {}
  {\bibfield  {journal} {\bibinfo  {journal} {Nature}\ }\textbf {\bibinfo
  {volume} {408}},\ \bibinfo {pages} {158} (\bibinfo {year}
  {2000})}\BibitemShut {NoStop}%
\bibitem [{\citenamefont {Gureyev}\ \emph {et~al.}(2004)\citenamefont
  {Gureyev}, \citenamefont {Stevenson}, \citenamefont {Nesterets},\ and\
  \citenamefont {Wilkins}}]{DeblurByDefocus}%
  \BibitemOpen
  \bibfield  {author} {\bibinfo {author} {\bibfnamefont {T.~E.}\ \bibnamefont
  {Gureyev}}, \bibinfo {author} {\bibfnamefont {A.~W.}\ \bibnamefont
  {Stevenson}}, \bibinfo {author} {\bibfnamefont {{\relax Ya}.~I.}\
  \bibnamefont {Nesterets}},\ and\ \bibinfo {author} {\bibfnamefont {S.~W.}\
  \bibnamefont {Wilkins}},\ }\bibfield  {title} {\bibinfo {title} {Image
  deblurring by means of defocus},\ }\href@noop {} {\bibfield  {journal}
  {\bibinfo  {journal} {Opt. Commun.}\ }\textbf {\bibinfo {volume} {240}},\
  \bibinfo {pages} {81} (\bibinfo {year} {2004})}\BibitemShut {NoStop}%
\bibitem [{\citenamefont {Gureyev}\ \emph {et~al.}(2026)\citenamefont
  {Gureyev}, \citenamefont {Paganin}, \citenamefont {Pakzad},\ and\
  \citenamefont {Quiney}}]{Gureyev2026}%
  \BibitemOpen
  \bibfield  {author} {\bibinfo {author} {\bibfnamefont {T.~E.}\ \bibnamefont
  {Gureyev}}, \bibinfo {author} {\bibfnamefont {D.~M.}\ \bibnamefont
  {Paganin}}, \bibinfo {author} {\bibfnamefont {A.}~\bibnamefont {Pakzad}},\
  and\ \bibinfo {author} {\bibfnamefont {H.~M.}\ \bibnamefont {Quiney}},\
  }\bibfield  {title} {\bibinfo {title} {On optimisation of {P}aganin's method
  for propagation-based {X}-ray phase-contrast imaging and tomography},\
  }\href@noop {} {\bibfield  {journal} {\bibinfo  {journal} {J. Microsc.}\
  }\textbf {\bibinfo {volume} {303}},\ \bibinfo {pages} {37} (\bibinfo {year}
  {2026})}\BibitemShut {NoStop}%
\bibitem [{\citenamefont {Beltran}\ \emph {et~al.}(2018)\citenamefont
  {Beltran}, \citenamefont {Paganin},\ and\ \citenamefont
  {Pelliccia}}]{Beltran2018}%
  \BibitemOpen
  \bibfield  {author} {\bibinfo {author} {\bibfnamefont {M.~A.}\ \bibnamefont
  {Beltran}}, \bibinfo {author} {\bibfnamefont {D.~M.}\ \bibnamefont
  {Paganin}},\ and\ \bibinfo {author} {\bibfnamefont {D.}~\bibnamefont
  {Pelliccia}},\ }\bibfield  {title} {\bibinfo {title} {Phase-and-amplitude
  recovery from a single phase-contrast image using partially spatially
  coherent x-ray radiation},\ }\href@noop {} {\bibfield  {journal} {\bibinfo
  {journal} {J. Opt.}\ }\textbf {\bibinfo {volume} {20}},\ \bibinfo {pages}
  {055605} (\bibinfo {year} {2018})}\BibitemShut {NoStop}%
\bibitem [{\citenamefont {Dirac}(1931)}]{Dirac1931}%
  \BibitemOpen
  \bibfield  {author} {\bibinfo {author} {\bibfnamefont {P.~A.~M.}\
  \bibnamefont {Dirac}},\ }\bibfield  {title} {\bibinfo {title} {Quantised
  singularities in the electromagnetic field},\ }\href@noop {} {\bibfield
  {journal} {\bibinfo  {journal} {Proc. Roy. Soc. A}\ }\textbf {\bibinfo
  {volume} {133}},\ \bibinfo {pages} {60} (\bibinfo {year} {1931})}\BibitemShut
  {NoStop}%
\bibitem [{\citenamefont {Nye}(1999)}]{Nye1999}%
  \BibitemOpen
  \bibfield  {author} {\bibinfo {author} {\bibfnamefont {J.~F.}\ \bibnamefont
  {Nye}},\ }\href@noop {} {\emph {\bibinfo {title} {Natural Focusing and Fine
  Structure of Light}}}\ (\bibinfo  {publisher} {Institute of Physics
  Publishing},\ \bibinfo {address} {Bristol},\ \bibinfo {year}
  {1999})\BibitemShut {NoStop}%
\bibitem [{\citenamefont {Gori}\ \emph {et~al.}(1993)\citenamefont {Gori},
  \citenamefont {Santarsiero},\ and\ \citenamefont {Guattari}}]{Gori1993}%
  \BibitemOpen
  \bibfield  {author} {\bibinfo {author} {\bibfnamefont {F.}~\bibnamefont
  {Gori}}, \bibinfo {author} {\bibfnamefont {M.}~\bibnamefont {Santarsiero}},\
  and\ \bibinfo {author} {\bibfnamefont {G.}~\bibnamefont {Guattari}},\
  }\bibfield  {title} {\bibinfo {title} {Coherence and the spatial distribution
  of intensity},\ }\href {https://doi.org/10.1364/JOSAA.10.000673} {\bibfield
  {journal} {\bibinfo  {journal} {J. Opt. Soc. Am. A}\ }\textbf {\bibinfo
  {volume} {10}},\ \bibinfo {pages} {673} (\bibinfo {year} {1993})}\BibitemShut
  {NoStop}%
\bibitem [{\citenamefont {Kress}(2014)}]{KressBook}%
  \BibitemOpen
  \bibfield  {author} {\bibinfo {author} {\bibfnamefont {R.}~\bibnamefont
  {Kress}},\ }\href@noop {} {\emph {\bibinfo {title} {Linear Integral
  Equations}}},\ \bibinfo {edition} {3rd}\ ed.\ (\bibinfo  {publisher}
  {Springer},\ \bibinfo {address} {New York},\ \bibinfo {year}
  {2014})\BibitemShut {NoStop}%
\bibitem [{\citenamefont {Gabor}(1961)}]{Gabor1961}%
  \BibitemOpen
  \bibfield  {author} {\bibinfo {author} {\bibfnamefont {D.}~\bibnamefont
  {Gabor}},\ }\bibfield  {title} {\bibinfo {title} {Light and information},\
  }\href@noop {} {\bibfield  {journal} {\bibinfo  {journal} {Prog. Opt.}\
  }\textbf {\bibinfo {volume} {1}},\ \bibinfo {pages} {109} (\bibinfo {year}
  {1961})}\BibitemShut {NoStop}%
\bibitem [{\citenamefont {Paganin}\ \emph {et~al.}(2004)\citenamefont
  {Paganin}, \citenamefont {Barty}, \citenamefont {McMahon},\ and\
  \citenamefont {Nugent}}]{QPAMIII}%
  \BibitemOpen
  \bibfield  {author} {\bibinfo {author} {\bibfnamefont {D.}~\bibnamefont
  {Paganin}}, \bibinfo {author} {\bibfnamefont {A.}~\bibnamefont {Barty}},
  \bibinfo {author} {\bibfnamefont {P.~J.}\ \bibnamefont {McMahon}},\ and\
  \bibinfo {author} {\bibfnamefont {K.~A.}\ \bibnamefont {Nugent}},\ }\bibfield
   {title} {\bibinfo {title} {Quantitative phase-amplitude microscopy. {III}.
  {T}he effects of noise},\ }\href@noop {} {\bibfield  {journal} {\bibinfo
  {journal} {J. Microsc.}\ }\textbf {\bibinfo {volume} {214}},\ \bibinfo
  {pages} {51} (\bibinfo {year} {2004})}\BibitemShut {NoStop}%
\bibitem [{\citenamefont {Nesterets}\ and\ \citenamefont
  {Gureyev}(2014)}]{Nesterets2014}%
  \BibitemOpen
  \bibfield  {author} {\bibinfo {author} {\bibfnamefont {Y.~I.}\ \bibnamefont
  {Nesterets}}\ and\ \bibinfo {author} {\bibfnamefont {T.~E.}\ \bibnamefont
  {Gureyev}},\ }\bibfield  {title} {\bibinfo {title} {Noise propagation in
  x-ray phase-contrast imaging and computed tomography},\ }\href
  {https://doi.org/10.1088/0022-3727/47/10/105402} {\bibfield  {journal}
  {\bibinfo  {journal} {J. Phys. D: Appl. Phys.}\ }\textbf {\bibinfo {volume}
  {47}},\ \bibinfo {pages} {105402} (\bibinfo {year} {2014})}\BibitemShut
  {NoStop}%
\bibitem [{\citenamefont {Gureyev}\ \emph
  {et~al.}(2014{\natexlab{b}})\citenamefont {Gureyev}, \citenamefont {Mayo},
  \citenamefont {Nesterets}, \citenamefont {Mohammadi}, \citenamefont {Lockie},
  \citenamefont {Menk}, \citenamefont {Arfelli}, \citenamefont {Pavlov},
  \citenamefont {Kitchen}, \citenamefont {Zanconati}, \citenamefont {Dullin},\
  and\ \citenamefont {Tromba}}]{Gureyev2014}%
  \BibitemOpen
  \bibfield  {author} {\bibinfo {author} {\bibfnamefont {T.~E.}\ \bibnamefont
  {Gureyev}}, \bibinfo {author} {\bibfnamefont {S.~C.}\ \bibnamefont {Mayo}},
  \bibinfo {author} {\bibfnamefont {Y.~I.}\ \bibnamefont {Nesterets}}, \bibinfo
  {author} {\bibfnamefont {S.}~\bibnamefont {Mohammadi}}, \bibinfo {author}
  {\bibfnamefont {D.}~\bibnamefont {Lockie}}, \bibinfo {author} {\bibfnamefont
  {R.~H.}\ \bibnamefont {Menk}}, \bibinfo {author} {\bibfnamefont
  {F.}~\bibnamefont {Arfelli}}, \bibinfo {author} {\bibfnamefont {K.~M.}\
  \bibnamefont {Pavlov}}, \bibinfo {author} {\bibfnamefont {M.~J.}\
  \bibnamefont {Kitchen}}, \bibinfo {author} {\bibfnamefont {F.}~\bibnamefont
  {Zanconati}}, \bibinfo {author} {\bibfnamefont {C.}~\bibnamefont {Dullin}},\
  and\ \bibinfo {author} {\bibfnamefont {G.}~\bibnamefont {Tromba}},\
  }\bibfield  {title} {\bibinfo {title} {Investigation of the imaging quality
  of synchrotron-based phase-contrast mammographic tomography},\ }\href
  {https://doi.org/10.1088/0022-3727/47/36/365401} {\bibfield  {journal}
  {\bibinfo  {journal} {J. Phys. D: Appl. Phys.}\ }\textbf {\bibinfo {volume}
  {47}},\ \bibinfo {pages} {365401} (\bibinfo {year}
  {2014}{\natexlab{b}})}\BibitemShut {NoStop}%
\bibitem [{\citenamefont {Gureyev}\ \emph {et~al.}(2017)\citenamefont
  {Gureyev}, \citenamefont {Nesterets}, \citenamefont {Kozlov}, \citenamefont
  {Paganin},\ and\ \citenamefont {Quiney}}]{Gureyev2017Unreasonable}%
  \BibitemOpen
  \bibfield  {author} {\bibinfo {author} {\bibfnamefont {T.~E.}\ \bibnamefont
  {Gureyev}}, \bibinfo {author} {\bibfnamefont {Y.~I.}\ \bibnamefont
  {Nesterets}}, \bibinfo {author} {\bibfnamefont {A.}~\bibnamefont {Kozlov}},
  \bibinfo {author} {\bibfnamefont {D.~M.}\ \bibnamefont {Paganin}},\ and\
  \bibinfo {author} {\bibfnamefont {H.~M.}\ \bibnamefont {Quiney}},\ }\bibfield
   {title} {\bibinfo {title} {On the ``unreasonable'' effectiveness of
  transport of intensity imaging and optical deconvolution},\ }\href
  {http://josaa.osa.org/abstract.cfm?URI=josaa-34-12-2251} {\bibfield
  {journal} {\bibinfo  {journal} {J. Opt. Soc. Am. A}\ }\textbf {\bibinfo
  {volume} {34}},\ \bibinfo {pages} {2251} (\bibinfo {year}
  {2017})}\BibitemShut {NoStop}%
\bibitem [{\citenamefont {Kitchen}\ \emph {et~al.}(2017)\citenamefont
  {Kitchen}, \citenamefont {Buckley}, \citenamefont {Gureyev}, \citenamefont
  {Wallace}, \citenamefont {Andres-Thio}, \citenamefont {Uesugi}, \citenamefont
  {Yagi},\ and\ \citenamefont {Hooper}}]{Kitchen2017}%
  \BibitemOpen
  \bibfield  {author} {\bibinfo {author} {\bibfnamefont {M.~J.}\ \bibnamefont
  {Kitchen}}, \bibinfo {author} {\bibfnamefont {G.~A.}\ \bibnamefont
  {Buckley}}, \bibinfo {author} {\bibfnamefont {T.~E.}\ \bibnamefont
  {Gureyev}}, \bibinfo {author} {\bibfnamefont {M.~J.}\ \bibnamefont
  {Wallace}}, \bibinfo {author} {\bibfnamefont {N.}~\bibnamefont
  {Andres-Thio}}, \bibinfo {author} {\bibfnamefont {K.}~\bibnamefont {Uesugi}},
  \bibinfo {author} {\bibfnamefont {N.}~\bibnamefont {Yagi}},\ and\ \bibinfo
  {author} {\bibfnamefont {S.~B.}\ \bibnamefont {Hooper}},\ }\bibfield  {title}
  {\bibinfo {title} {{CT} dose reduction factors in the thousands using {X}-ray
  phase contrast},\ }\href@noop {} {\bibfield  {journal} {\bibinfo  {journal}
  {Sci. Rep.}\ }\textbf {\bibinfo {volume} {7}},\ \bibinfo {pages} {15953}
  (\bibinfo {year} {2017})}\BibitemShut {NoStop}%
\bibitem [{\citenamefont {Saleh}\ and\ \citenamefont
  {Teich}(1991)}]{SalehTeichBook}%
  \BibitemOpen
  \bibfield  {author} {\bibinfo {author} {\bibfnamefont {B.~E.~A.}\
  \bibnamefont {Saleh}}\ and\ \bibinfo {author} {\bibfnamefont {M.~C.}\
  \bibnamefont {Teich}},\ }\href@noop {} {\emph {\bibinfo {title}
  {{Fundamentals of Photonics}}}}\ (\bibinfo  {publisher} {Wiley},\ \bibinfo
  {address} {New York},\ \bibinfo {year} {1991})\BibitemShut {NoStop}%
\bibitem [{\citenamefont {Gureyev}(2003)}]{GureyevCompositeTechniques2003}%
  \BibitemOpen
  \bibfield  {author} {\bibinfo {author} {\bibfnamefont {T.~E.}\ \bibnamefont
  {Gureyev}},\ }\bibfield  {title} {\bibinfo {title} {Composite techniques for
  phase retrieval in the {F}resnel region},\ }\href
  {https://doi.org/https://doi.org/10.1016/S0030-4018(03)01353-1} {\bibfield
  {journal} {\bibinfo  {journal} {Opt. Commun.}\ }\textbf {\bibinfo {volume}
  {220}},\ \bibinfo {pages} {49} (\bibinfo {year} {2003})}\BibitemShut
  {NoStop}%
\bibitem [{\citenamefont {Peskin}\ and\ \citenamefont
  {Schroeder}(1995)}]{PeskinSchroeder1995}%
  \BibitemOpen
  \bibfield  {author} {\bibinfo {author} {\bibfnamefont {M.~E.}\ \bibnamefont
  {Peskin}}\ and\ \bibinfo {author} {\bibfnamefont {D.~V.}\ \bibnamefont
  {Schroeder}},\ }\href@noop {} {\emph {\bibinfo {title} {{An Introduction to
  Quantum Field Theory}}}}\ (\bibinfo  {publisher} {CRC Press},\ \bibinfo
  {address} {Boca Raton, FL, USA},\ \bibinfo {year} {1995})\BibitemShut
  {NoStop}%
\bibitem [{\citenamefont {Magyar}\ and\ \citenamefont
  {Mandel}(1963)}]{MagyarMandel1963}%
  \BibitemOpen
  \bibfield  {author} {\bibinfo {author} {\bibfnamefont {G.}~\bibnamefont
  {Magyar}}\ and\ \bibinfo {author} {\bibfnamefont {L.}~\bibnamefont
  {Mandel}},\ }\bibfield  {title} {\bibinfo {title} {Interference fringes
  produced by superposition of two independent maser light beams},\ }\href@noop
  {} {\bibfield  {journal} {\bibinfo  {journal} {Nature}\ }\textbf {\bibinfo
  {volume} {198}},\ \bibinfo {pages} {255} (\bibinfo {year}
  {1963})}\BibitemShut {NoStop}%
\bibitem [{\citenamefont {Paganin}\ and\ \citenamefont {{S\'anchez del
  R\'{\i}o}}(2019)}]{paganinsanchezdelrio2019}%
  \BibitemOpen
  \bibfield  {author} {\bibinfo {author} {\bibfnamefont {D.~M.}\ \bibnamefont
  {Paganin}}\ and\ \bibinfo {author} {\bibfnamefont {M.}~\bibnamefont
  {{S\'anchez del R\'{\i}o}}},\ }\bibfield  {title} {\bibinfo {title} {Speckled
  cross-spectral densities and their associated correlation singularities for a
  modern source of partially coherent x rays},\ }\href
  {https://doi.org/10.1103/PhysRevA.100.043813} {\bibfield  {journal} {\bibinfo
   {journal} {Phys. Rev. A}\ }\textbf {\bibinfo {volume} {100}},\ \bibinfo
  {pages} {043813} (\bibinfo {year} {2019})}\BibitemShut {NoStop}%
\bibitem [{\citenamefont {Bale}\ and\ \citenamefont
  {Schmidt}(1984)}]{BaleSchmidt1984}%
  \BibitemOpen
  \bibfield  {author} {\bibinfo {author} {\bibfnamefont {H.~D.}\ \bibnamefont
  {Bale}}\ and\ \bibinfo {author} {\bibfnamefont {P.~W.}\ \bibnamefont
  {Schmidt}},\ }\bibfield  {title} {\bibinfo {title} {Small-angle
  x-ray-scattering investigation of submicroscopic porosity with fractal
  properties},\ }\href {https://doi.org/10.1103/PhysRevLett.53.596} {\bibfield
  {journal} {\bibinfo  {journal} {Phys. Rev. Lett.}\ }\textbf {\bibinfo
  {volume} {53}},\ \bibinfo {pages} {596} (\bibinfo {year} {1984})}\BibitemShut
  {NoStop}%
\bibitem [{\citenamefont {Teixeira}(1988)}]{Teixeira1988}%
  \BibitemOpen
  \bibfield  {author} {\bibinfo {author} {\bibfnamefont {J.}~\bibnamefont
  {Teixeira}},\ }\bibfield  {title} {\bibinfo {title} {{Small-angle scattering
  by fractal systems}},\ }\href {https://doi.org/10.1107/S0021889888000263}
  {\bibfield  {journal} {\bibinfo  {journal} {J. Appl. Cryst.}\ }\textbf
  {\bibinfo {volume} {21}},\ \bibinfo {pages} {781} (\bibinfo {year}
  {1988})}\BibitemShut {NoStop}%
\bibitem [{\citenamefont {Jakeman}\ and\ \citenamefont
  {Ridley}(2006)}]{JakemanRidleyBook}%
  \BibitemOpen
  \bibfield  {author} {\bibinfo {author} {\bibfnamefont {E.}~\bibnamefont
  {Jakeman}}\ and\ \bibinfo {author} {\bibfnamefont {K.~D.}\ \bibnamefont
  {Ridley}},\ }\href@noop {} {\emph {\bibinfo {title} {Modeling Fluctuations in
  Scattered Waves}}}\ (\bibinfo  {publisher} {Taylor \& Francis},\ \bibinfo
  {address} {New York},\ \bibinfo {year} {2006})\BibitemShut {NoStop}%
\bibitem [{\citenamefont {Pearson}(1916)}]{PearsonVII}%
  \BibitemOpen
  \bibfield  {author} {\bibinfo {author} {\bibfnamefont {K.}~\bibnamefont
  {Pearson}},\ }\bibfield  {title} {\bibinfo {title} {{IX}. {M}athematical
  contributions to the theory of evolution. {XIX}. {S}econd supplement to a
  memoir on skew variation},\ }\href@noop {} {\bibfield  {journal} {\bibinfo
  {journal} {Phil. Trans. R. Soc. A}\ }\textbf {\bibinfo {volume} {216}},\
  \bibinfo {pages} {429} (\bibinfo {year} {1916})}\BibitemShut {NoStop}%
\bibitem [{\citenamefont {Wolf}(1982)}]{Wolf1982}%
  \BibitemOpen
  \bibfield  {author} {\bibinfo {author} {\bibfnamefont {E.}~\bibnamefont
  {Wolf}},\ }\bibfield  {title} {\bibinfo {title} {New theory of partial
  coherence in the space-frequency domain. {P}art {I}: spectra and
  cross-spectra of steady-state sources},\ }\href@noop {} {\bibfield  {journal}
  {\bibinfo  {journal} {J. Opt. Soc. Am.}\ }\textbf {\bibinfo {volume} {72}},\
  \bibinfo {pages} {343} (\bibinfo {year} {1982})}\BibitemShut {NoStop}%
\bibitem [{\citenamefont {Thorne}\ and\ \citenamefont
  {Blandford}(2017)}]{ThorneBlandfordBook}%
  \BibitemOpen
  \bibfield  {author} {\bibinfo {author} {\bibfnamefont {K.~S.}\ \bibnamefont
  {Thorne}}\ and\ \bibinfo {author} {\bibfnamefont {R.~D.}\ \bibnamefont
  {Blandford}},\ }\href@noop {} {\emph {\bibinfo {title} {Modern Classical
  Physics: Optics, Fluids, Plasmas, Elasticity, Relativity, and Statistical
  Physics}}}\ (\bibinfo  {publisher} {Princeton},\ \bibinfo {address}
  {Princeton University Press},\ \bibinfo {year} {2017})\ pp.\ \bibinfo {pages}
  {183--184}\BibitemShut {NoStop}%
\bibitem [{\citenamefont {McCauley}(1993)}]{McCauleyChaosBook}%
  \BibitemOpen
  \bibfield  {author} {\bibinfo {author} {\bibfnamefont {J.~L.}\ \bibnamefont
  {McCauley}},\ }\href@noop {} {\emph {\bibinfo {title} {Chaos, Dynamics, and
  Fractals: An Algorithmic Approach to Deterministic Chaos}}}\ (\bibinfo
  {publisher} {Cambridge University Press, Cambridge},\ \bibinfo {year}
  {1993})\ pp.\ \bibinfo {pages} {105--108}\BibitemShut {NoStop}%
\bibitem [{\citenamefont {Alperin}\ \emph {et~al.}(2019)\citenamefont
  {Alperin}, \citenamefont {Grotelueschen},\ and\ \citenamefont
  {Siemens}}]{Alperin2019}%
  \BibitemOpen
  \bibfield  {author} {\bibinfo {author} {\bibfnamefont {S.~N.}\ \bibnamefont
  {Alperin}}, \bibinfo {author} {\bibfnamefont {A.~L.}\ \bibnamefont
  {Grotelueschen}},\ and\ \bibinfo {author} {\bibfnamefont {M.~E.}\
  \bibnamefont {Siemens}},\ }\bibfield  {title} {\bibinfo {title} {Quantum
  turbulent structure in light},\ }\href
  {https://doi.org/10.1103/PhysRevLett.122.044301} {\bibfield  {journal}
  {\bibinfo  {journal} {Phys. Rev. Lett.}\ }\textbf {\bibinfo {volume} {122}},\
  \bibinfo {pages} {044301} (\bibinfo {year} {2019})}\BibitemShut {NoStop}%
\bibitem [{\citenamefont {Goodman}(1985)}]{GoodmanStatisticalOpticsBook}%
  \BibitemOpen
  \bibfield  {author} {\bibinfo {author} {\bibfnamefont {J.~W.}\ \bibnamefont
  {Goodman}},\ }\href@noop {} {\emph {\bibinfo {title} {Statistical Optics}}}\
  (\bibinfo  {publisher} {John Wiley \& Sons},\ \bibinfo {address} {New York},\
  \bibinfo {year} {1985})\BibitemShut {NoStop}%
\bibitem [{\citenamefont {Goodman}(2007)}]{GoodmanSpeckleBook}%
  \BibitemOpen
  \bibfield  {author} {\bibinfo {author} {\bibfnamefont {J.~W.}\ \bibnamefont
  {Goodman}},\ }\href@noop {} {\emph {\bibinfo {title} {Speckle Phenomena in
  Optics}}}\ (\bibinfo  {publisher} {Roberts and Company},\ \bibinfo {address}
  {Englewood Colorado},\ \bibinfo {year} {2007})\BibitemShut {NoStop}%
\bibitem [{\citenamefont {Alonso}(2011)}]{Alonso2011}%
  \BibitemOpen
  \bibfield  {author} {\bibinfo {author} {\bibfnamefont {M.~A.}\ \bibnamefont
  {Alonso}},\ }\bibfield  {title} {\bibinfo {title} {Wigner functions in
  optics: describing beams as ray bundles and pulses as particle ensembles},\
  }\href@noop {} {\bibfield  {journal} {\bibinfo  {journal} {Adv. Opt.
  Photon.}\ }\textbf {\bibinfo {volume} {3}},\ \bibinfo {pages} {272} (\bibinfo
  {year} {2011})}\BibitemShut {NoStop}%
\bibitem [{\citenamefont {Marchand}\ and\ \citenamefont
  {Wolf}(1974)}]{MarchandWolf1974}%
  \BibitemOpen
  \bibfield  {author} {\bibinfo {author} {\bibfnamefont {E.~W.}\ \bibnamefont
  {Marchand}}\ and\ \bibinfo {author} {\bibfnamefont {E.}~\bibnamefont
  {Wolf}},\ }\bibfield  {title} {\bibinfo {title} {Walther's definitions of
  generalized radiance},\ }\href {https://doi.org/10.1364/JOSA.64.001273}
  {\bibfield  {journal} {\bibinfo  {journal} {J. Opt. Soc. Am.}\ }\textbf
  {\bibinfo {volume} {64}},\ \bibinfo {pages} {1273} (\bibinfo {year}
  {1974})}\BibitemShut {NoStop}%
\bibitem [{\citenamefont {Papoulis}(1974)}]{Papoulis1974}%
  \BibitemOpen
  \bibfield  {author} {\bibinfo {author} {\bibfnamefont {A.}~\bibnamefont
  {Papoulis}},\ }\bibfield  {title} {\bibinfo {title} {Ambiguity function in
  {F}ourier optics},\ }\href {https://doi.org/10.1364/JOSA.64.000779}
  {\bibfield  {journal} {\bibinfo  {journal} {J. Opt. Soc. Am.}\ }\textbf
  {\bibinfo {volume} {64}},\ \bibinfo {pages} {779} (\bibinfo {year}
  {1974})}\BibitemShut {NoStop}%
\bibitem [{\citenamefont {Erni}(2015)}]{Erni2015}%
  \BibitemOpen
  \bibfield  {author} {\bibinfo {author} {\bibfnamefont {R.}~\bibnamefont
  {Erni}},\ }\href@noop {} {\emph {\bibinfo {title} {Aberration-Corrected
  Imaging in Transmission Electron Microscopy: An Introduction}}},\ \bibinfo
  {edition} {2nd}\ ed.\ (\bibinfo  {publisher} {Imperial College Press,
  London},\ \bibinfo {year} {2015})\BibitemShut {NoStop}%
\bibitem [{\citenamefont {Anderson}\ \emph {et~al.}(2009)\citenamefont
  {Anderson}, \citenamefont {McGreevy},\ and\ \citenamefont
  {Bilheux}}]{NeutronImagingAndItsApplications}%
  \BibitemOpen
  \bibinfo {editor} {\bibfnamefont {I.~S.}\ \bibnamefont {Anderson}}, \bibinfo
  {editor} {\bibfnamefont {R.~L.}\ \bibnamefont {McGreevy}},\ and\ \bibinfo
  {editor} {\bibfnamefont {H.~Z.}\ \bibnamefont {Bilheux}},\ eds.,\ \href@noop
  {} {\emph {\bibinfo {title} {Neutron Imaging and Applications: A Reference
  for the Imaging Community}}}\ (\bibinfo  {publisher} {Springer, Boston},\
  \bibinfo {year} {2009})\BibitemShut {NoStop}%
\bibitem [{\citenamefont {Cloetens}\ \emph {et~al.}(1996)\citenamefont
  {Cloetens}, \citenamefont {Barrett}, \citenamefont {Baruchel}, \citenamefont
  {Guigay},\ and\ \citenamefont {Schlenker}}]{Cloetens}%
  \BibitemOpen
  \bibfield  {author} {\bibinfo {author} {\bibfnamefont {P.}~\bibnamefont
  {Cloetens}}, \bibinfo {author} {\bibfnamefont {R.}~\bibnamefont {Barrett}},
  \bibinfo {author} {\bibfnamefont {J.}~\bibnamefont {Baruchel}}, \bibinfo
  {author} {\bibfnamefont {J.-P.}\ \bibnamefont {Guigay}},\ and\ \bibinfo
  {author} {\bibfnamefont {M.}~\bibnamefont {Schlenker}},\ }\bibfield  {title}
  {\bibinfo {title} {Phase objects in synchrotron radiation hard x-ray
  imaging},\ }\href@noop {} {\bibfield  {journal} {\bibinfo  {journal} {J.
  Phys. D: Appl. Phys.}\ }\textbf {\bibinfo {volume} {29}},\ \bibinfo {pages}
  {133} (\bibinfo {year} {1996})}\BibitemShut {NoStop}%
\bibitem [{\citenamefont {Klein}\ and\ \citenamefont
  {Opat}(1976)}]{KleinOpat1976}%
  \BibitemOpen
  \bibfield  {author} {\bibinfo {author} {\bibfnamefont {A.~G.}\ \bibnamefont
  {Klein}}\ and\ \bibinfo {author} {\bibfnamefont {G.~I.}\ \bibnamefont
  {Opat}},\ }\bibfield  {title} {\bibinfo {title} {Observation of 2$\pi$
  rotations by {F}resnel diffraction of neutrons},\ }\href@noop {} {\bibfield
  {journal} {\bibinfo  {journal} {Phys. Rev. Lett.}\ }\textbf {\bibinfo
  {volume} {37}},\ \bibinfo {pages} {238} (\bibinfo {year} {1976})}\BibitemShut
  {NoStop}%
\bibitem [{\citenamefont {Pelliccia}\ and\ \citenamefont
  {Paganin}(2025)}]{PellicciaPaganin2025}%
  \BibitemOpen
  \bibfield  {author} {\bibinfo {author} {\bibfnamefont {D.}~\bibnamefont
  {Pelliccia}}\ and\ \bibinfo {author} {\bibfnamefont {D.~M.}\ \bibnamefont
  {Paganin}},\ }\href@noop {} {\emph {\bibinfo {title} {Synchrotron Light}}}\
  (\bibinfo  {publisher} {Oxford University Press, Oxford},\ \bibinfo {year}
  {2025})\BibitemShut {NoStop}%
\bibitem [{\citenamefont {Reimer}\ and\ \citenamefont
  {Kohl}(2008)}]{ReimerKohlBook}%
  \BibitemOpen
  \bibfield  {author} {\bibinfo {author} {\bibfnamefont {L.}~\bibnamefont
  {Reimer}}\ and\ \bibinfo {author} {\bibfnamefont {H.}~\bibnamefont {Kohl}},\
  }\href@noop {} {\emph {\bibinfo {title} {Transmission Electron Microscopy:
  Physics of Image Formation}}},\ \bibinfo {edition} {5th}\ ed.\ (\bibinfo
  {publisher} {Springer},\ \bibinfo {address} {New York},\ \bibinfo {year}
  {2008})\BibitemShut {NoStop}%
\bibitem [{\citenamefont {Klein}\ and\ \citenamefont
  {Werner}(1983)}]{KleinWerner1983}%
  \BibitemOpen
  \bibfield  {author} {\bibinfo {author} {\bibfnamefont {A.~G.}\ \bibnamefont
  {Klein}}\ and\ \bibinfo {author} {\bibfnamefont {S.~A.}\ \bibnamefont
  {Werner}},\ }\bibfield  {title} {\bibinfo {title} {Neutron optics},\ }\href
  {https://doi.org/10.1088/0034-4885/46/3/001} {\bibfield  {journal} {\bibinfo
  {journal} {Rep. Prog. Phys.}\ }\textbf {\bibinfo {volume} {46}},\ \bibinfo
  {pages} {259} (\bibinfo {year} {1983})}\BibitemShut {NoStop}%
\bibitem [{\citenamefont {Fomin}\ \emph {et~al.}(2022)\citenamefont {Fomin},
  \citenamefont {Fry}, \citenamefont {Pattie},\ and\ \citenamefont
  {Greene}}]{SpallationSources}%
  \BibitemOpen
  \bibfield  {author} {\bibinfo {author} {\bibfnamefont {N.}~\bibnamefont
  {Fomin}}, \bibinfo {author} {\bibfnamefont {J.}~\bibnamefont {Fry}}, \bibinfo
  {author} {\bibfnamefont {R.~W.}\ \bibnamefont {Pattie}},\ and\ \bibinfo
  {author} {\bibfnamefont {G.~L.}\ \bibnamefont {Greene}},\ }\bibfield  {title}
  {\bibinfo {title} {Fundamental neutron physics at spallation sources},\
  }\href {https://doi.org/https://doi.org/10.1146/annurev-nucl-121521-051029}
  {\bibfield  {journal} {\bibinfo  {journal} {Annu. Rev. Nucl. Part. Sci.}\
  }\textbf {\bibinfo {volume} {72}},\ \bibinfo {pages} {151} (\bibinfo {year}
  {2022})}\BibitemShut {NoStop}%
\bibitem [{\citenamefont {Wilkins}\ \emph {et~al.}(1996)\citenamefont
  {Wilkins}, \citenamefont {Gureyev}, \citenamefont {Gao}, \citenamefont
  {Pogany},\ and\ \citenamefont {Stevenson}}]{WilkinsFish}%
  \BibitemOpen
  \bibfield  {author} {\bibinfo {author} {\bibfnamefont {S.~W.}\ \bibnamefont
  {Wilkins}}, \bibinfo {author} {\bibfnamefont {T.~E.}\ \bibnamefont
  {Gureyev}}, \bibinfo {author} {\bibfnamefont {D.}~\bibnamefont {Gao}},
  \bibinfo {author} {\bibfnamefont {A.}~\bibnamefont {Pogany}},\ and\ \bibinfo
  {author} {\bibfnamefont {A.~W.}\ \bibnamefont {Stevenson}},\ }\bibfield
  {title} {\bibinfo {title} {Phase-contrast imaging using polychromatic hard
  {X}-rays},\ }\href@noop {} {\bibfield  {journal} {\bibinfo  {journal}
  {Nature}\ }\textbf {\bibinfo {volume} {384}},\ \bibinfo {pages} {335}
  (\bibinfo {year} {1996})}\BibitemShut {NoStop}%
\bibitem [{\citenamefont {Wilkins}\ \emph {et~al.}(2014)\citenamefont
  {Wilkins}, \citenamefont {Nesterets}, \citenamefont {Gureyev}, \citenamefont
  {Mayo}, \citenamefont {Pogany},\ and\ \citenamefont
  {Stevenson}}]{Wilkins2014}%
  \BibitemOpen
  \bibfield  {author} {\bibinfo {author} {\bibfnamefont {S.~W.}\ \bibnamefont
  {Wilkins}}, \bibinfo {author} {\bibfnamefont {{\relax Ya}.~I.}\ \bibnamefont
  {Nesterets}}, \bibinfo {author} {\bibfnamefont {T.~E.}\ \bibnamefont
  {Gureyev}}, \bibinfo {author} {\bibfnamefont {S.~C.}\ \bibnamefont {Mayo}},
  \bibinfo {author} {\bibfnamefont {A.}~\bibnamefont {Pogany}},\ and\ \bibinfo
  {author} {\bibfnamefont {A.~W.}\ \bibnamefont {Stevenson}},\ }\bibfield
  {title} {\bibinfo {title} {On the evolution and relative merits of hard
  {X}-ray phase-contrast imaging methods},\ }\href@noop {} {\bibfield
  {journal} {\bibinfo  {journal} {Phil. Trans. R. Soc. A}\ }\textbf {\bibinfo
  {volume} {372}},\ \bibinfo {pages} {20130021} (\bibinfo {year}
  {2014})}\BibitemShut {NoStop}%
\bibitem [{\citenamefont {Larsson}\ \emph {et~al.}(2011)\citenamefont
  {Larsson}, \citenamefont {Takman}, \citenamefont {Lundstr\"{o}m},
  \citenamefont {Burvall},\ and\ \citenamefont
  {Hertz}}]{LiquidMetalJetSources}%
  \BibitemOpen
  \bibfield  {author} {\bibinfo {author} {\bibfnamefont {D.~H.}\ \bibnamefont
  {Larsson}}, \bibinfo {author} {\bibfnamefont {P.~A.~C.}\ \bibnamefont
  {Takman}}, \bibinfo {author} {\bibfnamefont {U.}~\bibnamefont
  {Lundstr\"{o}m}}, \bibinfo {author} {\bibfnamefont {A.}~\bibnamefont
  {Burvall}},\ and\ \bibinfo {author} {\bibfnamefont {H.~M.}\ \bibnamefont
  {Hertz}},\ }\bibfield  {title} {\bibinfo {title} {A 24 ke{V} liquid-metal-jet
  x-ray source for biomedical applications},\ }\href
  {https://doi.org/10.1063/1.3664870} {\bibfield  {journal} {\bibinfo
  {journal} {Rev. Sci. Instrum.}\ }\textbf {\bibinfo {volume} {82}},\ \bibinfo
  {pages} {123701} (\bibinfo {year} {2011})}\BibitemShut {NoStop}%
\bibitem [{\citenamefont {Carminati}\ and\ \citenamefont
  {Schotland}(2021)}]{CarminatiSchotlandBook}%
  \BibitemOpen
  \bibfield  {author} {\bibinfo {author} {\bibfnamefont {R.}~\bibnamefont
  {Carminati}}\ and\ \bibinfo {author} {\bibfnamefont {J.~C.}\ \bibnamefont
  {Schotland}},\ }\href@noop {} {\emph {\bibinfo {title} {Principles of
  Scattering and Transport of Light}}}\ (\bibinfo  {publisher} {Cambridge
  University Press},\ \bibinfo {address} {Cambridge},\ \bibinfo {year}
  {2021})\BibitemShut {NoStop}%
\end{thebibliography}
\end{document}